\documentstyle[epsfig,axodraw]{aipproc2}

\renewcommand{\theequation}{\arabic{section}.\arabic{equation}}

\def\beq{\begin{equation}}
\def\eeq{\end{equation}}
\def\eq{\end{equation}}
\def\bea{\begin{eqnarray*}}
\def\eea{\end{eqnarray*}}

\def\ap  #1 #2 #3 #4 {Ann.~Phys.         {\bf  #1}, #2 (#3)#4 }
\def\aplb#1 #2 #3 #4 {Acta Phys.~Pol.    {\bf B#1}, #2 (#3)#4 }
\def\cpc #1 #2 #3 #4 {Comp.~Phys.~Comm.  {\bf  #1}, #2 (#3)#4 }
\def\jetp#1 #2 #3 #4 {JETP Lett.         {\bf  #1}, #2 (#3)#4 }
\def\npb #1 #2 #3 #4 {Nucl.~Phys.        {\bf B#1}, #2 (#3)#4 }
\def\mpla#1 #2 #3 #4 {Mod.~Phys.~Lett.   {\bf A#1}, #2 (#3)#4 }
\def\plb #1 #2 #3 #4 {Phys.~Lett.        {\bf B#1}, #2 (#3)#4 }
\def\pr  #1 #2 #3 #4 {Phys.~Rep.         {\bf  #1}, #2 (#3)#4 }
\def\prd #1 #2 #3 #4 {Phys.~Rev.         {\bf D#1}, #2 (#3)#4 }
\def\prl #1 #2 #3 #4 {Phys.~Rev.~Lett.   {\bf  #1}, #2 (#3)#4 }
\def\ptp #1 #2 #3 #4 {Prog.~Theor.~Phys. {\bf  #1}, #2 (#3)#4 }
\def\zpc #1 #2 #3 #4 {Zeit.~Phys.        {\bf C#1}, #2 (#3)#4 }

\def\centeron#1#2{{\setbox0=\hbox{#1}\setbox1=\hbox{#2}\ifdim
\wd1>\wd0\kern.5\wd1\kern-.5\wd0\fi
\copy0\kern-.5\wd0\kern-.5\wd1\copy1\ifdim\wd0>\wd1
\kern.5\wd0\kern-.5\wd1\fi}}
\def\ltap{\;\centeron{\raise.35ex\hbox{$<$}}{\lower.65ex\hbox{$\sim$}}\;}
\def\gtap{\;\centeron{\raise.35ex\hbox{$>$}}{\lower.65ex\hbox{$\sim$}}\;}
\def\gsim{\mathrel{\gtap}}
\def\lsim{\mathrel{\ltap}}

\def\slashchar#1{\setbox0=\hbox{$#1$}           
   \dimen0=\wd0                                 
   \setbox1=\hbox{/} \dimen1=\wd1               
   \ifdim\dimen0>\dimen1                        
      \rlap{\hbox to \dimen0{\hfil/\hfil}}      
      #1                                        
   \else                                        
      \rlap{\hbox to \dimen1{\hfil$#1$\hfil}}   
      /                                         
   \fi}                                        %

\def\singleandabitspaced{\baselineskip=\normalbaselineskip\multiply
    \baselineskip by 120\divide\baselineskip by 100}

\def\singlespaced{\baselineskip=\normalbaselineskip}

\newcommand{\newc}{\newcommand}
\newc{\Nmess}{{N}}
\newc{\Mmess}{{M_m}}

\newc{\GG}{{\widetilde G}}
\newc{\goldstino}{{\widetilde G}}
\newc{\mesino}{{\cal M}}
\newc{\sbaryon}{{\cal B}}
\newc{\stopq}{{\tilde t}}
\newc{\qbar}{{\bar{q}}}
\newc{\Rhadron}{{\widetilde R}}

\newc{\NI}{{\widetilde \chi_1^0}}
\newc{\NII}{{\widetilde \chi_2^0}}
\newc{\CIplusminus}{{\widetilde \chi_1^\pm}}
\newc{\CI}{{\widetilde \chi_1^\pm}}
\newc{\CIplus}{{\widetilde \chi_1^+}}
\newc{\CIminus}{{\widetilde \chi_1^-}}
\newc{\slepton}{{\widetilde\ell}}
\newc{\selectron}{{\widetilde e}}
\newc{\smuon}{{\widetilde \mu}}
\newc{\cG}{{C_G}}
\newc{\missET}{{\slashchar{E}_T}}
\newc{\stauI}{{\widetilde \tau_1}}
\newc{\stau}{{\widetilde \tau}}
\newc{\squark}{{\widetilde Q}}
\newc{\wino}{{\widetilde W}}
\newc{\bino}{{\widetilde B}}
\newc{\None}{{\widetilde \chi_1^0}}
\newc{\Ntwo}{{\widetilde \chi_2^0}}
\newc{\Cone}{{\widetilde \chi_1^\pm}}
\newc{\nni}{{\widetilde \chi_i^0}}
\newc{\nnone}{{\widetilde \chi_i^0}}
\newc{\cci}{{\widetilde \chi_i^{\pm}}}
\newc{\gluino}{{\widetilde g}}

\def\Rslash{\not \! \! R}
\def\D0{D\O}
\def\USA{{}}

\newcommand{\ldt}{${\cal L}$}
\newcommand{\nlsp}{$\tilde\chi^0_1$}
\newcommand{\tlsp}{$\tilde\tau_1$}

\newcommand{\cn}{$\tilde\chi^\pm_1\tilde\chi^0_2$}

\newcommand{\met}{\rlap{\kern0.25em/}E_T}
\newcommand{\gmet}{$\gamma\rlap{\kern0.25em/}E_T$}
\newcommand{\ggmet}{$\gamma\gamma\rlap{\kern0.25em/}E_T$}
\newcommand{\gjjmet}{$\gamma jj \rlap{\kern0.25em/}E_T$}
\newcommand{\gbjmet}{$\gamma bj \rlap{\kern0.25em/}E_T$}
\newcommand{\gbbmet}{$\gamma b\bar{b} \rlap{\kern0.25em/}E_T$}
\newcommand{\gdjjmet}{$\gamma^\prime jj \rlap{\kern0.25em/}E_T$}
\newcommand{\lljjmet}{$\ell^\pm\ell^\pm jj \rlap{\kern0.25em/}E_T$}
\newcommand{\gmetjj}{$\gamma\rlap{\kern0.25em/}E_T+\ge 2\ {\rm jets}$}
\newcommand{\llmet}{$\ell\ell\rlap{\kern0.25em/}E_T$}
\newcommand{\lldedx}{$\ell\ell+dE/dx$}
\newcommand{\lllj}{$\ell\ell\ell j\rlap{\kern0.25em/}E_T$}
\newcommand{\lljj}{$\ell^\pm\ell^\pm jj\rlap{\kern0.25em/}E_T$}
\newcommand{\rsb}{$N_s/\delta N_b$}

\newcommand{\Eslash}{\rlap{\kern0.25em/}E_T}

\def\eslt{E\llap/_T}
\def\to{\rightarrow}
\def\Re{{\cal R \mskip-4mu \lower.1ex \hbox{\it e}}\,}
\def\Im{{\cal I \mskip-5mu \lower.1ex \hbox{\it m}}\,}

\def\ttau{\tilde \tau}
\def\tg{\tilde g}

\def\tw{\tilde \chi^{\pm}}

\def\tz{\tilde \chi^0}

\def\tG{\tilde G}

\catcode`@=11\def\tableofcontents{\section*{\contentsname
\@mkboth{\uppercase{\contentsname}}{\uppercase{\contentsname}}}%
\@starttoc{toc}}
\catcode`@=12
\begin{document}
\pagenumbering{arabic}


\begin{flushright}
FERMILAB-Pub-00/251-T\\
SLAC-PUB-8643
\end{flushright}
\begin{center}
{\LARGE\bf Low-Scale and Gauge-Mediated Supersymmetry}

\vspace{0.05in} 

{\LARGE\bf Breaking at the Fermilab Tevatron Run II}
\end{center}

\vspace{-0.26in}

\author{{\sc Conveners:}}

\vspace{-0.34in}

\author{{Ray Culbertson}$^1$(CDF), 
{Stephen P.~Martin}$^{2,3}$,
{Jianming Qian}$^4$(D$\emptyset$), 
{Scott Thomas}$^5$}

\author{{\sc{Contributors:}}
Howard~Baer$^6$,
Wasiq~Bokhari$^7$(CDF),
Sailesh~Chopra$^8$(\D0),
Chih-Lung~Chou$^9$,
Amy~Connolly$^{10}$(CDF),
Ray~Culbertson$^1$(CDF),
Dave~Cutts$^{11}$(\D0),
Regina~Demina$^{12}$(CDF,\D0),
Bhaskar~Dutta$^{13}$,
Gary~Grim$^{14}$(CDF),
Greg~Landsberg$^{11}$(\D0),
Stephen~P.~Martin$^{2,3}$,
Konstantin~Matchev$^3$,
P.G.~Mercadante$^{6,15}$,
D.J.~Muller$^{16}$,
S.~Nandi$^{16}$,
Michael~Peskin$^{9}$,
Jianming~Qian$^4$(\D0),
Uri~Sarid$^{17}$,
David~Stuart$^1$(CDF),
Benn~Tannenbaum$^{18}$(CDF),
Xerxes~Tata$^{15}$,
Scott~Thomas$^5$,
Randy~Thurman-Keup$^{19}$(CDF),
Ming-Jer~Wang$^{20}$(CDF),
Yi-li~Wang$^{15}$}
\address{$^1$MS 318, Fermi National Accelerator Laboratory, Batavia IL
60510 \USA\\
$^2$Physics Department, Northern Illinois University, DeKalb IL 60115
\USA\\
$^3$Theoretical Physics, MS 106, Fermilab, Batavia IL 60510 \USA\\
$^4$Physics Department, University of Michigan, Ann Arbor MI 48109 \USA\\
$^5$Physics Department, Stanford University, Stanford CA 94305 \USA\\
$^6$Department of Physics, Florida State University, Tallahassee FL 32306 
\USA\\
$^7$Department of Physics, University of Pennsylvania, Philadelphia
PA 19104 \USA\\
$^8$Brookhaven National Laboratory, Upton NY 11973 \USA\\
$^9$Stanford Linear Accelerator Center, Stanford University,
Stanford CA 94309 \USA\\
$^{10}$Ernest Orlando Lawrence Berkeley National Laboratory, Berkeley CA
94720 \USA\\
$^{11}$Physics Department, Brown University, Providence RI 02912 \USA\\
$^{12}$Physics Department, Kansas State University, Manhattan KS 66506
\USA\\
$^{13}$Center for Theoretical Physics, Dept.~of Physics, Texas A\&M
University, College Station TX 77843 \USA\\
$^{14}$Physics Department, University of California, Davis CA
95616 \USA\\
$^{15}$Dept. of Physics and Astronomy, University of Hawaii, Honolulu
HI 96822 \USA\\
$^{16}$Department of Physics, Oklahoma State University, Stillwater OK
74078 \USA\\
$^{17}$Department of Physics, University of Notre Dame, Notre Dame IN
46556 \USA\\
$^{18}$Department of Physics, University of California, Los Angeles CA 
90095 \USA\\
$^{19}$Argonne National Laboratory, Argonne IL 60439 \USA\\
$^{20}$Institute of Physics, Academia Sinica, Taipei, Taiwan 11529
Republic of China\\
}

\maketitle

\vspace{-0.25in}

\begin{abstract} 
\singleandabitspaced 
\noindent 
The prospects for discovering and studying signals of low-scale
supersymmetry breaking models at the Tevatron Run II and beyond are
explored.  These models include gauge-mediated supersymmetry breaking as
the most compelling and concrete realization, but more generally are
distinguished by the presence of a nearly massless Goldstino as the
lightest supersymmetric particle.  The next-lightest supersymmetric
particle(s) (NLSP) decays to its partner and the Goldstino.  Depending on
the supersymmetry breaking scale, these decays can occur promptly or on a
scale comparable to or larger than the size of a detector. A systematic
analysis based on a classification in terms of the identity of the NLSP
and its decay length is presented.  The various scenarios are discussed in
terms of signatures and possible event selection criteria. The Run II and
beyond discovery and exclusion reaches, including the effects of
background, are detailed for the most compelling cases.  In addition to
standard event selection criteria based on missing energy and photons,
leptons, jets, taus, tagged $b$-jets, or reconstructed $Z$-bosons, more
exotic signals of metastable NLSPs such as displaced photons, large
negative impact parameter tracks, kink tracks, both opposite and same-sign
highly ionizing tracks, time of flight measurements, charge-changing
tracks, charge-exchange tracks, and same-sign di-top events are
investigated.  The interesting possibility of observing a Higgs boson
signal in events that are efficiently ``tagged'' by the unique signatures
of low-scale supersymmetry breaking is also considered.  \end{abstract}

\setcounter{footnote}{1}
\setcounter{section}{0}
\setcounter{subsection}{0}
\setcounter{subsubsection}{0}
\singlespaced
\pagestyle{plain}

\vspace{-0.1in}
\tableofcontents


\section{Introduction}\label{sec:intro}
\setcounter{equation}{0}
\setcounter{footnote}{1}
\indent

Supersymmetry has emerged as the most promising candidate solution to the
hierarchy problem associated with the large separation between the
electroweak and Planck scales. Supersymmetry (SUSY) stabilizes the Higgs
boson mass against potentially dangerous quantum contributions from
ultraviolet physics. Supersymmetry requires that for each known particle
there exists a superpartner which differs in spin by ${1 \over 2}$ unit of
angular momentum. Spontaneous SUSY breaking in general splits the particle
and superpartner masses, consistent with the non-observation of Bose-Fermi
degeneracy in the physical spectrum at low energies. In supersymmetric
extensions of the Standard Model with spontaneously broken supersymmetry,
quantum effects of the top squark scalar superpartners of the top quark
lead naturally, through the large top quark Yukawa coupling, to the
observed electroweak symmetry breaking. In this way the
electroweak scale is {\it determined} by the superpartner masses.

The search for superpartners with electroweak scale masses constitutes a
major effort at present and future high energy colliders. This report
presents studies of interesting and unique signatures of low-scale
supersymmetry breaking which can be probed at the Fermilab Tevatron Run
II. The results of these studies suggest a wide range of new analysis
which should be implemented in the search for supersymmetry. In
some channels the Tevatron has a very significant discovery reach for
supersymmetry in Run II. All the studies presented here were completed as
part of the Supersymmetry and Higgs Workshop in preparation for Run II.

The manner and scale at which supersymmetry is spontaneously broken
has crucial implications for the phenomenology of collider searches for
the superpartners. The spontaneous breaking of any global symmetry implies
the existence of a Nambu-Goldstone particle to realize the symmetry in the
broken phase. Since supersymmetry is a fermionic symmetry which relates
Bosons and Fermions, the Nambu-Goldstone particle is a fermion, the
Goldstino.

In order to preserve supersymmetry in a theory which includes gravity,
supersymmetry must be promoted to a local symmetry. For any local symmetry
realized in the broken phase, the gauge particle becomes massive by eating
the Nambu-Goldstone particle. With spontaneously broken local
supersymmetry, the spin ${1 \over 2}$ Goldstino becomes the longitudinal
components of the spin ${3 \over 2}$ gravitino superpartner of the
graviton. The gravitino-Goldstino gains a mass by this super-Higgs
mechanism of
\beq
m_{\GG} = {F \over \sqrt{3} M_P} \simeq ~ 2.4~
\left( { \sqrt{F} \over 100~{\rm TeV}} \right)^2~{\rm eV} ,
\label{gravmass}
\eq
where 
$M_P=2.4 \times 10^{18}$ GeV is the reduced Planck mass, and
$F$ is the order parameter for spontaneous supersymmetry
breaking with units of [mass]$^2$.
$F$ is a vacuum
expectation value (of an auxiliary field) which measures the magnitude of
supersymmetry breaking in the vacuum state. 
Supersymmetry is restored for $F \rightarrow 0$.
Couplings of the helicity $\pm 1/2$ Goldstino components
discussed below
are only directly relevant to accelerator phenomenology for a gravitino
which is much lighter than the energy scale of a collider
experiment.
In addition, the helicity $\pm 3/2$ components only couple
with gravitational strength and are never relevant to accelerator
phenomenology.
It is therefore most appropriate to consider only the
essentially massless Goldstino, although
gravitino and Goldstino are used interchangeably in the literature.
Use of the gravitino mass rather than the
supersymmetry breaking scale, $\sqrt{F}$,
to characterize accelerator signatures is also not very appropriate
since this mass scale does not appear in any relevant process.

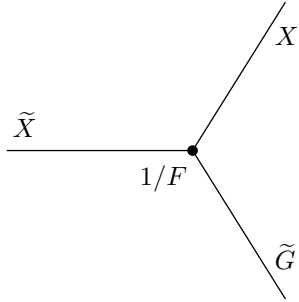
\begin{figure}[t]
\centering
\vspace{.2cm}
\begin{picture}(350,100)(-200,-75)
\Line(-70,0)(0,0)
\put (-68,5){$\widetilde X$}
\Line(35,56)(0,0)
\put (31,40){$X$}
\Line(35,-56)(0,0)
\Vertex(0,0)2
\put (31,-45){$\widetilde G$}
\put (-20,-13){$1/F$}
\end{picture}
\vspace{-.1cm}
\caption{The Goldstino $\GG$ derivatively couples each particle $X$ to its
superpartner $\widetilde X$, with an interaction strength inversely
proportional to $F$.}
\label{fig:goldstinocoupling}
\vspace{.25cm}
\end{figure}

All the signatures studied in this report follow either directly
or indirectly from the existence of the Goldstino.
The analog of the Goldberger-Treiman relation implies that
each particle is derivatively coupled to its superpartner
through the Goldstino with a strength inversely proportional $F$,
as illustrated in Fig.~\ref{fig:goldstinocoupling}.
If the scale of supersymmetry breaking is not too high, these interactions
can be relevant to accelerator physics. In particular a massive
superpartner, $\widetilde X$, can decay to its
partner particle, $X$, by emitting an essentially massless
Goldstino \cite{fayet} with decay rate:
\beq
\Gamma (\widetilde X \rightarrow X\GG) =
{\kappa m_{\tilde X}^5 \over 16 \pi F^2}
\left ( 1 - {m^2_X \over m^2_{\widetilde X}} \right )^4 ,
\label{goldstinodecaywidth}
\eeq
where $\kappa$ is a mixing parameter to be evaluated for the particular
model parameters.
If $X$ and $\widetilde{X}$ are unmixed states within the same supermultiplet,
such as for slepton decay to a lepton and Goldstino,
$\kappa=1$.
However, for superpartner mass eigenstates
which are mixtures of superpartners in different supermultiplets,
$\kappa < 1$ is possible.
For a pure $U(1)_Y$
Bino decay to a photon and Goldstino, $\kappa=\cos^2\theta_W$.
A complete list of the decay rates for the particles of the
minimal supersymmetric standard model (MSSM) through Goldstino
emission is given in Appendix B.
The decay rate (\ref{goldstinodecaywidth})
corresponds to a decay length of
\beq
c \tau  (\widetilde X \rightarrow X\GG)
\simeq \left( { 100 ~\mu{\rm m} \over \kappa} \right)
\left( 100~{\rm GeV} \over m_{\tilde{X}} \right)^5
\left( \sqrt{F} \over 100 ~{\rm TeV} \right)^4
\left( 1 -  {m^2_X \over m^2_{\widetilde X}} \right )^{-4}
\label{ctaueq}
\eq
For a supersymmetry breaking scale, $\sqrt{F}$, much larger
than a few 1000 TeV, superpartner decays through the Goldstino
take place well outside a collider detector, and are
not directly relevant to accelerator physics.
However, for supersymmetry breaking scales below a few 1000 TeV,
decays through the Goldstino can occur within a detector,
and therefore have a very direct impact on the types of
SUSY signatures which can be observed \cite{prlgoldstino}.

\begin{figure}[tpb]
\centering \epsfxsize=4.0in
\hspace*{0in}
\epsffile{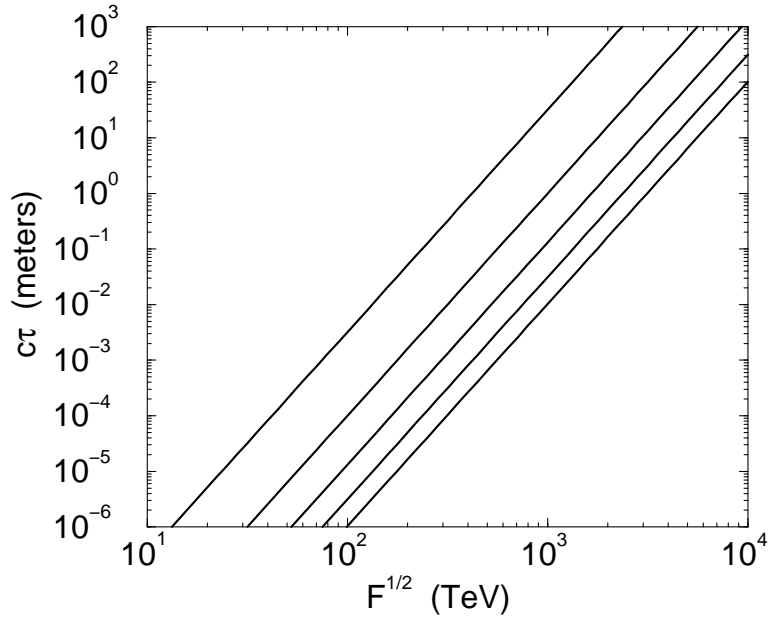}
\caption{NLSP decay length, $c \tau(\widetilde X \to X \widetilde G)$, in
meters, as a function of the supersymmetry
breaking scale, $\sqrt{F}$, in TeV.
From top to bottom the lines are for an NLSP mass $m_{\tilde X}$ =
50, 100, 150, 200, 250 GeV and with $m_X=0$ and $\kappa =1$.}
\label{ctaufig}
\end{figure}

Supersymmetric particles are generally unstable to decays to lighter
superparticles. These decays can take place through interactions related
by supersymmetry to the ordinary strong, electromagnetic, or weak
couplings. If kinematically allowed, these decays are generally much more
rapid than decay through the Goldstino for any reasonable $\sqrt{F}$ above
the electroweak scale. Heavy superpartners therefore generally rapidly
cascade decay to the lightest standard model superpartner. If $R$-parity
is conserved, the lightest standard model superpartner is stable with
respect to all the MSSM couplings. However, as discussed above, if decay
through the Goldstino can take place within a detector, the Goldstino is
essentially massless, and is therefore the lightest supersymmetric
particle (LSP). The lightest standard model superpartner is therefore the
next to lightest superpartner (NLSP). Because of the Goldstino coupling,
the NLSP is only meta-stable, and can decay to its partner through LSP
Goldstino emission with decay rate eq.~(\ref{goldstinodecaywidth}) or
decay length eq.~(\ref{ctaueq}). The NLSP decay length is shown in
Fig.~\ref{ctaufig} as a function of the SUSY breaking scale $\sqrt{F}$ for
a number of different NLSP masses.
Because all cascades pass through the NLSP, the identity of the NLSP and
the SUSY breaking scale, which determines the NLSP decay length, are the
crucial parameters in classifying the types of SUSY signatures which may
arise \cite{prlgoldstino}.

It is important to note that more than one superpartner can effectively
act as the NLSP. This can occur if the lightest standard model
superpartners are nearly degenerate, with any allowed decays between these
states taking place well outside the detector. Each of these superpartners
can however decay to the Goldstino, and so each is effectively an NLSP.
This situation is most natural if there is some (perhaps approximate)
symmetry which enforces the near degeneracy of some of the superpartners.
As an example, the approximate lepton chiral flavor symmetry, indicated by
the small lepton Yukawa couplings, can be sufficient to enforce the near
degeneracy of the lightest slepton in each of the three generations. If
the splitting between these sleptons is smaller than the associated lepton
mass, then conservation of individual lepton number forbids decays between
the slepton states which involve charged leptons. Decays involving
neutrinos are allowed, but are slow enough to take place well outside a
detector. The lightest slepton in each generation then form an effective
slepton co-NLSP \cite{phenimp,mgm,Ambrosanio:1997rv,threebody}.

The range of possibilities for the NLSP and SUSY breaking
scale present many unique and challenging experimental signatures
which must be classified in order to fully cover all discovery modes
for supersymmetry.
In principle any of the MSSM particles may be the NLSP.
Some of possibilities for the NLSP, along with the
decay modes to the Goldstino, are listed in Table \ref{NLSPtable}.
A neutralino NLSP is in general a mixture of gaugino and Higgsino
eigenstates, and so can decay to either the $\gamma$ or $Z$ gauge
bosons, or the $h$ Higgs boson,
$\NI \to (\gamma, Z, h) \GG$.
A slepton NLSP decays to its partner lepton and the Goldstino,
$\slepton \to \ell \GG$.
Likewise, a squark NLSP decays to its partner quark and the
Goldstino, $\squark \to q \GG$.
In the case of a stop-like squark lighter than the top quark,
decay to a $b$-quark, $W$ boson, and Goldstino,
$\tilde t \to b W \GG$ (or $\tilde t \to c \GG$) results.
Finally, a gluino NLSP would decay to the gluon and Goldstino,
$\gluino \to g \GG$.
\begin{table}[htbp]
\caption{Decay modes to the Goldstino in various NLSP scenarios.}
\label{NLSPtable}
\renewcommand{\arraystretch}{1.8}
\begin{tabular}{ll}
\hline \hline 
\multicolumn{1}{c}{NLSP} & \multicolumn{1}{c}{Decay to the Goldstino}  \\
\hline  
Bino-like Neutralino     & $ \NI \to \gamma ~\GG$ \\     
Higgsino-like Neutralino~~~~ & $ \NI \to (h,Z,\gamma) ~\GG$ \\ 
Stau                     & $ \stau \to \tau ~\GG$   \\   
Slepton Co-NLSP          & $ \slepton \to \ell~\GG$  \\ 
Squark                   & $ \squark \to (q, q^{\prime} W) ~\GG$ \\ 
Gluino                   & $ \gluino \to g  ~\GG$   \\   
\hline \hline
\end{tabular}
\end{table}

The NLSP decay lengths can generally be divided into three relevant
ranges. The first are prompt decays which can not be resolved as secondary
vertices, and therefore appear to originate from the interaction region.
Since SUSY particles are produced in pairs the signatures are then two
hard partons coming from the NLSP decays, significant missing energy
carried by the Goldstino pair, and possibly other partons from cascade
decays to the NLSPs. Since the Goldstino is essentially massless, the
partons arising from the NLSP decays can be very hard. A search for each
NLSP possibility may be implemented by an inclusive search for the
appropriate hard parton pairs in association with missing energy.

The second range of decay lengths are macroscopic but within the detector.
In this case the partons arising from NLSP decay do not necessarily point
back to the interaction region. For a neutralino decay $\NI \to \gamma
\GG$ this leads to displaced photons. This presents the experimental
challenge of accurate photon pointing. For a neutralino NLSP displaced
decay $\NI \to h \GG$ or $\NI \to Z \GG$ secondary vertices arise with
decay product invariant mass which can be identified with the parent Higgs
or $Z$ boson. This generally requires a special analysis to identify high
momentum partons which form secondary vertices. In some cases this can
greatly reduce background. In the case of a displaced decay $\NI \to h
\GG$, with $h \to bb$, a search for high momentum large negative impact
parameters (LNIPs) greatly reduces potential backgrounds for Standard
Model $b$-quark production. For a slepton NLSP, displaced decays $\slepton
\to \ell \GG$ inside the inner tracking region will yield charged particle
tracks which do not point back to the interaction region. A
non-relativistic slepton which traverses at least part of the tracking
region before decay $\slepton \to \ell \GG$ gives a highly ionizing track
(HIT) with a kink to a minimum ionizing track (MIT) or tau jet. In either
case identifying
these tracks is very challenging and probably requires that such events
have some other characteristic on which to trigger and identify in
analysis. A strongly interacting squark or gluino NLSP can hadronize as
either a neutral or charged bound state. In the neutral case macroscopic
decay to a jet $\squark \to q \GG$ or $\gluino \to g \GG$ gives a
displaced jet which can be searched for using an LNIP analysis. In the
case of a charged squark or gluino bound state the additional possibility
arises of observing the charged bound state as a highly ionizing track
which decays to a jet. Observation of a macroscopic decay length for any
NLSP type would be a smoking gun for low scale supersymmetry breaking. In
addition, a measure of the decay length distribution, along with the
superpartner mass and identity, would give an essentially model
independent measure of the supersymmetry breaking scale.

The final range of NLSP decay lengths is well outside
the detector.
A neutralino NLSP which decays well outside the detector
appears as missing energy.
This is the conventional form of missing energy considered
in standard phenomenological studies of SUSY signatures,
and will not be considered in this report.
A study of these signatures at the Tevatron Run II is
presented in the SUGRA working group report \cite{SUGRAreport}.
For the other possible types of NLSPs,
even if decay to the Goldstino takes place
well outside a detector, the eventual decay
can evade cosmological constraints on the nature of the NLSP.
So in this case the existence of the Goldstino, while
not directly relevant to accelerator physics, does allow
the possibility for the NLSP to be of any type listed in
Table \ref{NLSPtable} and
to be
effectively stable on the scale of a detector.
A slepton NLSP which traverses the detector is generally
not ultra-relativistic and therefore appears as a HIT.
Since slepton pairs result for any SUSY production,
the resulting signature is HIT pairs often {\it without} significant
missing energy, and possibly other partons from cascade decays.
Time of flight (TOF) can also be utilized in the search 
for slow moving sleptons. 
A squark NLSP hadronized with light
quark(s) can exchange isospin and charge through strong interactions
with background material as it 
traverses the detector. 
This gives rise to intermittent charge exchange in the calorimeters
associated with highly ionizing
tracks in the vertex and tracking volume (CE-HITs). 
Because of the reduced average ionization, and intermittent nature
of the charged tracks, CE-HIT momenta would generally be mismeasured,
leading to an apparent missing energy.
Identification of CE-HITs is likely to be challenging and requires specialized
analysis.
A squark NLSP hadronized with light antiquark(s) cannot 
as readily charge exchange with matter. 
Charged bound states of this type however can still have soft
hadronic activity along a highly ionizing track (H-HIT). 
Both CE-HITs and H-HITs are likely to appear as HITs in the
inner tracking region. 

Some of the distinctive signatures associated with the NLSP 
decay to the Goldstino may be useful in obtaining samples of Higgs bosons. 
In the case of a Bino-like neutralino NLSP, $\NI$, 
cascade decays to $\NI$ can often include a Higgs boson, $h$. 
In the case of a Higgsino-like neutralino NLSP, $\NI$, 
direct decay of $\NI$ to the Goldstino can yield a Higgs boson, 
$\NI \to h \GG$. 
The unique low scale supersymmetry breaking signatures 
can then be used as a method 
to tag for events which contain a Higgs boson. 

The existence of the Goldstino clearly presents many unique experimental
challenges and opportunities, many of which have been overlooked in the
past. This report contains a nearly comprehensive compendium of the
relevant signatures which can be probed at Run II.

The overall scale of SUSY breaking, which is so crucial in determining the
experimental signatures, can tell us much about the manner in which SUSY
is broken.
Weakly-coupled spontaneous supersymmetry breaking that occurs directly in
the visible MSSM sector leads to an unacceptable superpartner mass
spectrum, with some of the superpartners very light. Supersymmetry must
therefore be broken in some other sector of nature, generally referred to
as the SUSY breaking sector. This spontaneous breaking may result as the
consequence of non-perturbative gauge dynamics, but might also be the
result of non-supersymmetric boundary conditions on an internal space, or
result geometrically from a non-BPS ground state of extended branes on
which the MSSM degrees of freedom reside. Regardless of the ultimate
source, a messenger sector must couple the SUSY breaking sector to the
MSSM visible sector superpartners. The messenger sector is then said to
``transmit" SUSY breaking to the visible sector. The MSSM visible sector
superpartner masses, $\widetilde m$, are related to the intrinsic SUSY
breaking scale $\sqrt{F}$, and the messenger sector mass scale, $M_m$, by
\beq
\widetilde m \propto {F \over M_m} .
\label{googlymoogly}
\eq
The messenger interactions must be at least as strong as gravity,
implying an upper limit on the messenger scale of $M_m \lsim M_p$.
The limit in which the messenger interactions are of gravitational
strength, $M_m \sim M_P$, is generally referred to as high-scale or
gravity-mediated SUSY breaking.
In this case the intrinsic SUSY breaking scale is the intermediate
scale $\sqrt{F} \sim 10^{11}$ GeV.
In principle, however, the messenger scale can be anywhere between
just above the electroweak scale up to the Planck scale.
A messenger scale significantly below the Planck scale,
$M_m \ll M_P$, is generally referred to as low-scale SUSY breaking.
Note that with low-scale SUSY breaking, the gravitino mass
eq.~(\ref{gravmass})
is well below the MSSM
superpartner masses and electroweak
scale eq.~(\ref{googlymoogly});
$m_{\GG} \ll \widetilde m$ for $M_m \ll M_p$.
So with low-scale SUSY breaking the gravitino is naturally the LSP,
and NLSP superpartner of an MSSM particle is unstable to decay
to the Goldstino as discussed above.

If the messenger scale is in fact well below the Planck scale, it is
likely that the usual $SU(3)_C \times SU(2)_L \times U(1)_Y$ Standard Model
gauge interactions play some role in the messenger sector \cite{phenimp,mgm}.
This is because the structure of supersymmetric gauge theories
dictates that gauginos couple at the renormalizable
level only through gauge interactions.
If the MSSM scalars, including the Higgs bosons which determine
the electroweak scale, received mass
predominantly from non-gauge (and therefore
non-renormalizable and hence suppressed) interactions, the gauginos would
be unacceptably light.
It is therefore natural within low-scale SUSY breaking, to consider
theories of gauge-mediated SUSY
breaking (GMSB) \cite{lsgauge,hsgauge,dnmodels}.
In general, GMSB arises if some massive fields which couple to
the SUSY breaking sector, and therefore have a non-supersymmetric
mass spectrum, also transform under the Standard Model gauge groups.
These heavy fields are referred to as messengers, with
the messenger masses determining the messenger scale $M_m$.

In GMSB theories the MSSM squarks, sleptons, and gauginos obtain
mass radiatively from gauge interactions with the massive
messengers.
Since the messenger interactions are the usual gauge interactions,
the superpartners generally acquire a mass in proportion
to associated gauge coupling squared or equivalently fine
structure constant
\beq
{\widetilde m} \sim {\alpha_a \over 4\pi} {F\over \Mmess}
\>
\label{gmsbestimate}
\eeq
where $M_m$ is the characteristic messenger mass, and
$\alpha_1, \alpha_2 \alpha_3$ are the fine structure constants
for the $U(1)_Y$, $SU(2)_L$, and $SU(3)_C$ gauge interactions.
The precise definition of the minimal model of gauge-mediation
(MGM) is given in Appendix A1.
The dependence of the fine structure constants
generally leads to a hierarchy between the masses of
the strongly interacting squarks and gluino,
$\squark$ and $\gluino$, which couple to $SU(3)_C$,
the Wino and left handed sleptons, $\wino$ and $\slepton_L$,
which couple to $SU(2)_L$, and
the Bino and right handed sleptons, $\bino$ and
$\slepton_R$, which couple to $U(1)_Y$.
The minimal expectation for the
mass ordering of the superpartners with GMSB is then
$m_{\squark}, m_{\gluino} \gg m_{\wino}, m_{\slepton_L} >
m_{\bino},m_{\slepton_R}$.
Based on this mass ordering, either the Bino or right handed slepton,
$\bino$ and $\slepton_R$, are natural candidates within GMSB
for the NLSP,
which is crucial in determining the phenomenology.
However, it is worth noting that this mass ordering
is only representative of the
minimal expectations for GMSB.
The structure and representations of the messengers as
well as their couplings to the SUSY breaking sector need
not be universal.
Almost any mass ordering can in fact be obtained from sufficiently
general models of GMSB \cite{mgm,Martin:1997zb,gmsb}.
So it is important when considering phenomenological signatures
not to focus too closely on any one particular class of
underlying model.

Gauge-mediated supersymmetry breaking, even as an effective
phenomenological theory of SUSY breaking, is by itself incomplete. The
MSSM Higgs sector possesses certain global symmetries in the
supersymmetric limit which must be broken in order to obtain acceptable
electroweak symmetry breaking without the appearance of an unacceptable
Peccei-Quinn axion or very light Higgsino. These symmetries, however,
commute with the gauge interactions at leading order, and are not broken
at this order by gauge couplings with the messenger fields. Additional
interactions between the Higgs and messenger sectors are therefore
required, beyond the minimal gauge interactions \cite{dnmodels}. The
existence of these additional interactions has the possible
phenomenological consequence of making the relative gaugino and Higgsino
mixtures of the neutralinos uncertain \cite{mgm}, as described in Appendix
A2i.

Perhaps the most appealing theoretical feature of GMSB is the natural lack
of SUSY contributions to lepton flavor or quark flavor violating processes
such as $\mu \to e \gamma$ decay, $K \leftrightarrow \bar{K}$
oscillations, or $b \to s \gamma$ decay. This arises because the leading
contributions to visible sector soft SUSY breaking involving the squark
and slepton superpartners depend only on gauge couplings. All soft
SUSY-breaking parameters are then automatically flavor independent or
aligned with the quark or lepton Yukawa couplings. Because of decoupling,
this is generally possible if the messenger scale is well below the flavor
scale at which the standard model Yukawa couplings are determined. But
with high-scale SUSY breaking arising from Planck scale operators no
separation of the messenger and flavor scales is possible and it is
difficult to enforce a symmetry in the high energy theory that can prevent
flavor violation in the visible sector soft SUSY breaking parameters. Even
with GMSB, however, if the messenger scale is not too far below the flavor
scale, sub-leading flavor violating effects can persist. In particular,
this presents the possibility of observing small lepton flavor violation
with NLSP sleptons \cite{sleptonflavor}.

In order to assess discovery reaches for a collider experiment it is
useful to introduce the notion of a Model Line (rather than a model point)
in which 
dimensionless parameters 
and 
ratios of the dimensionful model parameters 
are fixed, but with the
overall superpartner mass scale varying. This allows a discovery reach to
be deduced along a Model Line for a given type of representative model in
terms of the overall superpartner mass scale, or more conveniently in terms
of the mass of a given superpartner. Since with low-scale SUSY breaking it
is likely that gauge-mediation plays some role in transmitting SUSY
breaking to the visible sector, it is reasonable to employ the MGM to
define some of the Model Lines. Representative Model Lines for each
type of NLSP were developed for the Run II workshop, and the results
presented in this report are based on these Model Lines. The NLSP decay
length can be varied at each point on the Model Line by varying the SUSY
breaking scale, holding the other ratios of parameters fixed. The
phenomenological
importance of the parameters which determine the Model Lines are described
in Section \ref{sec:mgm}, and the specific parameters of the individual
Model Lines are presented in subsequent sections.

The results of the Low-Scale Supersymmetry Breaking Working
Group for the SUSY-Higgs Run II workshop
are presented in the remainder of this report.
Section \ref{sec:objectid} contains a summary of the object identification
and acceptances employed for the studies presented in this
report.
Separate studies were undertaken by the CDF and \D0 detector
collaborations, and by
independent groups referred to as ISAJET studies, and
PYTHIA-SHW studies for convenience.
The remaining sections present results organized by the
various scenarios for the NLSP.
Sections \ref{sec:bino} through 
\ref{sec:squarknlsp}
contain results for Bino-like neutralino NLSP,
Higgsino-like neutralino NLSP, stau NLSP, slepton co-NLSP,
and squark NLSP respectively.
The types of triggers, cuts, and
analysis required to probe each NLSP scenario are
detailed, as well as Run II discovery reaches along the defined
Model Lines.
When applicable the cases of prompt NLSP decay,
macroscopic NLSP decay, and NLSP decay well outside the detector
are each considered.
The CDF, \D0, ISAJET, and PYTHIA-SHW studies are complementary in the
sense that they are often based on different event selection criteria,
and are
presented separately
within each section.
Section \ref{sec:waco} discusses the signatures associated
with other NLSP scenarios.
Section \ref{sec:directGG} presents results for direct
Goldstino pair production.
The Appendices contain the precise definition of the MGM,
phenomenological implications of variations of the MGM,
decay rates for MSSM particles to the Goldstino, a list of the Model
Lines studied, a short glossary of acronyms peculiar to our analyses, and
an estimate of the charge exchange length for massive
strongly interacting particles in matter.

Contributions from the large number of participants in the working group
are dispersed throughout the report. The CDF studies include contributions
from W. Bokhari, A. Connolly, R. Culbertson, R. Demina, G. Grim, R.
Thurman-Keup, D. Stuart, B. Tannenbaum and Ming-Jer Wang. The \D0 studies
were carried out by J. Qian, S. Chopra, D. Cutts, and G. Landsberg. The
ISAJET studies were carried out by H.
Baer, P. Mercadante, X. Tata, and Y. Wang, who presented these and related
studies in \cite{BMTW}. The PYTHIA-SHW studies were carried out by K.
Matchev and S. Thomas. The work on prompt squark decay was contributed by
C.-L. Chou and M. Peskin, and on sbaryon and mesino signals by U. Sarid
and S. Thomas. 
Additional work on Bino-like neutralino NLSP, stau NLSP,
and slepton co-NLSPs was contributed at the workshop meetings by B. Dutta,
D. Muller, and S. Nandi and has also been reported on together with
related studies in \cite{DuttaNandi} and \cite{Nanditwo}. 
The appendices and introductory material in each
section were prepared by S. Martin and S. Thomas.


\section{Model Lines for Run II Studies and Minimal Gauge 
Mediation}\label{sec:mgm}
\setcounter{equation}{0}
\setcounter{footnote}{1}
\indent

Since gauge mediation is likely to play some role in low-scale SUSY
breaking it is reasonable to employ the minimal model of gauge mediation
(MGM) to define representative Model Lines for the studies presented in
this report. This provides a well-defined framework to assess discovery
reach and to study specific signatures for each possible type of NLSP. Any
specific model is special in some way. However, the Model Lines developed
for this working group were chosen to be as representative as possible.
Most of the important features of each Model Line depend only the NLSP and
superpartner mass orderings which determine the cascade decays. In most
cases the specific cascades are not important, so the results of the
studies presented in this report should be fairly robust.

The MGM is the simplest phenomenological model of gauge
mediated SUSY breaking (GMSB).
This model is useful for phenomenological studies since any given
model is specified in terms of six parameters
\beq
\Lambda~,~\Nmess~,~\Mmess~,~\tan\beta~,~ {\rm sgn}(\mu)~,~C_G
\eq
defined below.
In addition, either a Bino-like neutralino NLSP, stau NLSP, or
slepton co-NLSP naturally arise a functions of the model
parameters, allowing these possibilities to be covered within
the MGM.

The MGM assumes $N$ generations of messenger fields in the
${\bf 5} \oplus {\bf \overline 5} \in SU(5) \supset
SU(3)_C \times SU(2)_L \times U(1)_Y$.
The messengers have an overall supersymmetric mass
$M_m$, with the scalar and fermion split by a SUSY breaking
auxiliary order parameter $F_S$ with units of [mass]$^2$.
The MSSM gaugino masses arise from one-loop coupling to the
messengers,
\beq
M_a =  k_a N \Lambda ~{\alpha_a \over  4 \pi}
\label{textgauginomass}
\eq
where 
\beq
\Lambda = F_S / M_m
\eeq 
is the effective visible sector
SUSY breaking parameter, and
$a=1,2,3$ for the Bino, Wino, and gluino respectively,
and $k_1={5 \over 3}$, $k_2=k_3=1$.
The MSSM scalar masses arise from two-loop coupling
to the messengers
\beq
m_{\phi}^2 = 2 N \Lambda^2 \left[
  { 5 \over 3} \left( {Y \over 2} \right)^2
       \left( {\alpha_1 \over  4 \pi} \right)^2
 + C_2 \left( {\alpha_2 \over  4 \pi} \right)^2
 + C_3 \left( {\alpha_3 \over  4 \pi} \right)^2
  \right]
  \label{textscalarmass}
\eq
where $Y$ is the ordinary weak hypercharge normalized as
$Q=T_3 + {1 \over 2} Y$,
$C_2={3\over 4}$ for weak isodoublet scalars and zero
for weak isosinglets, and
$C_3={4 \over 3}$ for squarks and zero for other scalars.
The gaugino masses eq.~(\ref{textgauginomass}) which arise at
one loop, and the scalar masses eq.~(\ref{textscalarmass}) which arise
at two loops are each specific realizations of the
estimate eq.~(\ref{gmsbestimate}) for GMSB.
Note that the gaugino masses scale like $N$ while the scalar
masses scale like $\sqrt{N}$.

The MSSM superpartner masses (\ref{textgauginomass}) and
(\ref{textscalarmass}) are evolved from the messenger scale
to the electroweak scale by renormalization group evolution.
This is the only way in which the messenger scale $M_m$
directly enters, so the MSSM parameters only depend
weakly (logarithmically) on $M_m$.
The constraint of electroweak symmetry breaking
with the known $Z$ boson mass
is imposed on the electroweak scale parameters.
These constraints may be determined in terms of the ratio
of Higgs expectation values,
$\tan \beta = \langle H_u^0 \rangle/\langle H_d^0 \rangle$,
and ${\rm sgn}(\mu)$ where $\mu$ is supersymmetric Higgs and
Higgsino mass parameter.
A Bino-like neutralino or slepton is generally the NLSP.
Many groups have developed computer code to calculate electroweak
scale MSSM parameters in terms of the MGM parameters,
including the ISAJET code in versions 7.34 and later \cite{isajet}.

The effective SUSY breaking order parameter $F_S$ felt by
the messengers may not coincide with the
ultimate underlying SUSY breaking order parameter $F$ which
determines the Goldstino coupling.
To account for this a dimensionless factor $C_G \gsim 1$ may be
introduced
relating $F$ and $F_S$ by
\beq
F = \cG F_S 
\label{relateFtoFS}
\eeq
With all other MGM parameters fixed, $C_G$ may be used to
control the NLSP decay length.

The phenomenological meaning and importance of the MGM parameters can be
summarized as follows:

\begin{itemize}

\item[$\bullet$] $\Lambda$:  This effective visible sector
SUSY breaking
parameter sets the overall mass scale for all the MSSM superpartners.
For electroweak scale superpartners
$\Lambda \sim {\cal O}(100~{\rm TeV})/\sqrt{\Nmess}$.
To first approximation, all of the MSSM superpartner masses
scale linearly with $\Lambda$.

\item[$\bullet$] $\Nmess$:
The gaugino masses scale like the number of messenger
generations, $N$, while the
squark and slepton masses scale like $\sqrt{\Nmess}$.
For low values of $N$ a Bino-like neutralino, $\NI$, is the
NLSP, while for larger values a right-handed
slepton, $\slepton_R$, is the NLSP.

\item[$\bullet$] $\Mmess$: The messenger scale enters
as the scale at which the boundary conditions for renormalization
group evolution of the MSSM parameters are imposed.
The electroweak scale and all of the sparticle masses depend only on the
logarithm of $\Mmess$.
The lower limit $\Mmess > \Lambda$ is required in order to avoid
color and charge breaking in the messenger sector, and $\Mmess \lsim
10^{16}$ GeV to satisfy the defining criterion for low-scale SUSY
breaking.

\item[$\bullet$] $\tan\beta$: The ratio of the MSSM Higgs vacuum
expectation values is in a range $1.5 \lsim \tan\beta \lsim 60$.
The lower
limit leads to a light CP-even Higgs scalar, which is presently being
confronted at LEP.
Large values of $\tan\beta$
yield a $\stau$ slepton which is significantly lighter than the other
sleptons.

\item[$\bullet$] sgn$(\mu)$: The sign of Higgs and Higgsino supersymmetric
mass parameter $\mu$ appears in
the chargino and neutralino mass matrices.
For a Higgsino-like
neutralino $\NI$ NLSP with
low to moderate values of $\tan \beta$,
${\rm sgn} (\mu)$ is crucial in determining
the relative strength of the $\NI$ coupling
to Higgs and $Z$ bosons through the Goldstino.
Our convention for the sign of $\mu$ follows that of \cite{haberkane}
and \cite{primer}. (In general, one could allow $\mu$ to have a
complex phase once the other parameters are fixed to be real. However,
in general this would require CP violation and we will not consider it
here.)

\item[$\bullet$] $\cG$: The ratio of the
messenger sector SUSY breaking order parameter to the
intrinsic SUSY breaking order parameter
controls the coupling to the Goldstino.
The NLSP decay length scales like $C_G^2$.

\end{itemize}

For each Model Line, $\Lambda$ is varied with the other parameters or
ratios held fixed.
Since the superpartner masses scale nearly linearly
with $\Lambda$, it is straightforward and natural to find a discovery or
exclusion reach just by varying $\Lambda$.
However, since the scale $\Lambda$ does not appear directly
in any relevant process accessible at an accelerator,
it is best to quote
the discovery reach in terms of physical masses such as the
NLSP mass or the mass of the superpartner which has the largest
production cross-section (typically the lightest chargino).

The parameters for MGM Model Lines studied in the Run II workshop for a
Bino-like neutralino NLSP, stau NLSP, and slepton co-NLSP are given in
Sections \ref{sec:bino}, \ref{sec:stauNLSP} and \ref{sec:sleptonconlsp}
respectively. The parameters for the Higgsino-like Neutralino NLSP Model
Lines studied are given in Section \ref{sec:higgsino}. These two Model
Lines are modified MGM Lines with fixed ratios of $\mu / M_1$ in order to
ensure that the NLSP $\NI$ has a sizeable Higgsino component. This is
equivalent to modifying the Higgs scalar soft masses at the messenger
scale to be different from the left-handed slepton masses. In the MGM
these fields have the same gauge quantum numbers and are degenerate at the
messenger scale. 
Since a squark NLSP is likely to have the largest SUSY
production cross section, the squark NLSP Model Line studied in Section
\ref{sec:squarknlsp} is simply defined to be a single squark with varying
mass. 
For each Model Line, the Goldstino decay constant parameter $\cG$ is a
free parameter, and can be varied in order to adjust the NLSP decay
length.


\section{Object Identification and Acceptances}
\label{sec:objectid}
\setcounter{equation}{0}
\setcounter{footnote}{1}
\indent

The identification and acceptance of objects within events generated for
the results presented in this report differ slightly for the CDF, \D0, and
ISAJET studies. This section presents some information on how objects such
as electrons, muons, taus, photons, jets, $b$-jets, etc.~are identified
and accepted. Some differences in identification and acceptance between
Run I and Run II are also discussed. In addition, the general procedures
are indicated for how signal efficiencies are derived for the present
studies.

A 2~TeV Tevatron center-of-mass energy is assumed throughout.
All event generators used in the studies presented
in this report employ leading order tree-level
cross sections for signal processes.
Inclusion of next to leading order QCD $K$-factor corrections
can increase production cross sections for electroweak states
by 15-20\% \cite{Kfactor}. This can have an important impact on the
interpretation of any observed signal.
However, because the cross section is a rapidly falling function
of superpartner mass, these $K$-factor corrections would only increase
the discovery reach in mass for signals originating from chargino or slepton
production by roughly 5 GeV.

\subsection {CDF study object identification and acceptance parameters}

Both current limits on low-scale SUSY breaking based on Run I data
and projections of sensitivities for Run II are presented as part
of the CDF contribution to the working group.
Although some Run I results
are based on the PYTHIA event generator, most results, including all Run
II projections, are based on the ISAJET event generator. The efficiencies
for Run I analysis were found with the standard full CDF Monte Carlo
detector simulation.  However, at the time of this study, a full Run II
Monte Carlo was not available so the efficiencies for the Run II
projections are estimated using the highly--parameterized
SHW Monte Carlo \cite{SHWreport},
developed for the SUSY-Higgs Run II workshop.
The SHW object identification cuts and efficiencies are employed.
These are
based on Run I experience with corrections for detector changes in Run II.

High-$p_T$ electron identification begins with clusters of energy in the
electromagnetic calorimeters.
A track is required to be found in the
tracking chambers with a $p_T$ consistent with the cluster $E_T$.
The cluster is required to be isolated from other energy in the calorimeter.
Typically, for high-$E_T$ electrons, the additional $E_T$ in a cone of
radius
${\cal R} \equiv \sqrt{(\Delta \phi)^2 + (\Delta \eta)^2} = 0.4$ around
the
electron cluster is required to be less than 10\% of the cluster's energy.
However, a cut
requiring less than 2~GeV in the cone is used for the low-$p_T$ electrons
in the trilepton analysis. Triggers for electrons over 20~GeV are on the
order of 97\% efficient.

Photons are found as electrons with no tracks pointing to them. Without
the track, the trigger has more background and it is necessary to require
that the
photon (with $23<E_T<$50~GeV, $|\eta|<1.0$) is isolated in the trigger,
passes a tight fiducial cut (clusters are far from calorimeter cracks),
and has a cluster in the shower max detector, the combination of which is
not very efficient.  Above 50~GeV, there is no isolation or shower max
requirement and the trigger is much more efficient.

In Run II, we expect to use looser fiducial cuts for electrons and
photons. A new, more sophisticated and efficient isolation trigger will be
installed and fiducial cuts are expected to be loosened.  The Central Outer
Tracker (COT) will provide tracking coverage for $|\eta|<1$, similar to
Run I. However the Intermediate Silicon Layers (ISL) will provide new
tracking coverage out to $|\eta|<2$ in Run II. Run I studies indicate the
$1<|\eta|<2$ region has signal--to--noise comparable to the central region
so with the additional tracking coverage, it is expected to be nearly equal
to the central region.  Therefore all electrons and photons
with $|\eta|<2$ are included in Run II projections.  A fiducial cut
efficiency of approximately 90\% and a identification cut efficiency of
approximately 85\% are expected for Run II.

Muons are identified by matching tracks in the central tracker
to tracks in the
muon chambers, which are outside steel shielding. In Run I, the coverage
was broken up between a central region, a central extension and a forward
region with significant gaps in between. In Run II, the far forward region
will be removed but the central region coverage is greatly expanded.
Overall, muon coverage in CDF will be increased by approximately 15\%. In
the region $|\eta|<0.6$ we will have 100\% coverage, for $0.6<|\eta|<1.0$,
95\% coverage, and for $1.0<|\eta|<1.5$, 75\% coverage. For
muons that are contained in the muon chambers, the identification efficiency
is expected to be approximately 90\%.
The ISL again provides tracking at larger $\eta$,
making these regions similar to the central region.

The $\tau$ lepton is identified as an $e$ or $\mu$ in its leptonic
decays. They can also be identified in their hadronic decays when there
are one or three tracks in a narrow cone.
These tracks are also required to
be isolated from other tracks and the corresponding energy cluster to be
thin, consistent with a low-mass object.  For $\tau$'s passing $E_T$ and
$\eta$ cuts, the identification efficiency is approximately 55\%. In Run
I, it was possible to identify $\tau$'s with $|\eta|<1$ and
it should be possible to extend the coverage for
$\tau$ identification to larger $|\eta|$ by using the ISL.
In projections for this report, it will not be assumed that
this can be done.

In Run I in CDF, jets were clustered using a cone algorithm. The cone has
a radius of 0.4, 0.7 or 1.0 in $\eta - \phi$ space; the smaller radii
being used in to separate jets in a busier environment.  In Run II there
are likely to be other algorithms available, including improved energy
measurement due to analyzing tracks in the jet. Jet identification is
approximately 100\% efficient, apart from the 10-15\% energy resolution.

The tagging of jets as heavy flavor ($b$ or $c$ quarks) will improve
significantly in Run II.  The Silicon Vertex Detector (SVX)  coverage of
the long interaction region ($\sigma=30$~cm) will go from 65\% in Run I to
90\% in Run II. Overall, the probability for tagging one of the two $b$
jets in top events will go from 40\% to 65\%. In addition the new SVX will
be double-sided, improving background rejection. The ISL will also be able
to contribute to tagging, with somewhat worse signal--to--noise compared
to the SVX.

The listing of identifiable objects here is not complete. Of course there
is missing $E_T$, the only object that might be measured more poorly in
Run II (due to additional interactions in an event), but even this may be
compensated by improved understanding.  Unless noted otherwise, the
total momentum of the generated non-interacting particles (usually the
$\GG$ and possibly neutrinos) is employed as the missing $E_T$ for the
signal samples. This is adequate
since the signal events have such large missing $E_T$ that the resolution
smearing doesn't significantly affect the efficiency estimate.

A heavy charged particle may be identified through its
$dE/dx$ and/or its time of flight.  Objects which decay to jets,
leptons and even photons, close to the interaction vertex or far from it
can also be identified.
Charged objects that decay in flight leaving a ``kink'' in a
track can also be found.
The detector has good coverage with powerful, versatile devices,
leaving us limited by the amount of time the collaboration can put into
analysis.

In all cases, backgrounds are projected from Run I data or very
conservative background assumptions are made.  Using data for the
projections correctly includes all sources of ``physics'' background
(Standard Model sources which yield the same signature as the SUSY model)
and ``detector'' backgrounds (mismeasurements and misidentifications)
which are very difficult to model thoroughly. Unless there are obvious
improvements or degradations, using Run I data is a conservative
assumption.  The main degradation will come from multiple vertices in
events, but that is balanced by a detector improved in almost all
respects.  It is also balanced by the enormous experience gained in Run I,
which should be matched by similar innovation and improvement in Run
II.  The history of the accelerator is to exceed projections in delivered
luminosity and the history of the collaborations is to exceed projected
sensitivities.

\subsection{\rm {\D0} study object identification and acceptance
parameters}

All \D0 studies described in this report, except those extrapolated from
Run~I analysis, are carried out at the particle level using ISAJET
\cite{isajet}. Due to a large number of Monte Carlo events generated, no
detector simulation is done for supersymmetry signals.

Leptons, $\ell=e,\mu$, and photons, $\gamma$,  are `reconstructed' from the
generated particle list by requiring them to have transverse energy, $E_T$, or
momentum, $p_T$, greater than 5~GeV and to be within the pseudorapidity
ranges:
\begin{itemize}
  \item[] $e$:      $|\eta|<1.1$ or $1.5<|\eta|<2.0$;
  \item[] $\mu$:    $|\eta|<1.7$;
  \item[] $\gamma$: $|\eta|<1.1$ or $1.5<|\eta|<2.0$.
\end{itemize}
These fiducial ranges are dictated by the coverages of the electromagnetic
calorimeter and the central tracker of the D\O\  detector. Furthermore,
the leptons and photons must be isolated. Additional energy in a cone with
a radius ${\cal R}\equiv\sqrt{(\Delta\phi)^2+(\Delta\eta)^2}=0.5$ in
$\eta-\phi$ space around the lepton/photon is required to be
less than 20\% of its energy.

Jets are reconstructed using a cone algorithm with a radius ${\cal R}=0.5$
in $\eta-\phi$ space and are required to have $E^j_T>20$~GeV and
$|\eta^j|<2.0$. All particles except neutrinos, the lightest supersymmetric
particles (LSP), and the identified leptons and photons are used in
the jet reconstruction. The transverse momentum imbalance ($\met$) is defined
to be the total transverse energy of neutrinos and the LSPs.

Energies or momenta of leptons, photons and jets of Monte Carlo events are
taken from their particle level values without any detector effect. Smearing
of energies or momenta of leptons, photons and jets according to their
expected resolution typically changes signal efficiencies by less than 10\%
relatively and therefore has negligible effect on the study.

The reconstruction efficiencies are assumed to be 90\% for leptons and photons.
For the purpose of background estimations, the probability for
a jet to be misidentified as a lepton
or a photon are assumed to be
${\cal P}(j\to\ell) = 10^{-4}$ and
${\cal P}(j\to\gamma) = 10^{-4}$ respectively.
The probability for an electron
to be misidentified as a photon
is also assumed to be ${\cal P}(e\to\gamma) = 10^{-4}$.
These probabilities are slightly smaller than those obtained in Run~I.
For comparison, the typical misidentification probabilities
determined in Run I are
${\cal P}(j\to e ) = 5 \times 10^{-4}$,
${\cal P}(j\to\gamma) = 7 \times 10^{-4}$, and
${\cal P}(e\to\gamma) = 4 \times 10^{-3}$.
With a new magnetic central tracking system, the improved misidentification
rates given above should be achievable
in Run~II.

In Run~I, tagging of b-jets was limited to the use of soft muons in D\O.
Secondary vertex tagging of b-jets will be a powerful addition in Run~II.
For the studies described below, a tagging efficiency of 60\% is assumed
for those b-jets with $E_T>20$~GeV and $|\eta|<2.0$. The probability
${\cal P}(j\to b)$ for a light-quark or gluon jet to be tagged as a b-jet is
assumed to be $10^{-3}$. These numbers are optimistic extrapolations of
what CDF achieved in Run~I.

Heavy stable charged particles can be identified~\cite{gll,dedx} using the
expected large ionization energy losses, $dE/dx$, in the silicon detector,
fiber tracker, preshower detectors and calorimeter. Based on Ref.~\cite{gll},
a generic $dE/dx$ cut is introduced with an efficiency of
68\% for heavy stable charged particles and a rejection factor of 10
for the minimum ionization particles~(MIP). Note that the
efficiency for identifying at least one such particle in events with two
heavy stable charged particles is 90\%.

With the addition of preshower detectors, D\O\  will be able
to reconstruct the distance of the closest approach~(DCA) of a photon
with a resolution $\sim 1.5$~cm~\cite{gll}. Here the DCA is defined as
the distance between the primary event vertex and the reconstructed
photon direction. Thereby it will enable identification
of photons produced at secondary vertices. In the following, a photon
is called displaced if its DCA is greater than 5.0~cm and is
denoted by $\gamma^\prime$. The further assumption
is made that the probability for a
photon produced at the primary vertex to have the measured DCA$>5$~cm is
${\cal P}(\gamma\to\gamma^\prime)=2\times 10^{-3}$ (about $3\sigma$).

All final states studied have large $E_T$ ($p_T$) leptons/photons with or
without large $\met$. A minimum $p_T$ of 50~GeV of the hard scattering is
applied for all signal processes. Triggering on these events is not
expected to pose any problem. Nevertheless, a 90\% trigger efficiency is
assume for all the final states.

\begin{figure}[tpb]
  \centerline{\epsfysize=3.5in\epsfbox{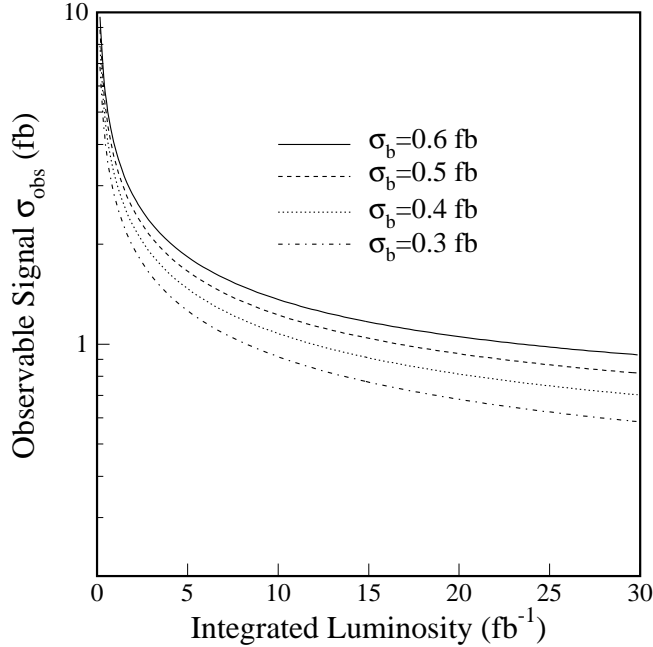}}
  \caption{The minimum observable signal cross section $\sigma_{obs}$ for
           a $5\sigma$ discovery as a function of integrated luminosity
           for four different values of background cross sections.}
  \label{fig:obs}
\end{figure}

In order to find the reach in mass scale for a given supersymmetric
signature, a significance for the signal must be defined.
The significance $N_s/\delta N_b$ is defined as the ratio
between the number of expected signal events, $N_s$, and the error,
$\delta N_b$, on the estimated number of background events.
Here a 20\% systematic uncertainty is assumed for all estimated
observable background cross sections. Therefore,
\beq
\delta N_b=\sqrt{{\cal L}\cdot\sigma_b + (0.2\cdot{\cal
L}\cdot\sigma_b)^2}
\eeq
The sensitivity is characterized using the minimum signal cross section
$\sigma_{dis}$ for a 5 standard deviation ($5\sigma$) discovery:
\beq
\frac{N_s}{\delta N_b} = \frac{{\cal L}\cdot\sigma_{dis}\cdot\epsilon}
                                {\delta N_b} = 5
\eeq
where $\epsilon$ is the efficiency for the signal. The minimum observable
signal cross section $\sigma_{obs}$ defined as $\sigma_{dis}\cdot\epsilon$
for the discovery is therefore independent of signal processes.
The $\sigma_{obs}$ as a function of \ldt\ for several different values of
$\sigma_b$ are shown in Fig.~\ref{fig:obs}. 
It decreases dramatically as \ldt\   increases for small \ldt\ 
values and flattens out for large \ldt\   values. Clearly, the sensitivity
can be improved for large \ldt\   values by tightening the cuts to reduce
backgrounds further.

\subsection{ISAJET study object identification and acceptance parameters }

Signals for the ISAJET studies presented in this report are generated
with ISAJET \cite{isajet}.
A bug in the program which resulted in an
underestimate of the chargino production cross section has also been
corrected.
To model the experimental conditions at the Tevatron,
the toy calorimeter package ISAPLT is
interfaced with ISAJET.
Calorimetry coverage of
$-4 \leq \eta \leq 4$ is assumed with a cell size given by $\Delta\eta \times
\Delta\phi= 0.1 \times 0.087$.
The hadronic (electromagnetic)
calorimeter resolution is taken to be $0.7/\sqrt{E}$ ($0.15/\sqrt{E}$).
Jets are defined as hadronic clusters with $E_T > 15$~GeV within a cone of
$\Delta R= \sqrt{\Delta\eta^2+\Delta\phi^2} = 0.7$ with $|\eta_j| \leq
3.5$.
Muons and electrons with $E_T > 7$~GeV and $|\eta_{\ell}| < 2.5$
are considered to be isolated if the scalar sum of
electromagnetic and hadronic $E_T$ (not including the lepton, of course)
in a cone with $\Delta R=0.4$ about the lepton is smaller than
${\rm max}(2~{\rm GeV}, {1 \over 4} E_T(\ell))$.
Isolated leptons are also required to be
separated from one another by $\Delta R \geq 0.3$.
Photons are identified within $|\eta_{\gamma}|< 1$ if $E_T > 15$~GeV, and
are considered to be isolated if the additional $E_T$ within a cone of
$\Delta R = 0.3$ about the photon is less than 4~GeV.
$\tau$ leptons are
identified as narrow jets with just one or three charged prongs with
$p_T>2$~GeV within $10^{\circ}$ of the jet axis and no other charged
tracks in a 30$^{\circ}$ cone about this axis.
The invariant mass of
these tracks is required to be $\leq m_{\tau}$ and the net charge of the
three prongs required to be $\pm 1$. QCD jets with $E_T =15 (\geq
50)$~GeV are misidentified as taus with a probability of 0.5\% (0.1\%)
with a linear interpolation in between.

\subsection{PYTHIA-SHW study object identification and acceptance
parameters}

Signal and background rates for the PYTHIA-SHW studies
presented in this report were generated with the
PYTHIA \cite{PYTHIA} option of the SUSY-Higgs Workshop (SHW) v.2.2
detector simulation package \cite{SHW,TAUOLA,STDHEP}
developed for the SUSY-Higgs Run II workshop.
The definition of SHW v.2.2 objects is given in the
SHW description in the SUSY-Higgs workshop report \cite{SHWreport}.
The following modifications in the SHW--TAUOLA package are made
for the results presented in this report:
(1) TAUOLA is modified to account for the correct (on average)
polarization of tau leptons coming from decays of supersymmetric particles.
(2) The tracking coverage is extended to $|\eta|<2.0$, which increases
the electron and muon acceptance, as is expected in Run II \cite{CDFTDR}.
For muons with $1.5<|\eta|<2.0$, the same fiducial efficiency
as for $1.0<|\eta|<1.5$ is applied.
(3) The existing electron isolation requirement is retained and a
muon isolation requirement $I<2$ GeV is added, where $I$ is the total
transverse energy contained in a cone of size $\Delta
R=\sqrt{\Delta\phi^2+\Delta\eta^2}=0.4$ around the muon.
(4) The jet cluster $E_T$ cut is increased to 15 GeV and the
jet energy is corrected for muons.
A simple electron/photon rejection cut of
$E_{em}/E_{had}<10$  is also added
to the jet reconstruction algorithm, where
$E_{em}$ ($E_{had}$) is the cluster energy from the
electromagnetic (hadronic) calorimeter.
(5) The calorimeter $\met$ is corrected for muons.
The addition of the muon isolation cut and the jet $E_{em}/E_{had}$
cut allows the occasional ambiguity between
jet and muon objects in SHW v. 2.2 to be uniquely resolved.

\section{Bino-like Neutralino NLSP}\label{sec:bino}
\setcounter{equation}{0}
\setcounter{footnote}{1}
\indent

Neutralinos are in general mixtures of the gauginos and
Higgsinos.
Since the gauginos are superpartners of the gauge bosons,
a gaugino-like neutralino NLSP decays to the Goldstino
predominantly by emission of a $\gamma$ or $Z$ boson.
In many models of supersymmetry breaking, including
the MGM, the gaugino masses are related
by gaugino mass unification relations,
$M_1 \simeq 0.5 ~M_2$,
which imply that
a gaugino-like neutralino NLSP is mostly Bino,
the superpartner of the $U(1)_Y$ hypercharge gauge boson.
Since the projection of the hypercharge gauge boson
is larger in the photon than in the $Z$ boson and because of the more
favorable
kinematics,
a Bino-like neutralino
NLSP decays to the Goldstino
predominantly by emission of a photon:
\beq
\NI \rightarrow \gamma \GG
\label{photonmode}
\, .
\eeq
Supersymmetric particles are produced in pairs, with
all cascades passing through the NLSP.
For a Bino-like neutralino NLSP which decays by eq.~(\ref{photonmode}),
all
supersymmetric final states include two hard photons,
large missing energy carried off by the Goldstinos, and possibly
other hard partons from cascade decays to the NLSP,
$\gamma \gamma X \met$.
If the supersymmetry breaking scale $\sqrt{F}$ is
smaller than a few 100 TeV, the decay length eq.~(\ref{ctaueq}) for
the decay eq.~(\ref{photonmode}) is short enough that
the two hard photons appear to originate from the interaction
point.
In this case the photons are said to be prompt.
However, for $\sqrt{F}$ between a few 100 and a few 1000 TeV,
the decay eq.~(\ref{photonmode}) can take
place over a macroscopic distance, but within the detector.
In this case the photons are said to be non-prompt or displaced,
with a finite distance of closed approach (DCA) to the interaction
point.
For $\sqrt{F}$ greater than a few 1000 TeV, the decays
eq.~(\ref{photonmode}) take place outside the detector.
In this case $\NI$ is essentially stable on the scale of
the experiment and escapes as missing energy.
The resulting signatures are then qualitatively similar
to traditional SUSY missing energy signatures with a
stable $\NI$, and will
not be considered further in this report.
The experimental signatures that are unique to a Bino-like neutralino
NLSP with low scale supersymmetry breaking are therefore:
\beq
\begin{array}{llll}
\bullet~~{\rm Prompt~decays} & \NI \rightarrow \gamma \GG~~: &
      \gamma\gamma X \missET,
                 &   X= {\rm leptons~and~jets} \\
& & & \\
\bullet~~{\rm Macroscopic~decays} & \NI \rightarrow \gamma \GG~~: &
      \gamma\gamma X \missET,
                 &   X= {\rm leptons~and~jets} \\
    & & {\rm Displaced~photons} & \\
\end{array}
\nonumber
\eeq
Observation of either of these signatures would yield
interesting information about the superpartners and supersymmetry breaking.
The final state $\gamma \gamma X  \met$, interpreted as arising from
decay to Goldstino pairs, would immediately
imply that the supersymmetry breaking scale is low.
A large branching ratio for $\NI \rightarrow \gamma \GG$ would imply
the NLSP is mostly Bino.
Finally, with displaced photons,
the decay length distribution would yield the neutralino
life time, and give an essentially model independent measure
of the SUSY breaking scale.

\begin{figure}[tpb]
\centering
\epsfxsize=4.5in
\hspace*{0in}
\epsffile{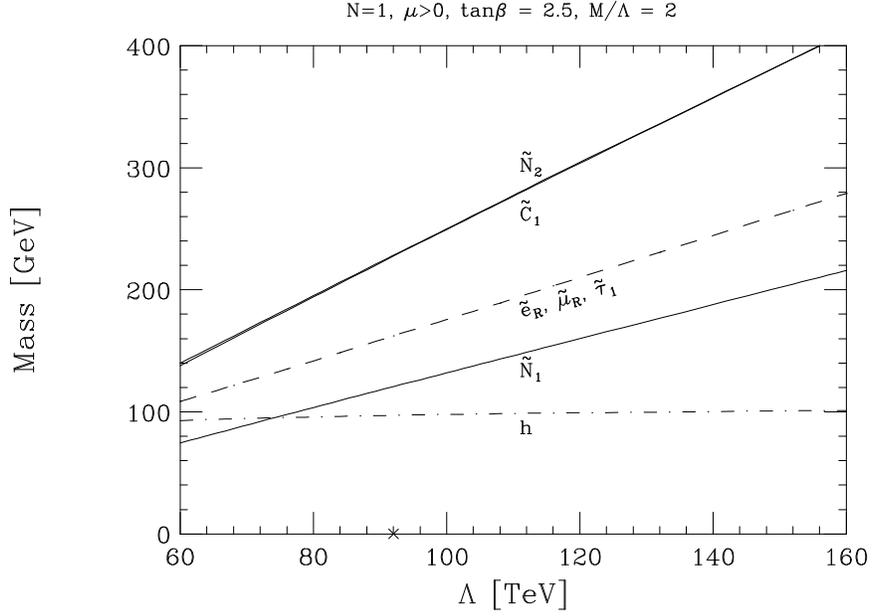}
\caption{The masses of the lightest neutralinos, charginos, sleptons,
and lightest CP-even Higgs boson as a function
of the overall scale $\Lambda$ along the Bino-Like
Neutralino NLSP Model Line.}
\label{fig:n1line_mass}
\end{figure}

For the quantitative studies presented below, a Model Line
within the MGM is defined in which the NLSP is Bino-like with
nearly 100\% branching ratio $\NI \to \gamma \GG$.
The fixed parameters that define the Model Line are:
\beq
{\rm Bino-like~Neutralino~NLSP~Model~Line:}~~~~~~
\Nmess = 1,\>\> 
{\Mmess \over \Lambda} = 2,\>\> \tan\beta = 2.5, \>\> \mu > 0,
\label{binolineparameters}
\eeq
with the overall superpartner mass scale defined by $\Lambda$, which is
allowed to vary.
The mass spectrum of the phenomenologically important superpartners and
the lightest CP-even Higgs boson $h^0$, is shown in Figure
\ref{fig:n1line_mass} as a function of $\Lambda$ in the range 60-160 TeV.
Over the entire model line $\NI$ is the NLSP, and the mass ordering of the
superpartner spectrum is $m_{\tilde \chi_1^0} < m_{\slepton_R} <
m_{\tilde \chi_2^0,
\tilde \chi_1^\pm}$. The
left-handed sleptons, the mostly-Higgsino neutralino and chargino states,
the squarks, and the heavy Higgs bosons, are all too heavy to be produced
at the Tevatron. The lightest neutralino, $\NI$, is mostly $U(1)_Y$ Bino,
while $\NII$ and $\CI$ are mostly $SU(2)_L$ Wino, and nearly degenerate.
The light right-handed sleptons, $\selectron_R$ and $\smuon_R$, are
effectively degenerate with $\stauI$ which is mostly $\stau_R$ with a
small $\stau_L$ component from left-right mixing. The Higgs mass varies
very slowly along the model line, due mainly to the varying virtual
effects of the massive stop squarks.

\begin{figure}[tpb]
\centering
\vspace{1cm}
\epsfxsize=4.5in
\hspace*{0in}
\epsffile{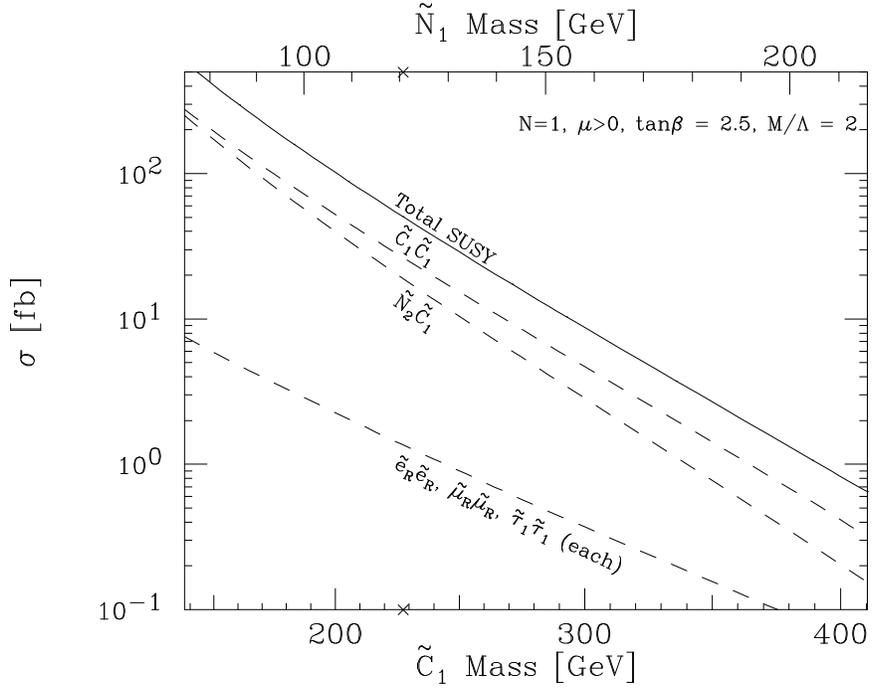}
\caption{The most significant supersymmetric total production
cross-sections in $p\protect\overline p$ collisions with $\protect\sqrt{s}
= 2$ TeV, as a function of $m_{\tilde \chi_1^\pm}$ and $m_{\tilde
\chi_1^0}$ along the
Bino-like
Neutralino NLSP Model Line.}
\label{fig:n1line_sigma}
\end{figure}

The total SUSY cross sections for the light states
are shown in Fig.~\ref{fig:n1line_sigma}.
The largest cross sections are for
$\widetilde\chi_1^+ \widetilde\chi_1^-$ and $\Cone \NII$ production.
These arise predominantly in the $S$-wave through
off-shell $\gamma^*$, $Z^*$, and $W^*$ couplings to the
$SU(2)_L$ Wino components.
Even though the right-handed sleptons are lighter,
$\slepton_R^+ \slepton_R^-$ production cross sections are
smaller because of $P$-wave suppression and smaller
$U(1)_Y$ hypercharge coupling.
Since the two largest production cross-sections both involve the chargino,
$m_{\tilde \chi_1^\pm}$ is
probably the best
figure of merit for the discovery reach along this Model Line.

\begin{figure}[tpb]
\centering
\epsfxsize=4.5in
\hspace*{0in}
\epsffile{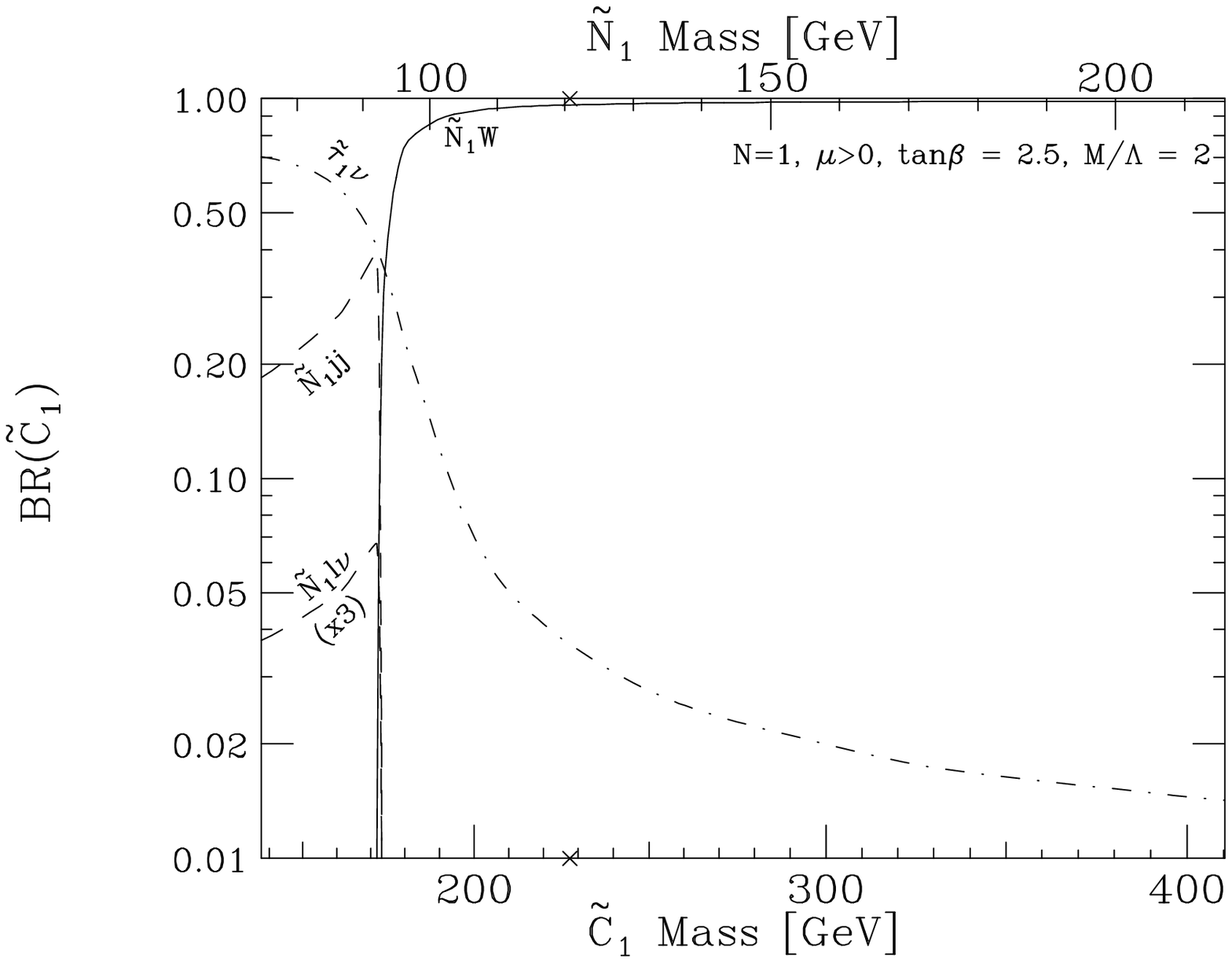}
\caption{Important branching ratios for
$\CI$ decay as a function of $m_{\tilde \chi_1^\pm}$ and $m_{\tilde
\chi_1^0}$ along
the Bino-like
Neutralino NLSP Model Line.}
\label{fig:n1line_brc1}
\end{figure}

\begin{figure}[tpb]
\centering
\vspace{1cm}
\epsfxsize=4.5in
\hspace*{0in}
\epsffile{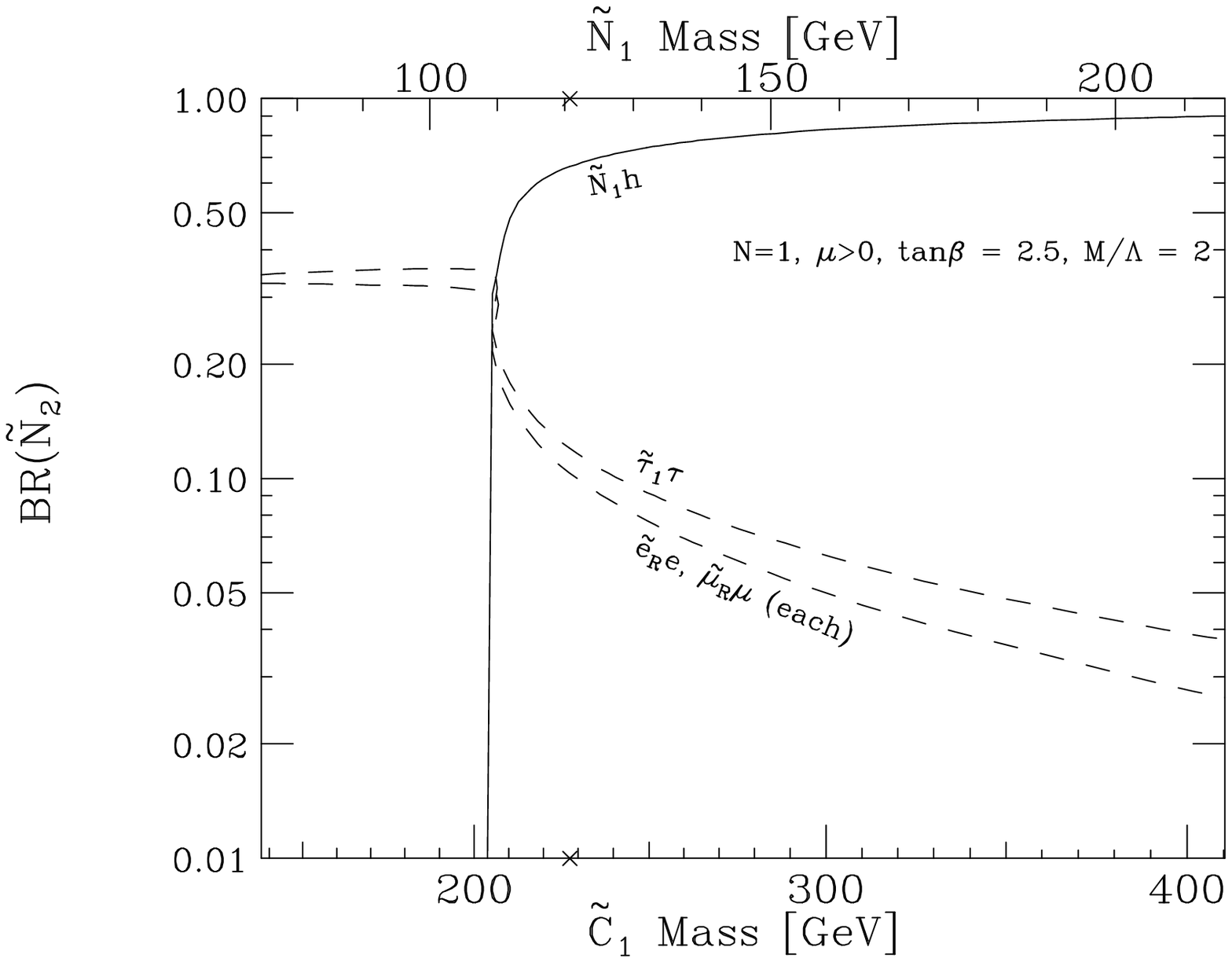}
\caption{Important branching ratios for
$\NII$ decay as a function of $m_{\tilde \chi_1^\pm}$ and $m_{\tilde
\chi_1^0}$ along
the Bino-like
Neutralino NLSP Model Line.}
\label{fig:n1line_brn2}
\end{figure}

The specific final states which arise from $\widetilde\chi_1^+
\widetilde\chi_1^-$ and $\CI
\NII$ production depend on the $\CI$ and $\NII$ decay modes which give
rise to various possible cascades to the NLSP $\NI$. The $\CI$ and $\NII$
branching ratios are shown in Figs. \ref{fig:n1line_brc1} and
\ref{fig:n1line_brn2} as a function of mass.
Since $\CI$ and $\NII$ are highly Wino-like, the dominant decay modes are
dictated primarily by which states with $SU(2)_L$ quantum numbers are
open, and therefore depend on the overall superpartner mass scale. This
leads to three regions along the model line which have qualitatively
different cascades and final states. In the low mass region, $m_{\tilde
\chi_1^\pm}
\lsim 175$ GeV, only three-body decays are open. For intermediate masses,
175 GeV $ \lsim m_{\tilde \chi_1^\pm} \lsim$ 205 GeV, the two-body mode
$\CI \to W^{\pm}
\NI$ is open and dominates. And for larger masses, $m_{\tilde \chi_1^\pm}
\gsim 205$
GeV, the two-body mode $\NII \to h \NI$ is open and dominates. The
dominance of the Higgs mode if kinematically open is generic if the
lightest two neutralinos are gaugino-like, since the Higgs coupling is
only
singly suppressed by the small gaugino-Higgsino mixing, while other two
body modes, such as $\NII \to Z \NI$, are doubly suppressed in this
mixing. This presents the interesting possibility of obtaining Higgs
bosons from supersymmetric cascades, and is discussed further in
subsection \ref{higgsgamgam}.

\begin{figure}[tpb]
  \centerline{\epsfysize=3.5in\epsfbox{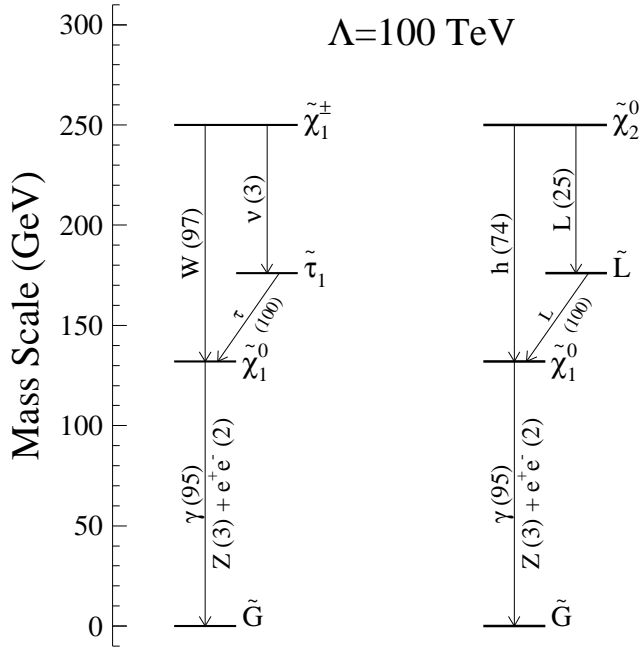}}
  \caption{Superpartner spectrum and important cascade decays for the
  Bino-like Neutralino NLSP Model Line with $\Lambda=100$~TeV.
  Branching ratio percentages are shown in parentheses.}
  \label{fig:p1br}
\end{figure}

A representative spectrum and decay chains with some branching ratios are
shown in Fig.~\ref{fig:p1br} for $\Lambda = 100$ TeV.
For internal comparisons between different studies and detailed Monte
Carlo studies, a reference Model Point on the model line is defined by
$\Lambda = 90$ TeV, corresponding to $m_{\tilde \chi_1^\pm} = 225$ GeV.
The reference
Model Point is indicated by an $\times$ in Figs. (\ref{fig:n1line_mass})
through (\ref{fig:n1line_brn2}).


\subsection{Prompt Decays to Photons}

Prompt decays $\NI \to \gamma \GG$ give rise to spectacular events with
two hard photons and significant missing energy. A search for $\gamma
\gamma X \met$ events provides a very sensitive discovery reach for SUSY
in this channel for a number of reasons. The $\gamma \gamma$ branching
ratio is nearly 100\%, which gives a large advantage over other channels
which require, for example, leptonic decay of $W$ and/or $Z$. There is
essentially no standard model background to $\gamma \gamma X \met$,
although there is significant background from jets faking photons or
mismeasured $\met$. An inclusive search for $\gamma \gamma X \met$,
independent of $X$, is possible. So the specific form of cascade decays do
not affect the discovery reach which depends mainly on the production
cross section. Finally, the detectors have a relatively large coverage and
detection efficiency for photons. CDF and \D0~have searched for this mode
in Run I, and a brief review of those results is included below.
Improvements in triggering, coverage, efficiency, and the projected Run II
sensitivity in this channel are also presented.

\subsubsection{\rm CDF study of promptly-decaying Bino-like NLSP}

CDF searched for the signature of two photons and $\missET$ in Run I and
reported no excess of events beyond the one unusual candidate event
mentioned below \cite{gmsbcdfdiphoton}. The search utilizes 85~pb$^{-1}$
of data and requires two central ($|\eta|<1$) photons with $E_T>12$~GeV
and 35~GeV of $\missET$.  Photons are separated from electrons and jets by
requiring the EM clusters to be isolated in the calorimeter and isolated
from tracks in the central tracking chamber. The sample comes from two
triggers: a high-threshold diphoton trigger with loose requirements and a
low-threshold trigger with tighter fiducial and isolation requirements.  
Events that have both photons with $|\eta|<1$ and over the $E_T$
threshold, and come on the high-threshold trigger have a per--photon
acceptance of 87\% and an efficiency of 84\%.  However the corresponding
numbers for the low-threshold trigger are 73\% and 68\% due to the tighter
trigger requirements. The fraction of diphoton events which actually
contain two photons as opposed to jets faking photons, is low, 15\%.

\begin{figure}[tpb]
\centering
\epsfysize=3.5in
\hspace*{0in}
\epsffile{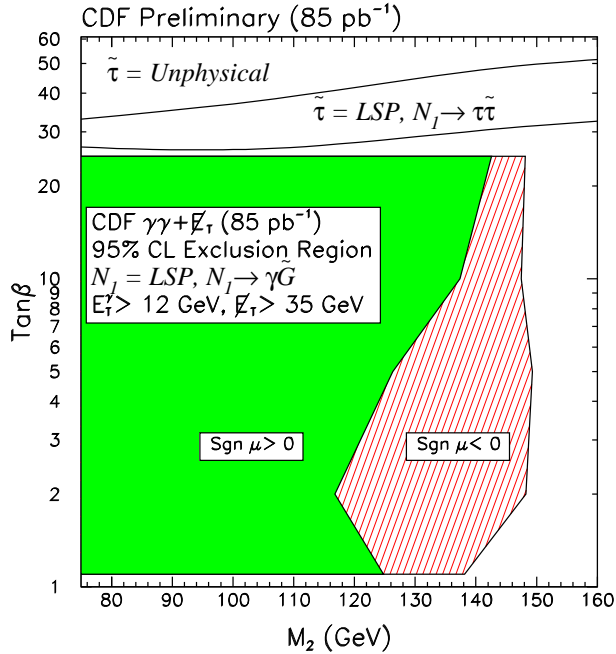}
\caption{Limits on the MGM as a function of $\tan \beta$ and the Wino
mass, $M_2$, with $N=1$ and $M/\Lambda=3$ from $\gamma \gamma X \met$ CDF
Run I data.}
\label{fig:gmsbcdfggi}
\end{figure}

A model for the $\missET$ distribution in these events is derived from
$Z\rightarrow ee$ events which have a similar topology and should have no
true $\missET$.  The $\missET$ is measured as a function of the event
scalar $\Sigma E_T$, excluding the EM clusters.  This function is then
applied to the diphoton sample to produce an expected $\missET$
distribution which agrees well with the observed distribution.  Above
$\missET$ of 35~GeV, only 0.5 events are expected and only one event
survives, the $ee\gamma\gamma\missET$ candidate event. This event contains
two photons, a well-measured electron, a second, unreliably identified,
electron candidate, and $\missET$. The event, which has an estimated
standard model background many orders of magnitude less than one, is
analyzed and discussed elsewhere \cite{gmsbcdfdiphoton}. As a candidate
for low scale supersymmetry breaking, the electrons would come from
cascade decays of $\widetilde\chi$'s or sleptons and the photons and
$\missET$ would come from the $\NI\rightarrow \gamma\GG$ decay. To set
limits, the MGM is employed with parameters $N=1$, $M/\Lambda = 3$ with
$\tan\beta$, sign$(\mu)$, and $\Lambda$ varied (similar to the Model Line
eq.~(\ref{binolineparameters})). The resulting excluded region is shown in
Fig.~\ref{fig:gmsbcdfggi}.

In Run II, many proposed signatures will be investigated. Techniques are
required which cover as many theoretical model possibilities as feasible
in a simple way. One such technique is to investigate a baseline signature
(such as $\gamma\gamma$) and look for other objects in those events. The
Run I $\gamma\gamma$ search includes not only the search with $\missET$
but also with 4 or more jets, $b$-quark jets, electrons, muons, and
hadronically decaying $\tau$'s\cite{gmsbcdfdiphoton}. The results are
shown in Table~\ref{tab:cdfgmsbggruni} - no excesses are found. Note that
this technique requires a sophisticated set of tools from the
collaboration to calculate instrumental backgrounds and an equally
sophisticated set of simulations from theorists for physics backgrounds.
\begin{table}[htbp]
\caption{Summary of the results from searches for
identifiable objects in events with two photons with $E_T>12$~GeV
in the Run Ib CDF data.}
\label{tab:cdfgmsbggruni}
\renewcommand{\arraystretch}{1.75}
\begin{tabular}{|l|c|c|} \hline
Event Selection & Events Observed & Background Estimate\\ \hline
$\missET > 35$~GeV, $|\Delta \phi_{\missET - jet}|>10^\circ$
                                       &1& $0.5\pm 0.1$ \\ \hline
$N_{jet}>4$, $E_T>10$~GeV, $|\eta|<1.0$ &2& $1.6\pm 0.4$ \\ \hline
b-tag, $E_T>25$~GeV, $|\eta|<2.0$      &2& $1.3\pm 0.7$ \\ \hline
$\gamma$, $E_T>10$~GeV, $|\eta|<1.0$    &0& $0.1\pm 0.1$ \\ \hline
Central $e$ or $\mu$, $E_T>25$~GeV     &3& $0.3\pm 0.1$ \\ \hline
$\tau$, $E_T>25$~GeV, $|\eta|<1.2$     &1& $0.2\pm 0.1$ \\ \hline
\end{tabular}
\end{table}

To project the sensitivity in Run II, the signal efficiency and
backgrounds must be estimated. To do this the Model Line
eq.~(\ref{binolineparameters}) and various approximations for estimating
the efficiency are employed. First, in Run II the plug region of the CDF
calorimeter will have very similar properties to the central region in Run
I.  The addition of the ISL tracking silicon detector makes isolation cuts
for plug clusters very similar to the central region. Since no detailed
simulation is available, it is assumed that the plug region will have the
same efficiency and background as the central. This is also justified by
preliminary studies of the plug in Run I which show that the plug is
almost as effective as the central region even using the VTX (a
vertex detector which doesn't measure $p_T$) instead of
the ISL. The region $|\eta|<2$ will therefore be assumed to be available
in Run II.  For the model points investigated here, this improves the
efficiency by 60\%.

Trigger isolation cuts will be made much more efficient and it is assumed
that the fiducial cuts on the low-threshold trigger will be loosened. This
means the Run I high-threshold acceptance and efficiency for all of the
projected Run II data may be used, with an improvement in efficiency of a
few percent (most events will pass the high-threshold trigger anyway).
Since the expected $\missET$ is large, the resolution on the $\missET$ is
not expected to have much of an effect on the efficiency, and the Monte
Carlo generated $\missET$ is used for the event's total $\missET$.

To estimate backgrounds, note that the main backgrounds are QCD with
either one real photon and one jet faking a photon or two fake photons.
The Run II projection will be based on the Run I measured background
distribution with a correction for the increase in the center--of--mass
energy from 1.8~TeV to 2~TeV. QCD Monte Carlo indicates that the dijet
cross section above a threshold $E_T$ cut typically increases by 20\%, so
this factor is included in the background estimate. The $E_T$ distribution
also tends to increase an average of 5\% and this effect is included as
well. To make the projection, these factors are applied to the measured
photon $E_T$ spectrum. These factors are then also applied to the measured
$\Sigma E_T$ distribution, along with the conversion between $\Sigma E_T$
and $\missET$, to arrive at the projected $\missET$ distribution.

With a background and signal distribution, a new set of cuts can be
optimized (with the constraint that the projected background is not to be
reduced
far below 1 event).
The Run I analysis optimized to 12~GeV photon $E_T$ and 35~GeV
$\missET$. Here the projected Run II data optimizes at $E_T>14$~GeV and
$\missET>40$~GeV.

\begin{figure}[tpb]
\centering
\epsfysize=3.5in
\hspace*{0in}
\epsffile{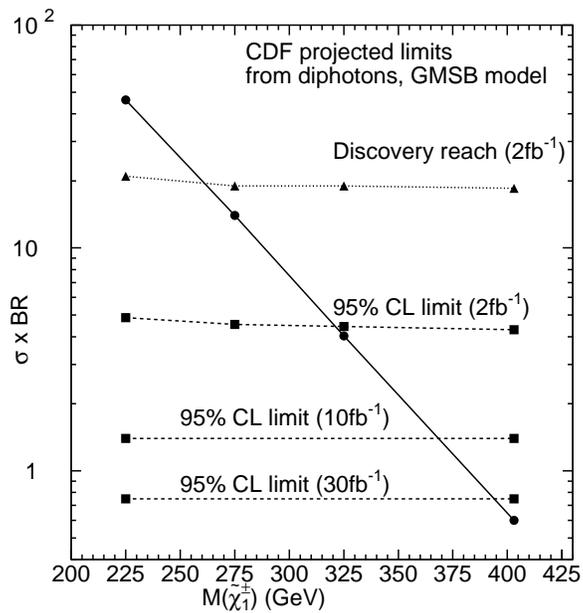}
\caption{Projected CDF limits on the total SUSY cross section
for the Bino-like Neutralino NLSP Model Line from the $\gamma \gamma X
\met$
Run II analysis.  The total SUSY cross section in fb as a function
of the $\Cone$ mass is also indicated.}
\label{fig:gmsbcdfggii}
\end{figure}

With 2~fb$^{-1}$, 1 background event and 20\% systematics, we can expect
to exclude more than 4 signal events. For a $5\sigma$ discovery about 17
events would be required. Table \ref{tab:gmsbcdfdipholim} summarizes the
sensitivity along the Model Line (\ref{binolineparameters}). Figure
\ref{fig:gmsbcdfggii} shows that the reach for CDF alone is approximately
a $\CI$ mass of 320~GeV for a limit or 250~GeV for a discovery.  With 10
fb$^{-1}$, and no further optimization, these masses are estimated to be
are 370 and 310~GeV, and with 30 fb$^{-1}$, 390 and 330~GeV.
\begin{table}[htbp]
\caption{Projected limits and discovery potential for $\gamma \gamma
X \met$
events along the Bino-like Neutralino NLSP Model Line with 2~fb$^{-1}$ in
the CDF study.}
\label{tab:gmsbcdfdipholim}
\renewcommand{\arraystretch}{1.75}
\begin{tabular}{|c|c|c|c|c|} \hline
$\Lambda$~~(TeV)          & 92  & 110    & 128  & 157 \\ 
$m_{\tilde \chi_1^\pm}$~~(GeV)          & 225   & 275    & 325  & 403 \\
$m_{\tilde \chi_1^0}$~~(GeV)          & 121   & 146    & 171  & 212 \\
\hline
$\sigma\times$BR (fb)    & 46  & 14.0  & 4.0 &0.60   \\ \hline
Total $A\cdot \epsilon$  (\%) & 41    & 44  & 45 &  46 \\ \hline
Signal events       &  38    &  12    &    3.6  & 0.55  \\ \hline
95\% C.L. limit (fb)     & 4.9 & 4.5 & 4.4 & 4.3 \\ \hline
$5\sigma$ discovery (fb) & 21.0 & 19.0 & 19.0 & 19.0 \\ \hline
\end{tabular}
\end{table}


\subsubsection{\rm {\D0} study of promptly-decaying Bino-Like NLSP}

Motivated by supersymmetric models with low scale supersymmetry breaking
with a Goldstino, the \D0 Collaboration reported a search~\cite{ggmet} in
Run I for di-photon events with large $\met$~($\gamma\gamma X \met$
events) from a data
sample with an integrated luminosity of $106.3\pm 5.6$~pb$^{-1}$. The
$\gamma\gamma X \met$ events were selected by requiring two identified
photons, one with
$E_T^\gamma>20$~GeV and the other with $E_T^\gamma>12$~GeV, each within
pseudorapidity $|\eta^{\gamma}|<1.1$ or $1.5<|\eta^{\gamma}|<2.0$, and a
$\rlap{\kern0.25em/}E_T$ greater than 25~GeV. Two events satisfied all
requirements.

\begin{figure}[tpb]
  \centerline{\epsfysize=3.5in\epsfbox{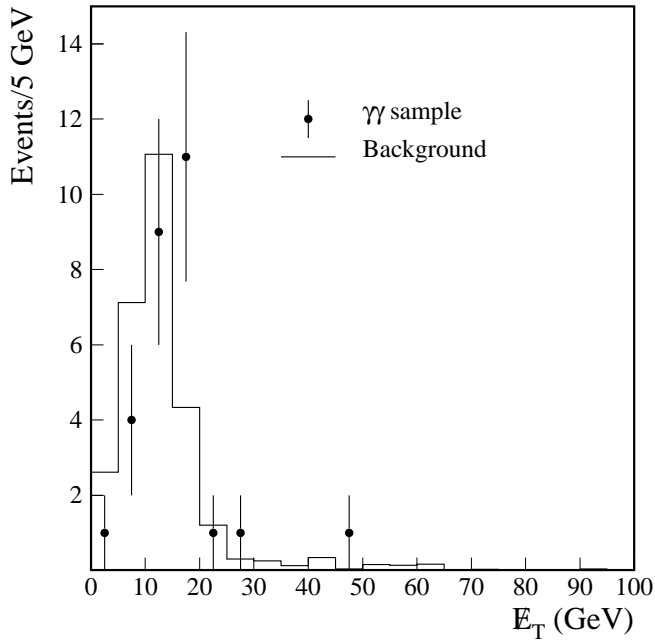}}
  \caption{The \protect$\rlap{\kern0.25em/}E_T$\  distributions of the
           $\gamma\gamma$ and background samples for \D0\ in
                   Run I. The number of
           events with \protect$\rlap{\kern0.25em/}E_T<20$~GeV in the
           background sample is normalized to that in the $\gamma\gamma$
           sample. Note that there is a 14~GeV $\met$ requirement
           in the trigger. The $\met$ values plotted here are
           calculated off-line and therefore may differ from their values
           at the trigger level.}
  \label{fig:run1_ggmet_met}
\end{figure}

The principal backgrounds were multijet, direct photon, $W+\gamma$,
$W+{\rm jets}$, $Z\rightarrow ee$, and $Z\rightarrow\tau\tau\rightarrow ee$
events from Standard Model processes with misidentified photons and/or
mismeasured $\rlap{\kern0.25em/}E_T$. The numbers of estimated background
events were $2.1\pm 0.9$ from $\met$ mismeasurement~(QCD) and $0.2\pm 0.1$ from
misidentified photons~(fakes). This led to an observed background cross
section of 20~fb from QCD and of 2~fb from fakes in Run~I. The $\met$
distributions before the $\met$ cut for both candidates and background events
are shown in Fig.~\ref{fig:run1_ggmet_met}. Note that events with large $\met$
are rare.

Since the backgrounds are dominated by the $\met$ mismeasurement, they can
be significantly reduced by raising the $\met$ cut. Therefore, the following
selection criteria are used for the Run II studies:
\begin{itemize}
  \item[1)] At least two photons with $E^\gamma_T > 20$ GeV;
  \item[2)] $\met>50$ GeV.
\end{itemize}
The backgrounds with this set of selection criteria are expected to be
significantly reduced by the increased cutoffs on $\met$ and photon $E_T$
and by the improved photon identification. The total observable background
cross section in Run II
is estimated to be $\sigma_b = 0.4 ({\rm QCD}) + 0.2 ({\rm fakes})=0.6$~fb
assuming reduction factors of 5 from the raised $\met$ cutoff, 
4 from the improved ${\cal P}(j \to \gamma)$,
3 from the
higher photon $E_T$ requirement, and 10 from the decreased
${\cal P}(e\to\gamma)$ fake probability.

\begin{figure}[tpb]
\centerline{\epsfysize=3.0in\epsfbox{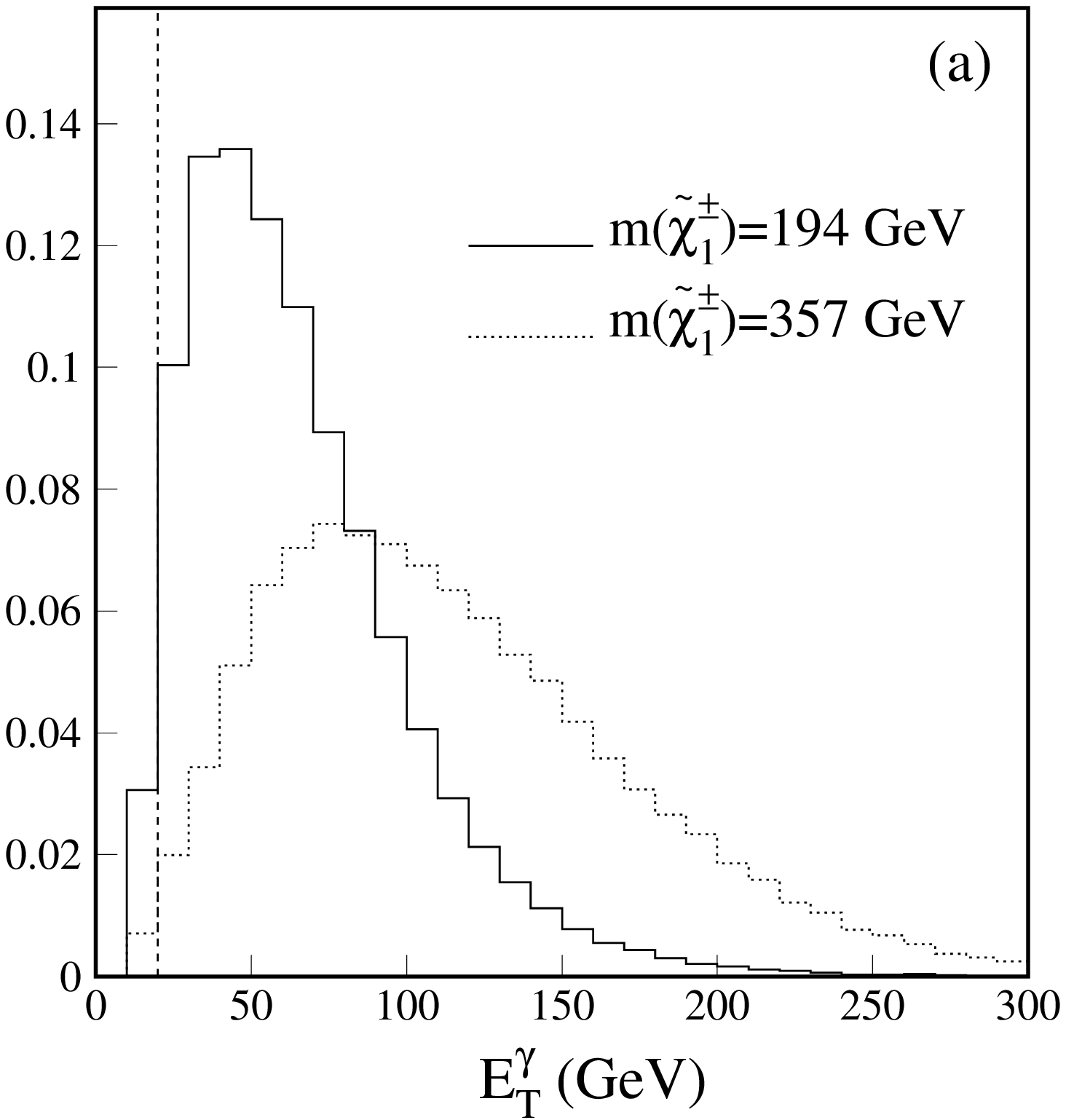}
              \epsfysize=3.0in\epsfbox{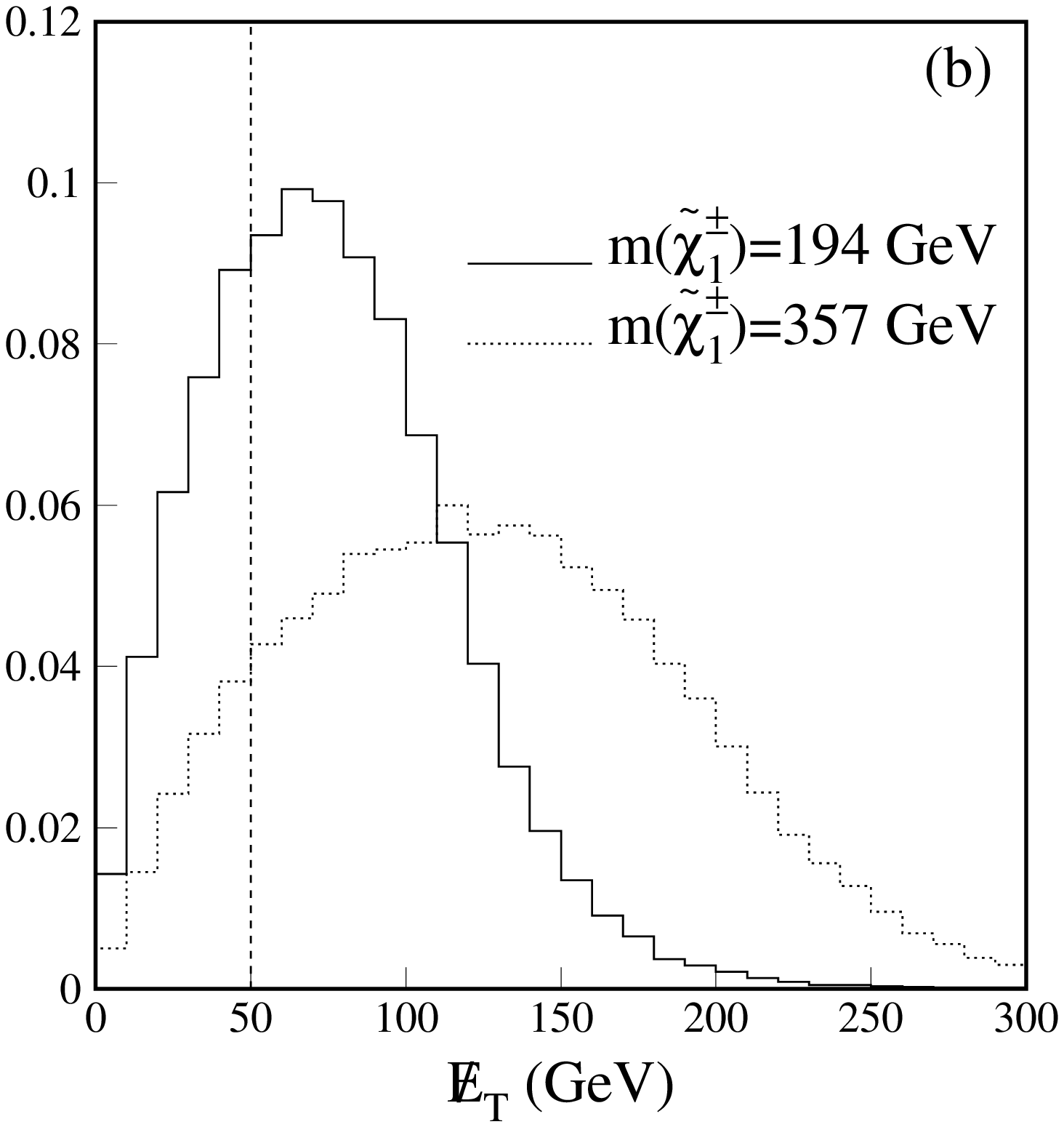}}
\caption{Distributions of (a) photon $E_T$ and (b) event $\met$ for
$\Lambda=80,\ 140$~TeV ($m_{\tilde\chi^\pm_1}=194,\ 357$~GeV) for the
Bino-like Neutralino NLSP Model Line with prompt decay in the {\D0} study.
The vertical dashed lines indicate the cutoffs. All distributions are
normalized to have unit area.}
\label{fig:p1}
\end{figure}

\begin{figure}[tpb]
  \centerline{\epsfysize=3.5in\epsfbox{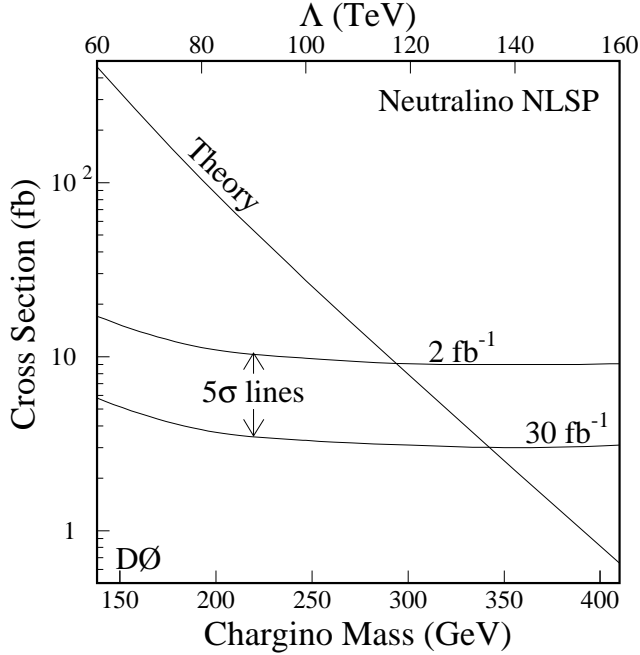}}
  \caption{The $5\sigma$ discovery cross section curves as functions of
the chargino mass along with the SUSY cross sections for the Bino-like
Neutralino Model Line in the \D0 studies. The two curves correspond to
integrated luminosities of 2 and 30~fb$^{-1}$.}
\label{fig:p1lim}
\end{figure}

If the $\NI \to \gamma \goldstino$ decays are prompt, $\gamma\gamma
X \met$ events are
expected. The distributions of photon $E_T$ and event $\met$ for
$\Lambda=80,140$~TeV, corresponding to $m_{\tilde \chi_1^\pm}= 194,357$
GeV, are shown
in Fig.~\ref{fig:p1} for Run II. 
These events typically have high $E_T$
photons together with large transverse momentum imbalances, and therefore
can be selected using the $\gamma\gamma X \met$ criteria discussed above.
Table~\ref{tab:p1} shows the detection
efficiencies, significances, along with the total theoretical SUSY cross
sections, and chargino and neutralino masses for different values of
$\Lambda$ along the Bino-like Neutralino Model Line.
Figure~\ref{fig:p1lim} compares the $5\sigma$ discovery cross sections
$\sigma_{dis}$ with the theoretical SUSY cross sections for two different
values of \ldt\ as functions of the chargino mass $m_{\tilde\chi^\pm_1}$.
The discovery reach in chargino mass with \ldt=2, 30~fb$^{-1}$ is 290, 340
GeV respectively.
\begin{table}[htbp]
\caption{The supersymmetry cross section ($\sigma_{th}$),
$\tilde\chi^\pm_1$ and $\tilde\chi^0_1$ masses, detection efficiency of
the $\gamma\gamma X \met$ selection, and significances for different
values of
$\Lambda$ along the Bino-like Neutralino NLSP Model Line in the {\D0}
study. The relative statistical error on the efficiency is typically 2\%.
The observable background cross section is assumed to be 0.6~fb with a
20\% systematic uncertainty.}
\label{tab:p1}
\renewcommand{\arraystretch}{1.75}
  \begin{tabular}{|c|cccccc|} \hline
  $\Lambda$ (TeV)              &  60 &   80 &  100 &  120 &  140 &  160 \\
  $m_{\tilde\chi^\pm_1}$ (GeV) & 138 &  194 &  249 &  304 &  357 &  410
\\
  $m_{\tilde\chi^0_1}$ (GeV)   &  75 &  104 &  132 &  160 &  188 &  216 \\
\hline
  $\sigma_{th}$ (fb)           & 464 &  105 &   27 &  7.7 &  2.2 &  0.7 \\
\hline
  $\epsilon$ (\%)              & 16.1& 24.3 & 28.2 & 30.1 & 30.6 & 30.2 \\
  \protect\rsb\ (2 fb$^{-1}$)  & 136 &  46  &  14  &  4.2 &  1.2 &  0.4 \\
  \protect\rsb\ (30 fb$^{-1}$) & 400 & 137  &  41  &  12  &  3.6 &  1.1 \\
 \end{tabular}
\end{table}


\subsubsection{ISAJET Studies of Bino-like NLSP}\label{binoisajet}

Tevatron signals for the MGM Bino-like Neutralino Model Line parameters
eq.~(\ref{binolineparameters}) have been simulated using
ISAJET~\cite{isajet}.
This is an extension of a previous study \cite{bbct} of hadron collider
signatures of the MGM with a Bino like NLSP. This analysis has been
repeated for acceptances and cuts more appropriate to Run II. A bug in the
program which resulted in an underestimate of the chargino production
cross section has also been corrected.

The ISAJET simulations are used
to classify the supersymmetric signal events primarily by
the number of isolated photons --- events with $<2$ photons arise when
one or more of the photons is outside the geometric acceptance, has too
low an $E_T$, or happens to be close to hadrons.
The signal events are separated further
into clean and jetty events and classified by the number of
isolated leptons ($e$ and $\mu$).
In addition to the acceptance cuts
described above, an additional global requirement  of $\eslt >
40$~GeV is imposed, which together with the presence of jets,
leptons or photons may also serve as a trigger for these events.

Before describing the results of the computation it is useful to
consider non-supersymmetric Standard Model backgrounds to the signal events.
The backgrounds
are expected to be smallest in the two photon channel,
which will be the main focus for the purpose of assessing the reach.
A detailed estimate of the background has not been attempted
because the recent analysis by the \D0 collaboration
\cite{dzero}, searching for charginos and neutralinos in the GMSB
framework, points out that the major portion of the background arises
from mismeasurement of QCD jets and for yet higher values of $\eslt$
from misidentification of jets/leptons as photons.
In other words, this
background is largely instrumental, and
hence rather detector-dependent.
From Fig.~1 of Ref.~\cite{dzero},
the inclusive $\gamma\gamma X \eslt > 40$~GeV (60~GeV) background
level (for $E_T(\gamma_1,\gamma_2)>$ (20~GeV, 12~GeV))
is estimated to correspond to
$\sim$~0.9 (0.1) event in the \D0 data sample of $\sim 100$ pb$^{-1}$.
The background from jet mismeasurement, of course, falls steeply with
$\eslt$.  The inclusive $\gamma\gamma X\eslt$ background is also
sensitive to the minimum $E_T$ of the photon.

\begin{figure}[t]
\centering
\epsfxsize=4.1in
\hspace*{0in}
\epsffile{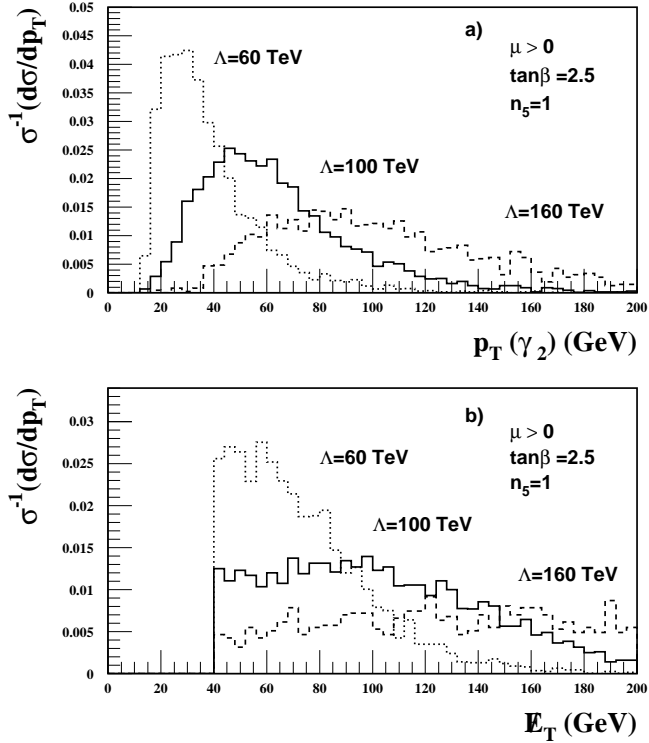}
\caption{a) Transverse energy distributions for the softer
photon, and b)
missing energy distributions for the inclusive
mode $\gamma \gamma X \met$
after cuts along the Bino-like
Neutralino Model Line in the ISAJET study. Results are shown for  $\Lambda
= 60,100,160$ TeV
corresponding to $m_{\tilde \chi_1^\pm}= 135,250,410$ GeV.}
\label{tatafig1}
\end{figure}

Changing the photon and $\eslt$ requirements alter the SUSY signal. To
assess this, the signal distributions of ({\it a})~$E_T(\gamma_2)$, the
transverse energy of the softer photon in two photon events, and ({\it
b})~$\eslt$ in $\gamma\gamma+\eslt$ events that pass the above cuts, are
shown in Fig.~\ref{tatafig1} for three values of $m_{\CIplusminus}$.
The following features are worth noting.
\begin{itemize}
\item For $m_{\tilde \chi_1^\pm} \simeq 250$ GeV
corresponding to
$\Lambda \simeq 100$~TeV
(which is within the Run II reach),
reducing the $E_T(\gamma)$ cut does not increase
the signal. In fact, it may be possible to further harden this cut to
reduce the residual backgrounds.
Although not shown explicitly here, it has been
checked that increasing the cut on the hard photon to
$E_T(\gamma_1) > 40$~GeV results in very little loss of signal for
$m_{\CIplusminus} > 250$ GeV or
$\Lambda \gsim 100$~TeV.
\item In view of the discussion of Standard Model backgrounds, it is clear that
requiring $\eslt > 60$~GeV greatly reduces the background with modest
loss of signal. Indeed, it may be possible to reduce the
background to negligible levels by optimizing the cuts on the photon and
on $\eslt$.
\end{itemize}

\begin{figure}[tpb]
\centering
\epsfxsize=5.2in
\hspace*{0in}
\epsffile{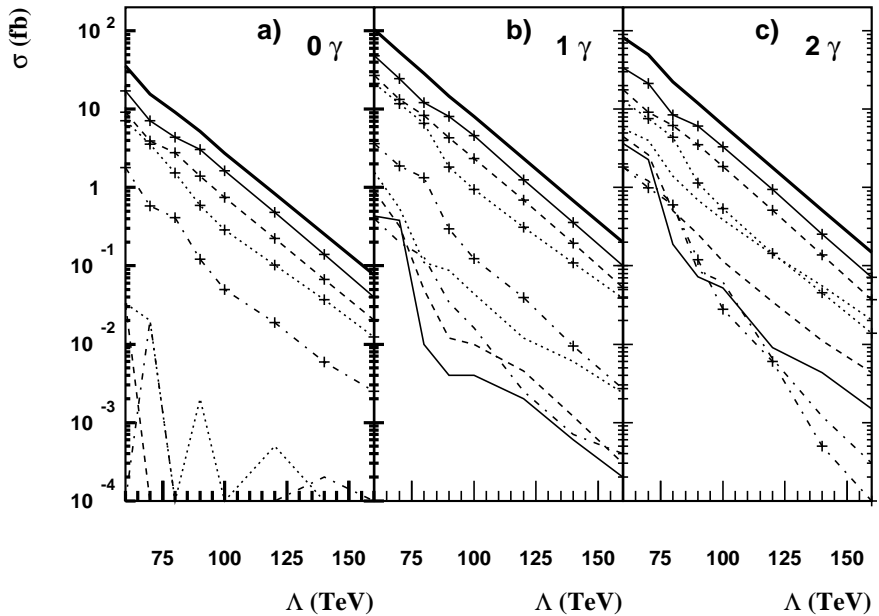}
\caption{Inclusive topological cross sections
for a) $0\gamma X \met$, b) $\gamma X \met$, and c)
$\gamma \gamma X \met$
after cuts along the Bino-like Neutralino
Model Line as a function of $\Lambda$ in the ISAJET study.
The heavy solid lines are the totals.
The solid, dash, dot, and dot-dash lines with crosses are for
events with at least one jet and with 0,1,2,3 leptons respectively.
The solid, dash, dot, and dot-dash lines without crosses are likewise
for events with no jets and with 0,1,2,3 leptons respectively.}
\label{tatafig2}
\end{figure}

The results of the computation of various topological cross sections for a
2 TeV $p\bar{p}$ collider after cuts are shown in Fig.~\ref{tatafig2} for
({\it a}) 0 photon, ({\it b})~one photon, and ({\it c})~two photon events.
For this figure a cut of $\eslt > 60$~GeV has been employed. As mentioned,
this reduces the cross section by just a small amount, especially for the
larger $\CIplusminus$ masses in this figure. The lines with the crosses
correspond to events with at least one jet, while those without correspond
to events free of jet activity. The solid, dashed, dotted and
dashed-dotted lines correspond to 0,1,2 and $\geq 3$ lepton events,
respectively. Finally, the heavy solid line represents the sum of all the
topologies, {\it i.e.} the inclusive SUSY cross section after cuts. Some
features are worth noting:
\begin{itemize}
\item The signal cross sections in $1\gamma$ and $2\gamma$
channels after cuts are comparable.
Since the background in the $2\gamma$ channel is considerably smaller
(recall that a significant portion of the background arises
from fake photons), the maximum reach is obtained in this channel.
\item Events with at least one jet dominate clean events, irrespective
of the number of photons.
\item Over much of the Model Line sparticle production is dominated
by $\tw_1\tw_1$ and $\tw_1\tz_2$ processes -- squarks and gluinos are
generally too heavy to be produced. The event topologies are thus qualitatively
determined by chargino and neutralino decay patterns.
\end{itemize}

A conservative estimate of the reach may be obtained by assuming an
inclusive $\gamma\gamma X \eslt \geq 60$~GeV background level of 0.1 event
per 100 pb$^{-1}$; {\it i.e.} assuming a background effective
cross section of 1~fb.
This corresponds to a ``$5\sigma$ reach''
for the signal cross section of 3.5 fb (1 fb)
for an integrated luminosity of 2 fb$^{-1}$ (25 fb$^{-1}$) at the
Tevatron Run II (TeV33).
This translates to a reach in chargino mass of
$m_{\CIplusminus} \leq 280 (330)$ GeV or equivalently
to $\Lambda \leq 110(130)$~TeV at Run II (TeV33).
As mentioned above, it may be possible to further
reduce the background by hardening the $E_T(\gamma)$ and $\eslt$
requirements with only modest loss of signal. The background may also be
reduced if jet/lepton misidentification as a photon is considerably
smaller than in Run I~\cite{dzero}.
Optimistically assuming that the
reach is given by the 5 (10) event level in Run II (TeV33),
leads to the conclusion that the experiments may probe
chargino masses as high as $m_{\CIplusminus} < 300 (375)$ GeV
corresponding to $\Lambda \lsim 118(145)$ TeV.

\subsection{Non-prompt Decays and Photon Pointing}
\label{subsec:npphoton}

The decay rate for $\NI\rightarrow \gamma\GG$ depends on the supersymmetry
breaking scale, and may take place over a macroscopic distance, as given
in eq.~(\ref{ctaueq}) and illustrated in Fig.~\ref{ctaufig}. Since the
$\NI$ are not ultra-relativistic, the decay photons are roughly uniformly
distributed in solid angle in the lab frame. Decay over a macroscopic
distance, but within the detector volume, therefore gives a displaced
photon with finite impact parameter or finite distance of closest approach
(DCA) to the beam axis. The calorimeters have multiple layers of position
measurements which allow some degree of photon pointing by determining the
line along which the EM shower develops. It is therefore possible to
identify a displaced photon arising from a secondary vertex, and possibly 
to determine the decay length by using time-of-flight information
\cite{ChenGunion}. A measurement
of the decay length distribution would yield an essentially model
independent measure of the supersymmetry breaking scale.

\subsubsection{\rm CDF study of Displaced Photons}

CDF has some power to resolve a photon shower that doesn't point back
to the interaction region.
The calorimeter has a preradiator that measures in the $r-\phi$ direction
with a resolution of 2.5~cm and a shower max detector with a resolution
of about 0.6~cm and 16~cm lever arm between them.
A finite $\NI$ lifetime
could be detected as a broadening of the distribution of $\Delta\phi$
between these $r-\phi$ measurements.  The resulting lifetime measurement
is best if $\NI$ has a lifetime of about 1~ns where the uncertainty
would be about 3~ns$/\sqrt{{N_{\gamma}}}$ and is susceptible to systematics.
This resolution can be characterized as about 10 times worse than
\D0's, completely due to the difference in resolution of the preradiators.

There are also other ways in which a finite $\NI$ lifetime could
be measured.
If there were many signal events the
one-photon excess to the two-photon excess
could be compared to derive a probability
of decay outside the detector and therefore a lifetime.
Also with enough statistics, the 10\% of diphoton events with
one or both photons converting could be studied.
The conversion could be reconstructed in the
silicon systems or central trackers and the
impact parameters and lifetime accurately measured.

\subsubsection{\rm {\D0} study of Displaced Photons}

\begin{figure}[tpb]
  \centerline{\epsfysize=3.0in\epsfbox{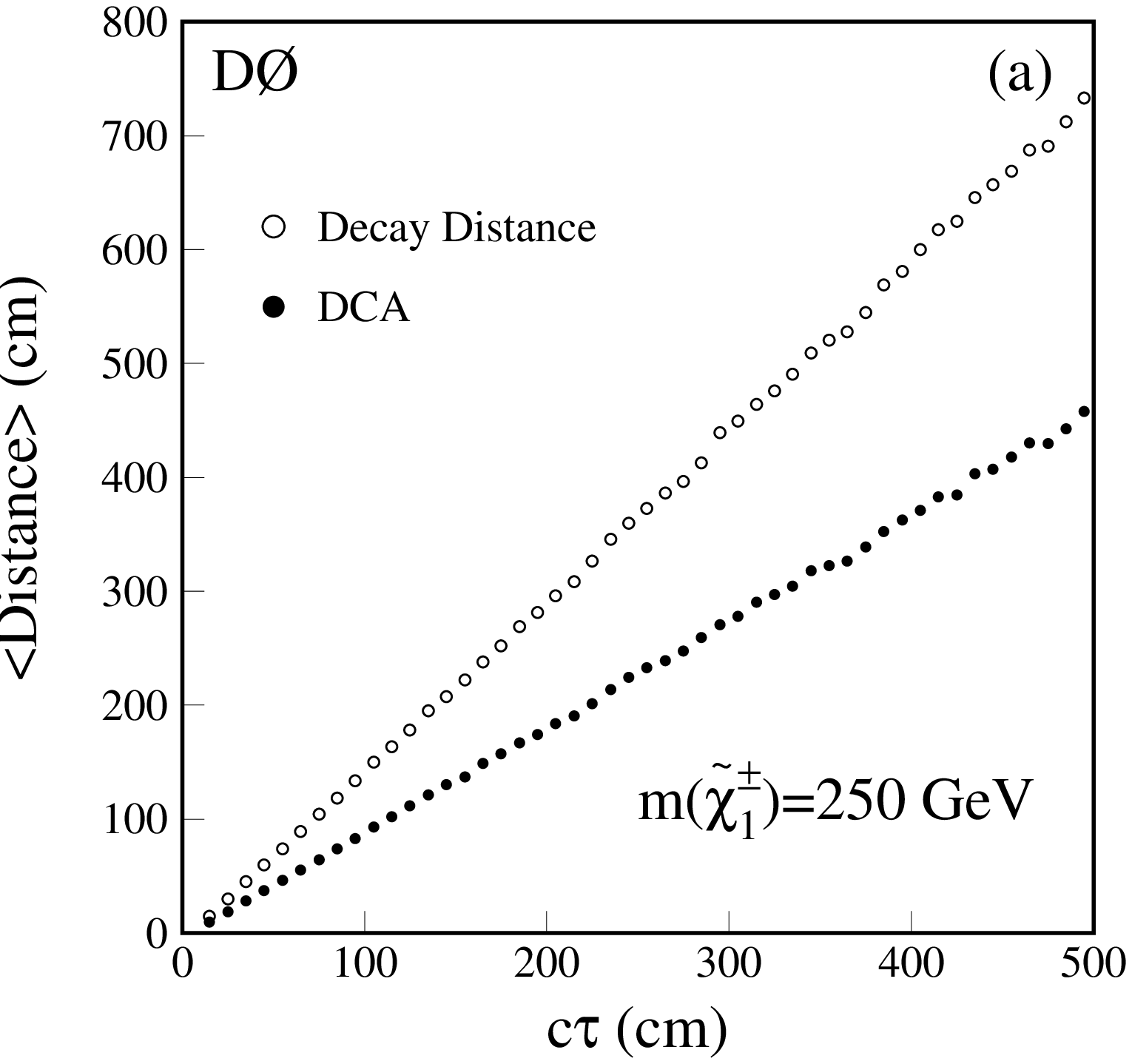}
              \epsfysize=3.0in\epsfbox{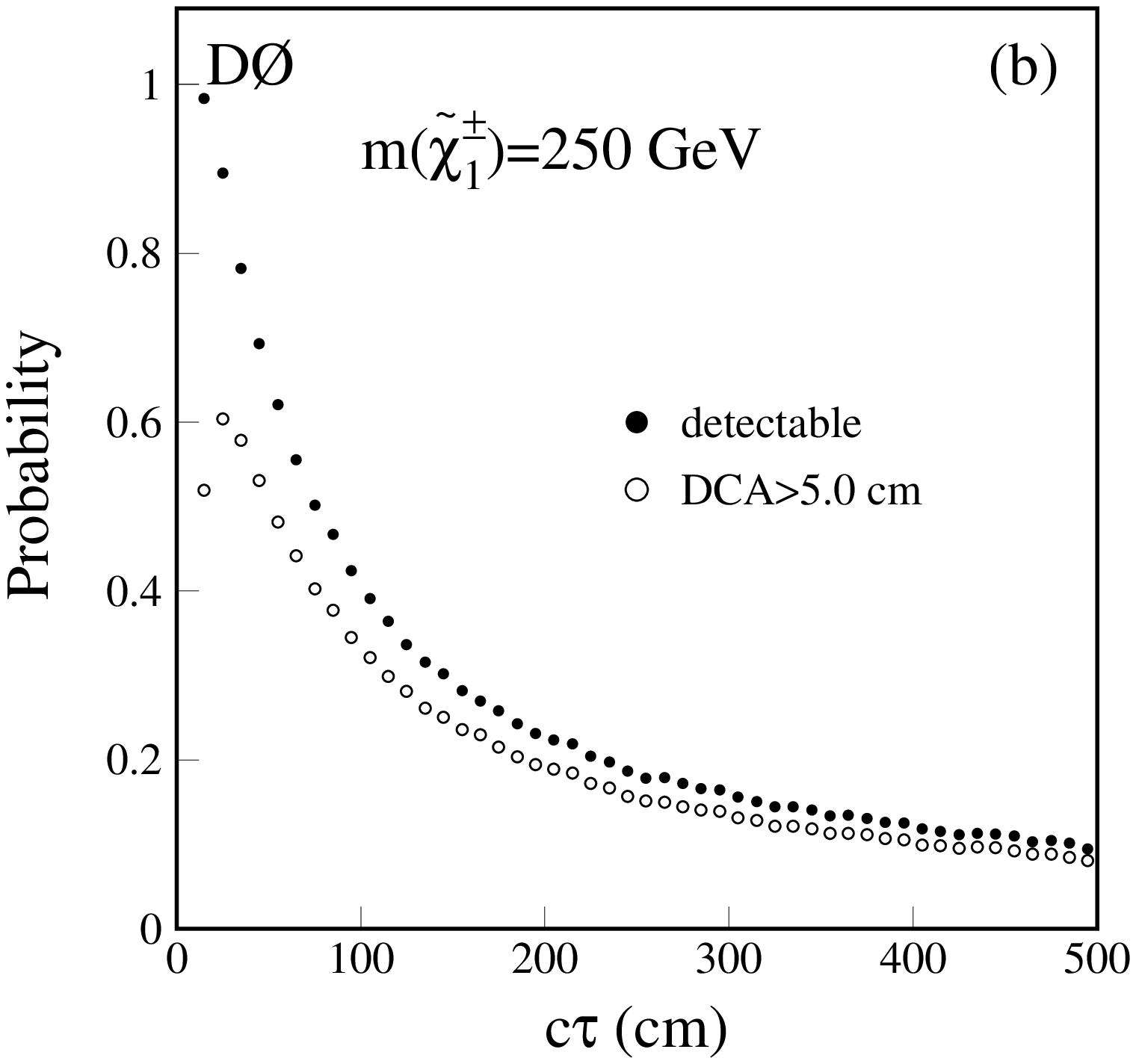}}
\caption{(a) Average lab frame decay distance and photon DCA as a function
of the proper decay distance $c \tau ( \NI \to \gamma \goldstino)$ in cm.
(b) The probability that a displaced photon appears within the tracking
volume, defined by $r <$ 50 cm and $|z| <$ 120 cm, and the probability
that such a photon has a DCA$>5$~cm as functions of $c \tau ( \NI \to
\gamma \goldstino)$ in cm. In both plots $m_{\tilde\chi^\pm_1}\simeq
250$~GeV and $m_{\tilde \chi_1^0} \simeq  130$ GeV corresponding to
$\Lambda=100$~TeV
on the Bino-like Neutralino NLSP Model Line in the {\D0} study.}
\label{fig:p1dd}
\end{figure}

If a non-prompt $\NI \to \gamma \goldstino$ decay occurs inside the
tracking volume of the D\O\ 
detector, the photon is expected to traverse
standard electromagnetic detectors (the preshower detectors and the
electromagnetic calorimeter). It can, therefore, be identified. However,
if the decay occurs outside the tracking detector, the photon
identification is problematic. For this study, the photon is assumed to be
identifiable if it is produced inside a cylinder defined by the D\O\
tracking volume, $r<50$~cm and $|z|<120$~cm, and is lost if it is produced
outside the cylinder. Figure~\ref{fig:p1dd}(a) shows the average decay
distance and photon distance of closest approach as functions of the
proper decay length, $c \tau ( \NI \to \gamma \goldstino)$, for
$\Lambda=100$~TeV corresponding to $m_{\CIplusminus}\simeq 250$ GeV and
$m_{\tilde \chi_1^0} \simeq 130$ GeV.
Due to its heavy mass, the Lorentz boost for the \nlsp\ is typically
small, $\gamma\sim 1.5$. The probabilities that a photon is identifiable
within the tracking volume as defined above, and that the identifiable
photon has DCA$>5$~cm as functions of the $\tilde\chi^0_1$ proper decay
distance $c \tau ( \NI \to \gamma \goldstino)$ are shown in
Fig.~\ref{fig:p1dd}(b), again for $\Lambda=100$~TeV. Photons with
DCA$>5$~cm could be identified as displaced. Distributions for other
superpartner masses are similar.

\begin{figure}[tpb]
  \centerline{\epsfysize=3.0in\epsfbox{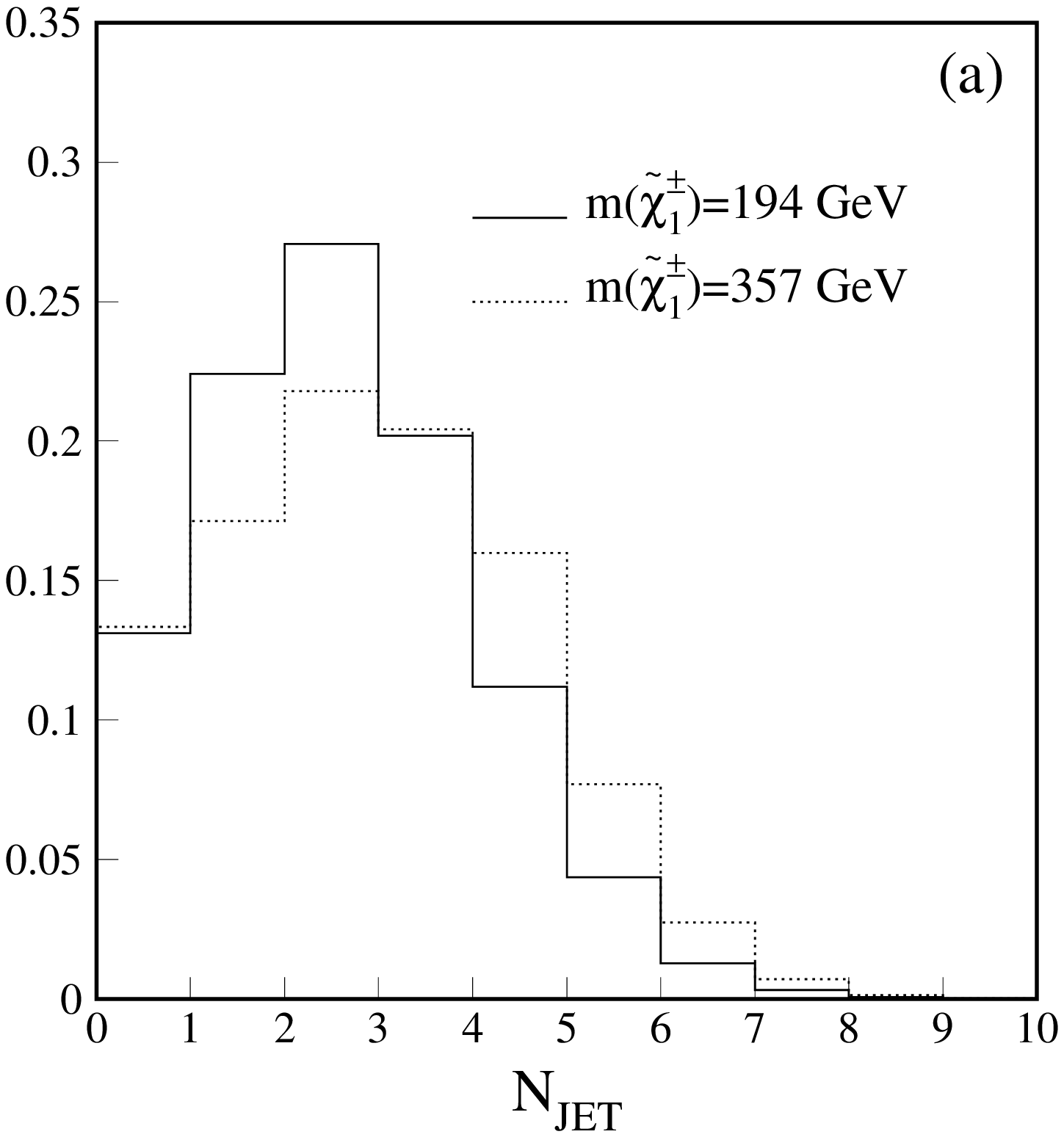}
              \epsfysize=3.0in\epsfbox{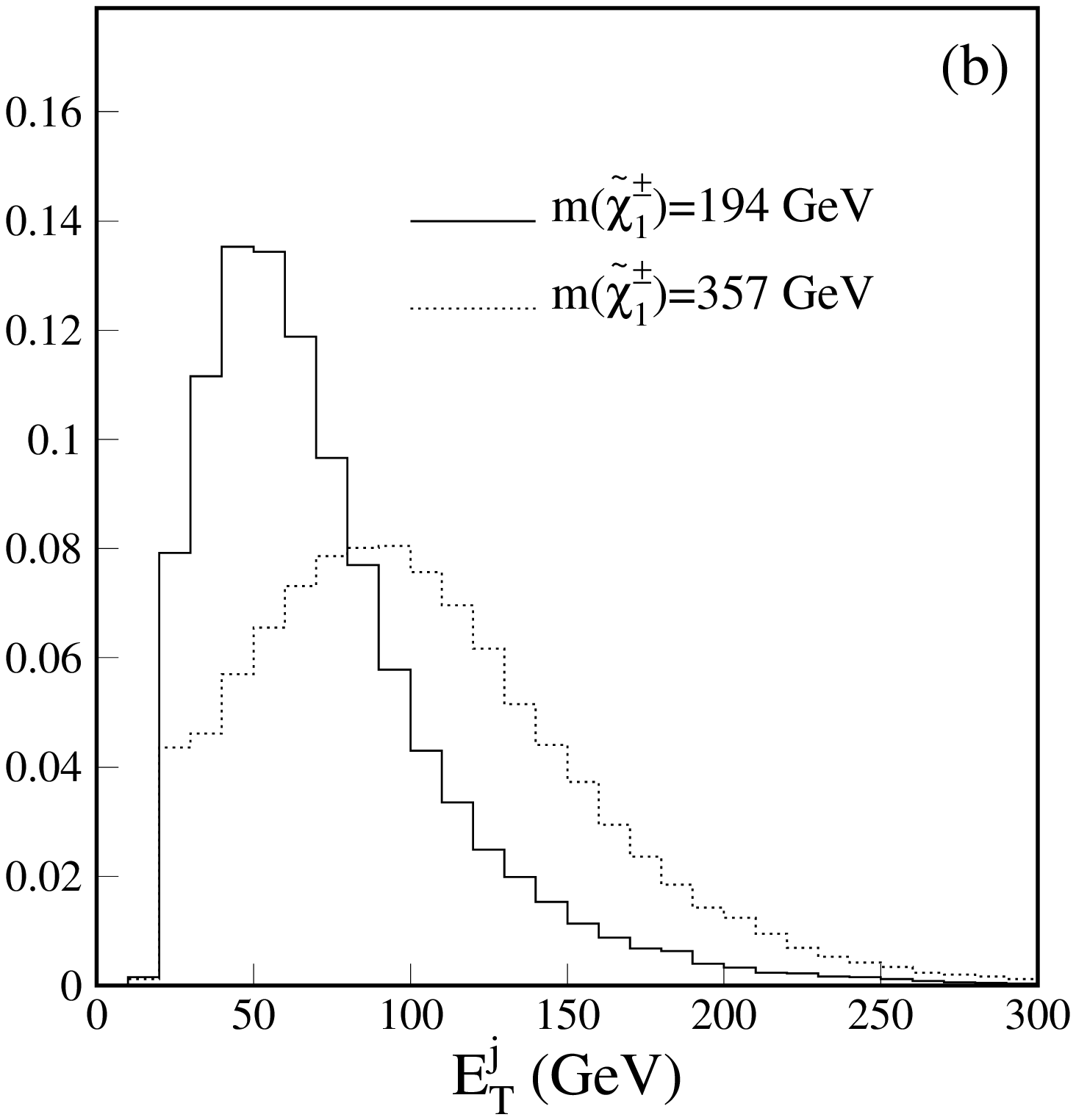}}
  \caption{Distributions of (a) jet multiplicity and (b) jet $E_T$
           for $m_{\tilde\chi^\pm_1}=194,\ 357$~GeV,
           corresponding to  $\Lambda=80,\ 140$~TeV on the
           Bino-like Neutralino NLSP Model Line in the {\D0} study.  All
           distributions are normalized to have unit area.}
  \label{fig:p1jet}
\end{figure}

\begin{figure}[tpb]
  \centerline{\epsfysize=3.5in\epsfbox{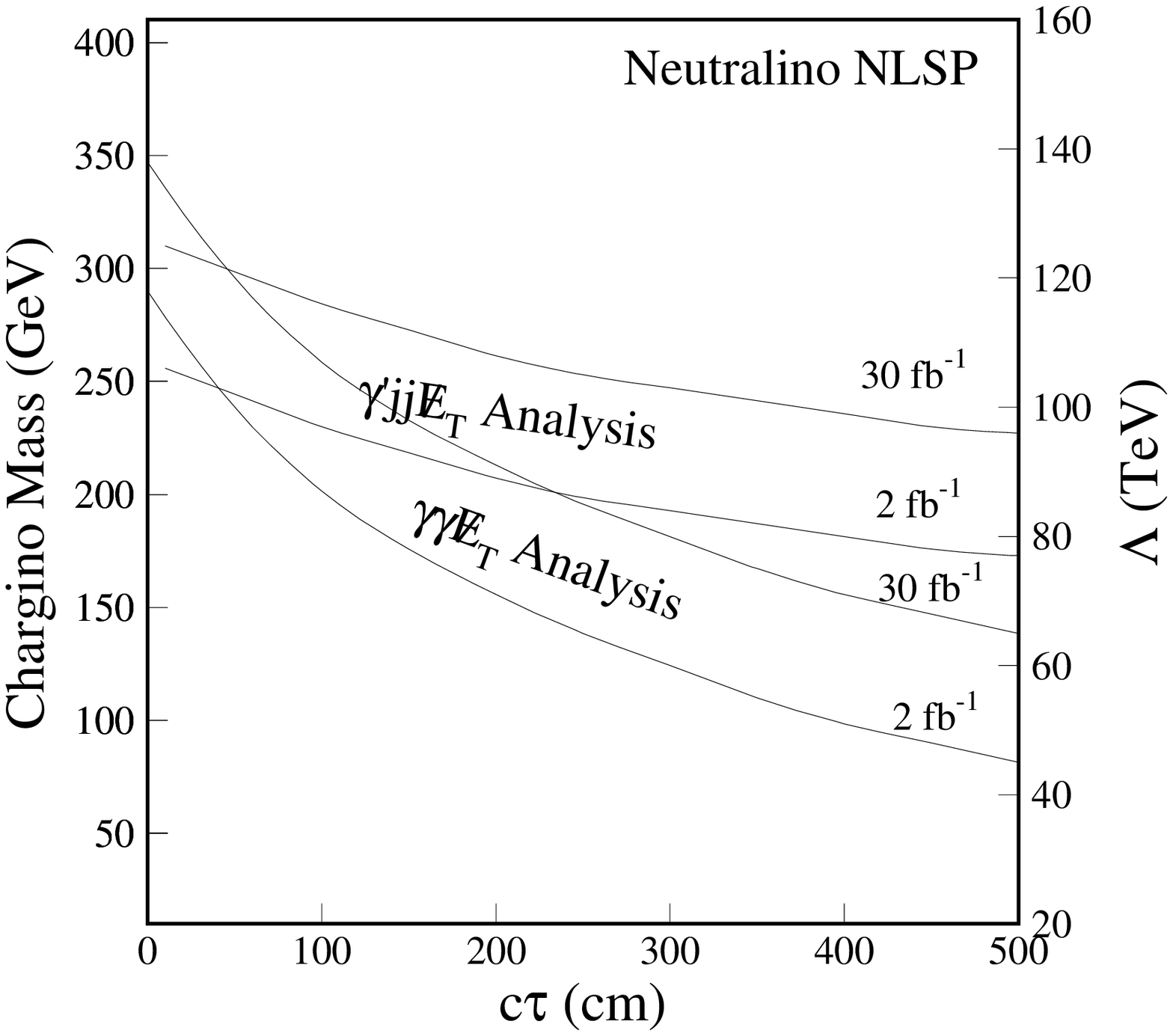}}
  \caption{The $5\sigma$ discovery reaches of the \protect\gdjjmet\ and
           the $\gamma\gamma X \met$ analysis in 
           chargino mass in GeV and $\Lambda$ in TeV
           as a function  $c\tau(\NI \to \gamma \goldstino)$ in cm
           for \protect\ldt=2, 30~fb$^{-1}$ on the Bino-like Neutralino
           NLSP Model Line in the {\D0} study.}
  \label{fig:p1dlim}
\end{figure}

Displaced photon events with secondary vertices are in principle very
dramatic. However, tagging on displaced photons alone is unlikely to
sufficiently reduce backgrounds from cosmic rays or mismeasurements. So
some other characteristic must be employed to tag these events.
Figure~\ref{fig:p1jet} shows jet multiplicity and $E_T$ distributions for
$m_{\CIplusminus} =194,357$ GeV corresponding to $\Lambda=80,140$~TeV on
the Bino-like Neutralino NLSP model line.
Most of the events have large $E_T$ jets.
Events are therefore selected
with displaced photons accompanied by jets and
large $\met$:
\begin{itemize}
  \item[1)] At least one displaced photon with $E^{\gamma^\prime}_T >20$~GeV;
  \item[2)] At least two jets with $E^j_T>20$~GeV;
  \item[3)] $\met>50$~GeV.
\end{itemize}
This gives a \gdjjmet\  sample, where $\gamma^{\prime}$ indicates
a displaced photon. The dominant backgrounds are the same as
those for \gjjmet\   events,
namely QCD direct photon and multijet events or $W$ events with an
electron misidentified as photon, with mismeasured $\met$ and a real or
fake photon, with in this case a vertex-pointing photon being 
misidentified as a displaced photon. Using ${\cal P} (\gamma \to
\gamma^\prime) = 2 \times 10^{-3}$ (about $3 \sigma$),
the observable background cross section from QCD and W events is estimated to
be 0.6~fb. The detection efficiencies and the expected significances of
the \gdjjmet\   selection for
$c\tau(\NI \to \gamma \goldstino)=50$~cm are tabulated in 
Table~\ref{tab:p1h} as an example. The estimated $5\sigma$ discovery
reaches in chargino mass and $\Lambda$ are shown in
Fig.~\ref{fig:p1dlim} as a function of $c\tau(\NI \to \gamma \goldstino)$
along with those expected from the $\gamma\gamma X \met$ 
analysis. 
As expected, the $\gamma\gamma X \met$   analysis has a stronger
dependence on the decay length than the \gdjjmet\  analysis.
Further discussion of the photon pointing capabilities of \D0
is given in the Beyond the MSSM Subgroup report \cite{zooper}.
\begin{table}[htbp]
\caption{The detection efficiency of the \protect\gdjjmet\ selection, and
the significances for different values of $\Lambda$ on the Bino-like
Neutralino NLSP Model Line with displaced photons $c\tau(\NI \to \gamma
\goldstino)=50$~cm in the {\D0} study. The relative statistical error on
the efficiency is typically 2\%. The observable background cross section
is assumed to be 0.6~fb with a 20\% systematic uncertainty.}
\label{tab:p1h}
\renewcommand{\arraystretch}{1.7}
  \begin{tabular}{|c|cccccc|}\hline
  $\Lambda$ (TeV)              &  60  &   80 &  100 &  120 &  140 &  160 \\
  $m_{\tilde\chi^\pm_1}$ (GeV) & 138  &  194 &  249 &  304 &  357 &  410 \\
  $m_{\tilde\chi^0_1}$ (GeV)   &  75  &  104 &  132 &  160 &  188 &  216 \\
\hline
  $\sigma_{th}$ (fb)           & 464  &  105 &   27 &  7.7 &  2.2 &  0.7 \\
\hline
  $\epsilon$ (\%)              & 11.2 & 23.6 & 31.1 & 33.3 & 33.2 &  32.1\\
  \protect\rsb\ (2 fb$^{-1}$)  & 93  &  44  & 15   & 4.6  & 1.3  & 0.4
\\
  \protect\rsb\ (30 fb$^{-1}$) & 278  & 133  & 45   &  14  & 3.9  & 1.2
\\
 \end{tabular}
\end{table}

\subsection{Associated Higgs Production from Supersymmetric Cascades}
\label{higgsgamgam}

An inclusive search for $\gamma \gamma X \met$ is very efficient
in identifying a Bino-like Neutralino NLSP which decays to the
Goldstino by emission of a photon.
The precise nature of the extra partons, $X$, would provide information
about the superpartner mass orderings and couplings.
As discussed at the beginning of this section,
if the two lightest neutralinos, $\NI$ and $\NII$, are gaugino-like,
the second
lightest neutralino decays predominantly to the lightest
one by emission of a Higgs boson, $\NII \to h \NI$, when this
mode is open.
This presents the exciting possibility of obtaining a sample of
Higgs bosons in
$\gamma \gamma X \met$ events.
In this case SUSY could be used to tag Higgs events.
Identification of the Higgs boson
in the mode $h \to b\bar{b}$ would require $b$-tag(s)
of the resulting jets and observation of a peak in the
di-jet invariant mass distribution.
This possibility of using SUSY to tag for Higgs bosons
has the great advantage over Standard Model Higgs channels
of being rate-limited rather than background-limited.

\subsubsection{\rm CDF study of Associated Higgs production in SUSY
events}

\begin{figure}[tbp]
\centering
\epsfysize=3.4in
\hspace*{0in}
\epsffile{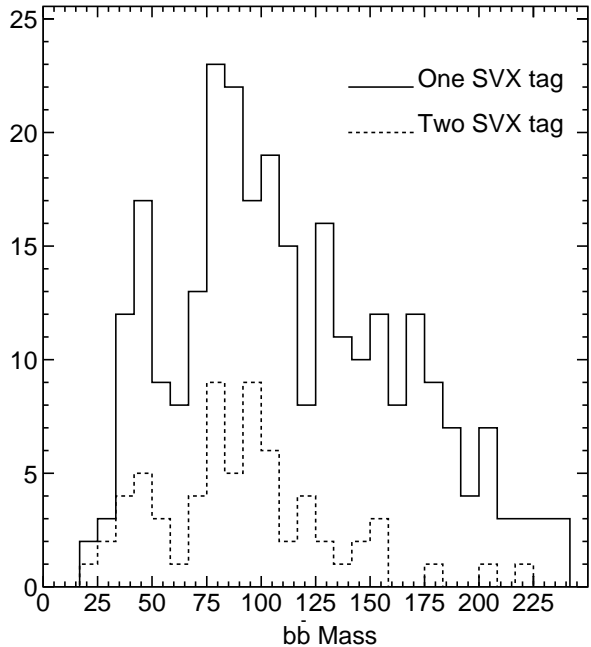}
\caption{The $b\bar{b}$ invariant mass distribution in
$\gamma \gamma X \met$ events for the reference
Model Point on the Bino-like Neutralino NLSP Model Line,
corresponding to $m_{\tilde \chi_1^\pm}=225$~GeV, $m_h  = 97$~GeV in the
CDF study
with 40 fb$^{-1}$.
Essentially no Standard Model
background is expected, and the background here is combinatoric.
In this crude analysis, a peak in the distribution is apparent --
a more sophisticated analysis would increase the signal to noise.}
\label{fig:gmsbcdfggbb}
\end{figure}

For the Bino-like Neutralino Model Line, $\CIplusminus \NII$ pair
production with the dominant decays $\CIplusminus \to W^{\pm} \NI$ with
$\NII \to h \NI$ and $h \to b \bar{b}$ gives the final states $\gamma
\gamma bbjj\met$ and $\gamma \gamma bb l \met$. Requiring an SVX $b$-tag
leaves no Standard Model background. At the reference Model Point on the
Model Line, with a $\Cone$ mass of 225~GeV, about 15 events with at least
one $b$-tag and three events with two $b$-tags are expected in 2
fb$^{-1}$. The
$b\bar{b}$ invariant mass with one and two SVX $b$-tags is shown in
Fig.~\ref{fig:gmsbcdfggbb} for the reference Model Point. 
Because of the large jet multiplicity (c.f.~Fig.~\ref{fig:p1jet}) the
combinatoric background within a SUSY event is significant. However, a
broad peak can be seen in this simple analysis, and a stronger Higgs
signal would be apparent with a more sophisticated analysis.


\section{Higgsino-like Neutralino NLSP}
\label{sec:higgsino}
\setcounter{equation}{0}
\setcounter{footnote}{1}
\indent

Neutralinos are in general mixtures of both gaugino and Higgsino
eigenstates.
A general NLSP neutralino can therefore decay to the Goldstino by
emission of either a Higgs boson, $Z$ boson, or photon
\beq
\NI \to (h,Z,\gamma)~\GG
\label{hZgmode}
\eq
Pair production of superpartners which cascade decay to a general NLSP
neutralino then gives rise to the di-boson final states
$(hh,h\gamma,hZ,Z \gamma, ZZ, \gamma \gamma) X \met$,
where $X$ represents additional partons from the cascade
decays \cite{phenimp,higgsinotalks,hshort}.
Di-boson signatures which include Higgs and $Z$ bosons and
$\met$ are quite novel discovery modes for SUSY at the Tevatron.
In conventional SUSY signatures, in which the lightest neutralino,
$\NI$, is assumed to escape the detector without
decay to the Goldstino, the mass splittings between supersymmetric
particles required in order for $h$ or $Z$ to arise in a cascade
decay typically imply the superpartners are too heavy to be
produced in sufficient numbers at the Tevatron.
For this reason events with reconstructed $Z$ bosons are in
fact generally rejected in present SUSY searches.
However, since the Goldstino is essentially massless, sufficient
phase space is available for the $\NI \to h \goldstino$ and
$\NI \to Z \goldstino$ modes.
The Higgs final states also present the
exciting possibility of discovering and studying the
Higgs boson in association with supersymmetry.

If the supersymmetry breaking scale $\sqrt{F}$ is
smaller than a few 100 TeV, the decay length (\ref{ctaueq}) for
the decays (\ref{hZgmode}) is short enough that
the decay products appear to originate from the interaction
point.
For the photon mode $\NI \to \gamma \GG$ the photon is then prompt,
while for  $\NI \to h \GG$ with $h \to b\bar{b}$ or
$\NI \to Z \GG$ with $Z \to \ell \ell, jj$, the final state
partons are prompt with invariant masses associated
to the parent Higgs or $Z$ boson.
However, for $\sqrt{F}$ between a few 100 and a few 1000 TeV,
the decays (\ref{hZgmode}) can take
place over a macroscopic distance, but within the detector.
In this case the di-bosons are non-prompt or displaced.
For displaced decays $\NI \to h \GG$ with $h \to bb$ or
$\NI \to Z \GG$ with $Z \to \ell \ell, jj$ secondary
vertices arise, again with decay product invariant masses
which can be associated to the parent Higgs or $Z$ boson.
For the hadronic decay modes, especially $h,Z \to b\bar{b}$,
Standard Model backgrounds from heavy quark production with
displaced secondary vertices are possible, but
may be controlled by use of invariant mass and angular distributions.
For $\sqrt{F}$ greater than a few 1000 TeV, the decays
eq.~(\ref{hZgmode}) take place outside the detector.
The resulting signatures are then qualitatively similar
to traditional SUSY missing energy signatures with a
stable $\NI$.
The experimental signatures for a general neutralino
NLSP with low scale supersymmetry breaking are therefore:
\beq
\begin{array}{llll}
\bullet~~{\rm Prompt~decays} & \NI \rightarrow (h,Z,\gamma)
         \GG~~: &
        (hh,h\gamma,hZ,Z\gamma,ZZ,\gamma\gamma) X  \missET,
                 &   X= {\rm leptons~and~jets} \\
& & & \\
\bullet~~{\rm Macroscopic~decays} & \NI \rightarrow (h,Z,\gamma)
         \GG~~:  &
                 (hh,h\gamma,hZ,Z\gamma,ZZ,\gamma\gamma) X  \missET,
                  &  X= {\rm leptons~and~jets} \\

    & & ~{\rm Displaced~photons}, ~b\bar{b}~{\rm pairs}, & \\
        & & ~~~{\rm or}~ \ell^+ \ell^-~{\rm pairs}
      & \\
\end{array}
\nonumber
\eeq
All the possible final state signatures for a general
neutralino NLSP with observable decay to the Goldstino are given
in Table \ref{higgsinosignatures},
including the heavy boson decays $h \to b\bar{b}$ and
$Z \to \ell^+ \ell^-, \nu \bar{\nu}, jj$.
Observation of any of these signatures would yield
interesting information about the $\NI$  composition, and would imply
the SUSY breaking scale is low.
Observation of a displaced $h$ or $Z$ decay would be very dramatic.
A measure of the decay length distribution would give an
indirect measure of the SUSY breaking scale.
Certain of the di-boson final states could be an interesting source
of Higgs bosons.
\begin{table}[htbp]
\caption{Final state signatures for a general Higgsino-gaugino
neutralino NLSP with observable decay to the Goldstino,
$\NI \to (h,Z,\gamma) \GG$.
Partons resulting from a common $h$ or $Z$ parent, with the associated
invariant mass, are grouped together.
$X\equiv$ additional partons from cascade decays to the $\NI$ NLSP.
\label{higgsinosignatures}}
\renewcommand{\arraystretch}{1.5}
\begin{tabular}{||c||c||c||c|c|c||}
\hline\hline
& $\gamma$
  & $h\rightarrow b\bar{b}$
    & $Z\rightarrow \ell^+\ell^-$
      & $Z\rightarrow \nu\bar{\nu}$
        & $Z\rightarrow jj$   \\ \hline\hline
$\gamma$
  & $\gamma\gamma X \met$
    & $\gamma (b\bar{b})_h X \met$
      & $\gamma (\ell^+\ell^-)_ZX \met$
        & $\gamma X \met$
          & $\gamma (jj)_Z X \met$ \\ \hline\hline
$h\rightarrow b\bar{b}$
  &
    & $(b\bar{b})_h(b\bar{b})_h X \met$
      & $(b\bar{b})_h(\ell^+\ell^-)_Z X \met$
        & $(b\bar{b})_h X \met$
          & $(b\bar{b})_h(jj)_Z X \met$ \\ \hline\hline
$Z\rightarrow l^+l^-$
  &
    &
      & $(\ell^+\ell^-)_Z (\ell^+\ell^-)_Z X \met$
        & $(\ell^+\ell^-)_Z X \met$
          & $(\ell^+\ell^-)_Z (jj)_Z X \met$ \\ \hline
$Z\rightarrow \nu\bar{\nu}$
  &
    &
      &
        & $X \met$
          & $(jj)_Z X \met$ \\ \hline
$Z\rightarrow jj$
  &  &  &  &  & $(jj)_Z(jj)_Z X \met$\\ \hline\hline
\end{tabular}
\end{table}

\begin{figure}[tpb]
\begin{center}
\epsfysize=3.5in
\epsffile{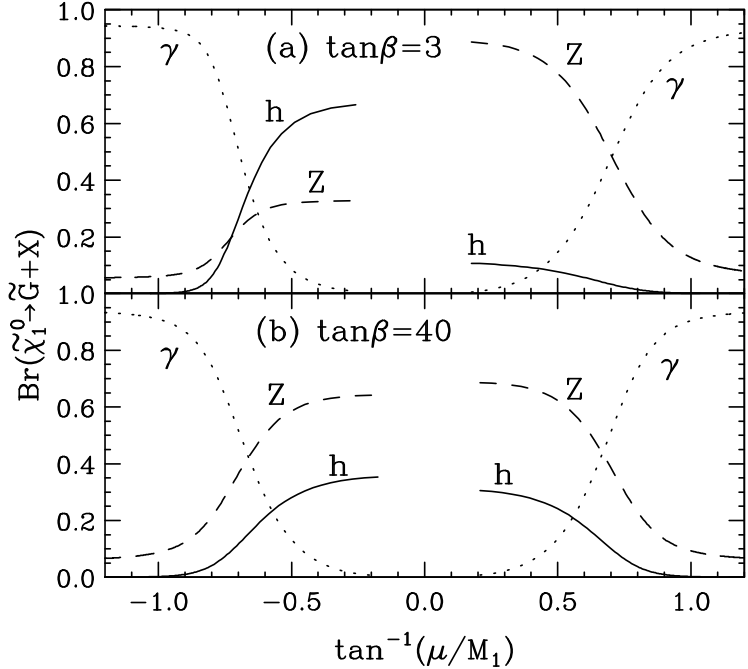}
\parbox{6.0in}{
\caption[]{Branching ratios of the lightest neutralino
to the Goldstino,
${\rm Br}(\widetilde \chi^0_1 \rightarrow \tilde G + (\gamma,h,Z))$,
as a function of the neutralino mixing angle $\tan^{-1}(\mu/M_1)$,
for a fixed mass $m_{\tilde \chi^0_1}=160$ GeV
and $m_h=105$ GeV for (a) $\tan\beta=3$ and (b) $\tan\beta=40$.
\label{higgsinobr}}}
\end{center}
\end{figure}

The branching ratios
${\rm Br}(\tilde \chi^0_1 \rightarrow \tilde G + (\gamma,h,Z))$
are determined by the relative Higgsino and gaugino content
of $\widetilde \chi_1^0$.
The general expressions for the decay widths are given
in Appendix B.
The Higgsino-gaugino content dependence is
illustrated in Fig.~\ref{higgsinobr} as a function of
the neutralino mixing angle $\tan^{-1}(\mu/M_1)$ for
$\tan \beta = 3$ and $40$ with fixed
$\tilde \chi_1^0$ mass, where $\mu$ and $M_1$ are the Higgsino
and Bino mass parameters respectively.
For definiteness the Higgs decoupling limit
in which decays to the heavy scalar and pseudoscalar Higgs bosons,
$H$ and $A$, are kinematically blocked is employed throughout.
For $|\tan^{-1}(\mu/M_1)|$ large, $\NI$ is
gaugino-like and $\NI \to \gamma \GG$ dominates,
while for  $|\tan^{-1}(\mu/M_1)|$ small, $\NI$ is
Higgsino-like and $\NI \to h \GG$ and $\NI \to Z \GG$ are important.
The dependence of the branching ratios
on ${\rm sgn}(\mu)$ and $\tan \beta$
apparent in Fig.~\ref{higgsinobr} can be understood
in terms of the $\widetilde \chi_1^0$ quantum numbers and
couplings \cite{hlong}.
For small $\tan \beta$ a Higgsino-like $\NI$ is predominantly
an $SU(2)_L$ triplet(singlet) for ${\rm sgn}(\mu)=+(-)$, and couples
to the Goldstino predominantly through the $Z$ (Higgs) boson.
For larger $\tan \beta$ a Higgsino-like $\NI$ is equal mixtures
of $SU(2)_L$ triplet and singlet and couples through the
Goldstino with equal
strength to the $Z$ and Higgs bosons.

Because of the large range of possibilities for combinations of
$\NI$ NLSP branching ratios illustrated in Fig.~\ref{higgsinobr},
and the many associated possible final states listed in
Table \ref{higgsinosignatures}, two Higgsino-like Neutralino
NLSP Model Lines are defined for the Run II workshop.
The first Model Line is defined such that the branching ratios
in each mode $\NI \to (h,Z,\gamma) \GG$ are roughly of the same
order for a $\NI$ mass in the range 150-200 GeV.
This is realized for the
fixed MGM parameters
\beq
{\rm Higgsino-like~Neutralino~NLSP~Model~Line~I:}~~~~~~
\Nmess = 2,\>\> {\Mmess \over \Lambda} = 3,\>\> \tan\beta = 3, \>\>
\mu = -{3 \over 4} ~M_1
\label{hIlineparameters}
\eeq
with the overall superpartner mass scale defined by
$\Lambda$ varying.
The ratio $\mu / M_1 = -3/4$ fixes the neutralino
Higgsino-gaugino mixing angle
$\tan^{-1}(\mu/M_1) \simeq -0.64$
along Model Line I.
This is a modification of the usual relation implied
by electroweak symmetry breaking with minimal MGM boundary
conditions for the Higgs soft masses at the messenger scale.
As described in Appendix A2i,
modification of the Higgs soft masses may in fact arise
from additional interactions between the Higgs and
messenger sectors which are required in any realistic model.
Because all the $\NI \to (\gamma, h , Z) \GG $
branching ratios are of the same order, the
Higgsino-like Neutralino NLSP Model Line I is useful
for study of the di-boson final states
$hhX \met$, $\gamma h X \met$, $\gamma Z X \met$,
and $hZ X \met$.

The second Model Line is defined such that the
$\NI \to Z \GG$ mode dominates.
This occurs for a Higgsino-like $\NI$ with low $\tan \beta$
and $\mu >0$.
The MGM Model Line fixed parameters are
\beq
{\rm Higgsino-like~Neutralino~NLSP~Model~Line~II:}~~~~~~
\Nmess = 2,\>\> {\Mmess \over \Lambda} = 3,\>\> \tan\beta = 3, \>\>
\mu = {1 \over 3} ~M_1
\label{hIIlineparameters}
\eeq
with the overall superpartner mass scale defined by
$\Lambda$ varying.
The ratio $\mu / M_1 = 1/3$ fixes the neutralino
Higgsino-gaugino mixing angle
$\tan^{-1}(\mu/M_1) \simeq 0.32$
along Model Line II, and again is
a modification of the usual MGM relation.
Because of the dominant $\NI \to Z \GG$ decay mode, the
Higgsino-like Neutralino NLSP Model Line II is most useful
for study of the many interesting signatures
associated with the $Z Z X \met$ and $\gamma Z X \met$
final states.

\begin{figure}[tpb]
\centering
\epsfxsize=4.5in
\hspace*{0in}
\epsffile{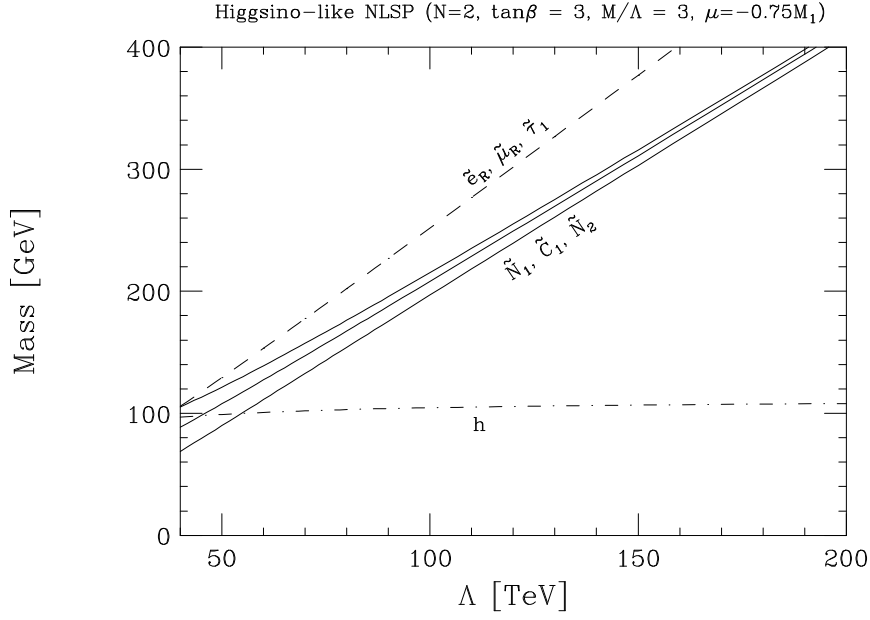}
\caption{Masses in the lightest neutralinos, chargino, sleptons, and
CP-even Higgs boson as a function of the overall scale $\Lambda$ along
the Higgsino-like Neutralino NLSP Model Line I.}
\label{fig:higgsinoline_mass}
\end{figure}

For both the Higgsino-like Neutralino NLSP Model Line I and Model Line II,
the two lightest neutralinos, $\NI$ and $\NII$, and the lightest chargino,
$\CI$, are all Higgsino-like and approximately degenerate. The masses of
the lightest neutralinos, chargino, right-handed sleptons, and lightest
CP-even Higgs boson as a function of the overall scale $\Lambda$ along
Model Line I are shown in Fig.~\ref{fig:higgsinoline_mass}.
The superpartner mass spectrum along Model Line II is similar. Even though
the right-handed sleptons are not much heavier than the Higgsino-like
neutralinos and chargino, the $\slepton_R^+ \slepton_R^-$ $s$-channel
production cross section through $\gamma^*$ and $Z^*$ is relatively small
because of $P$-wave suppression and pure $U(1)_Y$ hypercharge coupling. In
contrast, the $s$-channel production cross sections for $\widetilde
\chi_1^+ \widetilde \chi_1^-$, $\CI \widetilde \chi_i^0$, and $\widetilde
\chi_i^0 \widetilde \chi_j^0$ for $i,j=1,2$ through $\gamma^*, Z^*$ and
$W^*$ are $S$-wave through $SU(2)_L$ couplings. The total SUSY production
cross section turns out to be almost entirely from these Higgsino-like
states on both Model Line I and Model Line II. So the Higgsino-like
Neutralino Model Lines are effectively determined only by the $\mu$
parameter or equivalently the $\NI$ or $\CI$ mass, the ratio $\mu/M_1$ or
equivalently the neutralino mixing angle, and the Higgs and $Z$ boson
masses. Note in Fig.~\ref{fig:higgsinoline_mass} that the splitting
between the Higgsino-like states is small compared to the overall mass
scale. So the additional partons $X$ from cascade decays $\NII \to X \NI$
and $\CI \to X \NI$ may not be particularly useful at the trigger level,
but might be useful in disentangling the general magnitude of the mass
splittings if a signal in one of the di-boson channels were established.

\begin{figure}[tpb]
\centering
\epsfysize=3.1in
\hspace*{0in}
\epsfbox{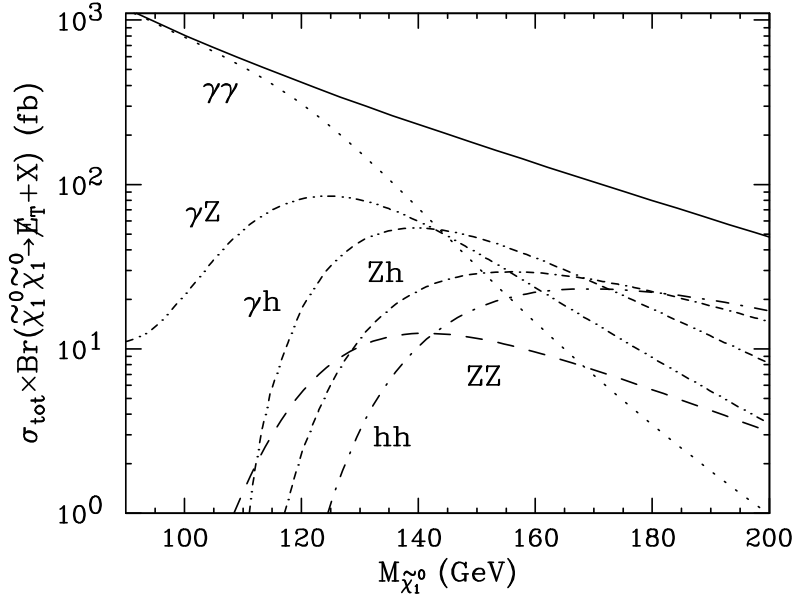}
\caption{Total SUSY production cross section
times branching ratios into various di-boson final states
in $p \bar{p}$ collisions
with $\sqrt{s}=2$ TeV, as a function of the $\NI$ mass along the
Higgsino-like Neutralino NLSP Model Line I with a fixed
ratio $\mu/M_1 = -3/4$.
The solid line indicates the total SUSY production cross section.}
\label{fig:HIcross}
\end{figure}

In addition to the Higgsino-gaugino content discussed above, the $\NI \to
(h,Z, \gamma) \GG$ branching ratios also depend on the $\NI$ mass through
the phase space available to the $h$ and $Z$ modes which suffer a
$\beta^4$ velocity suppression near threshold \cite{gwidth,mgm,baggermgm}.
So even a Higgsino-like $\widetilde \chi_1^0$ decays predominantly by
$\widetilde \chi_1^0 \rightarrow \gamma \GG$ for masses not too far above
the $h$ and $Z$ masses. The mass dependence of the branching ratios is
illustrated for Model Line I in Fig.~\ref{fig:HIcross} in which the total
SUSY cross section times branching ratio into the di-boson final states is
given as a function of the $\NI$ mass.
The $hh$ and $hZ$ di-boson modes
dominate for very large masses, while the $\gamma \gamma$ mode
dominates for smaller masses.
However, because of the strong phase space suppression near threshold
there is a transition region which extends over a significant
range of mass between these limits
in which the mixed di-boson mode $\gamma h$ is important.
This mode is particularly useful for masses in the transition
region since the photon is quite hard.
The existence of a transition region in mass in which the
$\gamma h$ mixed di-boson mode
is significant
is generic for Higgsino-like neutralino
NLSP with mostly $SU(2)_L$ singlet quantum numbers
(low to moderate $\tan \beta$ with $\mu <0$ or large $\tan \beta$
with either sign of $\mu$).
The major model dependence is the extent  of this transition
region in $\NI$ mass.

\begin{figure}[tpb]
\centering
\epsfysize=3.1in
\hspace*{0in}
\epsfbox{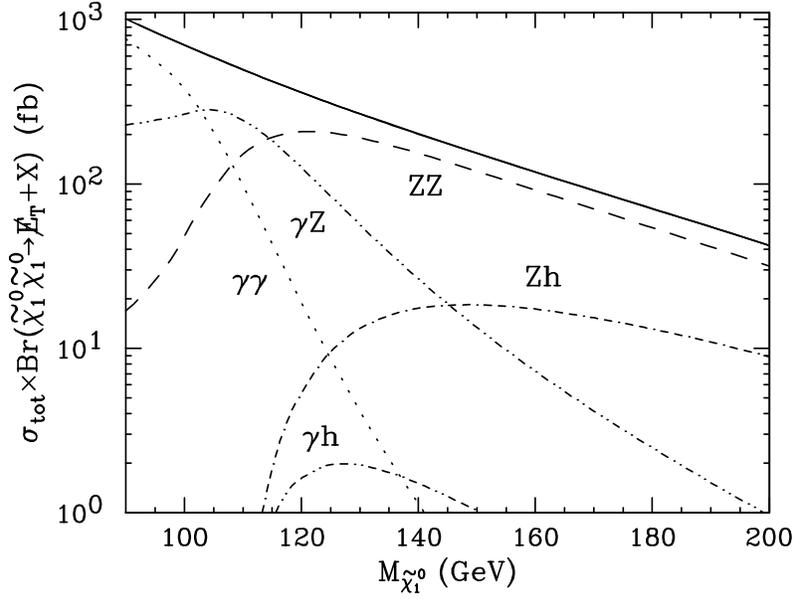}
\caption{Total SUSY production cross section
times branching ratios into various di-boson final states
in $p \bar{p}$ collisions
with $\sqrt{s}=2$ TeV, as a function of the $\NI$ mass along the
Higgsino-like Neutralino NLSP Model Line II with a fixed
ratio $\mu/M_1 = 1/3$.
The solid line indicates the total SUSY production cross section.}
\label{fig:HIIcross}
\end{figure}

The mass dependence of the $\NI$ branching ratios is illustrated
for Model Line II
in Fig.~\ref{fig:HIIcross}
in which the total SUSY cross section times
branching ratio into the di-boson final states
is given as a function of the $\NI$ mass.
The $ZZ$ mode dominates for moderate to large mass, while the
$\gamma \gamma$ mode dominates for very small mass.
The $\gamma Z$ mixed di-boson mode is important in a transition
region of mass between these limits.
The existence of this transition region in mass is generic
for a Higgsino-like neutralino NLSP with mostly $SU(2)_L$  triplet
quantum numbers (low to moderate $\tan \beta$ with $\mu >0$), but
the extent of the transition region is model dependent.
Because of the dominance of the neutralino coupling to the $Z$ boson
through the Goldstino on Model Line II,
the $\gamma Z$ mixed mode is important in a relatively narrow transition
region of masses.

For comparisons and detailed study of the di-boson final
states which contain Higgs bosons, a reference Model Point
is defined on Model Line I by $\Lambda = 80$ TeV
corresponding to $m_{\tilde \chi_1^0} = 154$ GeV, $m_{\tilde \chi_1^\pm} =
167$ GeV,
and $m_h = 104$ GeV.


\subsection{Prompt Decays to Higgs Bosons, $Z$ Bosons, and Photons}

Prompt decays $\NI \to (h,Z,\gamma)\GG$ in various combinations
for each $\NI$ yield the large number
of signatures listed in Table \ref{higgsinosignatures}, 
all with $\met$. 
Channels with tagged $b$-jets from Higgs decay, with a photon, 
or with a reconstructed leptonic $Z$ boson are particularly
useful, although hadronic channels may also be observable. 

\subsubsection{CDF study of Higgsino-like Neutralino NLSP}
\label{subsubcdfhiggsino}

\begin{figure}[tpb]
\centering
\epsfysize=3.5in
\hspace*{0in}
\epsfbox{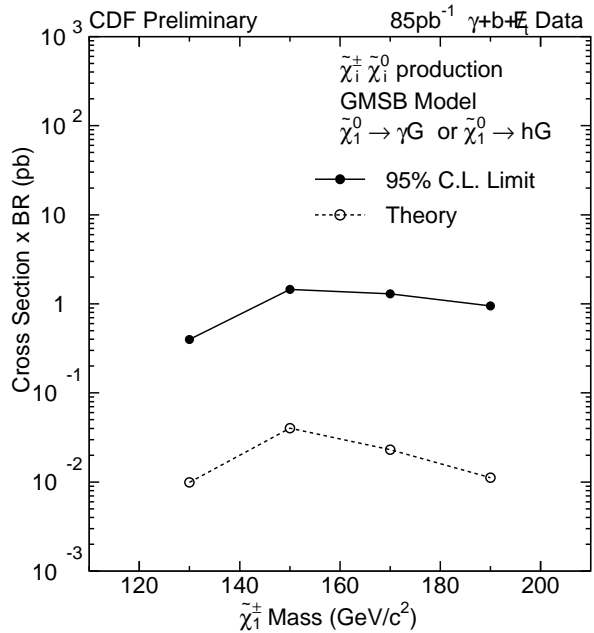}
\caption{CDF Run I limits with 85~pb$^{-1}$ of integrated luminosity
on the total SUSY cross section times branching ratio
in the $\gamma b X \met$ channel along the Higgsino-like Neutralino
Model Line I as a function of chargino mass.}
\label{fig:gmsbcdfgtmlim}
\end{figure}

CDF has investigated the Higgsino-like Neutralino Model Line I
in both the $\gamma b X \met$ and $b \bar{b} X \met$ channels
arising from the $\gamma h$ and $hh$ di-boson modes.
The $\gamma b X \met$ channel
has been studied in Run I and the same data has been used to
set limits on several models, including Model Line I
\cite{gmsbcdfrlc}.
The data is based on a single isolated, central photon trigger with a
threshold of 23~GeV.  Off-line, an isolated, central ($|\eta|<1$)
photon is required with $E_T>25$~GeV.
A standard SVX $b$--tag of a jet is required
with corrected $E_T>30$~GeV and $|\eta|<2$ and the event must have 40~GeV
of $\missET$.  Two events survive all
cuts while no reliable background prediction can be calculated.
More than 7 events of anomalous production are excluded --
Table \ref{tab:gmsbcdfgtm} and Fig.~\ref{fig:gmsbcdfgtmlim}
show the resulting limits.
\begin{table}[htbp]
\caption{Efficiencies and limits
in the $\gamma b X \met$ channel for the Higgsino-like
Neutralino NLSP Model Line I from
CDF in Run I with 85~pb$^{-1}$ of integrated luminosity.
Efficiencies do not include branching ratios.}
\label{tab:gmsbcdfgtm}
\renewcommand{\arraystretch}{1.75}
\begin{tabular}{|c|c|c|c|c|}  \hline
$\Lambda$ (TeV)              & 60.9 &70.3&81.0&89.5 \\ 
$m_{\tilde \chi_1^\pm}$ (GeV)              & 130 &147&170&186 \\ 
$m_{\tilde \chi_1^0}$ (GeV)              & 113 &132&156&174 \\ \hline
$A\cdot \epsilon (\%)$       & 20.9&5.7&6.4&8.7 \\ \hline
BR (\%)                      & 3&20&23&18 \\ \hline
$\sigma_{th}\times BR$ (fb)       & 10.0&40.2&23.0&11.0 \\ \hline
$\sigma_{95\%~lim}\times BR$ (fb) & 400&1450&1300&950 \\ \hline
\end{tabular}
\end{table}

Projected limits in the $\gamma b X \met$ channel may also be obtained for
Run II. To estimate the background, the Run I histograms of photon $E_T$
and $\missET$ are scaled up in energy by 5\% and the overall normalization
is increased by a factor of 1.2 to correct for the increase to 2~TeV and
for the 2~fb$^{-1}$ of integrated luminosity. Comparing the signal and
background photon $E_T$ and $\missET$ distributions it is found that the
$\missET$ distribution shows the greater discrimination power. So the
$\missET$ cut is increased to 60~GeV where the one event background level
can be maintained.

\begin{figure}[tpb]
\centering
\epsfysize=3.4in
\hspace*{0in}
\epsfbox{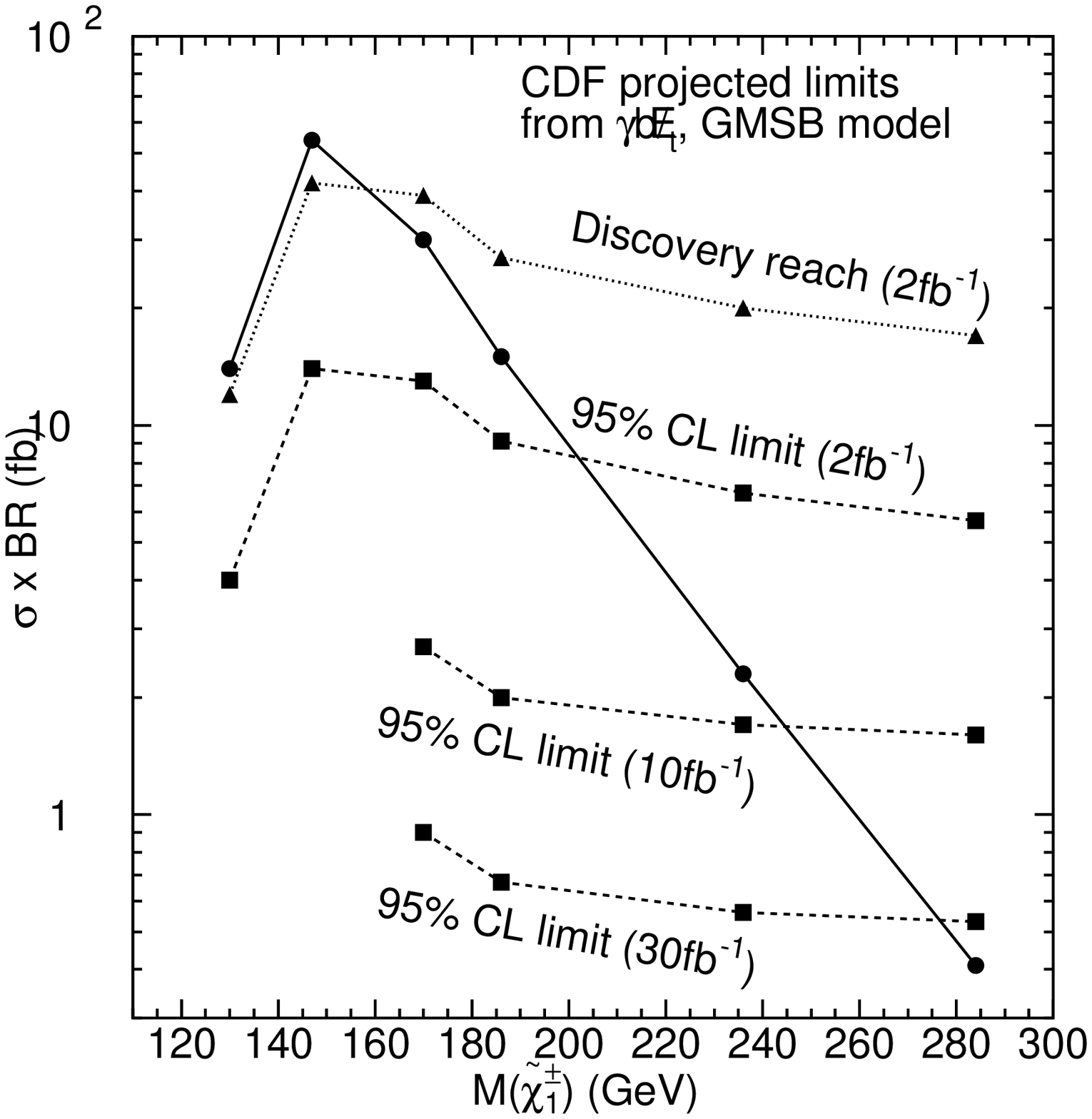}
\caption{CDF projected Run II limits
on the total SUSY cross section times branching ratio
in the $\gamma b X \met$ channel along the Higgsino-like Neutralino
Model Line I as a function of chargino mass. The solid line is the
theoretical expectation from the Higgsino-like Neutralino NLSP Model Line
I.} \label{fig:gmsbcdfgtmlimii}
\end{figure}

For the efficiency in Run II, the full efficiency of the Run I analysis is
used as a starting point. The b--tagging coverage improvement in Run II is
assumed to contribute a factor of 1.6 to the efficiency due to the
increase in the length of the SVX detector.  The improved photon coverage
contributes another factor of 1.6 due to improvements in the trigger
isolation and fiducial efficiency, and increasing the coverage from
$|\eta|<1$ to $|\eta |<2$. The projected Run II limits are presented in
Table \ref{tab:gmsbcdfgtmmodel} and Figure \ref{fig:gmsbcdfgtmlimii}. This
mixed di-boson signature requires that one neutralino NLSP decays by $\NI
\to \gamma \GG$ while the other decays by $\NI \to h \GG$. This mixed mode
is only significant in the transition region of $\NI$ mass, as discussed
above, and is less useful for very large masses. The CDF Run II exclusion
reach will be up to approximately 180~GeV, while a $5\sigma$ discovery may
or may not be possible. For larger luminosities, the signal and background
may be scaled without optimizing the analysis cuts again.
As shown in Figure \ref{fig:gmsbcdfgtmlimii} a sensitivity up to a $\CI$
mass of 250~GeV for 10~fb$^{-1}$ and  280~GeV for 30~fb$^{-1}$
are expected.
\begin{table}[htbp]
\caption{Summary of the Monte Carlo points used to investigate the
projected limits in the $\gamma b X \met$ channel for the Higgsino-like
Neutralino NLSP Model Line I for CDF in Run II with 2~fb$^{-1}$ of
integrated luminosity. Efficiencies do not include branching ratios.}
\label{tab:gmsbcdfgtmmodel}
\renewcommand{\arraystretch}{1.7}
\begin{tabular}{|c|c|c|c|c|c|c|} \hline
$\Lambda$ (TeV)              & 60.9 &70.3&81.0&89.5 & 110 & 134\\ 
$m_{\tilde \chi_1^\pm}$ (GeV)              & 130 &147&170&186 & 236 &
284\\ 
$m_{\tilde \chi_1^0}$ (GeV)              & 113 &132&156&174 & 226 & 275\\
\hline
BR(\%)                   & 3    &  20  & 23   &  18 &   11&   6\\ \hline
$\sigma\times$BR (fb)    & 14.0 & 54.0 & 30.0 & 15.0 & 2.3 &0.4 \\ \hline
$A\cdot \epsilon$ (\%)   &  50  &  14  &  16  &  22  & 30 & 35 \\ \hline
$\sigma\times$BR 95\% C.L. limit (fb)& 4.0& 14.0& 13.0& 9.1& 6.7 & 5.7 
\\  \hline
\end{tabular}
\end{table}

The Higgsino-like Neutralino NLSP Model Line I can also have be probed
with a $b \bar{b} X \met$ selection.
An analysis of this channel may be
borrowed from a SUGRA analysis which uses this channel
to search for sbottom squark pair production,
$\tilde b \tilde b$, with each sbottom decaying by
$\tilde b \to b \NI$ and with the $\NI$ escaping the detector
as missing energy \cite{SUGRAreport}.
In that analysis, several options
for the final cuts are offered which may be
evaluated for Model Line I.
The requirements placed on the events
are: two jets with $E_T>20$ and 30~GeV, one jet with $|\eta|<1$,
$\missET>50$~GeV, the jets must not be correlated with the $\missET$,
the jets must not be back--to--back,
and one jet must have a standard SVX $b$--tag.  These are the same as the
basic SUGRA analysis cuts with the exception of the requirement of no
isolated leptons in the event (to reject $W$'s). At this point the
efficiency of additional cuts can be checked and compared to the
background prediction, as shown in Table \ref{tab:cdfgmsbbbmopt}. The
no--leptons requirement is not a large loss in efficiency, requiring 2 or
3 jets is a clear loss, and either the sample with one tag or with two
tags may be used. For the present analysis the sample with two tags will
be employed.
\begin{table}[htbp]
\caption{Summary of the CDF optimization for the
Monte Carlo cuts for the $b\bar{b}X \missET$ channel for the
Higgsino-like
Neutralino NLSP Model Line I at the point with $m_{\tilde
\chi_1^\pm}=187$~GeV.
The efficiency for ``w/o lepton removal" is obtained without
requiring the absence of central or plug isolated
leptons (no corresponding background estimate was given in
\protect\cite{SUGRAreport}).}
\label{tab:cdfgmsbbbmopt}
\renewcommand{\arraystretch}{1.7}
\begin{tabular}{|c|c|c|c|c|c|c|} \hline
Cuts    & $\epsilon (\%)$  & background (fb) &
Signal$/\sqrt{{\rm background}}$
\\ \hline
basic, w/o lepton removal                           & 59  & -   & - \\
\hline
basic                   & 48  & 356 & 2.5 \\ \hline
basic, $n_{jet}\leq 3$ & 2.6 & 172 & 0.2 \\ \hline
basic, 2 $b$--tags     & 18  &  50 & 2.5 \\ \hline
\end{tabular}
\end{table}

\begin{figure}[tpb]
\centering
\epsfysize=3.4in
\hspace*{0in}
\epsfbox{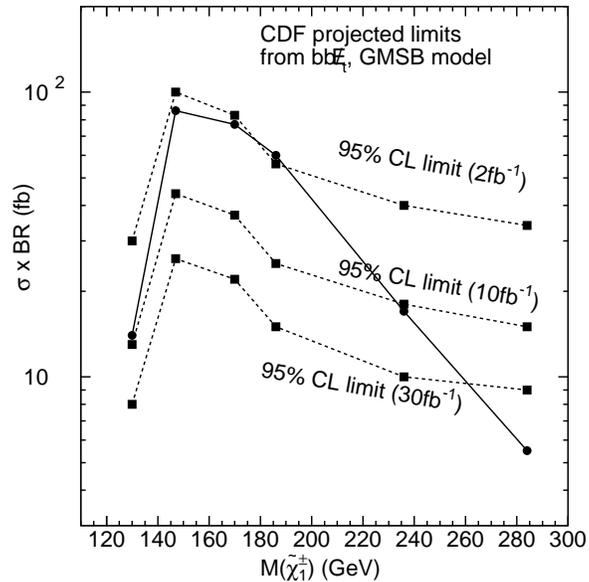}
\caption{CDF Projected Run II limits
on the total SUSY cross section times branching ratio
in the $bb X \met$ channel along the Higgsino-like Neutralino
Model Line I as a function of chargino mass. The solid line is the
theoretical expectation.}
\label{fig:cdfgmsbbbmlim}
\end{figure}

In Run II with 2~fb$^{-1}$ of integrated luminosity, assuming statistical
uncertainties dominate, a limit of approximately 20 events is expected or
a discovery sensitivity to 50 events or more. These limits translate to
cross section limits as indicated in Table~\ref{tab:cdfgmsbbbmlim} and
Fig.~\ref{fig:cdfgmsbbbmlim}.  For larger luminosities, the signal and
backgrounds are scaled with the results shown in
Fig.~\ref{fig:cdfgmsbbbmlim}.
\begin{table}[htbp]
\caption{Results for CDF projected Run II
limits on the Higgsino-like Neutralino NLSP Model Line I
in the $b \bar{b} X \missET$ channel with 2~fb$^{-1}$ of integrated
luminosity. Cross sections include branching ratios.}
\label{tab:cdfgmsbbbmlim}
\renewcommand{\arraystretch}{1.75}
\begin{tabular}{|c|c|c|c|c|c|c|} \hline
$\Lambda$ (TeV)              & 60.9 &70.3&81.0&89.5 & 110 & 134\\ 
$m_{\tilde \chi_1^\pm}$ (GeV)              & 130 &147&170&186 & 236 &
284\\ 
$m_{\tilde \chi_1^0}$ (GeV)              & 113 &132&156&174 & 226 & 275\\
\hline
BR(\%)                   & 3   &  32  & 58   & 70   &   81&  84\\ \hline
$\sigma\times$BR (fb)    & 14  & 86   & 77   & 60   & 17  &  6 \\ \hline
$A\cdot \epsilon$ (\%)   &  33 &  10  &  12  & 18   & 25  &  29 \\ \hline
$\sigma\times$BR 95\% C.L. limit (fb)
                         & 30  & 100  & 83   & 56   & 40  & 34 \\ \hline
\end{tabular}
\end{table}

Although it is clear from this estimate
that there is not a great deal of sensitivity,
this channel has a large potential for
more optimization for the parameters of Model Line I.
For example, in a significant fraction of events,
one of the NLSP neutralinos decays by
$\NI \to Z \GG$ in which the $Z$ boson could be
found as leptons or a second dijet peak.
In addition, there are jets and leptons in the cascade decays
and the $\missET$ and jet cuts are not optimized.

\subsubsection{\rm {\D0} study of Higgsino-like Neutralino NLSP}
\label{subsubd0higgsino}

The \D0 collaboration
has investigated the Higgsino-like Neutralino Model Line I
in the $\gamma b j \met$ channel arising from the $\gamma h$
di-boson mode.
In Run I a similar \gjjmet\ 
search~\cite{gjjmet} for single-photon
events with at least two jets and large $\met$ was carried out.
The \gjjmet\ 
events were selected by requiring at least one identified photon
with $E_T^\gamma>20$~GeV and within pseudorapidity ranges $|\eta^\gamma|<1.1$
or $1.5<|\eta^\gamma|<2.0$, two or more jets having $E^j_T>20$~GeV and
$|\eta^j|<2.0$, and $\rlap{\kern0.25em/}E_T>25$~GeV. A total of 318 events
were selected from a data sample corresponding to an integrated luminosity of
$99.4\pm 5.4$~pb$^{-1}$.

\begin{figure}[tpb]
  \centerline{\epsfysize=3.5in\epsfbox{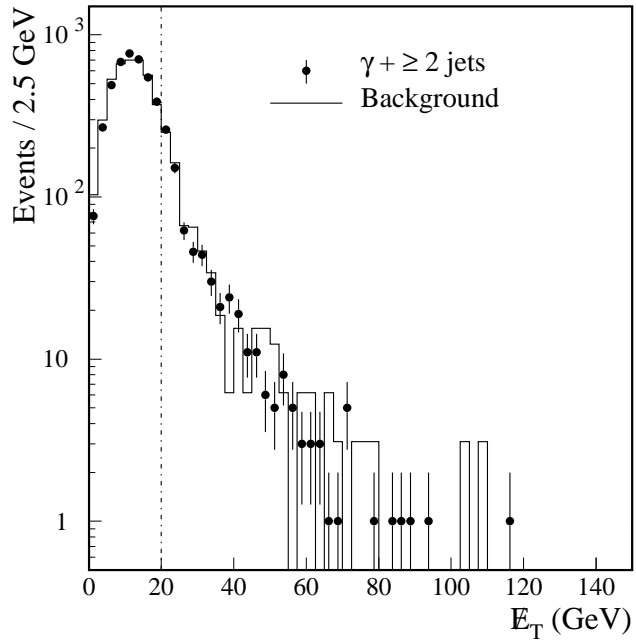}}
  \caption{The \D0 Run I $\met$ distributions of $\gamma jj$ and background
events. The number of events in the background is normalized to the
$\gamma jj$ sample for $\rlap{\kern0.25em/}E_T < 20$~GeV, the 
region left of the dot-dashed line.}
  \label{fig:run1_gjjmet_met}
\end{figure}

The principal backgrounds were found to be QCD direct photon and multijet
events, where there was mismeasured $\met$ and a real or fake photon. The
number of events from this source was estimated to be $315\pm 30$. Other
backgrounds such as those from $W$ with electrons misidentified as photons
were found to be small, contributing $5\pm 1$ events. This led to an
observed background cross section of 3200~fb from the $\met$
mismeasurement and of 50~fb from the fakes. The $\met$ distribution before
the $\met >25$~GeV cut is shown in Fig.~\ref{fig:run1_gjjmet_met}. The
backgrounds can be significantly reduced by raising the requirement on
$\met$, as is apparent in Fig.~\ref{fig:run1_gjjmet_met}.

Events with a high $E_T$ photon, $b$-jets and large $\met$ are expected in
several new physics models, including the Higgsino-like
Neutralino
NLSP Model Line I. These events, referred to as \gbjmet, are in many ways
similar to the \gjjmet\ events and thereby can be selected similarly:
\begin{itemize}
  \item[1)] At least one photon with $E^\gamma_T >20$~GeV;
  \item[2)] At least two jets with $E^j_T>20$~GeV;
  \item[3)] At least one jet is tagged as a b-quark jet with $E^b_T>20$~GeV;
  \item[4)] $\met >50$~GeV;
  \item[5)] No leptons with $E^\ell_T>20$~GeV.
\end{itemize}
The backgrounds from the QCD multijet events with real or misidentified
photons and from the W events with electrons faking photons are estimated to
be 0.63~fb, assuming background reduction factors of 5 from the raised $\met$
requirement, 2 from the improved photon identification and using the assumed
value of ${\cal P}(j\to b)=10^{-3}$.
The dominant background sources are expected to be $\gamma b\bar{b}$ and
$\gamma t\bar{t}$ events. These background sources cannot be reduced by
the tagging of $b$-jets. However, the $\gamma b\bar{b}$ contribution is
expected to be small due to the large $\met$ requirement. Monte Carlo studies
show that it is negligible. The $\gamma t\bar{t}$
with $t\bar{t}\to W^+W^-b\bar{b}$ contribution is reduced by the requirements
4) and 5) and is estimated using the cross section of Ref.~\cite{gtt}.
A total of 0.9~fb observable background cross section is assumed.

\begin{figure}[tpb]
  \centerline{\epsfysize=3.0in\epsfbox{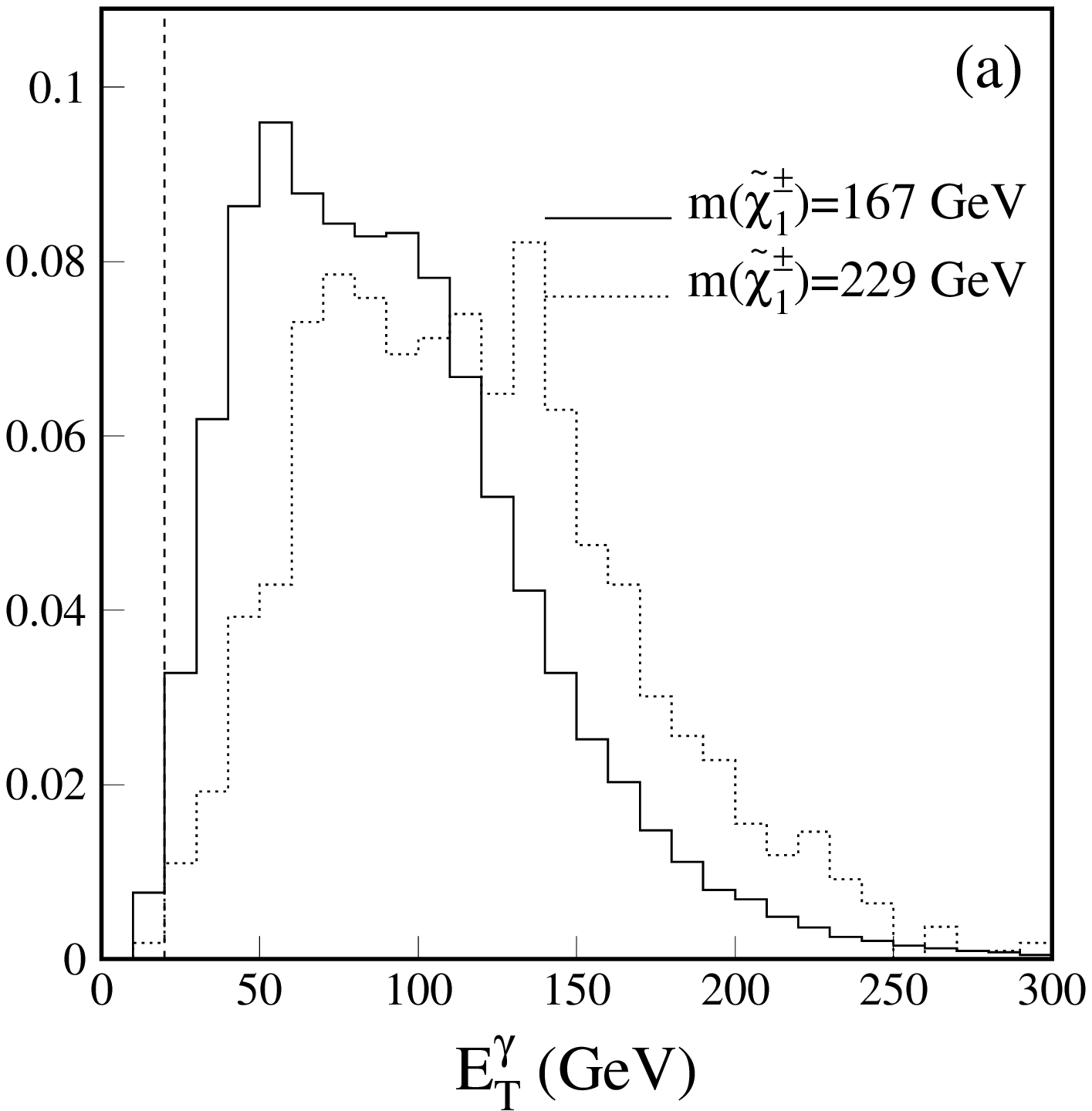}
              \epsfysize=3.0in\epsfbox{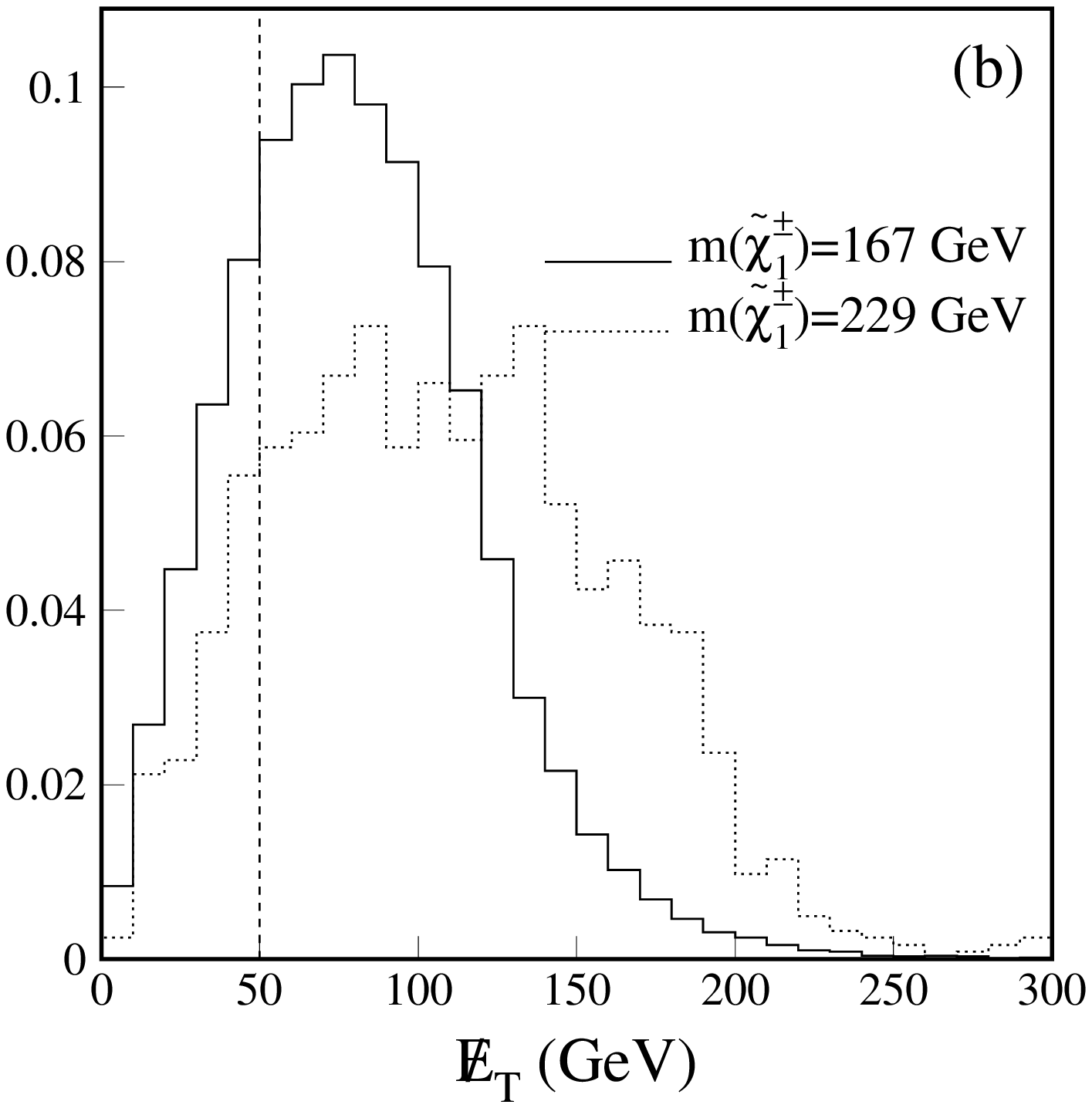}}
  \caption{The SUSY photon $E_T$ (a) and event $\met$ (b) distributions
for the Higgsino-like Neutralino NLSP Model Line I with $\Lambda=80,\
110$~TeV corresponding to $m_{\tilde\chi^\pm_1}=167,\ 229$~GeV, in the
{\D0} study. All distributions are normalized to have unit area.}
  \label{fig:p5}
\end{figure}

\begin{figure}[tpb]
  \centerline{\epsfysize=3.5in\epsfbox{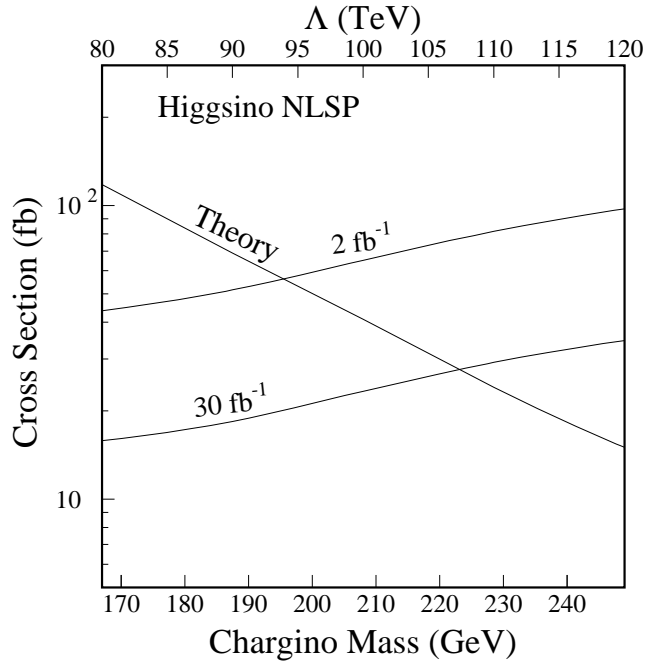}}
  \caption{The $5\sigma$ discovery cross section curves as functions of
           $\Lambda$ and the $\Cone$
           mass along the Higgsino-like Neutralino NLSP Model Line I
in the \D0\  study     along with the theoretical cross section.}
  \label{fig:p5lim}
\end{figure}

The SUSY events for Model Line I in the $\gamma b j \met$
channel are characterized by high $E_T$ photons and large $\met$ as
shown in the Fig.~\ref{fig:p5} for $\Lambda=80, 110$~TeV, corresponding
to $m_{\Cone}=167,229$ GeV and
$m_{\tilde \chi_1^0}=154,218$ GeV respectively.
These events can be selected using the \gbjmet\ selection criteria discussed
above. The detection efficiencies and $N_s/\delta N_b$
significances are shown in Table~\ref{tab:p5} for different values of
$\Lambda$ along Model Line I.
\begin{table}[htbp]
\caption{The $\tilde\chi^\pm_1$, 
$\tilde\chi^0_1$ and $h$
masses, theoretical SUSY cross section,
\protect\nlsp\   decay branching ratios, detection efficiency of
the \protect\gbjmet\   selection, and significances for different values
of $\Lambda$ along the Higgsino-like Neutralino Model Line I
in the \protect\D0\   study.
The relative statistical error on the efficiency is typically
4\%. The background cross section is assumed to be 0.9~fb with
a 20\% systematic error.}
\label{tab:p5}
\renewcommand{\arraystretch}{1.75}
  \begin{tabular}{|c|ccccc|}\hline
     $\Lambda$ (TeV)              &  80 &  90 & 100 & 110 & 120 \\ 
     $m_{\tilde\chi^\pm_1}$ (GeV) & 167 & 188 & 208 & 229 & 249 \\
     $m_{\tilde\chi^0_1}$ (GeV)   & 154 & 176 & 197 & 218 & 239 \\
     $m_h$ (GeV)                  & 103 & 104 & 105 & 105 & 106 \\ \hline
     $\sigma_{th}$ (fb)           & 118 &  69 &  41 & 24  & 15 \\ \hline
     ${\rm Br}(\tilde\chi^0_1\to\gamma\tilde G)$ & 0.38 & 0.21 & 0.15 & 0.10
  & 0.08 \\
     ${\rm Br}(\tilde\chi^0_1\to h\tilde G)$  & 0.38 & 0.52 & 0.59 & 0.63
& 0.66 \\ \hline
     $\epsilon$ (\%)               &  8.0 & 6.8 & 5.4 & 4.3 & 3.6 \\
     \protect\rsb\ (2 fb$^{-1}$)   &  13  & 6.7 & 3.2 & 1.5 & 0.8 \\
     \protect\rsb\ (30 fb$^{-1}$)  &  38  &  19 & 8.9 & 4.1 & 2.2 \\
  \end{tabular}
\end{table}
Most of the events selected are due to the $\gamma h$ di-boson
mode with $h\to b\bar{b}$. However, a non-negligible fraction of the
events are actually due to the $\gamma Z$ di-boson
mode with $Z\to b\bar{b}$.
The Run II discovery reach
in $\Lambda$ and $m_{\tilde\chi^\pm_1}$ along Model Line I
is shown in Fig.~\ref{fig:p5lim}
for integrated luminosities of \ldt=2,30~fb$^{-1}$.

\subsubsection{ISAJET studies of Higgsino-like Neutralino NLSP}
\label{subsubisahiggsino}

The $hh$ di-boson mode along the Higgsino-like Neutralino NLSP
Model Line I with $h \to b\bar{b}$ leads to the $bbbbX \met$ channel.
Signals in this channel with multiple tagged
$b$-jet plus $\met$ events, and perhaps other jets, leptons,
and possibly photons in the case that one of the neutralinos
decays as $\NI \to \gamma \GG$, have been simulated using
ISAJET \cite{isajet}.

The dominant Standard Model background to multi-$b$ events presumably comes from
$t\bar{t}$ production and is shown in Table \ref{bbbbtable}, where
the signal cross section for Model Line I with $\Lambda=100$~TeV, corresponding
to $m_{\Cone}= 208$ GeV and $m_{\tilde \chi_1^0}=197$ GeV, is also shown.
For events
with one or two tagged $b$-jets, the $t\bar{t}$ backgrounds come when
the $b$-jets from $t$ decay are tagged; {\it i.e.} the rate for events
where
other jets are mis-tagged as $b$-jets is just a few percent. This is also
true for signal events. On the other hand, in the $bbbX\met$ channel at least
one of the tagged $b$-jets in the $t\bar{t}$ background has to
come from a $c$ or light quark or gluon jet that is misidentified as a
$b$-jet, or from an additional $b$ produced by QCD radiation. This is
not, however, the case for signal events which contain up to four
$b$-jets.
In each of the last two columns of Table \ref{bbbbtable} where
the top background and the SUSY signal are shown, two
numbers are presented: the first of these is the cross section when all the tagged
jets come from real $b$'s, while the second number in parenthesis is the
cross section including $c$ and light quark or gluon jets that are
mistagged as $b$, with the
assumed probabilities
${\cal P}( c \rightarrow b) = 5$\% and
${\cal P}( q,g \rightarrow b) = 0.2$\%.
Indeed it can be see that the bulk of the $bbbX\met$ background is
reducible and comes from mistagging jets, whereas the signal is
essentially all from real $b$-jets.
\begin{table}[htbp]
\caption[]{The background cross section in fb for multiple tagged $b$-jets
plus lepton plus $\eslt$ events from $t\bar{t}$ production after basic
cuts and triggers in the ISAJET study. Also shown are the corresponding
SUSY signal cross sections for the Higgsino-like Neutralino NLSP Model
Line I with $\Lambda =100$~TeV corresponding to $m_{\Cone}= 208$ GeV and
$m_{\tilde \chi_1^0}=197$ GeV. The numbers in parenthesis for the
$bbbX\met$ channel
include events from charm or light quark jets faking a $b$-jet. Whereas
this fake rate dominates the background in the $3b$ channel, it is
negligible in the $bX\met$ and $bbX\met$ channels.} \label{bbbbtable}
\renewcommand{\arraystretch}{1.75}
\begin{tabular}{|c|cc|cc|cc|}
\tableline
 & \multicolumn{2}{c |} {$bX\met$ } &\multicolumn{2}{c |}{$bbX\met$ } &
\multicolumn{2}{c |} {$bbbX\met$ } \\
 &  $t\bar{t}$ & 
SUSY & $t\bar{t}$ &  
SUSY & $t\bar{t}$ &  
SUSY \\
\hline
~~~$0\ell$~~~ &  508 & 11.6 & 221 & 7.5 & 1.2 (8.1) & 2.6 (2.7)\\   \hline
~~~$1\ell$~~~ & 812 & 1.2 & 345 & 0.57 & 1.5 (9.3) & 0.11 (0.11)  \\
\hline
~~~$2\ell$~~~ & 132 & 0.49 & 56 & 0.31 & 0.13 (0.31) & 0.04 (0.05) \\
\hline
~~~$3\ell$~~~ & 0.22 & 0.03 & 0 & 0.04 & 0 (0) & 0 (0) \\ \hline
\end{tabular}
\end{table}

It is clear from Table \ref{bbbbtable} that the best 
signal to background ratio
is obtained in the $bbbX\met$ channel
for events with $\geq 3b$-jets. Our detailed analysis shows
that although the signal cross section is rather small,
$bbbX\met$ channel with a lepton veto (since top events with large $\met$
typically contain leptons) offers the best hope for identifying the
signal above Standard Model backgrounds. The
signal is seen to be
of similar magnitude as the background for $\Lambda =100$~TeV
on Model Line I,
corresponding to $m_{\Cone}= 208$ GeV,
a point beyond reach via the $\gamma\gamma$ channel. To further
enhance the signal relative to the background the additional
cuts are imposed,
\begin{itemize}
\item $\eslt \geq 60$~GeV, and
\item 60~GeV $\leq m_{bb} \leq 140$~GeV for at least one pair of tagged
$b$-jets in the event.
\end{itemize}
The first of these reduces the signal from 2.7 fb to 2.1 fb while
the background is cut by more than half to 3.4 fb. The mass cut was
motivated by the fact that for Model Line I, at least one pair of tagged
$b$'s comes via $h \to bb$ decay, with $m_h \sim 100$~GeV, while the
$b$'s from top decay form a continuum. However, this cut
leads to only a marginal improvement in the statistical significance and
the signal to background ratio.
This may be traced to the fact that, because
of top event kinematics, one $b$-pair is likely to fall in the `Higgs
mass window'. Reducing this window to $100 \pm 20$~GeV leads to a
slightly improved S/B but leads to too much loss of signal to improve
the significance.

\begin{figure}[tpb]
\epsfysize=3.5in
\epsffile[-170 50 180 500]{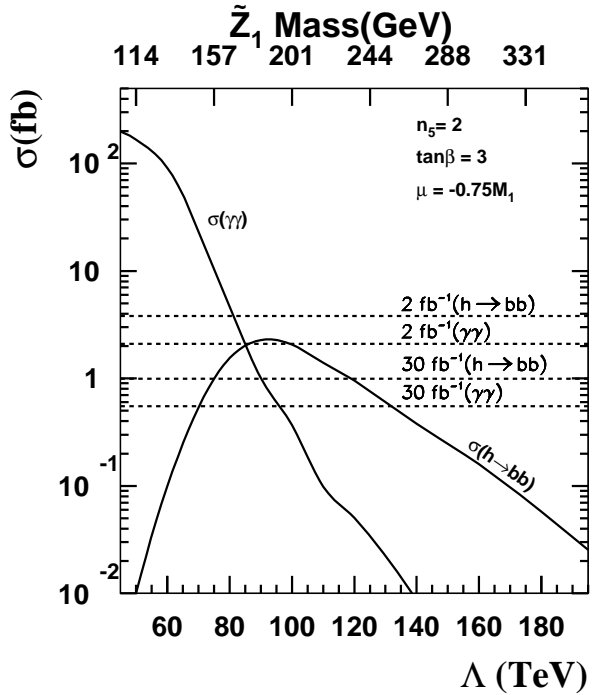}
\begin{center}
\caption[]
{SUSY signal cross sections for the Higgsino-like Neutralino Model Line I
in the $\gamma \gamma X \met$ channel
(labeled $\sigma(\gamma\gamma)$, and $bbbX\met$ channel
after all ISAJET study cuts described in the text.
The dashed horizontal lines denote the
minimum cross section for $3\sigma$ observation in Run II with
an integrated luminosity of 2 fb$^{-1}$ and 30 fb$^{-1}$.}
\label{mlD:reach}
\end{center}
\end{figure}

The signal cross section via the $bbbX\met$ channel after the basic cuts as
well as the additional $\eslt$ and $m_{bb}$ cuts introduced above is
shown by the solid curve labeled $h \to bb$ in Fig.~\ref{mlD:reach}.
For small $\NI$ masses, the signal is small because of the reduction in
the branching ratio $\NI \to h \GG$ apparent in Fig.~\ref{fig:HIcross}.
The corresponding dashed lines show the minimum cross section for a signal
to be observable at the $3\sigma$ level
at Run II with an integrated luminosity of 2 fb$^{-1}$ and 30 fb$^{-1}$.
For Run II with 2 fb$^{-1}$ there could just barely be an
observable signal in the $bbbX\met$ channel for the parameters
of Model Line I.
With 30 fb$^{-1}$ of integrated luminosity, the
signal exceeds the $3\sigma$ level for 75~TeV $\leq \Lambda \leq
120$~TeV, corresponding to a
$\NI$ mass of up to 240 GeV,
and significantly extends the reach obtained along Model Line I with,
for example, the
$\gamma\gamma X \met$ channel.
Furthermore there appears to be no window
between the upper limit of the $\gamma\gamma X \met$ channel and the lower
limit of the $bbbX\met$ channel.  A few points are worth noting.
\begin{enumerate}
\item Since the background dominantly comes from events where a $c$ or
light quark or gluon jet is mis-tagged as a $b$-jet, the reach via the
$bbbX\met$ channel is very sensitive to the assumptions about the $b$ mis-tag
rate. Indeed, if the mis-tag rate is twice as big as assumed here,
there will be no reach in this channel at Run II.

\item The $bbbX\met$ signal starts to become observable for
$m_{\tilde \chi_1^0} \sim 140$ GeV
where the branching ratio for $\NI \to h \GG$
becomes comparable to that for $\NI \to \gamma \GG$.
The $\NI$ mass for which $\NI \to h \GG$ becomes
dominant depends on $m_h$, which in turn is sensitive to
$\tan\beta$.


\end{enumerate}
The $bbbX \met$
channel provides a promising probe of the Higgsino-like Neutralino
Model Line I at Run II, and is rapidly helped by increasing
luminosity and by reducing the background from mis-tagged
charm (light quark or gluon jets to below 5\% (0.2\%).

Despite the fact that the top background alone is 50 to several hundred
times larger than the SUSY signal in all relevant one and two tagged $b$
plus multilepton channels in Table \ref{bbbbtable}, we have examined whether
it is possible to separate the signal from the background.
Attention was focused
on the $2b + 0\ell$ signal which has the best $S/B$ ratio, and required
in addition that $\eslt \geq 60$~GeV (which reduces the background by
almost 50\% with about a 20\% loss of signal) and further 60~GeV $\leq
m_{bb} \leq$ 140~GeV (which reduces the background by another factor of
half with a loss of 25\% of the signal).
Several other distributions were also studied including, jet
multiplicity, $\theta_{bb}$, $\delta\phi(bb)$ but none of these proved
useful to enhance the signal over the top background.
The signal is below the 5$\sigma$ discovery level over essentially
the entire Model Line I; only for $m_{\tilde \chi_1^0} \sim 155 \pm
10$~GeV does the signal cross section exceed the $5\sigma$ level of
7.7~fb.
Moreover the $S/B$ ratio never exceeds about 15\% which falls
below our detectability criterion $S/B \geq 20$\%.
While
it is possible to improve the $S/B$ ratio via additional cuts, these
typically degrade the statistical significance of the signal.

\subsubsection{PYTHIA-SHW studies of Higgsino-like Neutralino NLSP}
\label{sssec:phyhnlsp}

The $Z$ boson decay mode of a general neutralino NLSP,
$\NI \to Z \GG$, leads to the possibility of di-boson modes which have
one or two $Z$ bosons.
The existence of $Z$ boson(s) arising from decay to the
Goldstino gives many possible final state signatures,
depending on the $Z$ boson decay mode, $Z \to \ell \ell, jj, \nu \nu$.
The possible di-boson final state signatures
which include at least one $Z$ boson
were listed in the
last three rows and columns of Table \ref{higgsinosignatures}.
The sensitivity
of Run II to signatures associated with the $\gamma Z$ mode
for Model Line I and the $ZZ$ mode
for Model Line II have been
estimated based on some Run I analysis of similar final states.
The signatures associated with this have been
analyzed using the PYTHIA \cite{PYTHIA} option of the SHW detector
simulation package \cite{SHW,TAUOLA,STDHEP}.


For the parameters of Model Line I
the $\gamma Z \met$ di-boson mode dominates the total
cross section in the transition region of masses
as discussed at the beginning of Section \ref{sec:higgsino}, and illustrated
in Fig.~\ref{fig:HIcross}.
Leptonic decay of the $Z$ provides the cleanest final state,
$\gamma \ell^+ \ell^- \met$, which is similar to existing
Run I studies of Standard Model
$Z\gamma$ production without $\met$ \cite{Zgamma_CDF,Zgamma_D0}.
For a Higgsino-like neutralino NLSP search,
however, an additional large $p_T$ cut, made possible by decay to the
Goldstinos,
as well as a more stringent photon $E_T$ cut should reduce
the backgrounds to a negligible level.
In order to isolate a leptonically decaying $Z$ boson,
a hard photon, and missing energy, the following cuts are applied
\begin{enumerate}
\item At least one photon with $p_T>25$ GeV and $|\eta|<2.0$.
\item Two opposite-charge, same flavor leptons with
$p_T>(15,10)$ GeV, $|\eta|<2.0$ and
invariant mass $|m_{\ell\ell}-M_Z|<10$ GeV.
\item $\ \met>25$ GeV.
\end{enumerate}
Backgrounds necessarily include a fake lepton, fake photon, or
energy mismeasurement.
The probabilities for misidentifying an electron or jet
as a photon are conservatively assumed
to be
${\cal P}(e \to \gamma) = 10^{-3}$ and
${\cal P}(j \to \gamma) = 10^{-3}$.
The probability for jet energy mismeasurement to be greater than
or equal to $E_0$ is taken
to be
\beq
{\cal P}( j \to \met > E_0)=\exp(1.86-0.3E_0 / {\rm GeV})
\eeq
valid for
$E_0 >20$ GeV \cite{Greg}.
For the missing energy cut of 25 GeV given above this corresponds
to ${\cal P}( j \to \met > 25~{\rm GeV}) \simeq 0.0036$.
The main backgrounds are from
$WWj \to \ell \ell  j \met$ with the jet faking a photon,
$t \bar{t} j \to \ell \ell b \bar{b} j \met$ with the jet
faking a photon,
$Zjj \to \ell \ell jj$ with one jet faking $\met$ and the
other faking a photon,
and
$Z \gamma j$ with the jet faking $\met$ .
Other potential backgrounds from
$Z \to \ell \ell \tau \tau$ with an electron from
$\tau \to e \met$ decay misidentified as a photon,
$WZ \to e \ell \ell$ with the electron misidentified as a photon,
are significantly smaller.
Continuum $\gamma \ell \ell j $ with the jet faking $\met$
is very efficiently reduced to negligible levels by the invariant
mass cut given above, while
$ZZj\to \tau \tau j \to \ell \ell j\met$ with the jet
faking a photon is also reduced to negligible levels by
the combination of invariant mass and $\met$ cuts.
The total expected background from all these sources
to the $\gamma \ell \ell \met$ final state
with the cuts given is found to be 0.06 fb.
For the parameters of Model Line II with
a $\NI$ mass of 130 GeV, the acceptance times efficiency
for the signal in the $\gamma \ell \ell \met$ channel with the above cuts
(not including Br$(Z \to \ell \ell) \simeq 0.07$)
is found to be 0.39.
Using this along with the background quoted above,
the reach for a 3$\sigma$ observation with
2 (30) fb$^{-1}$ of integrated luminosity should be
155 (220) GeV for the $\NI$ mass along Model Line I, and
130 (165) GeV along Model Line II.
The reach is better along Model Line I since the $\NI \to \gamma \GG$
and $\NI \to Z \GG$ modes are comparable in the relevant mass range, while
for Model Line II $\NI \to Z \GG$ dominates above the narrow transition
region of mass.

Invisible decay of the $Z$ gives rise to the signature $\gamma \met$.
This channel has been studied in Run I as a probe for anomalous
$\gamma Z$ couplings with $Z \to \nu \nu$
\cite{photon_D0,diboson_review_D0}.
The largest background in Run I was
from single $W$ production with $W \rightarrow e \nu$ and
the electron misidentified as a photon.
This background can be substantially reduced by
raising the photon $E_T$ and $\met$ cuts
beyond the Jacobian peak for
$W\rightarrow \ell\nu$ \cite{Greg}.
This may be accomplished with the cuts
\begin{enumerate}
\item One photon with $p_T>50$ GeV and $|\eta|<2.0$.
\item $\ \met>50$ GeV.
\end{enumerate}
Monte Carlo simulation indicates that only roughly 1\% of
$W \to e \nu$ decays have both an
electron with $p_T > 50$ GeV and $\met > 50$ GeV.
With a total cross section $\sigma(p \bar{p} \to W) \simeq 6$ nb,
a misidentification rate ${\cal P}(e \to \gamma)=10^{-3}$, branching ratio
${\rm Br}(W \to e \nu) \simeq$ 11\%, and assuming a photon acceptance
of 0.8, this gives a $\gamma \met$ background from this source of
5 fb after cuts.
Additional backgrounds arise from
$\gamma j$ and $jj$ with one jet faking a photon
{\it and} in each case the remaining jet energy mismeasured to be below
the minimum pedestal.
The Run I $\gamma \met$ analysis \cite{photon_D0,diboson_review_D0}
estimated $< 45$ fb
from this source with a $\met$ cut of 35 GeV.
Raising the $\met$ cut to 50 GeV given above should reduce
this background to an insignificant level.
A final source for background is muon bremsstrahlung from cosmic
or beam halo muons.
With the combination of larger $\met$ cut,
improved photon pointing, and for cosmic rays a
larger signal to noise implied by larger instantaneous luminosity,
it should be possible to reduce the background from this source, estimated
to be 135 fb after cuts in the Run I $\gamma \met$ analysis
\cite{photon_D0,diboson_review_D0}, by a factor of roughly 30 \cite{Greg}.
The total $\gamma \met$ background with the above cuts is therefore
estimated to be of order 10 fb.
Note that with data, it should be possible
to accurately characterize the $\gamma \met$
backgrounds from these sources, as was the case in Run I.
With the cuts given above the acceptance time efficiency for the signal
in the $\gamma \met$ channel
(not including ${\rm Br}(Z \to \nu \nu) \simeq 0.20$)
is found to be 0.5.
With the estimated background given above, the reach for a $3 \sigma$
observation with 2 (30) fb$^{-1}$ of integrated luminosity should be
135 (165) GeV for the $\NI$ mass along Model Line I, and
130 (145) GeV along Model Line II.

Hadronic decay of the $Z$ in the $\gamma Z\met$ mode gives
rise to the signature $\gamma jj \met$.
Backgrounds are similar to those of the
$\gamma \met$ channel.  The $\gamma jj \met$
channel has been studied in Run I in order to place limits
on squark and gluino masses in very specific supersymmetric
models \cite{photon_jets_D0}.
Further background suppressions not included in 
the Run I study are possible with acoplanarity, sphericity and invariant
dijet mass cuts to reconstruct the $Z$ boson, and a lepton veto.
Although a detailed study has not been attempted,
the total background is expected to be smaller than
for the $\gamma\met$ channel, due to the presence of two additional
hard partons.
Given the significant $Z$ hadronic branching ratio, the $\gamma jj \met$
channel should therefore provide somewhat better reach than the
$\gamma \ell^+\ell^- \met$ or $\gamma \met$ channels in Run II.


For the parameters of Model Line II
the $ZZ \met $ di-boson mode dominates at
larger $\tilde \chi_1^0$ mass
as discussed at the beginning of Section \ref{sec:higgsino}, and illustrated
in Fig.~\ref{fig:HIIcross}.
Leptonic decay of each $Z$ boson gives rise to the spectacular
signature $\ell^+\ell^-\ell^{\prime +} \ell^{\prime -} \met$,
with the lepton pairs
reconstructing the $Z$ mass (in one choice of pairing for
$\ell=\ell^{\prime}$).
This gold plated channel is expected to be essentially background free,
but suffers from small leptonic branching ratio.
Because of this Run II with even 30 fb$^{-1}$  of integrated luminosity
will not be sensitive to the parameters of Model Line I
in this channel.
For Model Line II however, with the dominant $\NI \to Z \GG$ decay
mode, the reach for a 3$\sigma$ observation with 30 fb$^{-1}$
of integrated luminosity should approach 170 GeV for the $\NI$ mass.



\subsection{Higgs Bosons from Higgsino Decay}
\label{subsec:hfhiggsino}

The Higgs boson decay mode of a Higgsino-like neutralino NLSP, 
$\NI \to h \GG$, presents the possibility of collecting 
SUSY events which contain a real Higgs boson. 
The most promising channels which include a Higgs boson are
$h\gamma \met$ and $hh\met$ which yield the final state
signatures $\gamma b j \met$ discussed in sections \ref{subsubcdfhiggsino} 
and \ref{subsubd0higgsino}
and $bbbj\met$ discussed in section \ref{subsubisahiggsino}. 
It is then interesting to consider the reach in these 
channels as a general function of both the Higgsino and 
Higgs masses \cite{hshort}.

\begin{figure}[t!]
\epsfysize=3.0in
\epsffile[-50 230 160 560]{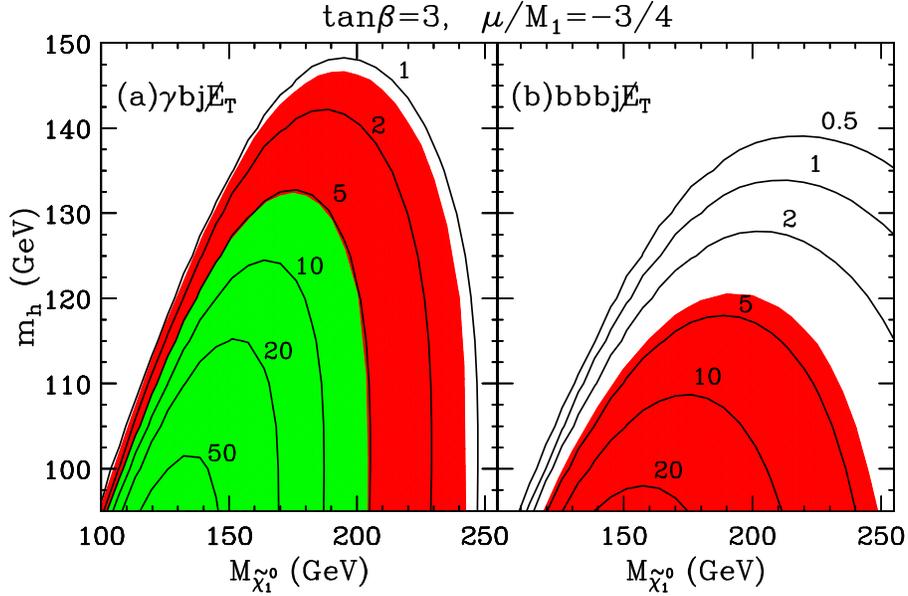}
\begin{center}
\parbox{5.5in}{
\caption[]{ Signal cross-section times branching ratio
contours in fb
for the (a) $\gamma bb\met$ and (b) $bbbb\met$ channels,
as a function of the neutralino mass
$M_{\tilde \chi^0_1}$, and the Higgs mass $m_h$,
for $\tan\beta=3$ and $\mu/M_1=-3/4$.
The Run IIa(b) reach with 2(30) fb$^{-1}$ of integrated luminosity
is indicated in light(dark) region. 
\label{reach}}}
\end{center}
\end{figure}

The total cross section times branching ratio contours for the
$\gamma bb \met$ and $bbbb \met$ channels as a function
of the $h$ and $\tilde \chi_1^0$  masses are shown in Fig.~\ref{reach}.
These contours include ${\rm Br}(\chi_1^0 \to (\gamma, h) \tilde G)$
for $\tan\beta=3$ and $\mu/M_1=-3/4$ and SM
values for ${\rm Br}(h \to bb)$.
The Run IIa 3$\sigma$ discovery reach with 2 fb$^{-1}$
discussed in section \ref{subsubd0higgsino} for the
$\gamma bj \met$ channel corresponds to a
signal times branching ratio cross section of 6 fb.
For the parameters of Fig. \ref{reach} this corresponds to a Higgs
mass of up to at least 120 GeV for $\tilde \chi_1^0$ masses in the
range 140-195 GeV, with a maximum reach in Higgs mass of just over
130 GeV. 
The Run IIa reach is indicated by the light shaded region in 
Fig.~\ref{reach}. 
This is to be contrasted with the search
for the SM Higgs from direct $Wh$ and $Zh$ production.
These SM channels are background limited, and no sensitivity
to a Higgs mass beyond current limits is expected in Run IIa.
So the $\gamma bj \met$ channel presents the interesting possibility
for Run IIa of a SUSY signal which contains real Higgs bosons.
The Run IIb 3$\sigma$ discovery reaches 
implied by the results of section \ref{subsubd0higgsino} and 
\ref{subsubisahiggsino} with 30 fb$^{-1}$ of integrated 
for the
$\gamma bj \met$ and $bbbj \met$ channels respectively correspond to
signal times branching ratio cross sections of
1.5 fb and 4 fb respectively. For the parameters of Fig.~\ref{reach}
the maximum reach in Higgs mass then corresponds to
almost 145 GeV and 115 GeV respectively.
The Run IIa reach is indicated by the dark shaded region in 
Fig.~\ref{reach}. 

\begin{figure}[t!]
\epsfysize=3.0in
\epsffile[-230 20 120 470]{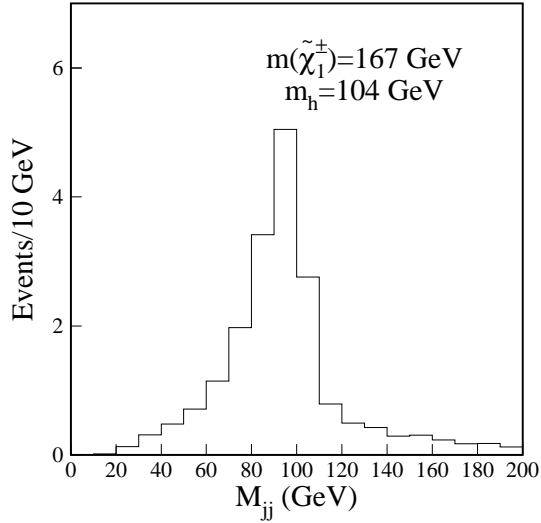}
\begin{center}
\caption[]
{\small The invariant mass distribution of the two leading jets
in the \D0 $\gamma bj\met$ analysis 
for the Model Point on the Higgsino-like Neutralino NLSP Model Line I with 
$m_{\tilde\chi^\pm_1}=167$ GeV ($\Lambda=80$ TeV) and with 
${\rm L}=2$ ${\rm fb}^{-1}$. Note that 19 signal and less
than 1 background events are expected.
\label{gbb}}
\end{center}
\end{figure}

In order to identify the Higgs boson directly in a sample
of events arising from Higgsino decays it is necessary to
observe a peak in the $bb$ invariant mass.
The identifiable di-boson final states and
large $\met$ carried by the Goldstinos
render the supersymmetric Higgs boson final states discussed
here relatively clean.
Reconstructing the Higgs mass peak should be
relatively straightforward
compared to SM $Wh$ and $Zh$ production modes
which suffer from much larger continuum $bb$ backgrounds.
Fig.~\ref{gbb} shows the invariant mass distribution of the
two leading jets 
in the \D0 $\gamma bj \met$ analysis described in section 
\ref{subsubd0higgsino}
for the Model Point on the Higgsino-like Neutralino NLSP Model Line I with 
$m_{\tilde\chi^\pm_1}=167$ GeV ($\Lambda=80$ TeV) and with 
${\rm L}=2$ ${\rm fb}^{-1}$. 


\subsection{Non-Prompt Decays to Higgs and $Z$ Bosons}
\label{subsec:nphiggisino}

The decay length for  $\NI \rightarrow (h,Z,\gamma) \GG$
may be macroscopic, as illustrated in Fig.~\ref{ctaufig}.
Decay over a macroscopic distance, but within the detector
volume gives displaced hard partons with finite impact
parameter. 
Detection of displaced hard photons is discussed in section 
\ref{subsec:npphoton}.

Decays of a metastable Higgsino-like neutralino with macroscopic
decay length by 
$\NI \to h \GG$ with $h \to bb$ or $\NI \to Z \GG$ with the 
$Z$ decaying hadronically give rise to large $E_T$ displaced jets 
with finite impact parameter and large $\met$. 
Since the metastable $\NI$ is non-relativistic the jets from 
$h$ or $Z$ decay have a roughly uniform angular distribution
in the lab frame. 
This is in contrast to potential high $E_T$ displaced jet 
background from heavy quark decay, which is highly boosted 
in the direction of motion of the relativistic heavy quark. 
So the angular distribution of displaced high $E_T$
jets can greatly aid in reducing background. 
A useful selection criterion in this regard is the 
large negative impact parameter
(LNIP) described in section \ref{subsec:dissquark} in the context of 
NLSP squark decay. 
LNIPs select displaced jets which are emitted towards rather than away 
from the beam axis. 
Also note that 
since each $h$ or $Z$ decays to a pair of jets, the displaced 
jets should reconstruct displaced vertices in pairs from the
point of each $h$ or $Z$ decay. 
Such pairs of displaced jets have an invariant mass appropriate
to the $h$ or $Z$, which could also be used as a further cut. 

Decays of a neutralino with macroscopic decay length 
contained within the tracking region by 
$\NI \to Z \GG$ with $Z \to \ell \ell$ where $\ell=e,\mu$
presents the possibility of very cleanly reconstructing the 
displaced $Z$. 
First, the $\ell \ell$ vertex can be accurately determined 
using tracking. 
Second the $\ell \ell$ invariant momentum can also be required
to reconstruct the $Z$ very accurately. 
Based in part on the possibility of observing a metastable neutralino
which decays to a $Z$ boson, CDF has carried out a search 
for displaced $Z \to ee$ decays in Run I \cite{longZ}.
Although the production cross section is too small to have 
been observed in Run I, as discussed in section \ref{sssec:phyhnlsp} 
some channels with a leptonic $Z$ could be observed in Run II. 
The search for displaced $Z$ bosons should therefore be extended
to Run II. 

It is important to note that the backgrounds for 
any of the final state signatures
for a Higgsino-like neutralino NLSP described in the previous
section are greatly reduced if the $\NI$ decay
is displaced but contained within the tracking region. 
A search for 
Non-prompt Higgsino-like neutralino decays can therefore 
be implemented simply by applying an LNIP or displaced leptonic
$Z$ search to an event sample derived from any of the searches 
described in the previous sections on prompt decays. 
This interesting combination of analysis for these 
spectacular signatures should not be overlooked in Run II. 
In the case of metastable $\NI \to h$ decays with $h \to bb$, this would
yield a very interesting sample of Higgs bosons. 


\section{Stau NLSP}\label{sec:stauNLSP}
\setcounter{equation}{0}
\setcounter{footnote}{1}
\indent

A stau NLSP arises if the sleptons are lighter than the 
other MSSM superpartners
and if the stau is more than 
1.8 GeV lighter than the selectron and smuon, so that these
states eventually cascade decay to $\stauI$, as discussed below. 
A stau NLSP decays to the Goldstino by 
\begin{equation} 
\stauI \to \tau \GG
\label{staudecay}
\end{equation}
If the supersymmetry breaking scale $\sqrt{F}$ is sufficiently low, 
then the decay \ref{staudecay} occurs
promptly,
and all supersymmetric events contain at least two high-$p_T$ $\tau$'s.
Since the decays into the $\stauI$ NLSP also typically contain $\tau$'s,
the predicted events can quite often have three or even four
$\tau$ leptons \cite{mgm,DuttaNandi,Nanditwo}.
They can be manifested either in purely leptonic or hadronic channels.
Since the heavy $\stauI$ decays to two essentially massless particles,
these events will have high $\met$ and the leptons or jets
from the $\tau$'s will typically have high $p_T$. Furthermore, the other
leptons
and jets produced in the supersymmetric decay chains can have quite
distinctive  profiles.

If the supersymmetry breaking scale is larger, the 
$\stauI$ is very long-lived.
It is then appears as a
weakly interacting,
massive, slowly-moving charged particle so it has unique characteristics.
It will appear as a high-$E_T$ track that penetrates the calorimeters
and the muon system.  If it is moving slowly enough, it will leave
significantly more $dE/dx$ energy than a minimum--ionizing particle.
In the GMSB models, we expect two of these objects in each SUSY event,
leading to a variety of spectacular signals
\cite{mgm,FengMoroi,MartinWells}.
We can consider triggering in several ways.
The first is the $\stauI$ as a muon,
since it will likely be a high-$E_T$ track that
is weakly interacting so it can traverse the muon chambers.  In this
case
there are two possible failure modes: the particle is moving too slow to
reach
the muon system within the timing gate, or the event is rejected
because a MIP calorimeter energy cut is placed on the muon trigger track.
The second trigger is a simple requirement on high-$E_T$ isolated tracks
as mentioned in the discussion of two prompt $\tau$'s.
Finally, these triggers are necessary for the direct production of the
sleptons, but the trigger could require
the objects from cascade decays of heavier sparticles.
Off-line the events can also be detected several ways.  First, since
they can appear as muons, they can be detected simply as an excess of muons.
Since most of the $\stauI$'s will have excessive $dE/dx$, they can
potentially be detected in any detector that records $dE/dx$.  Finally
they can be detected as slow particles by detectors with timing information.

For intermediate values of $\sqrt{F}$, the stau decays can occur inside
the
detector but at a macroscopic distance. In that case, the impact parameter
of the stau from the interaction region or the kink in the charged
particle track when the decay occurs can yield a unique signal and provide
information about the decay length. Note that the decay width formula
essentially depends only on the stau mass (which kinematic information
can reveal) and the supersymmetry breaking order parameter $\sqrt{F}$. So
measuring the physical decay length of the charged staus in this
scenario may yield direct information or constraints on the supersymmetry
breaking mechanism.

In the stau NLSP scenario, the heavier sleptons $\widetilde e_R$ and
$\widetilde \mu_R$ must have allowed three-body decays
\beq
\widetilde e_R \rightarrow e\tau^\pm\widetilde \tau_1^\mp
\qquad\qquad {\rm and}\qquad\qquad
\widetilde \mu_R \rightarrow \mu\tau^\pm\widetilde \tau_1^\mp .
\label{threebodydecays}
\eeq
These decays have been studied in some detail in Ref.~\cite{threebody}.
The charges of the final-state $\tau $ and $\widetilde
\tau_1$ NLSP are largely uncorrelated with the charge of the decaying
slepton.  This means that production of the heavier sleptons, either
directly or in other sparticle decays, tends
to
produce a pair of final-state stau NLSPs with the same charge almost half
of the time.  This is also true for any supersymmetric events which
involve a neutralino, either as one of the initially produced particles,
or as part of the decay chain. This is because of the Majorana nature
of neutralinos, which requires that they decay democratically into final
states with $\tau^+\widetilde \tau_1^-$ and $\tau^- \widetilde \tau_1^+$.
This observation, that many or most supersymmetric events
feature like-charge staus in the final stage of the decay chain
before $\stauI \rightarrow \tau\GG$,  may be
useful
for defining
reduced-background signals.  Furthermore, the ratio of
same-charge vs. opposite-charge stau events (particularly the deviation
of this ratio from 1)
can be an
important
observable in disentangling the model parameters from the data.

It is also worth noting that the three-body decays
eq.~(\ref{threebodydecays})
may also have a macroscopic length in the lab frame, if the total mass
difference between initial and final states is less than about 1 GeV.
This can occur for $\tan\beta$ in the range from 5 to 10, depending
of course on the other model parameters. In this case, the $\tau$ and
$e$ or $\mu$ will be extremely soft and could be missed. This results
in an interesting signal of a charge-changing HIT (CC-HIT), with the
decaying
slepton or smuon track turning into a stau track with the opposite
charge about half the time.

\begin{figure}[tpb]
\centering
\epsfxsize=4.5in
\hspace*{0in}
\epsffile{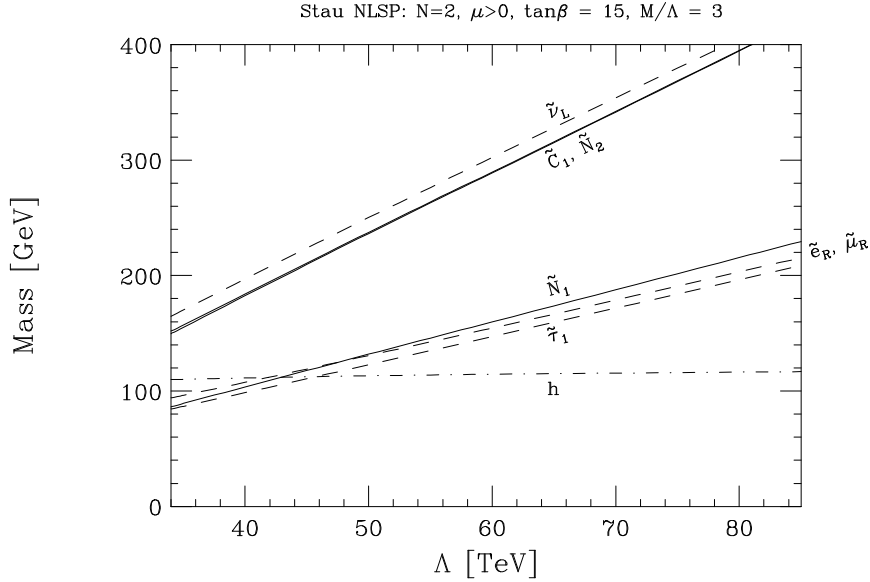}
\caption{The masses of the lightest neutralinos, charginos, sleptons
and CP-even Higgs boson in the Stau NLSP Model Line, as a function
of $\Lambda$.}
\label{fig:n2line_mass}
\end{figure}

\begin{figure}[tpb]
\centering
\epsfxsize=4.5in
\hspace*{0in}
\epsffile{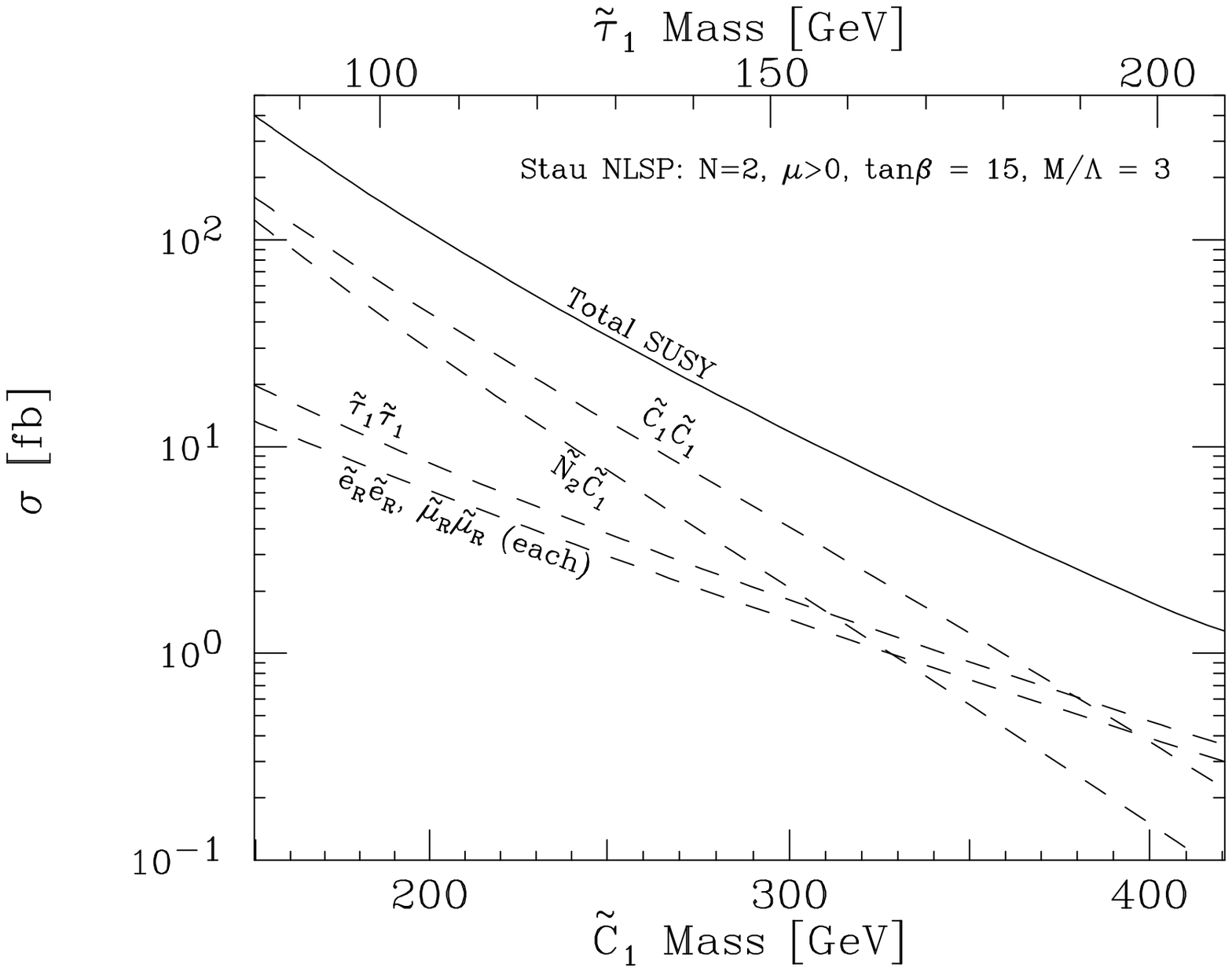}
\caption{Production cross-sections in
$p\protect\overline p$ collisions with $\protect\sqrt{s} = 2$ TeV, for
superpartner pairs in the Stau NLSP Model Line,
as a function of
$m_{\tilde \chi_1^\pm}$ and $m_{\protect\tilde \tau_1}$.}
\label{fig:n2line_sigma}
\end{figure}

For quantitative studies, we follow the general strategy outlined
in Section \ref{sec:mgm} and define a Stau
NLSP Model Line.
The fixed parameters that define this Model Line are:
\beq
{\rm Stau~NLSP~Model~Line:}~~~~~~
\Nmess = 2,\>\> {\Mmess \over \Lambda} = 3,\>\> \tan\beta = 15, \>\> \mu > 0,
\label{staulineparameters}
\eeq
with $\Lambda$ allowed to vary.
In Fig.~\ref{fig:n2line_mass}, we show the masses of the lightest
few superpartners and the lightest CP-even Higgs boson $h^0$,
as a function of 34 TeV $<\Lambda<$ 85 TeV.
The relevant sparticle masses scale almost linearly with $\Lambda$.
Note that
$m_{\tilde e_R} - m_{\tilde \tau_1}$ and
$m_{\tilde \mu_R} - m_{\tilde \tau_1}$ are greater than a few GeV
for all points along this Model Line, so that all sparticles do indeed
have kinematically-allowed decays to the stau NLSP.
Therefore, the possible
general types of experimental signatures are:
\begin{itemize}
\item[$\bullet$] Small $\cG$ with prompt NLSP decays:\\
Events with 2 or 3 or more high-$p_T$ $\tau$'s,
often with additional leptons and jets.
\item[$\bullet$] Intermediate $\cG$ with delayed NLSP decays:\\
Events with $\widetilde \tau_1 \rightarrow \tau$
decay kinks.
\item[$\bullet$] Large $\cG$ with a quasi-stable NLSP:\\
Events with heavy charged particle tracks featuring
anomalous ionization rate or time-of-flight from slow staus,
and/or fake ``muons" from fast staus.
\end{itemize}
In Figure \ref{fig:n2line_sigma}, we show the most important Run II
production cross-sections for superpartners in this model,
including $\CIplus \CIminus$, $\CIplusminus\NI$,
$\widetilde e^+_R \widetilde e^-_R$,
$\widetilde \mu^+_R \widetilde \mu^-_R$, and
$\widetilde \tau^+_1 \widetilde \tau^-_1$, as well as
the total inclusive cross-section for all supersymmetric particles,
as a function of $m_{\tilde \chi_1^\pm}$ and $m_{\tilde \tau_1}$.

\begin{figure}[tpb]
\centering
\epsfxsize=4.5in
\hspace*{0in}
\epsffile{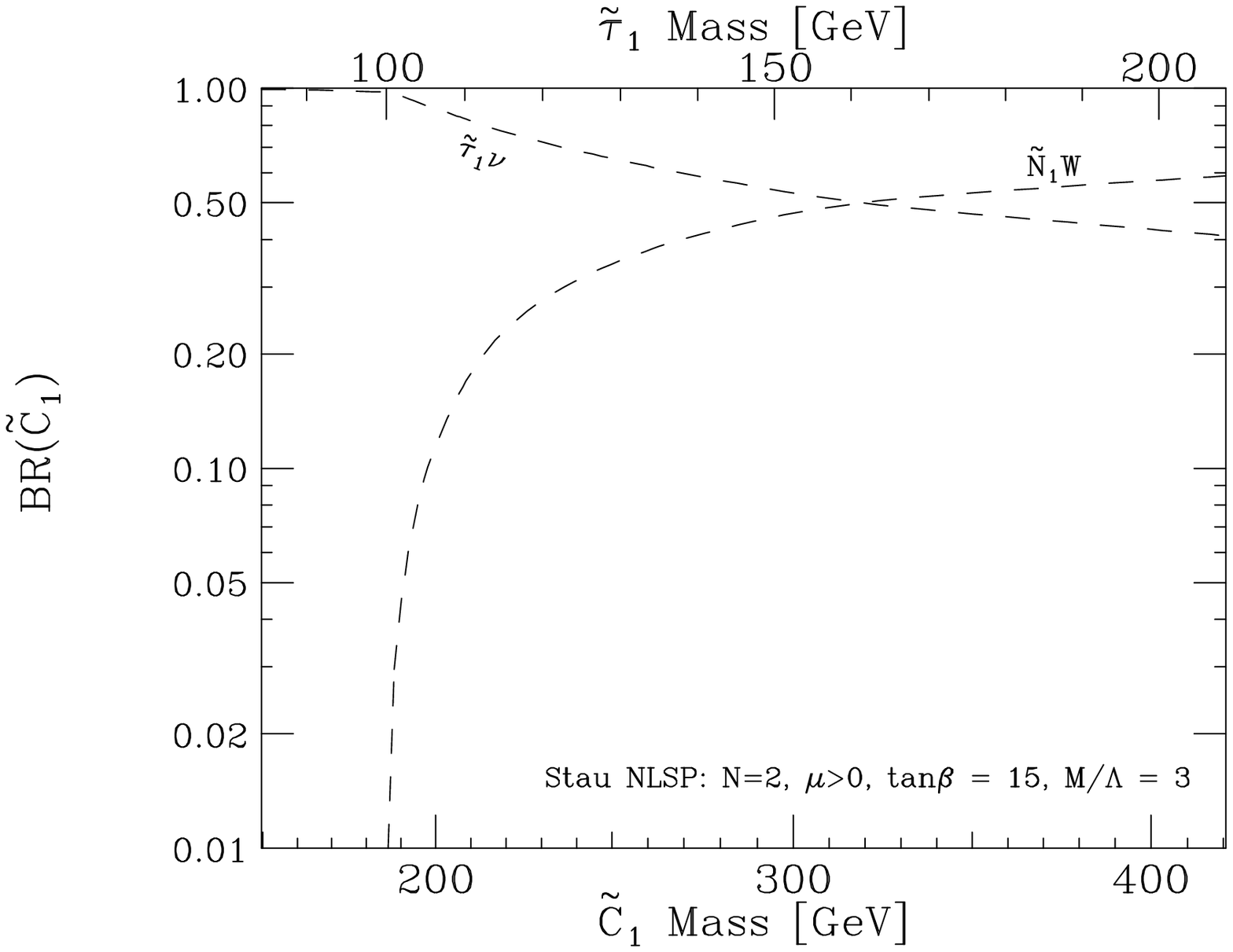}
\caption{Branching fractions for the decay of
$\CI$ in the Stau NLSP Model Line,
as a function of
$m_{\tilde \chi_1^\pm}$ and $m_{\protect\tilde \tau_1}$.}
\label{fig:n2line_brc1}
\end{figure}

\begin{figure}[tpb]
\centering
\epsfxsize=4.5in
\hspace*{0in}
\epsffile{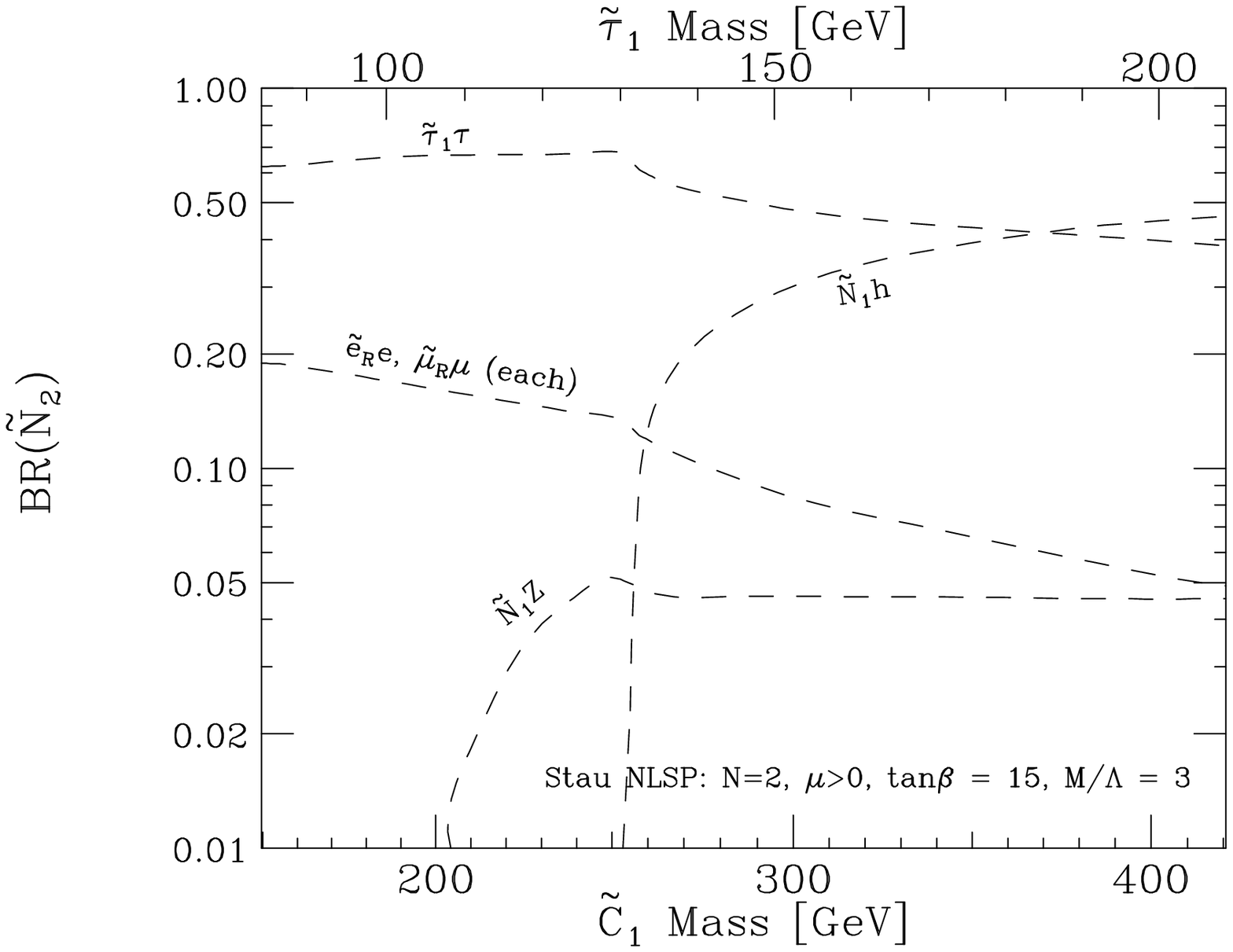}
\caption{Branching fractions for the decay of
$\NII$ in the Stau NLSP Model Line,
as a function of
$m_{\tilde \chi_1^\pm}$ and $m_{\protect\tilde \tau_1}$.}
\label{fig:n2line_brn2}
\end{figure}

\begin{figure}[tpb]
\centering
\epsfxsize=4.5in
\hspace*{0in}
\epsffile{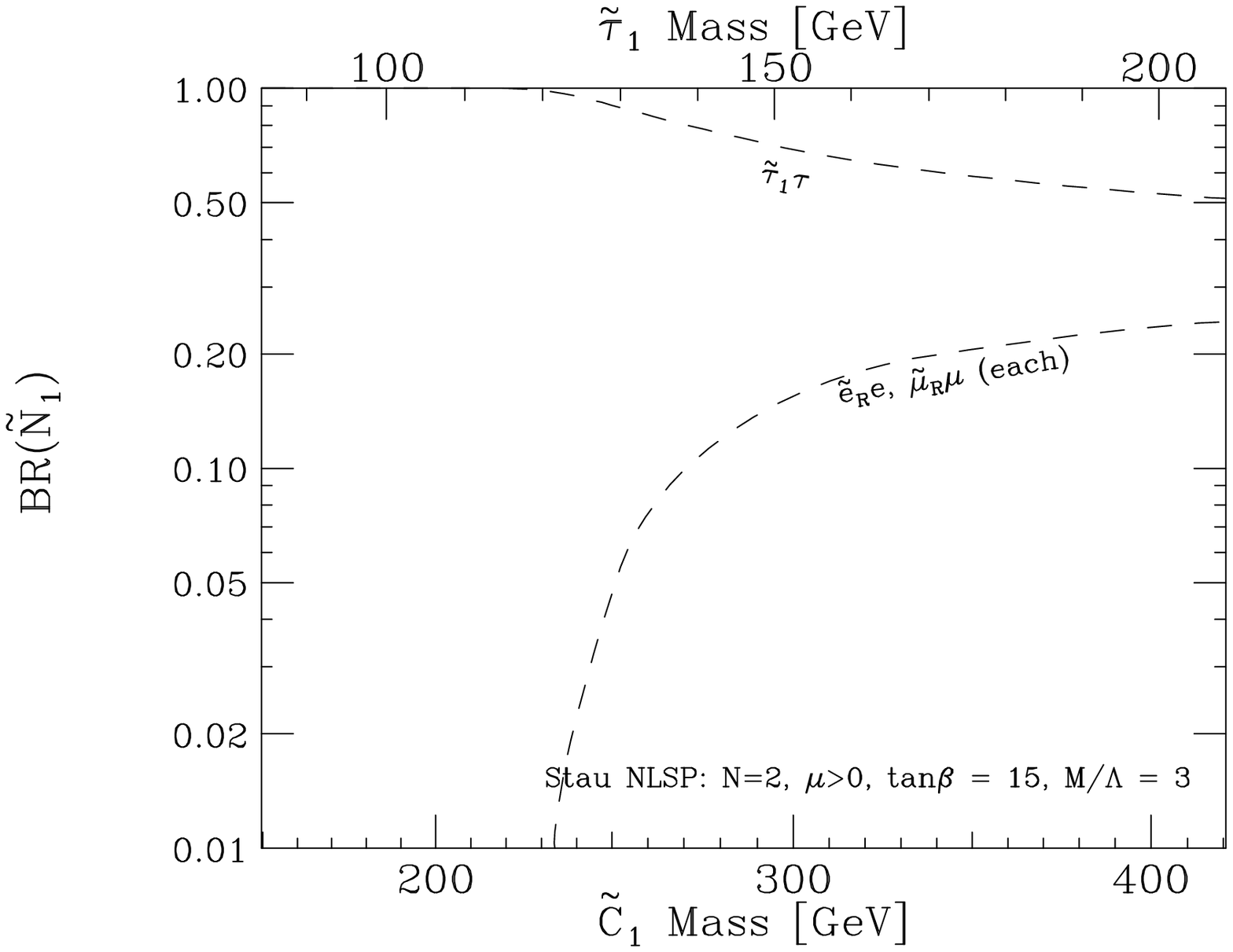}
\caption{Branching fractions for the decay of
$\NI$ in the Stau NLSP Model Line,
as a function of
$m_{\tilde \chi_1^\pm}$ and $m_{\protect\tilde \tau_1}$.}
\label{fig:n2line_brn1}
\end{figure}

\begin{figure}[tpb]
\centering
\epsfxsize=4.5in
\hspace*{0in}
\epsffile{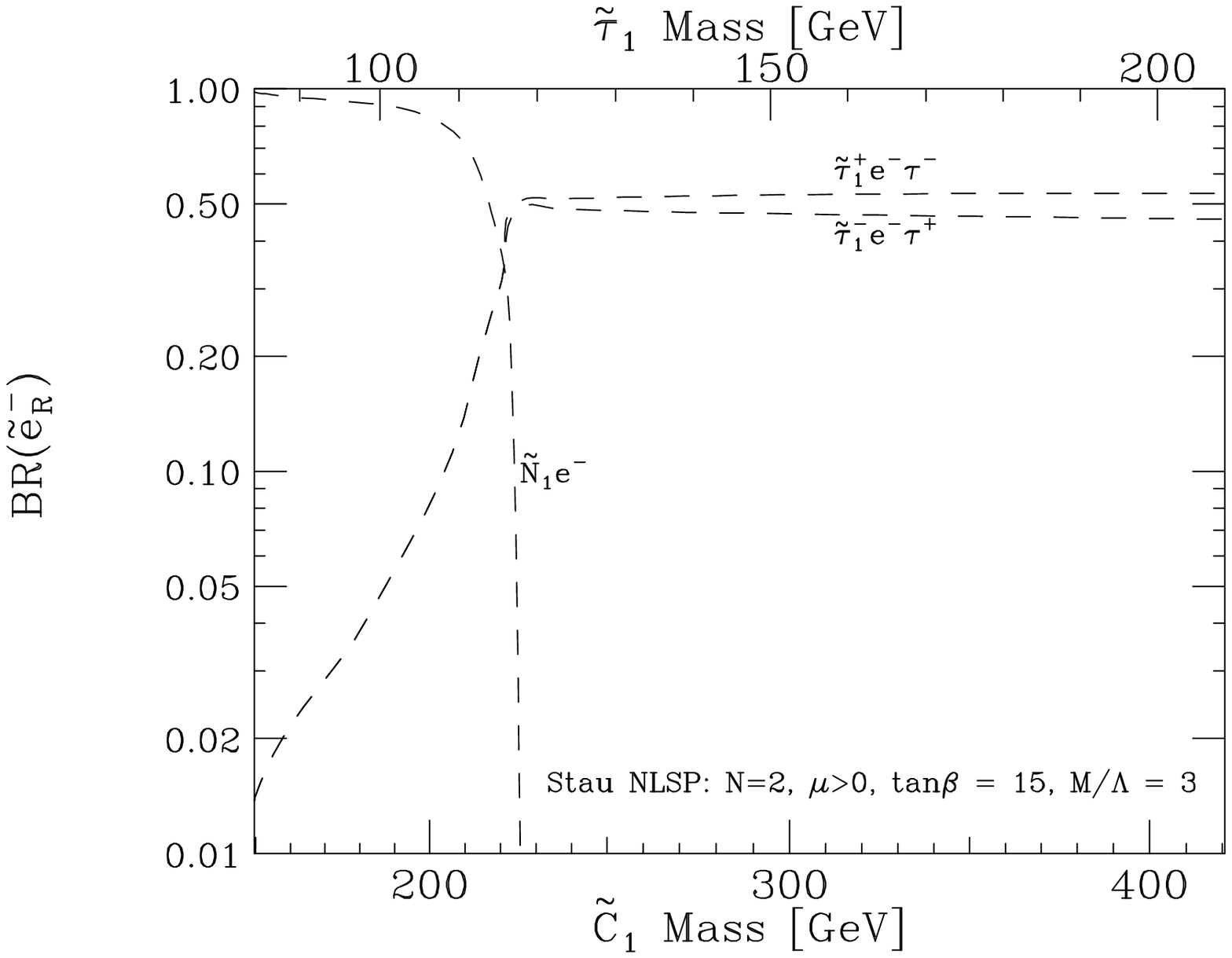}
\caption{Branching fractions for the decay of
$\widetilde e^-_R$ or $\widetilde \mu^-_R$ in the
Stau NLSP Model Line,
as a function of
$m_{\tilde \chi_1^\pm}$ and $m_{\protect\tilde \tau_1}$.}
\label{fig:n2line_brer}
\end{figure}

\begin{figure}[tpb]
  \centerline{\epsfysize=3.5in\epsfbox{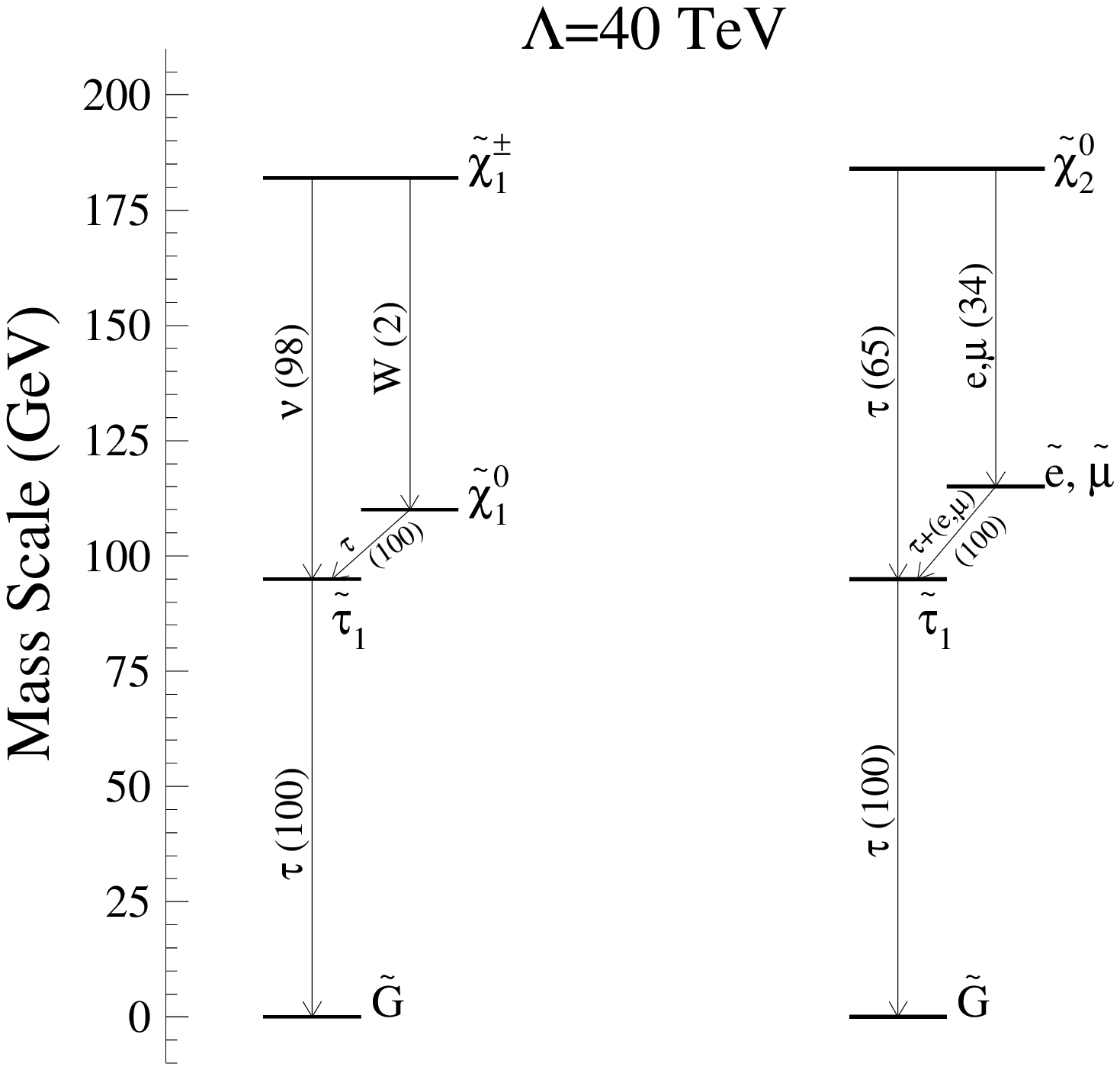}}
  \caption{Decay schematics of $\tilde\chi^\pm_1$ and $\tilde\chi^0_2$
           for $\Lambda=40$~TeV for the Model Line with a \protect\tlsp\
as
           the NLSP. Percentage branching ratios for main decay modes
           are shown in parentheses.
(The $\stauI$, $\widetilde e_R$, $\widetilde \mu_R$ and $\NI$
mass levels have been displaced slightly for labelling purposes.)}
  \label{fig:p2br}
\end{figure}

Again, $\CIplus\CIminus$ and \cn\ 
dominate the production
cross section for $\Lambda\lsim 75$~TeV, corresponding to
$m_{\tilde\tau_1} \lsim 185$ GeV and $m_{\tilde \chi_1^\pm} \lsim 370$
GeV.
For
larger
$\Lambda$
values, $\tilde\tau^+_1\tilde\tau^-_1$, $\tilde e^+_R\tilde e^-_R$, and
$\tilde\mu^+_R\tilde\mu^-_R$ productions become relatively more important.
Figures \ref{fig:n2line_brc1},
\ref{fig:n2line_brn2}, \ref{fig:n2line_brn1} and
\ref{fig:n2line_brer} show the most significant branching fractions
for $\CI$, $\NII$, $\NI$ and $\widetilde e_R$. As an example,
branching ratios of $\tilde\chi^\pm_1$ and $\tilde\chi^0_2$ for
$\Lambda=40$~TeV are graphically displayed in Fig.~\ref{fig:p2br}.
In the following,
signatures for short-lived and for quasi-stable $\tilde\tau_1$'s
will be discussed. It should be noted that the two analyses discussed
below also have some sensitivity to the case with an intermediate \tlsp\
lifetime.

\subsection{\rm CDF study of Stau NLSP Model Line with prompt $\stauI$
decay}

Prompt $\tau$ decays can be investigated though the $\ell\nu\nu$
leptonic decays (17\% each for $e$ or $\mu$),
where they often contribute to the efficiency
in leptonic analyses.  The other approach is to search for the
one-- (50\%) or three-- (15\%) charged--hadron decays
(which can contain additional neutrals)\cite{gmsbcdfgroer}.
For these ``hadronic'' $\tau$'s,
the signature is the one or three charged tracks, isolated from other
tracks, but associated with a calorimeter energy cluster with a narrow width.
The tracks must be within 10$^\circ$ of the jet centroid
and no other tracks are within 30$^\circ$.  The probably that a
randomly--selected jet pass these cuts is on the order of 1\%.
Note that this fake rate is, in general, too large to search for
$\tau$  decays in a data sample dominated by jets.
If selection criteria are imposed on other parts of the event,
then a good signal to noise can be achieved.
Note that a selection of a leptonically--decaying $\tau$ selected with
a hadronically--decaying $\tau$ has greater rejection than either
two leptons or two hadronic $\tau$'s because QCD and Drell--Yan can be
more effectively suppressed.

CDF has searched for the signature of two prompt $\tau$'s and $\missET$
in the context of a search for top decays to the charged Higgs
which in turn decays to $\tau\nu$ \cite{gmsbcdfHtau,gmsbcdfgroer}.
The data set was collected through the $\missET$ trigger.
The primary analysis cuts were $\missET>30$~GeV,
two hadronically decaying $\tau$'s with $E_T>30$~GeV and $|\eta|<1$,
and the $\tau$'s
must not be back--to--back, to remove $Z$ decays.
In 100~pb$^{-1}$, no events were observed
while $2.2\pm 1.3$ where expected from
Standard Model sources, mostly $W\rightarrow \tau\nu$ with a jet
faking the second $\tau$.

\begin{figure}[tpb]
\centerline{\epsfysize=3.5in\epsfbox{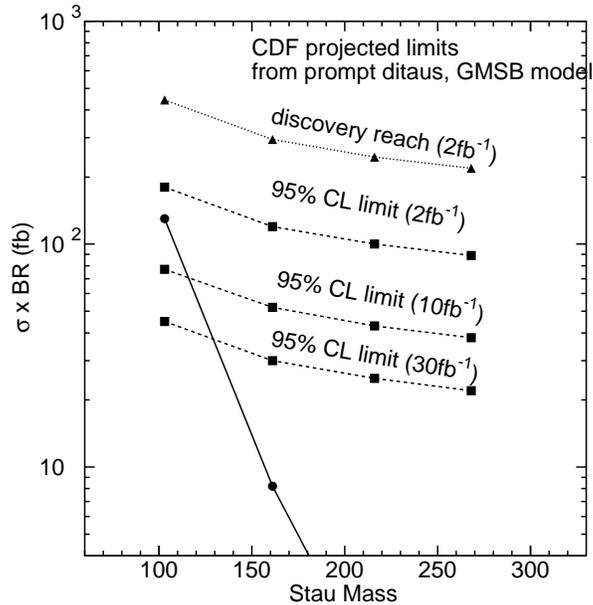}}
\caption{The limit on
cross section times branching ratio
from the CDF analysis of two $\tau$'s and $\missET$
applied to the Stau NLSP Model Line in Run II. The solid line is the
theoretical cross-section.}
\label{fig:gmsbcdftauii}
\end{figure}

Projecting this result in a straightforward way, we expect 40 events in
2~fb$^{-1}$ or a limit of 13 events at 95\% C.L. or 32 events at
$5\sigma$.
Using SHW, we find 3.6\% for the efficiency for a nominal model point
along the Stau NLSP Model Line which has $M_{\tilde \chi_1^\pm}=$193~GeV
and
$m_{\tilde\tau_1}=$103~GeV; see Table
\ref{tab:gmsbcdftauii}.
Due to the low efficiency for the $\tau$'s we do not expect to be
very sensitive to this model.  We might be able to set a limit
on the $\stauI$ near 100~GeV for 2~fb$^{-1}$,
similar to the sensitivity of the final LEP configuration.
For 30~fb$^{-1}$, we estimate we have a limit sensitivity near 130~GeV
and possibly a discovery sensitivity in the region of 100~GeV.
These results are shown in Figure \ref{fig:gmsbcdftauii}.
\begin{table}[htbp] 
\caption{The summary of the Monte Carlo points used to investigate
projected limits on the Stau NLSP Model Line for Run II (2~fb$^{-1}$) in
the CDF study.  Every event has two $\stauI$'s which decay promptly to
$\tau\GG$.}
\label{tab:gmsbcdftauii}
\renewcommand{\arraystretch}{1.75}
\begin{tabular}{|c|c|c|c|c|} \hline
$\Lambda$ (TeV)            & 42 & 66  & 88 & 109  \\ \
$m_{\tilde \chi_1^\pm}$ (GeV)            & 193 & 321  & 436 & 544  \\
$m_{\tilde \chi_1^0}$ (GeV)            & 109 & 176  & 237 & 296  \\
$m_{\tilde\tau_1}$ (GeV)         & 103 & 161  & 216 & 268  \\ \hline
$\sigma\times$BR (fb)      & 130 &  8.2 & 1.1 & .22  \\ \hline
$A\cdot \epsilon$ (\%)     & 3.6 &  5.4 & 6.5 & 7.3  \\ \hline
$\sigma\times$BR 95\% C.L. limit (fb)& 180 &  120 &  100 &   89  \\ \hline
\end{tabular}
\end{table}

This projection is based on data from a $\missET$ trigger and
one of the obvious potential improvements is to trigger on the $\tau$'s
directly.  Although we don't expect much more sensitivity from reducing
the $\missET$ requirements in this model, other searches, such as
$H^\pm\rightarrow \tau\nu$ and $A\rightarrow \tau^+\tau^-$, might
take advantage of this capability.  In addition we are attempting to
be sensitive to as many signatures as possible, with or without models.

\begin{figure}[tpb]
\centerline{\epsfysize=3.5in\epsfbox{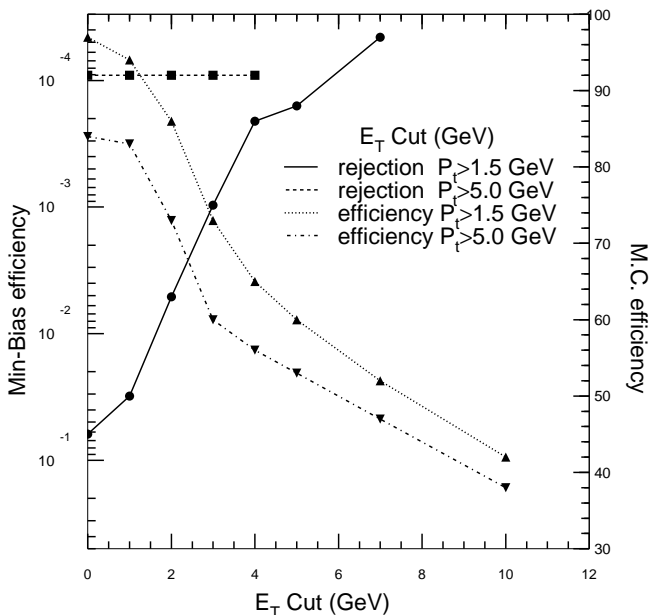}}
\caption{The rejection for CDF min--bias events and the efficiency
for the nominal stau model point ($m_{\tilde\tau_1}=103$~GeV).
The requirements are two trigger $\phi$ sectors with the indicated
requirements on minimum track $p_T$ and trigger tower $E_T$.
For a successful level 1 trigger, the rejection should be on the
order of $2 \times 10^{-4}$.}
\label{fig:gmsbcdftaut}
\end{figure}

A preliminary look at this trigger is shown in
Figure~\ref{fig:gmsbcdftaut}.  
This Figure displays the
rejection for min-bias events and efficiency for the nominal
point for this gauge-mediated model ($m_{\tilde\tau1}=$103~GeV).
The requirements are placed on the L1 calorimeter trigger.
For each trigger tower (twice as large in $\eta$ as a physical tower),
this trigger compares tracks, found in the $r-\phi$ plane, to
either the $EM$ or total tower energies.
We have plotted the rejection and efficiency
versus the tower energy requirement for different cases of
minimum track $p_T$.
We require two $\phi$ sectors pass the cuts
(indicating two $\tau$'s are present).
The L1 trigger must be a few percent of the level 1 trigger rate
to be acceptable, so the rejection should be near
$2\times 10^{-4}$.  We find we can achieve this rejection with the
requirement of two towers with $E_T>6$~GeV, each associated with tracks with
$p_T>1.5$~GeV.  Raising the $p_T$ requirement does not improve
the rejection rapidly.  These cuts are approximately 50\% efficient for
the gauge-mediated model.

Work is underway to understand how to achieve the needed rejection
at trigger level 2.  At this level we have the additional ability
to require that the jet is ``thin'', which is characteristic of
$\tau$'s.  We can require the tracks are isolated from other tracks
in the event, and that the $\tau$ candidates are not
back--to--back, reducing the QCD background.
We anticipate being able to keep the track and tower
thresholds near the L1 thresholds.

Further studies show that a single $\tau$ trigger will be
significantly more difficult. Here we find we approach the
required rejection with tower $E_T$ cuts of approximately 12~GeV,
and the track requirement doesn't improve the rejection.

A more inclusive trigger is to simply require two tracks above
a $p_T$ threshold.  This trigger could not only find more di-$\tau$'s,
it could find the long-lived $\stauI$'s (or, for example, $\tilde t_1$'s)
which do not trigger the
muon system.  We find we can achieve the necessary rejection when we
require two tracks with $p_T>5$~GeV.
At trigger level 2, we will need to require that the tracks are
well--isolated.  Work to plan these triggers is continuing.

\subsection{\rm {\D0} study of Stau NLSP Model Line with prompt
$\stauI$ decays}

If the \tlsp\ is short-lived and decays in the vicinity of the production
vertex ({\it i.e.} with a decay distance $\gamma c\tau\lsim 10$~cm),
anomalous $\tau$ production will provide a visible signal in the \D0
detector.
Combining the $\stauI \rightarrow \tau \GG$ decays with
the $W^*/Z^*$ productions from the cascade decays of primary supersymmetric
particles, these events will give rise to \lllj\ and \lljj\ 
final states
from leptonic $\tau$ decays.

\D0 searched for gaugino pair production using the tri-lepton
signature~\cite{lllj} in Run~I. The lepton $p_T$ cut was typically
15~GeV for the leading lepton and 5~GeV for the non-leading leptons.
The analysis also had a small $\met$ requirement. The observable background
cross section was estimated to be around 13~fb. Most of these backgrounds are
due to Drell-Yan processes. We select the \lllj\ events using the following
criteria:
\begin{itemize}
  \item[1)] $p^{\ell_1}_T>15$~GeV, $p^{\ell_2}_T>5$~GeV, $p^{\ell_3}_T>5$~GeV;
  \item[2)] $\met >20$~GeV;
  \item[3)] At least one jet with $E^j_T>20$~GeV.
\end{itemize}
The Drell-Yan production, a major background source for the Run~I analysis,
is significantly reduced by the new jet requirement. The total observable
background cross section is estimated to be 0.3~fb assuming background
reduction factors of 10 from the jet requirement, 2 from the improved particle
identification, 2 from the higher $\met$ cut.

Same-charge di-lepton events are expected from a variety of processes 
which produce slepton pairs either directly or in decays, or in any
processes which produce Majorana fermions such
as neutralino pair or gluino
pair
production. They are also expected from processes with three or more
leptons in the final states, but only two are identified. This
final state is expected to have small backgrounds. Again without a magnetic
tracker, D\O\  had no analysis of this nature in Run~I.
Based on Monte Carlo studies for several supersymmetric models, we select
\lljjmet\ events using the following criteria:
\begin{itemize}
  \item[1)] Two same-charge leptons  with $p^\ell_T>15$~GeV;
  \item[2)] At least two jets with $E^j_T>20$~GeV;
  \item[3)] $\met >25$~GeV.
\end{itemize}
Events with three or more identified leptons are removed to make the sample
orthogonal to the \lllj\ sample. Since leptons are relatively soft in
$p_T$ for the new physics model we investigated using this selection,
the effect of charge confusion due to a limited tracking resolution
is neglected in this study.
The major backgrounds are: $W+{\rm jets}$ events with one of the jets
misidentified as a lepton, $t\bar{t}$ events with energetic leptons from
b-quark decays, and Drell-Yan ($WZ$, $ZZ$) events. The $W+{\rm jets}$
background is estimated using the number of $W+ 3j$ events observed in
Run~I, folded with ${\cal P}(j\to\ell)$, to be 0.2~fb. The $t\bar{t}$ and
Drell-Yan backgrounds are estimated using Monte Carlo to be 0.1 and 0.1~fb
respectively. Adding the three background sources together yields
a total observable background cross section of 0.4~fb.

\begin{figure}[tpb]
  \centerline{\epsfysize=3.0in\epsfbox{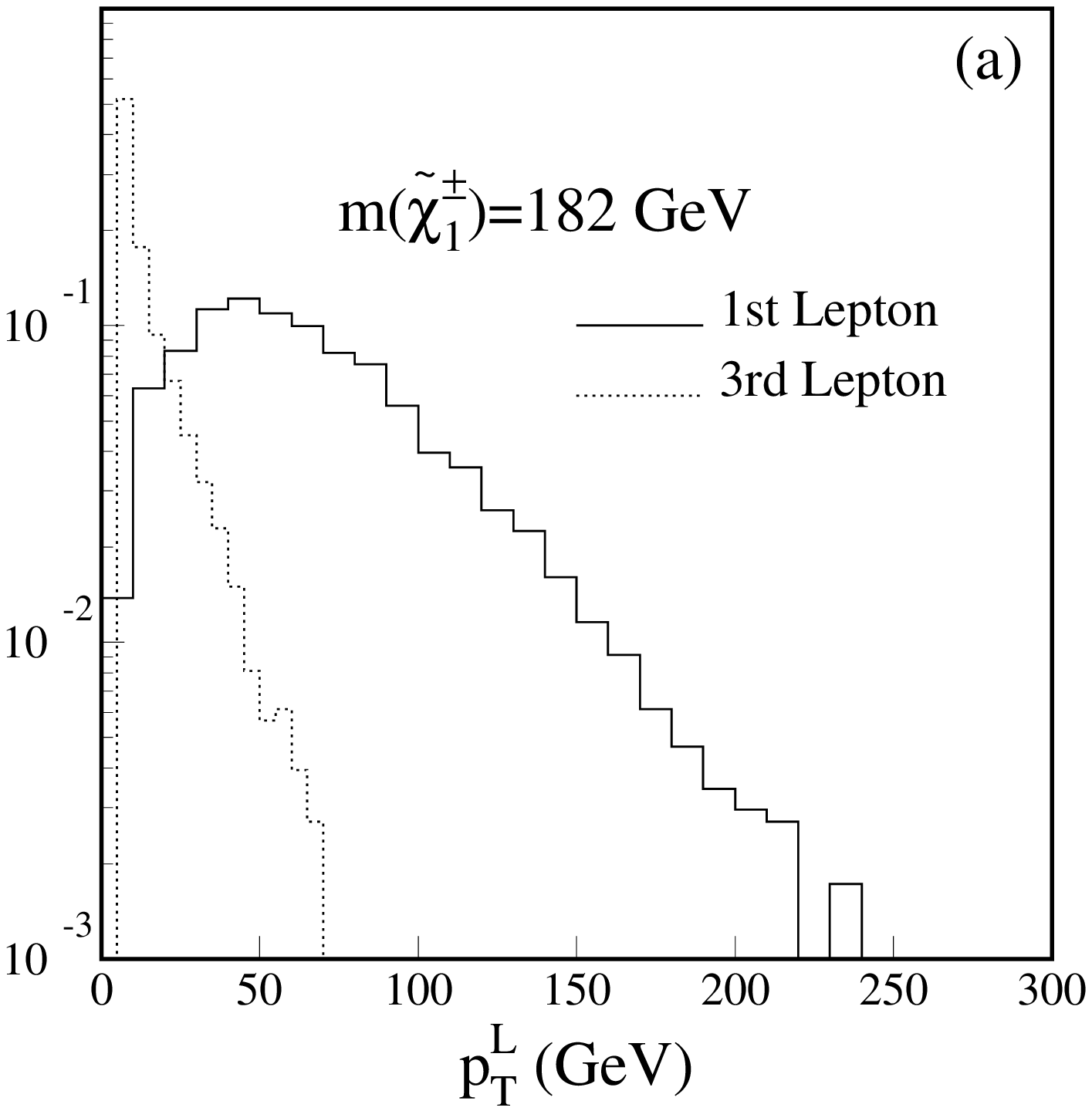}
              \epsfysize=3.0in\epsfbox{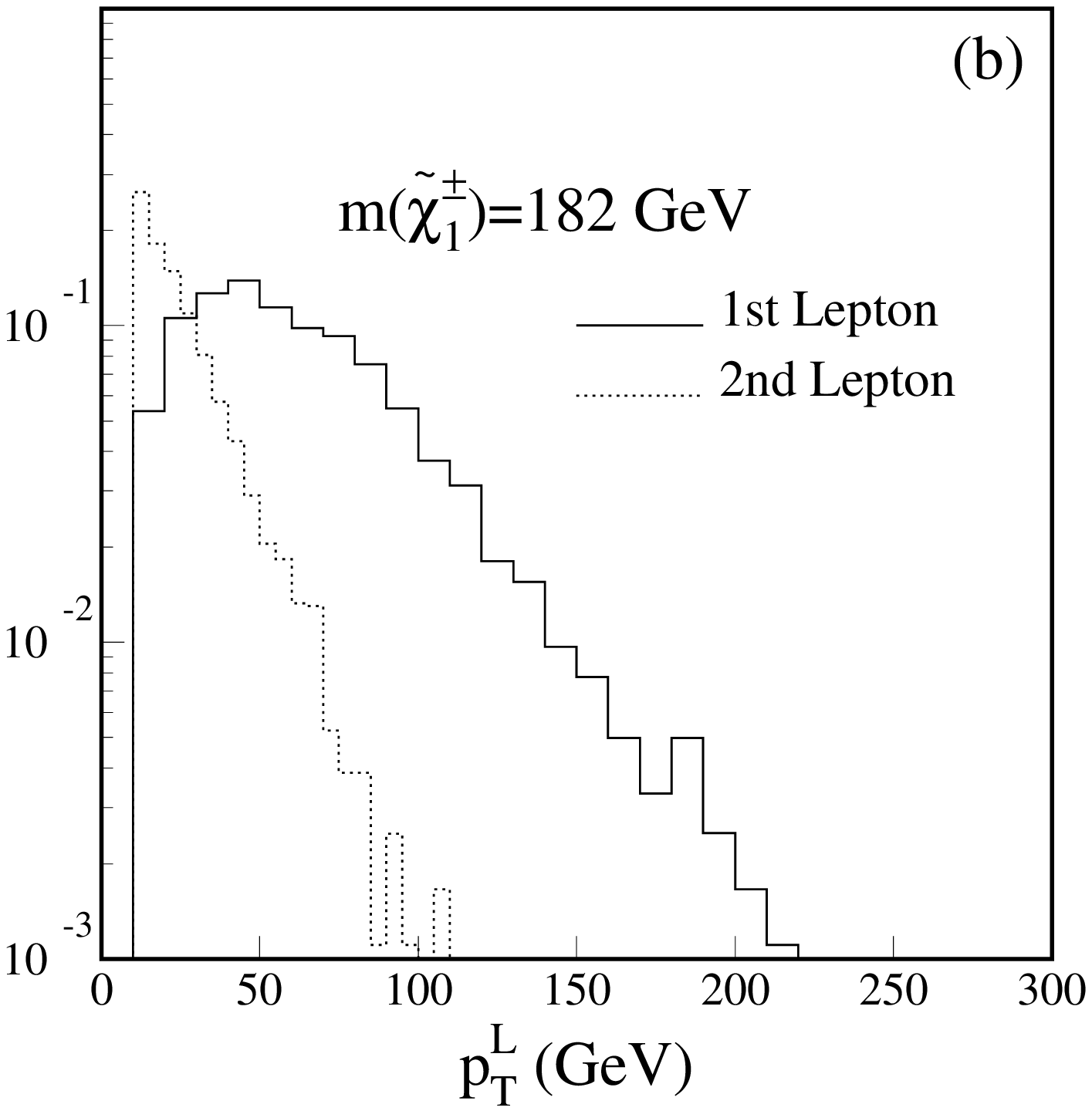}}
  \caption{Lepton $p_T$ distributions for (a) the \protect\lllj\ events and
           (b) the \protect\lljj\ 
events of the models with a short-lived
           \protect\tlsp\ for $m_{\tilde\chi^\pm_1}=182$~GeV
           ($\Lambda=40$~TeV) in the \D0 study of the Stau NLSP Model
Line. All distributions are
normalized to unit area.}
  \label{fig:p2}
\end{figure}

\begin{figure}[tpb]
  \centerline{\epsfysize=3.5in\epsfbox{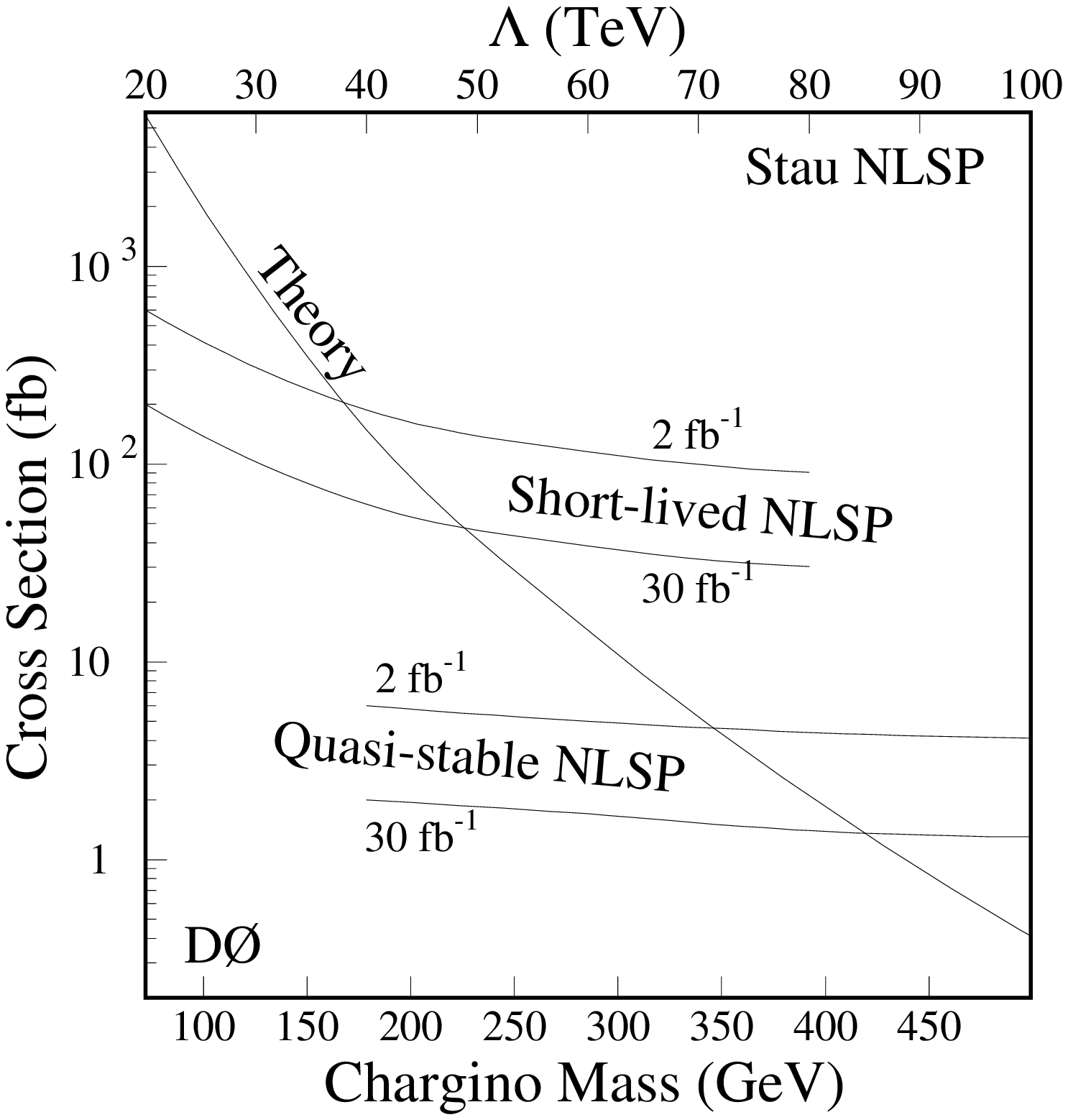}}
  \caption{The \D0 $5\sigma$ discovery cross section curves as functions
of
mass of the lighter chargino and the supersymmetry breaking scale
$\Lambda$ for the Stau NLSP Model Line,  along with the
theoretical cross sections. The $5\sigma$ curves are shown for both
short-lived NLSPs (combining \protect\lljj\ 
and \protect\lllj\ 
selections) and quasi-stable NLSP's
(\protect\lldedx\ 
selection) and for integrated luminosities of 2, 30~fb$^{-1}$.}
\label{fig:p2lim}
\end{figure}

The lepton $p_T$ distributions of the \lllj\  and \lljj\ 
events are shown in Fig.~\ref{fig:p2}.
Since most leptons are produced in $\tau$ decays, their
$p_T$'s are relatively soft. Table~\ref{tab:p2} shows the efficiencies of
the
\lllj\  and  \lljj\ 
selection criteria for these events along with the
theoretical cross sections, $\tilde\chi^\pm_1$ and $\tilde\tau_1$ masses.
Note that the \lllj\   and \lljj\ 
criteria are orthogonal. The efficiencies are relatively
small largely due to the small branching ratio of the events to tri-leptons.
We note that the total efficiencies shown in the table are somewhat
conservative. They do not take into account the migration of the \lllj\
events to the \lljj\ 
events due to inefficiency in the lepton identification.
The $5\sigma$ discovery curves are shown in Fig.~\ref{fig:p2lim}.
The lighter chargino with mass up to 160 and 230 GeV can be discovered for
\ldt=2, 30~fb$^{-1}$.
\begin{table}[htbp]
\caption{The $\tilde\chi^\pm_1$ and $\tilde\tau_1$ masses, theoretical 
cross sections, detection efficiencies of \protect\lljj\
and \protect\lllj\   selections, and significances for different values of
$\Lambda$ for the models with a short-lived $\tilde\tau_1$ as the NLSP in
the \D0 analysis. The relative statistical error on the efficiency is
typically 25\%. The combined \protect\lljj\   and  \protect\lllj\
background cross section is assumed to be 0.7~fb with a 20\% systematic
uncertainty.}
\label{tab:p2}
\renewcommand{\arraystretch}{1.65}
\begin{tabular}{|c|cccc|}\hline
     $\Lambda$ (TeV)              &   20 &  40 &  60  &  80 \\
     $m_{\tilde\chi^\pm_1}$ (GeV) &   72 & 182 & 289  & 394 \\
     $m_{\tilde\tau_1}$ (GeV)     &   54 &  99 & 147  & 196 \\ \hline
     $\sigma_{th}$ (fb)           & 5800 & 149 & 14.4 & 2.1 \\
\hline
     \protect\lljj\  $\epsilon$ (\%) &  --  & 0.6 & 1.0  & 1.3 \\
     \protect\lllj\  $\epsilon$ (\%) &  0.5 & 1.0 & 1.6  & 2.0 \\
     Total  $\epsilon$ (\%)         &  0.5 & 1.6 & 2.6  & 3.3 \\ \hline
     \protect\rsb\ (2 fb$^{-1}$)    &  48  & 4.0 & 0.6  & 0.1 \\
 ~~~~~    \protect\rsb\ (30 fb$^{-1}$) ~~~~~  & 140  &  12 & 1.8  & 0.3 \\
  \end{tabular}
\end{table}

It is often assumed that this analysis should benefit from a $\tau$
identification. However, it is not clear that it will have a dramatic
impact on the reach in the supersymmetry parameter space. Though a $\tau$
identification could improve the efficiency for the signal, it will
undoubtedly come with large backgrounds. Nevertheless, a $\tau$
identification is essential to narrow down theoretical models if an excess
is observed in the tri-lepton final state.

\subsection{\rm ISAJET study of Stau NLSP Model Line with prompt
$\stauI$ decays}

Tevatron signals for the Stau NLSP Model Line parameters
(\ref{staulineparameters}) have also been simulated using
ISAJET~\cite{isajet}. In this study, we only consider $\Lambda
\geq 35$~TeV.
[For $\Lambda \lsim 34$~TeV, $\tz_1\to \tau\ttau_1$ is kinematically
forbidden, and $\tz_1$ would decay via the four body decay $\tz_1 \to
\nu_{\tau}\ttau_1 W^*$ (which is not yet included in ISAJET) or via its
photon mode considered above.]
Gluinos and squarks are then too heavy to be produced at the
Tevatron, and sparticle production is dominated by chargino, neutralino
and slepton pair production. We expect that SUSY events will contain tau
leptons from sparticle cascade decays which are expected to result in a
tau slepton which then decays via $\ttau_1\to \tau \tG$. The observability
of SUSY realized as in this scenario thus will benefit from the capability
of experiments to identify hadronically decaying tau leptons, and further,
to distinguish these from QCD jets.  We classify SUSY events by the number
of identified taus, and further separate them into jetty and clean event
topologies labeled by the number of isolated leptons ($e$ and $\mu$). It
should be remembered that the efficiency for identifying taus is expected
to be smaller than for identifying photons. First, the tau has to decay
hadronically, and then the hadronic decay products have to form a jet.

We simulate this scenario using the toy ISAJET calorimeter described
above. To model the experimental conditions at the Tevatron the toy
calorimeter package ISAPLT is interfaced with ISAJET. The particle
identifications used in ISAPLT for photons, jets, electrons, and muons,
are described in section \ref{binoisajet}. In addition, $\tau$'s are
identified as narrow jets with just one or three charged prongs with
$p_T>2$~GeV within $10^{\circ}$ of the jet axis and no other charged
tracks in a 30$^{\circ}$ cone about this axis. The invariant mass of these
tracks is required to be $\leq m_{\tau}$ and the net charge of the three
prongs required to be $\pm 1$. QCD jets with $E_T =15 (\geq 50)$~GeV are
misidentified as taus with a probability of 0.5\% (0.1\%) with a linear
interpolation in between. In our analysis, we require $\eslt \geq 30$~GeV
together with at least one of the following which serve as a trigger for
the events: \begin{itemize} \item one lepton with $p_T(\ell) \geq 20$~GeV,
\item two leptons each with $p_T(\ell) \geq 10$~GeV, \item $\eslt \geq
35$~GeV. \end{itemize}

The dominant physics sources of
SM backgrounds to $n$-jet + $m$-leptons + $\eslt$ events, possibly
containing additional taus, are $W$,$\gamma^*$ or $Z$ + jet
production, $t\bar{t}$ production and vector boson pair
production. Instrumental backgrounds that we have attempted to estimate
are $\eslt$ from mismeasurement of jet energy and mis-identification of
QCD jets as taus.

In addition to our basic requirements above, we also impose:
\begin{itemize}
\item a veto on opposite-charge, same-flavor dilepton events with
$M_Z-10~{\rm GeV} \leq  m(\ell\bar{\ell}) \leq M_Z + 10~{\rm GeV}$ to
remove backgrounds from $WZ$ and $ZZ$ and high $p_T$ $Z$ production, and
\item for dilepton events, we require $\Delta\phi(\ell\bar{\ell'}) \leq
150^{\circ}$
($\ell,\ell'=e,\mu,\tau$) to remove
backgrounds from $Z \to \tau\bar{\tau}$ events.
\end{itemize}

We have checked that even after these cuts and triggers, SM backgrounds
from $W$ production swamp channels with no leptons or just one
identified lepton ($e$, $\mu$ or $\tau$). The former is the canonical
$\eslt$ signal, which after optimizing cuts, may be observable at Run 2
if gluinos are lighter than $\sim 400$~GeV. We do not expect that this
signal from gluino and squark production will be detectable since
$m_{\tg} =578$~GeV (with squarks somewhat
heavier) even for $\Lambda=35$~TeV. For this reason, and because there are
large single lepton
backgrounds from $W$ production, we focus on signals with two or more
leptons in our study. Also, because the presence of $\tau$'s is the
hallmark of this scenario, we mostly concentrate on leptonic events with
at least one identified $\tau$.

We begin by considering the signal and background cross section for
clean events. These are shown in Table \ref{clean}. 
Events are classified first by the number of identified taus, and then by
the lepton multiplicity; the $C$ in the topology column denotes
``clean'' events.  For
each topology, the first row of numbers denotes
the cross sections after the basic acceptance cuts and trigger
requirements along with the $Z$ veto and the $\Delta\phi$ cut discussed
above. 
\begin{table}[htbp]
\caption[]{SM background cross sections in fb for various clean
multilepton topologies from $W$, $Z \to \tau\tau$, $VV$ ($V=W,Z$) and
$t\bar{t}$ production at a 2~TeV $p\bar{p}$ collider, together with ISAJET
signal cross sections for the $\Lambda=40$~TeV and $\Lambda= 50$~TeV
points on the Stau NLSP Model Line described in the text. For each event
topology, the first number denotes the cross section after the basic
acceptance cuts and trigger requirements along with the $Z$ veto and the
$\Delta\phi$ cut discussed in the text. The second number is after the
additional cut, $p_{Tvis}(\tau_1) \geq 40$~GeV, for events at least one
identified $\tau$. The entries labeled $Total^*$ are the sum of all the
cross sections except those in the $1\tau 1\ell$ channel.  The last two
rows provide a measure of the statistical significance of the signal.}
\label{clean}
\smallskip
\renewcommand{\arraystretch}{1.55}
\begin{tabular}{|c|cccc|cc|}\hline
Topology & $W$ & $Z\to \tau\tau$ & $VV$ & $t\bar{t}$ & $\Lambda=40$~TeV
&$\Lambda=50$~TeV \\
& & & & & $m_{\tilde \chi_1^\pm}=183 $~GeV
&$m_{\tilde \chi_1^\pm} = 236$~GeV \\
\tableline
$C 3\ell$ & 0 & 0 & 0.39 & 0 & 0.68 & 0.24 \\
         & 0 & 0 & 0.39 & 0 & 0.68 & 0.24 \\
$C 1\tau 1\ell$ & 1045 & 4.2 & 36 & 0.044 & 8.6 & 1.96 \\
               &  43  & 2.0 & 10.8 & 0 & 5.3 & 1.27 \\
$C 1\tau 2\ell$ & 0 & 0.57 & 1.4 & 0 & 3.3 & 0.93 \\
               & 0 & 0.045 &    0.43 & 0  &  1.9  & 0.59 \\
$C 1\tau 3\ell$ & 0 & 0 & 0 & 0 & 0.31 & 0.16  \\
               & 0 & 0 & 0 & 0 & 0.21 & 0.10  \\
$C 2\tau 1\ell$ & 0 & 1.5 & 1.2 & 0 & 4.1 & 1.2 \\
               & 0    & 0.57 & 0.79 &0 & 3.3 & 1.02 \\
$C 2\tau 2\ell$ & 0 & 0 & 0 & 0 & 0.36 & 0.23 \\
               &  0 & 0 & 0 & 0 & 0.33 & 0.22 \\
\hline
$Total^*$      & 0 & 2.1 & 2.99 &  0 & 8.75 & 2.76 \\
               & 0   & 0.62 & 1.61 & 0 & 6.42 & 2.17 \\
\hline
$\sigma(sig)/\sqrt{\sigma(back)}$~(fb$^{1/2}$) & & & & &3.87 & 1.22\\
                        & & & & & 4.30 & 1.45
\end{tabular}
\end{table}
We see that there is still a substantial background in several of the
multilepton channels. This background can be strongly suppressed, with
modest loss of signal by
imposing an additional requirement,
\begin{itemize}
\item $p_{Tvis}(\tau_1) \geq 40$~GeV,
\end{itemize}
on the visible energy of the hardest tau in events with at least one
identified tau. In the background, the $\tau$'s typically come from
vector boson decays, while in the signal a substantial fraction of these
come from the direct decays of charginos and neutralinos that are
substantially heavier than $M_Z$
[even for $\Lambda =40$~(50)~TeV),
$m_{\tz_1}= 103$~(132)~GeV]. Thus signal taus
pass this cut more easily. A few points about this Table are worth
mentioning.
\begin{enumerate}
\item The signal cross sections in each channel are at most a few fb,
and with an integrated luminosity of 2~fb$^{-1}$, the individual signals
are below the $5\sigma$ level even for $m_{\tilde \chi_1^\pm} = 183$ GeV
($\Lambda=40$ TeV). It is clear
that with the luminosity expected at the MI we will be forced to add the
signal in various channels and see if this inclusive signal is
observable.

\item The sum of the signal in all the channels in Table \ref{clean}, except
the $1\tau 1\ell$ channel which has a very large background, is shown in
the next two rows with and without the $p_T$ cut on the $\tau$, while
in the last two rows we list $\sigma(sig)/\sqrt{\sigma(back)}$.
We see that a somewhat better significance is obtained after the
$p_{Tvis}(\tau_1) \geq 40$~GeV cut.

\item We see that the inclusive SUSY signal in the clean channels for
the $m_{\tilde \chi_1^\pm} = 183$ GeV ($\Lambda =40$ TeV) case should be
detectable with
the Run II
integrated luminosity, whereas for the
$m_{\tilde \chi_1^\pm} = 236$ GeV ($\Lambda =50$ TeV) case an
integrated luminosity of 12~fb$^{-1}$ is needed for a 5$\sigma$ signal.

\item We caution the reader that about 25-30\% of the $\tau$ background
comes from mis-tagging QCD jets as taus (except, of course, for the $W$
backgrounds and the backgrounds in the $C2\tau \ell$ channels
which are almost exclusively from these fake taus). Thus
our estimate of the background level is somewhat dependent on the $\tau$
faking algorithm we have used. The signal, on the other hand, almost
always contains only real $\tau$'s, so that improving the discrimination
between $\tau$ and QCD jets will lead to an increase in the projected
reach of these experiments.

\item In particular channels the background is completely dominated by
fake taus. For instance, after the $p_{Tvis}(\tau)$ cut, the
$C 1\tau 1\ell$ background from $W$ sources of just {\it
real taus} is only 1.9~fb, while the signal and other backgrounds
remain essentially unaltered from the cross sections in Table \ref{clean}.
Thus if fake $\tau$ backgrounds can be
greatly reduced, it may be possible to see signals in additional channels.

\end{enumerate}

Next, we turn to jetty signals for the Stau NLSP Model Line. Cross
sections
for
selected signal topologies together with SM backgrounds {\it after} the
$p_T(\tau_1) \geq 40$~GeV cut are shown in Table \ref{jetty}. The other
topologies appear to suffer from large SM backgrounds and we have not
included them here.
\begin{table}[htbp]
\caption{SM background cross sections in fb for various jetty multilepton
topologies from $W$, $Z\to \tau\tau$, $VV$ ($V=W,Z$) and $t\bar{t}$
production at a 2 TeV $p\bar{p}$ collider, together with ISAJET signal
cross sections for $m_{\tilde \chi_1^\pm} = 183$ GeV ($\Lambda=40$ TeV)
and $m_{\tilde
\chi_1^\pm} =
236$ GeV ($\Lambda= 50$ TeV) for the Stau NLSP Model Line. The cross
sections are with all the cuts including the $p_{Tvis}$ cut on the hardest
$\tau$.}
\label{jetty}
\bigskip
\renewcommand{\arraystretch}{1.55}
\begin{tabular}{|c|cccc|cc|}\hline
Topology & $W$ & $Z\to \tau\tau$ & $VV$ & $t\bar{t}$ & $\Lambda=40$~TeV
& $\Lambda=50$~TeV \\
& & & & & $m_{\tilde \chi_1^\pm}=183 $~GeV
&$m_{\tilde \chi_1^\pm} = 236$~GeV \\
\tableline
$J3\ell$ & 0 & 0.019 & 0.28 & 0.3 & 1.06 & 0.35 \\
$J1\tau 2\ell$ & 0 & 0.19 & 0.29 & 1.2 & 1.92 & 0.79 \\
$J2\tau 1\ell$ & 0.11 & 0.79 & 0.41 & 0.8 & 2.25 & 1.18 \\
$J2\tau 2\ell$ & 0 & 0 & 0 & 0 & 0.31 & 0.22 \\
\hline
$Total$    & 0.11  & 1.0 & 0.98 & 2.3 & 5.53& 2.54 \\
\hline
$\sigma(sig)/\sqrt{\sigma(back)}$~(fb$^{1/2}$) & & & & &2.64 & 1.21 \\
\hline
\end{tabular}
\end{table}

The following features are worth noting:
\begin{enumerate}
\item We see that Stau NLSP Model Line results in smaller cross
sections in
jetty channels. This should not be surprising since electroweak
production of charginos, neutralinos and sleptons are the dominant SUSY
processes, and because staus are light, branching fractions for hadronic
decays of $\tw$ and $\tz$ tend to be suppressed.

\item We see from Table \ref{jetty} that with the present set of cuts,
not only is the signal below the level of observability in any one of
the channels, the inclusive signal is not expected to be observable at
the MI even for the $m_{\tilde \chi_1^\pm} = 183$ GeV ($\Lambda = 40$ TeV)
case. With an
integrated luminosity
of 25 fb$^{-1}$ the signal for the $m_{\tilde \chi_1^\pm} = 236$ GeV
($\Lambda=50$ TeV
case is
observable at the $6\sigma$ level.

\item As for the clean lepton case, a significant portion of the
background comes from QCD jets faking a tau. The fraction of events with
a fake tau varies from channel to channel, but for the $2\tau 1\ell$
channel in Table \ref{jetty} almost 60\% of the background involves at
least one fake $\tau$ in contrast to essentially none of the signal.

\item A major background to the jetty signal comes from $t\bar{t}$
production. To see if we could enhance the signal relative to this
background we tried to impose additional cuts to selectively reduce the
top background. Since top events are expected to contain hard jets, we
first tried to require $E_T(j) \leq 50$~GeV. We also, independently,
tried vetoing events where the invariant mass of all jets exceeded
70~GeV. While both attempts lead to an improvement of the signal to
background ratio, the statistical significance of the signal is not
improved (and is even degraded). We do not present numbers for this for
the sake of brevity. It may be possible to reduce the top background by
vetoing events with identified $b$-jets, but we have not attempted to do
so here.

\end{enumerate}

\begin{figure}[tbp]
\centering
\epsfxsize=3.9in
\hspace*{0in}
\epsffile{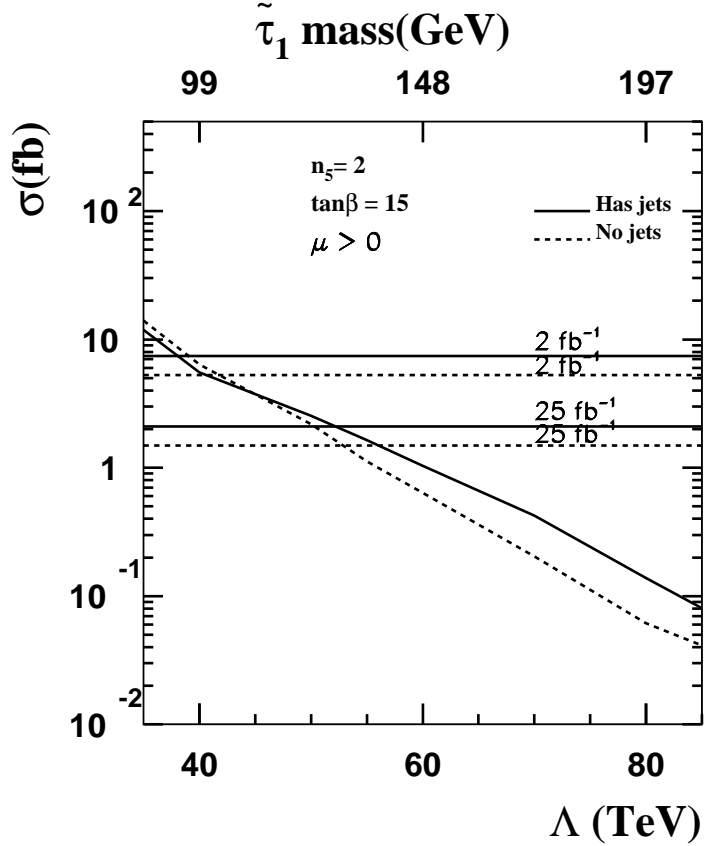}
\caption{Signal cross sections after all cuts versus $\Lambda$ and
$m_{\tilde\tau_1}$ for the
total signal in the clean (dashed line) and jetty (solid line) channels
in the ISAJET study of the Stau NLSP Model Line.
The corresponding horizontal lines denote the minimum cross
section for a $5\sigma$ signal for 2 fb$^{-1}$ and 25 fb$^{-1}$.}
\label{isajetstaunlspfig}
\end{figure}

For the Stau NLSP Model Line, it appears that experiments at the MI should
be able to probe $\Lambda$ values up to just beyond 40~TeV, corresponding
to $m_{\tilde \chi_1^\pm} = 183$ GeV, in the inclusive clean multilepton
channels. It
appears, however, that it will be essential to sum up several channels to
obtain a signal at the $5\sigma$ level. Confirmatory signals in inclusive
jetty channels may be observable at the $3.7\sigma$ level. Of course, for
an integrated luminosity of 25 fb$^{-1}$ it may be possible to probe
$\Lambda=50$~TeV ($m_{\tilde \chi_1^\pm} = 236$ GeV) even in the
unfavored jetty
channels, and somewhat beyond in the clean channels.  The situation is
summarized in Fig.~\ref{isajetstaunlspfig} where we show the signal cross
sections summed over the selected channels for events without jets
(dashed) and for events with jets (solid).
The horizontal lines denote the minimum cross section needed for the
signal to be observable at the $5\sigma$ level. We note that, in some
channels, a substantial fraction of background events come from QCD jets
faking a tau --- our assessment of the Run II reach is therefore sensitive
to our modeling of this jet mis-tag rate. By the same token, if this rate
can be reduced, the reach may be somewhat increased.

\subsection{\rm CDF study of quasi-stable staus in the Stau NLSP Model
Line}
\label{subsec:CDFqsstau}

CDF has searched for long-lived massive charged
particles in Run I \cite{gmsbcdfstuart,gmsbcdfrlc}.
The inclusive muon trigger data sample (90~pb$^{-1}$)
is searched for tracks with $p>35$~GeV, $|\eta|<1.0$, $dE/dx$
implying $\beta\gamma<0.85$,
and a mass, calculated from the $dE/dx$ and $p_T$,
greater than 60~GeV.
Since the $\stauI$ should be isolated,
cuts requiring less than 4~GeV in a cone of 0.4 in
the calorimeter and track isolation in the CTC are imposed.
Two events pass all cuts while $0.85\pm 0.25$
are expected.  Using a model quite similar to the Stau NLSP Model Line,
and allowing all SUSY production,
we find the limit is approximately a factor of 6 away from excluding
a 100~GeV $\stauI$.

For the Run II projection, we assume we are still using the muon triggers
so one of the $\stauI$'s must pass the trigger. The limit could
potentially be improved by including electron and $\missET$ triggers. For
the offline efficiency, we require $|\eta|<1.0$, $p>p_{\rm cut}$,
$\beta\gamma<0.85$ and isolation.  We take $p_{\rm cut} = 35$ GeV to start
with, and the effects of changing this cut will be discussed shortly. Once
the event passes the trigger, either $\tilde \tau$ may pass the offline
cuts. As defined by the Stau NLSP Model Line, we vary $\Lambda$ to find
the
mass dependence.

We estimate the cross section for the background will go up by 20\% due to
the higher energy. From the Run I data, we estimate the background
momentum distribution and for the 2~TeV projection, scale it up in
momentum by 5\%. The predicted background is 25 events for
$m_{\tilde\tau_1}=100$~GeV. The 95\% C.L. point is at $2\sigma$ or 10
events and
the $5\sigma$ discovery point is 25 events. We have assumed that the
larger $m_{\tilde\tau_1}$ is, the larger we can set the minimum mass
cut,
which greatly reduces the background for larger masses.

\begin{figure}[tpb]
\centerline{\epsfysize=3.5in\epsfbox{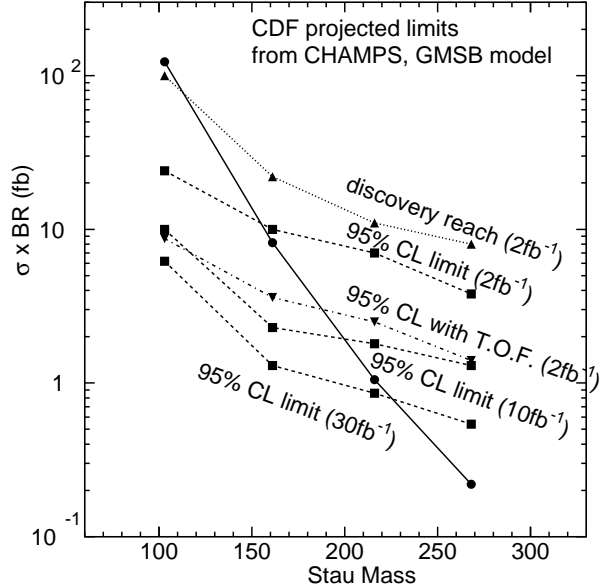}}
\caption{The projected CDF limits on the cross-section times branching
ratio for a long-lived stau in the Stau NLSP Model Line. The SUSY
production is gauginos and
sleptons. All sparticles decay to final states involving the $\stauI$
NLSP. The limit is the 95\% C.L. point assuming no signal and the
discovery reach is a signal 5$\sigma$ over background. The time-of-flight
projection assumes a 100 ps timing resolution. The other projections
assume only the $dE/dx$ in the COT and SVX and an isolation cut. Adding
timing information from other detectors or increasing the momentum cut
could increase sensitivity.}
\label{fig:gmsbcdfchamplimits}
\end{figure}

Table \ref{tab:gmsbcdfchampeff} and Figure \ref{fig:gmsbcdfchamplimits}
summarize our conclusions on the reach in Run II.
If we see no signal and set a limit it should be
on the order of 150~GeV in the $\stauI$ mass.
We should be sensitive to a $5\sigma$ signal out to $m_{\tilde
\tau_1} = 110$~GeV.
For higher luminosities, we scale the signal and backgrounds.
As shown in Figure \ref{fig:gmsbcdfchamplimits}, the limits
are extended to $m_{\tilde \tau_1} =  200$~GeV for 10~fb$^{-1}$ and
$m_{\tilde \tau_1} =  225$~GeV for 30~fb$^{-1}$.
\begin{table}[htbp]
\caption{Results of the CDF Monte Carlo study for four points in
the Stau NLSP Model Line with prompt $\stauI$ decays.
The cross section is the total SUSY cross section for
gauginos and sleptons. The integrated luminosity is assumed to be
2 fb$^{-1}$.}
\label{tab:gmsbcdfchampeff}
\renewcommand{\arraystretch}{1.75}
\begin{tabular}{|c|c|c|c|c|} \hline
$\Lambda$ (TeV)            & 42 & 66  & 88 & 109  \\ \
$m_{\tilde \chi_1^\pm}$ (GeV)            & 193 & 321  & 436 & 544  \\
$m_{\tilde \chi_1^0}$ (GeV)            & 109 & 176  & 237 & 296  \\
$m_{\tilde \tau_1 }$ (GeV)         & 103 & 161  & 216 & 268  \\ \hline
$\sigma$ (fb)          & 123& 8.2 & 1.05 & 0.22\\ \hline
$A\cdot \epsilon$ (\%)        & 21   & 25     & 31      & 40   \\ \hline
signal events   & 52   & 4      & .65     & .18  \\
\hline
$N_{bg}$     & 25  &  1.7 &   0.6 &   0.3  \\ \hline
95\% C.L. limit (fb)    & 24.0 & 10.0 & 7.0 & 4.0 \\ \hline
$5\sigma$ discovery (fb)& 100.0 & 22.0 & 11.0 & 8.0 \\ \hline
\end{tabular}
\end{table}

This projection is conservative since there are several methods
to improve the sensitivity.
We can easily check
if increasing the momentum cut can increase our sensitivity
and the results are in Table~\ref{tab:gmsbcdfpcut}.
We conclude that the sensitivity could be significantly increased
by increasing $p_{cut}$.  We do not include this result in the
prediction of the limits since it depends strongly on the
model of the background momentum distribution which could have tails
that degrade this result.
\begin{table}[htbp]
\caption{The effect of increasing the momentum cut on the $\stauI$
candidates in the CDF study of quasi-stable stau NLSPs, for $\Lambda =
42$ TeV and $m_{\tilde \chi_1^\pm} = 193$ GeV, with
2 fb$^{-1}$.}
\label{tab:gmsbcdfpcut}
\renewcommand{\arraystretch}{1.7}
\begin{tabular}{|c|c|c|c|} \hline
$p_{cut}$ (GeV) & $N_{\rm signal}$ & $N_{\rm bg}$ & $N_{\rm
signal}/\sqrt{N_{\rm bg}}$ \\ \hline 
35  & 52 &  25  & 10  \\ \hline
45  & 45 &   9  & 15  \\ \hline
55  & 35 &   3  & 20  \\ \hline
65  & 25 &   1  & 25  \\ \hline
75  & 12 & 0.4  & 8  \\ \hline
\end{tabular}
\end{table}

Another way to improve the sensitivity might be to require two heavy
charged particles in the event. While this might reduce the signal by a
factor of approximately three, the background would be zero (the fake
$dE/dx$ rate is $3\times 10^{-4}$ per track). Requiring the COT, hadron
calorimeter, or muon scintillator to show a late hit consistent with the
$dE/dx$ measurement could make dramatic improvements.

Since this model has light $\tilde \ell$'s and $W$'s in the cascade decays
of $\CI$ and $\NII$, there should be a significant number of leptons in
these events. Another way to improve the sensitivity would be to require
an $e$ or second $\mu$ in the event. The background would be greatly
reduced. As a simpler alternative, the additional lepton (or $\nu$ through
$\missET$) could supply the trigger to increase the efficiency.

The CDF Run II detector includes a new
time of flight (TOF) system. The system consists of
scintillators installed around the outside of the COT and read out with
precise timing.  The primary purpose is for $K/\pi$ separation in
$B$ physics but it will be very useful for this analysis too.

With 100~ps timing resolution (including the uncertainty on the
interaction time), we expect we could require $4\sigma$ separation at
$\approx$400~ps, which is $\beta\gamma<2.26$, or $p<$235~GeV for our
baseline 103~GeV $\stauI$.  This greatly increases the acceptance compared
to the $\beta\gamma<0.85$ we had to require in the baseline analysis.

\begin{figure}[tpb]
\centerline{\epsfysize=3.5in\epsfbox{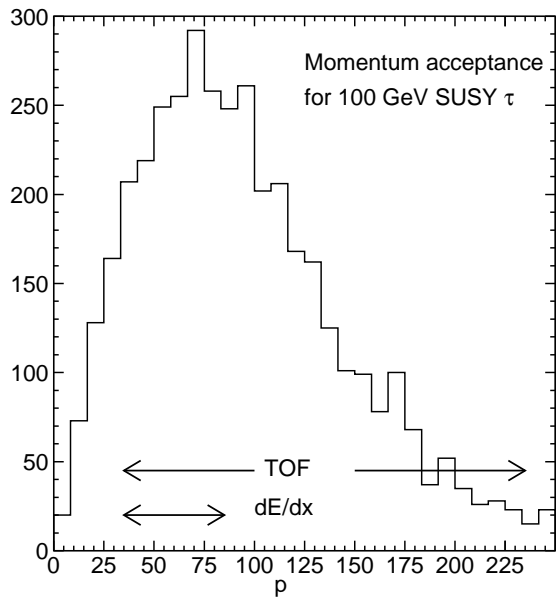}}
\caption{The large increase in momentum acceptance with the proposed CDF
time-of-flight system installed.}
\label{fig:gmsbcdftofp}
\end{figure}

The $\beta\gamma<2.26$ requirement has a total efficiency of 58\% compared
to the 21\% for the baseline analysis for the case $m_{\tilde
\chi_1^\pm} = 193$ GeV. We would expect 144 signal
events instead of 52.
Figure~\ref{fig:gmsbcdftofp} shows this significant increase in the
sensitive momentum range we could expect with a time-of-flight (T.O.F.)
system.

We can combine this result with the SUSY cross section to
arrive at the cross section limits show in
Figure \ref{fig:gmsbcdfchamplimits}.
The T.O.F. system could extend our limit sensitivity by 40~GeV
out to $m_{\tilde \tau_1} = 190$~GeV and our $5\sigma$ discovery
sensitivity out to $m_{\tilde \tau_1} =150$~GeV.
In this scenario our sensitivity is significantly greater than LEP.
However the limit for direct production of the $\stauI$ would
still be below the LEP reach.
The search for massive
stable strongly interacting 
charge particles, such as NLSP squarks discussed 
in section \ref{sec:squarknlsp}, will also be
significantly improved with the TOF detector.

\subsection{\rm {\D0} study of quasi-stable stau signals in the Stau NLSP
Model Line}
\label{subsec:D0qsstau}

If the \tlsp\ has a long lifetime (quasi-stable) and decays outside the
detector ($\gamma c\tau$ greater than $\sim 3$~m),
it can appear in the detector
as a slowly moving charged particle with
large ionization energy losses.
The signature
is, therefore, two high $p_T$ ``muons" with large $dE/dx$ values.

Though D\O\  had several di-lepton
analyses in Run~I, none of these can be extrapolated to Run~II,
thanks to the replacement
of the central tracker. Based on the expected signatures of several
supersymmetric models with heavy stable charged particles discussed below,
we select high $p_T$ di-lepton events (\lldedx) with large $dE/dx$ loss
using the following requirements:
\begin{itemize}
  \item[1)] At least two ``leptons" with $p^\ell_T>50$~GeV;
  \item[2)] $M_{\ell\ell}>150$ GeV;
  \item[3)] At least one ``lepton" passing the $dE/dx$ requirement.
\end{itemize}
The di-lepton mass requirement is intended to reduce Drell-Yan backgrounds.
The principal backgrounds are: QCD dijet events with jets misidentified as
leptons, $t\bar{t}$, and Drell-Yan events. Using ${\cal P}(j\to\ell)=10^{-4}$
and the assumed rejection factor of the $dE/dx$ cut for the MIP particles,
the observable background cross sections are estimated to
be 0.1~fb from QCD dijet, 0.2~fb from $t\bar{t}$ events,
and 0.2~fb from Drell-Yan processes. The QCD dijet cross section for
$p_T>50$~GeV is assumed to be 1~$\mu$b in the estimation.
The total observable cross section is therefore 0.5~fb for the above selection.

\begin{figure}[tpb]
  \centerline{\epsfysize=3.0in\epsfbox{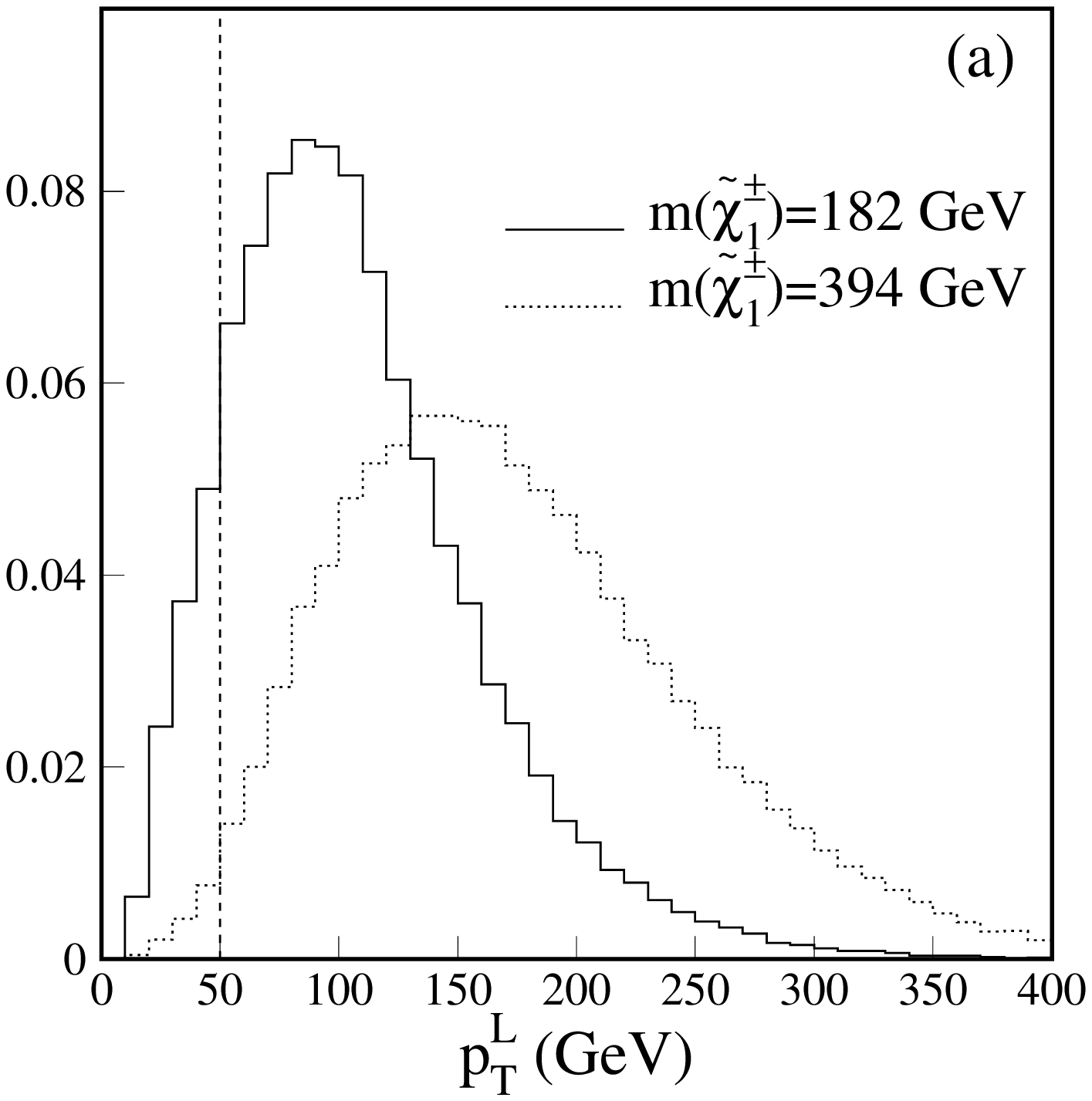}
              \epsfysize=3.0in\epsfbox{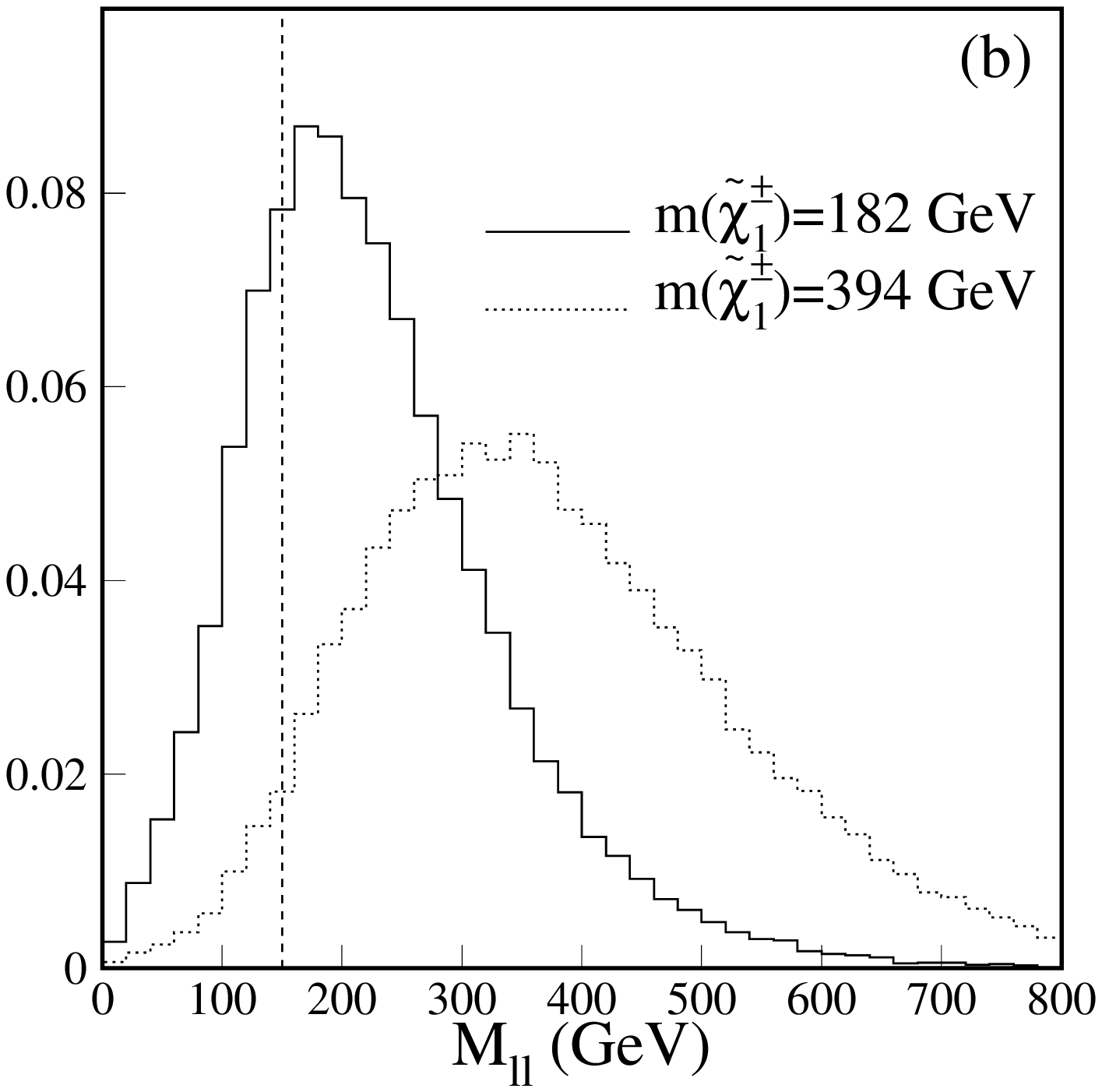}}
  \caption{The ``lepton" $p_T$ (a) and di-``lepton" mass $M_{\ell\ell}$
(b)
           distributions for the models with a quasi-stable \protect\tlsp\
           as the NLSP for $\Lambda=40,\ 80$~TeV
           ($m_{\tilde\chi^\pm_1}=182,\ 394$~GeV)
in the \D0 study. The ``leptons" are actually quasi-stable staus.
The vertical dashed lines indicate the cutoffs.
All distributions are normalized to unit area.}
  \label{fig:p2h}
\end{figure}

The expected $p_T$ distributions
of the \tlsp\ for two different values of $\Lambda$ are shown in
Fig.~\ref{fig:p2h}(a). The cut of $p_T>50$~GeV of the \lldedx\ selection
is efficient for the signal while it is expected to reduce backgrounds
significantly. The typical invariant mass of the two `muons' (assuming
massless) is very large as shown in Fig.~\ref{fig:p2h}(b).
A $M_{\ell\ell}>150$~GeV requirement does little harm to the signals.
Due to its large mass, the \tlsp\ is expected to move slowly. However since
most of the \tlsp's are produced in the decays of massive $\tilde\chi^\pm_1$s
and $\tilde\chi^0_2$s, the average speed $\beta(\equiv v/c)$ is relatively
large. It is around 0.7 for the $\Lambda$ values studied.

\begin{figure}[tpb]
  \centerline{\epsfysize=3.4in\epsfbox{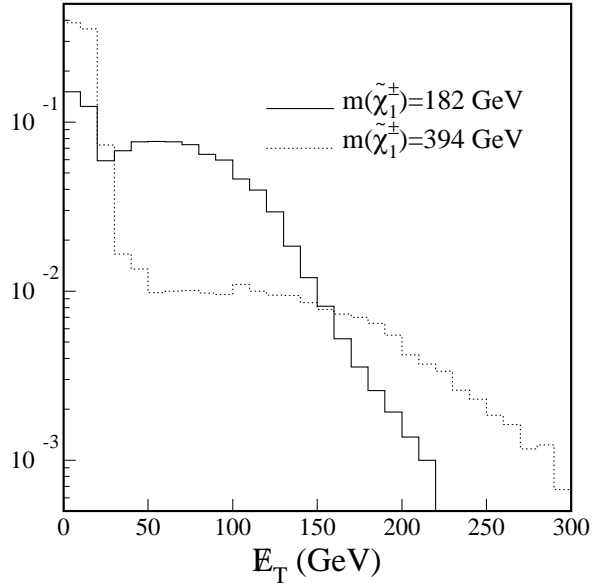}}
  \caption{The $\met$ distributions for the models with a quasi-stable
           \protect\tlsp\ as the NLSP for $\Lambda=40,\ 80$~TeV
          ($m_{\tilde\chi^\pm_1}=182,\ 394$~GeV) in the
\D0 study. The decays
          $\tilde\chi^\pm_1\to\tilde\chi^0_1 W^\pm\to\tilde\tau_1\tau W^\pm$
          and $\tilde\chi^0_2\to e\tilde e,\ \mu\tilde\mu$ contribute to events
          with small $\met$ while the decays
          $\tilde\chi^\pm_1\to\tilde\tau_1\nu$ and
          $\tilde\chi^0_2\to\tau\tilde\tau_1$ are the source
          for events with large $\met$.
          }
  \label{fig:p2hmet}
\end{figure}

The not-so-slow moving \tlsp's are expected to deposit large
ionization energies in the detector, differentiating them from other high
$p_T$ MIP particles. Since the backgrounds for the requirements $p_T>50$~GeV
and $M_{\ell\ell}>150$~GeV are already small, it pays to have a $dE/dx$
requirement with a relatively high efficiency for the signal and a
reasonable rejection for the MIP particles. The $\met$ distribution of
these events as shown in Fig.~\ref{fig:p2hmet} shows two distinct regions:
small and large $\met$. 
The decays $\tilde\chi^\pm_1\to\tilde\chi^0_1 W\to
\tilde\tau_1\tau W$ and $\tilde\chi^0_2\to e\tilde e,\ \mu\tilde\mu$
contribute to events with small $\met$.  The decays
$\tilde\chi^\pm_1\to \tilde\tau_1\nu$ and
$\tilde\chi^0_2\to\tau\tilde\tau_1$ are responsible for events 
with large $\met$.
The detection efficiencies and the expected significances of the
\lldedx\  selection for different values of $\Lambda$ are tabulated
in Table~\ref{tab:p2h}. The high efficiency is largely due to the high
momentum expected for the quasi-stable $\tilde\tau_1$.
The $5\sigma$ discovery curves are shown in Fig.~\ref{fig:p2lim} for two
values of \ldt. The lighter chargino with mass up to
340, 410~GeV and the \tlsp\  with mass up to 160, 200~GeV can be
discovered for the two integrated luminosities respectively.
\begin{table}[htbp]
\caption{The 
$\tilde\chi^\pm_1$ and 
$\tilde\tau_1$ masses, 
theoretical cross section, 
detection efficiencies of the \protect\lldedx\
selection, and significances in the \D0 study for different values of
$\Lambda$ for the models with a quasi-stable $\tilde\tau_1$ NLSP. The
relative statistical error on the efficiency is typically 1\%. The
background cross section is assumed to be 0.5~fb with an uncertainty of
20\%.}
\label{tab:p2h}
\renewcommand{\arraystretch}{1.7}
  \begin{tabular}{|c|cccc|}\hline
     $\Lambda$ (TeV)              &   40 &  60  &   80 &  100 \\
     $m_{\tilde\chi^\pm_1}$ (GeV) &  182 & 289  &  394 &  499 \\
     $m_{\tilde\tau_1}$ (GeV)     &   99 & 147  &  196 &  246 \\ \hline
     $\sigma_{th}$ (fb)           &  149 & 14.4 &  2.1 &  0.4 \\
\hline
     $\epsilon$ (\%)              & 37.4 & 44.6 & 51.6 & 54.9 \\
     \protect\rsb\ (2 fb$^{-1}$)  &  112 & 12   &  2.1 &  0.5 \\
     \protect\rsb\ (30 fb$^{-1}$) &  341 & 40   &  6.7 &  1.4 \\
  \end{tabular}
\end{table}


\section{Slepton Co-NLSP}\label{sec:sleptonconlsp}
\setcounter{equation}{0}
\setcounter{footnote}{1}
\indent

Slepton co-NLSPs result if the 
sleptons are lighter than the other MSSM superpartners and the 
slepton mass eigenstates
$\widetilde e_R$, $\widetilde \mu_R$, and $\widetilde \tau_1$ are
degenerate to within less than about 1.8 GeV, so that the
three body decays $\widetilde e_R \rightarrow e \tau \stauI$
and $\widetilde \mu_R \rightarrow \mu \tau \stauI$
are forbidden. Supersymmetric decay chains can therefore pass through
any of $\widetilde e_R$, $\widetilde \mu_R$ or $\stauI$.
As a result, the
sleptons ($\widetilde\ell\equiv \widetilde\tau_1,\widetilde
e_R,\widetilde\mu_R$) effectively
share the role of the NLSP.
The slepton co-NLSPs decay to the Goldstino by 
\begin{equation}
\widetilde \ell \rightarrow \ell \GG
\end{equation}

If the supersymmetry breaking scale 
$\sqrt{F}$ is in the lower part of the allowed range, then the slepton
decays can be prompt, so that supersymmetric events will be rich in high
$p_T$ leptons and $\tau$'s, with large $\missET$ from the Goldstinos
escaping the detector. The lepton flavors occurring in these decays
will often be nearly democratic because of the near-degeneracy
of staus, smuons, and selectrons in the slepton co-NLSP scenario, but
there is some preference for $\tau$'s, due to the effects of
$\widetilde\tau_R$-$\widetilde\tau_L$ mixing which provides a stronger
coupling of $\stauI$ to Winos.

For very large values of the supersymmetry breaking scale 
$\sqrt{F}$, the
slepton NLSPs will move through the detector before decaying. Since the
sleptons are required to be heavy from bounds at LEP, they may well often
be produced with velocities in the lab frame that are not
ultra-relativistic. This can cause them to be detected by virtue of their
high ionization rate ($dE/dx$) or by time-of-flight measurement,
just as discussed for the $\stauI$ in the Stau NLSP scenario of the
previous section.
Very energetic quasi-stable sleptons will penetrate all the way through
the detector with a near-minimal ionization rate, and may yield signals
that are identical or nearly identical to those of muons.

For intermediate values of $\sqrt{F}$, the slepton decays can occur inside
the
detector but at a macroscopic distance. Just as discussed above
for the stau NLSP case, the
impact parameter
of the slepton from the interaction region or the kink in the charged
particle track when the decay occurs can yield a unique signal and provide
information about the decay length. Measuring the physical decay length of
the charged sleptons in this
scenario may yield direct information or constraints on the supersymmetry
breaking mechanism, which would be one of the most theoretically
interesting
determinations one can make in this scenario.

In general, the signatures in the slepton co-NLSP scenario are
rather similar to those in the stau NLSP case. However,
the profiles of the leptons and jets coming from the sparticle
decays can be quite different, with the hallmark being a tendency
for lepton democracy rather than the preponderance of
$\tau$'s found in the stau NLSP scenario.

\begin{figure}[tpb]
\centering
\epsfxsize=4.5in
\hspace*{0in}
\epsffile{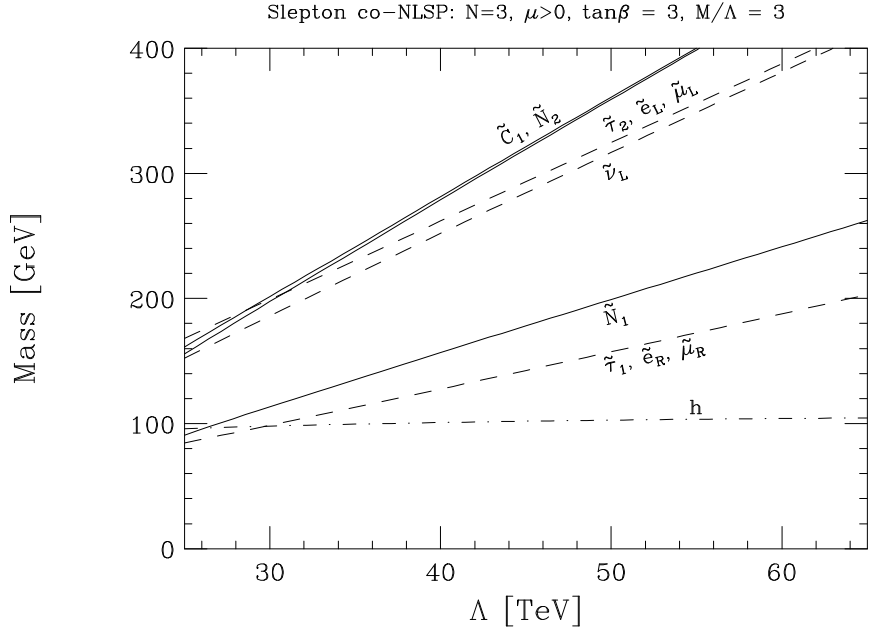}
\caption{The masses of the lightest neutralinos, charginos, sleptons
and CP-even Higgs boson in the Slepton co-NLSP Model Line, as a function
of $\Lambda$.}
\label{fig:n3slepton_mass}
\end{figure}

\begin{figure}[tpb]
\centering
\vspace{1cm}
\epsfxsize=4.5in
\hspace*{0in}
\epsffile{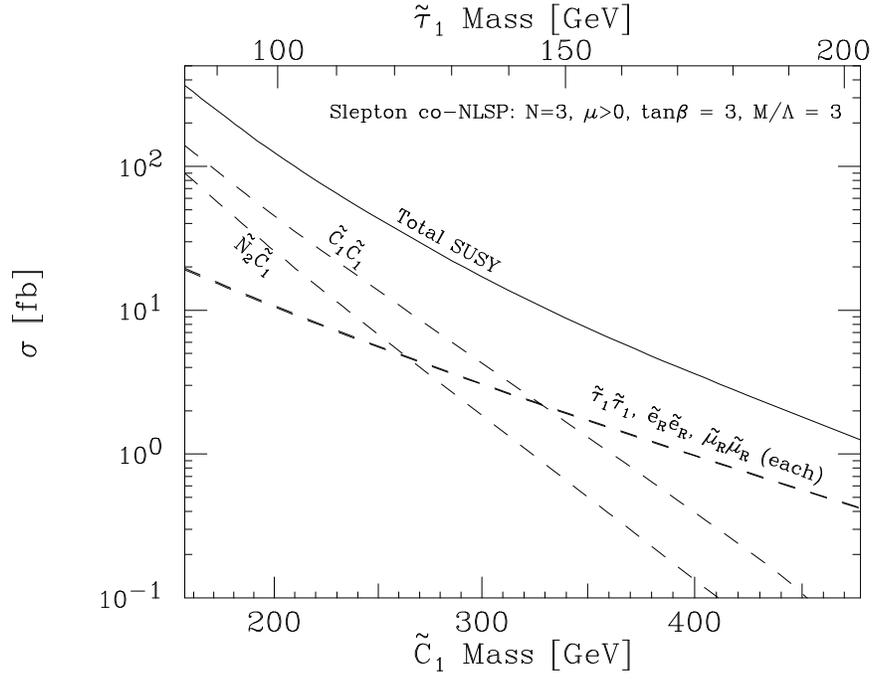}
\caption{Total production cross-sections in
$p\protect\overline p$ collisions with $\protect\sqrt{s} = 2$ TeV, for
superpartner pairs in the Slepton co-NLSP Model Line,
as a function of
$m_{\tilde \chi_1^\pm}$ and $m_{\protect\tilde \tau_1}$.}
\label{fig:n3slepton_sigma}
\end{figure}

\begin{figure}[tpb]
\centering
\epsfxsize=4.5in
\hspace*{0in}
\epsffile{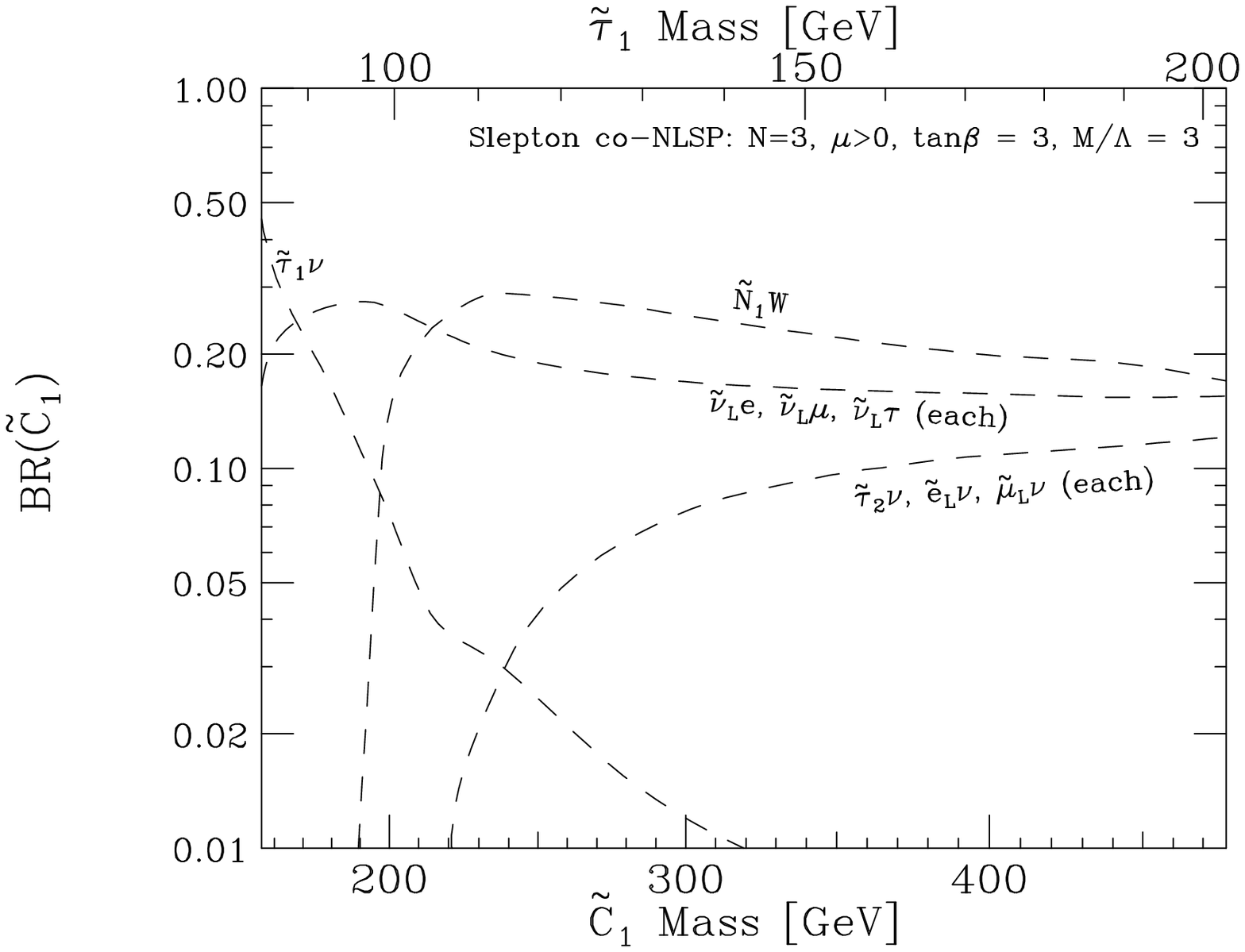}
\caption{Branching fractions for the decay of
$\CI$ in the Slepton co-NLSP Model Line,
as a function of
$m_{\tilde \chi_1^\pm}$ and $m_{\protect\tilde \tau_1}$.}
\label{fig:n3slepton_brc1}
\end{figure}

\begin{figure}[tpb]
\centering
\vspace{1cm}
\epsfxsize=4.5in
\hspace*{0in}
\epsffile{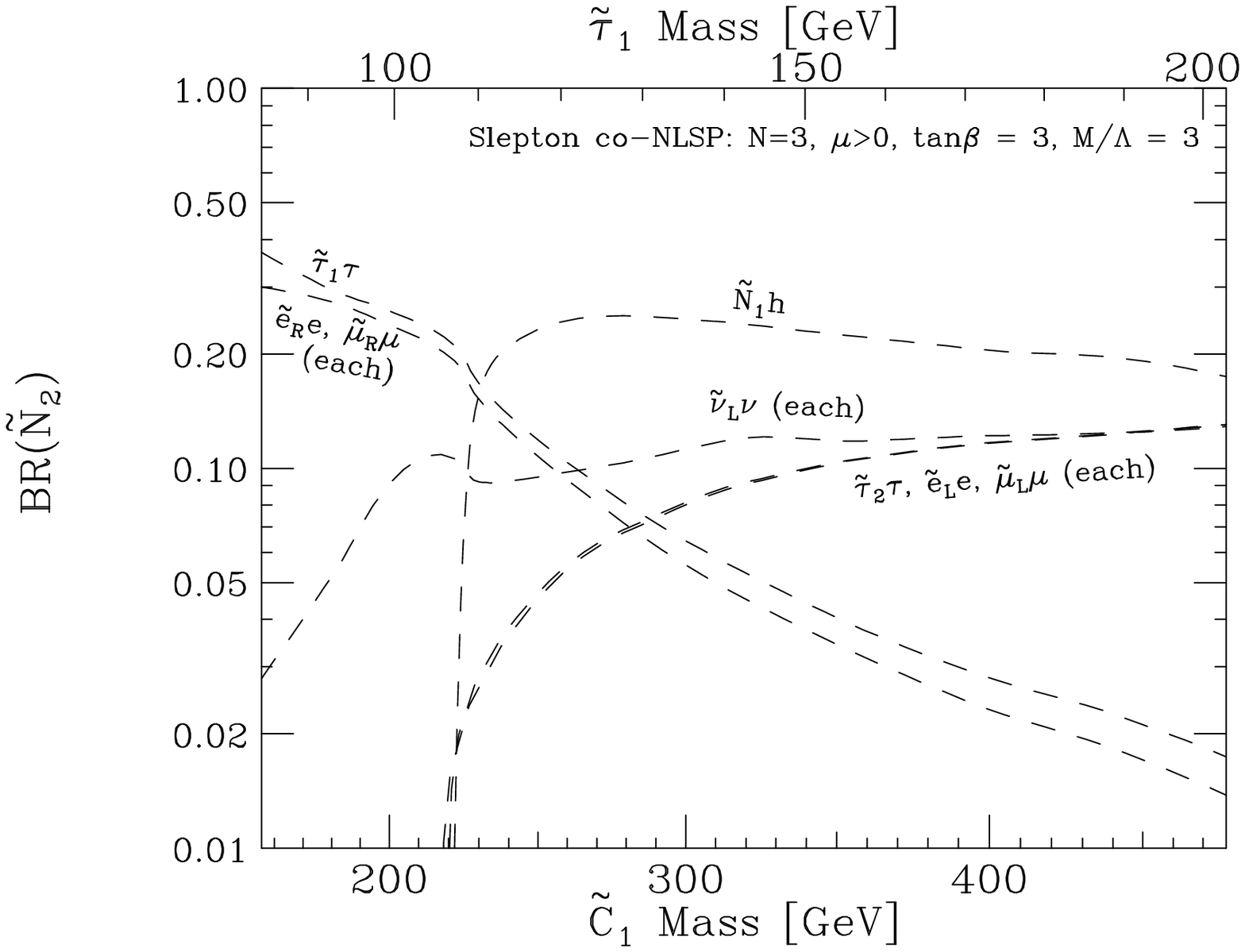}
\caption{Branching fractions for the decay of
$\NII$ in the Slepton co-NLSP Model Line,
as a function of
$m_{\tilde \chi_1^\pm}$ and $m_{\protect\tilde \tau_1}$.}
\label{fig:n3slepton_brn2}
\end{figure}

For quantitative studies, we define a Slepton co-NLSP Model Line
according to the general strategy outlined in Section \ref{sec:mgm}.
The fixed parameters  that define our Model Line are:
\beq
{\rm Slepton~Co-NLSP~Model~Line:}~~~~~~
\Nmess = 3,\>\> {\Mmess \over \Lambda} = 3,\>\> \tan\beta = 3, \>\> \mu > 0,
\label{sleptonlineparameters}
\eeq
and $\Lambda$ is allowed to vary. In Figure \ref{fig:n3slepton_mass}, we
show the masses of the lightest few superpartners and the lightest CP-even
Higgs boson $h^0$, as a function of 34 TeV $<\Lambda<$ 85 TeV.
In Figure \ref{fig:n3slepton_sigma}, we show the most important Run II
production cross-sections for superpartners in this Model Line,
including $\CIplus \CIminus$, $\CIplusminus\NI$,
$\widetilde e^+_R \widetilde e^-_R$,
$\widetilde \mu^+_R \widetilde \mu^-_R$, and
$\widetilde \tau^+_1 \widetilde \tau^-_1$, as well as
the total inclusive cross-section for all supersymmetric particles,
as a function of $m_{\tilde \chi_1^\pm}$ and $m_{\tilde \tau_1}$.
Figures \ref{fig:n3slepton_brc1} and \ref{fig:n3slepton_brn2} show the
most significant branching fractions for $\CI$ and $\NII$. 
The decay of $\NI$ for these models is almost
democratic into the three final states $e\widetilde e_R$,
$\mu \widetilde \mu_R$, and $\tau \widetilde\tau_1$, with a slight
preference for the last.

\begin{figure}[tpb]
  \centerline{\epsfysize=3.5in\epsfbox{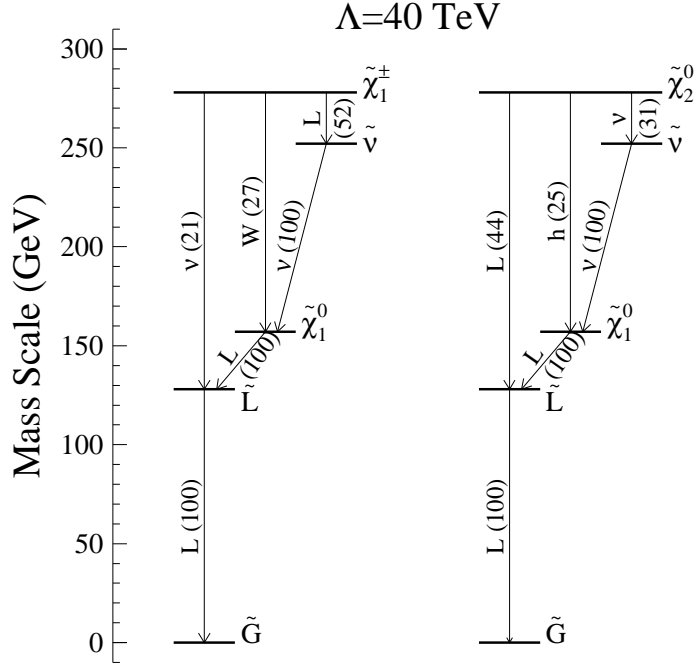}}
  \caption{Decay schematics of $\tilde\chi^\pm_1$ and $\tilde\chi^0_2$
           for $\Lambda=40$~TeV for the Slepton co-NLSP Model Line.
Percentage branching ratios for main
           decay modes are shown in parentheses.}
  \label{fig:p3br}
\end{figure}

In this Model Line, the three lightest mass eigenstates $\widetilde e_R$,
$\widetilde \mu_R$, and $\widetilde \tau_1$ are separated by less than 1
GeV. Therefore, they effectively share the role of the NLSP. For small
values of $\Lambda$, $\widetilde\chi_1^+ \widetilde \chi_1^-$ and \cn\
dominate the production cross section. As illustrated in
Fig.~\ref{fig:p3br}, $\widetilde\chi_1^+ \widetilde \chi_1^-$ and \cn\
production will yield events with multileptons in the final state. 
The slepton pair production in this Model Line surpasses
chargino-neutralino production if
$m_{\tilde \ell} \gsim 140$ GeV.

The lifetime of $\tilde\ell$ determines the event topology.
The general types of signals to be expected in this Model Line are:
\begin{itemize}
\item[$\bullet$] Small $\cG$ with prompt NLSP decays:\\
Events with 2 or 3 or more high-$p_T$ leptons ($e,\mu,$ or $\tau$),
often accompanied by two or more jets.
\item[$\bullet$] Intermediate $\cG$ with delayed NLSP decays:\\
Events with $\widetilde \ell \rightarrow \ell$
decay kinks.
\item[$\bullet$] Large $\cG$ with a quasi-stable NLSP:\\
Events with heavy charged particle tracks featuring
anomalous ionization rate or time-of-flight from slow slepton NLSPs,
and/or fake ``muons" from fast slepton NLSPs.
\end{itemize}
In the following, we discuss the cases with short-lived
and quasi-stable $\tilde\ell$'s. Again, the analyses should also be
sensitive
to the slepton co-NLSPs with a intermediate lifetime.

\subsection{\rm CDF study of prompt slepton decay signals in the Slepton
co-NLSP Model Line}

In the slepton co-NLSP scenario, supersymmetry events always produce two
high--$E_T$ leptons and may produce
other leptons or jets in the cascade decays.  One option for approaching
this model is through a trilepton and $\missET$ search which overlaps with
the SUGRA standard search.  Another option is dileptons with jets and
$\missET$, which also overlaps a standard SUGRA search.  Finally a new search
requiring only two leptons and $\missET$ is also potentially sensitive.
The only significant background would be $WW/WZ/ZZ$ and $t\bar{t}$ production.
This could be mitigated by harder cuts on the $E_T$ and $\missET$,
which should be more efficient
for SUSY events than the Standard Model backgrounds.
CDF has examined the trilepton approach by applying the results of the
SUGRA trilepton analysis to the GMSB model for both the Run I data
result\cite{gmsbcdftrilruni} and the Run II projection.

\begin{figure}[tpb]
\centerline{\epsfysize=3.7in\epsfbox{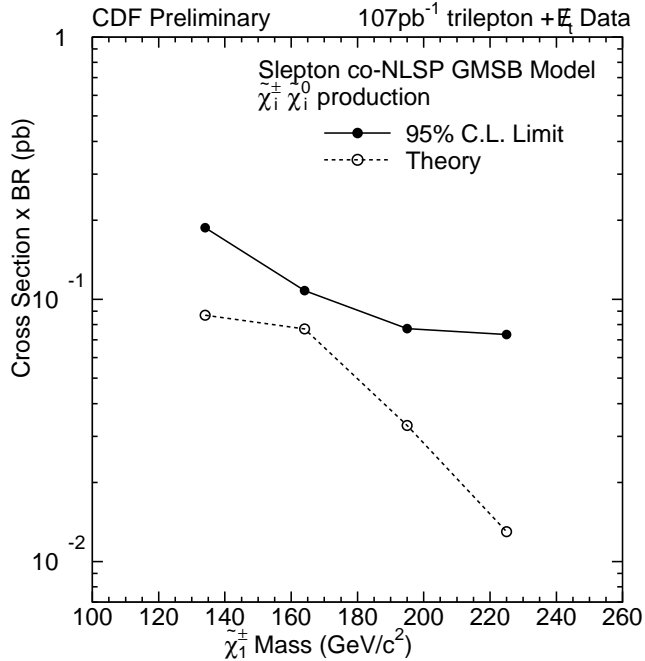}}
\caption{The CDF Run I limit on cross section times branching ratio
from the trilepton analysis applied to the Slepton co-NLSP Model Line.}
\label{fig:gmsbcdftrili}
\end{figure}

In 107~pb$^{-1}$ of Run I data, one lepton is required to be central,
have $E_T>11$~GeV and pass tight identification cuts.  Two other leptons
must have $E_T>5$~GeV and pass looser cuts.  A $\missET$ of more than 15~GeV
is then required.  No events pass all cuts and anything more than 3.2 events
of anomalous production is excluded.
Table \ref{tab:gmsbcdftrili} and Figure \ref{fig:gmsbcdftrili}
shows the resulting limits.
The branching ratio falls off when the $\CI$ mass is less than
about 150~GeV because the $\widetilde \nu$ becomes heavier than the $\CI$
and the leptonic feed--down decays are suppressed.
\begin{table}[htbp]
\caption{The summary of the Monte Carlo points used to investigate the
limits on the slepton co-NLSP model from the CDF Run I trilepton analysis.  
Cross sections include the branching ratios. The $A\cdot \epsilon$ is for
events that satisfy the branching ratio criteria which is any three
leptons at the generator level.}
\label{tab:gmsbcdftrili}
\renewcommand{\arraystretch}{1.7}
\begin{tabular}{|c|c|c|c|c|} \hline
$\Lambda$ (TeV)      & 23   & 26   & 31 & 35  \\ 
$m_{\tilde \chi_1^\pm}$ (GeV)       & 139   &  164   & 206 &  238 \\
$m_{\tilde\ell}$ (GeV)& 78   &  87  & 101   & 114  \\ \hline
BR(\%)                & 16 & 33 & 37 & 35 \\ \hline
$\sigma\times$BR (fb)      & 87   & 77  & 33  & 13  \\ \hline
$A\cdot \epsilon$ (\%)     &  15.9  &  27.7  &  38.7  & 40.7  \\ \hline
$\sigma\times$BR 95\% C.L. limit (fb)& 188   &
108 & 77.3 & 73.4 \\ \hline
\end{tabular}
\end{table}

\begin{figure}[tpb]
\centerline{\epsfysize=3.7in\epsfbox{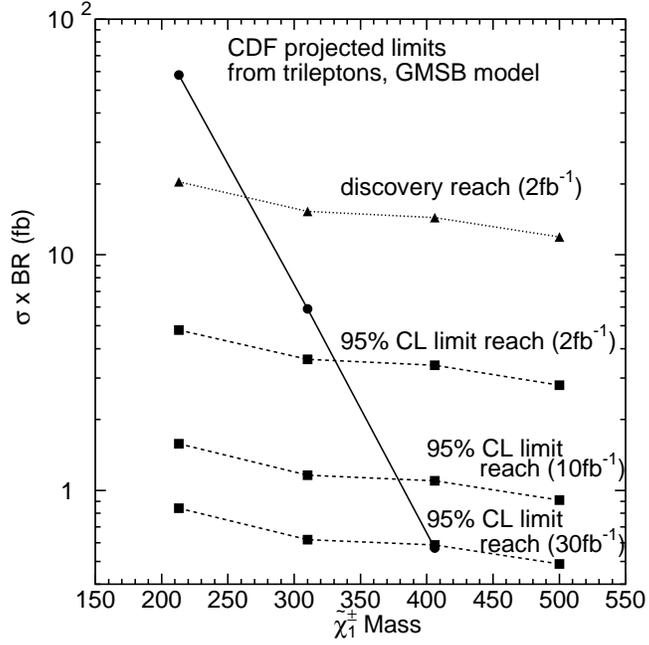}}
\caption{The projected Run II CDF limit on cross section times branching
ratio from the trilepton analysis applied to the Slepton co-NLSP Model
Line. The solid line is the theoretical prediction.}
\label{fig:gmsbcdftrilii}
\end{figure}

The SUGRA trilepton analysis has been projected to Run II and
it has been estimated that approximately one background event
will remain in 2~fb$^{-1}$ if the $\missET$ cut is raised to 25~GeV.
With 20\% systematics on the background and efficiency,
we expect limits of 4 events at 95\% C.L. and 17 events at 5$\sigma$.
Table \ref{tab:gmsbcdftrilii} and Figure \ref{fig:gmsbcdftrilii}
show the projected limits.
\begin{table}[htbp]
\caption{The summary of the Monte Carlo points used to investigate
projected limits on the slepton co-NLSP model for Run II (2~fb$^{-1}$) in
the CDF study.  Cross sections include the branching ratios. The $A\cdot
\epsilon$ is for events that satisfy the branching ratio criteria which is
any three leptons at the generator level.} 
\label{tab:gmsbcdftrilii}
\renewcommand{\arraystretch}{1.7}
\begin{tabular}{|c|c|c|c|c|} \hline
$\Lambda$ (TeV)           & 32    & 44    & 56     & 68    \\ 
$M_{\tilde \chi_1^\pm}$ (GeV)            & 213   & 310   & 406    & 500
\\ 
$M_{\tilde\ell}$ (GeV)     & 104   & 140   & 176    & 212   \\ \hline
BR(\%)                     & 61    & 39    & 16     & 5    \\ \hline
$\sigma\times$BR (fb)      & 58.0 & 5.9 & 0.57  & 0.056  \\ \hline
$A\cdot \epsilon$ (\%)     &  41 &  56 &  59  & 71   \\ \hline
$\sigma\times$BR 95\% C.L. limit (fb)& 4.8  & 3.6  & 3.4  & 2.8  \\ \hline
\end{tabular}
\end{table}

For 2~fb$^{-1}$ we expect to have a $5\sigma$ discovery reach of
280~GeV in chargino mass and a limit reach of 330~GeV.
For larger luminosities, we have not attempted to re--optimize the
cuts; we simply scale the signals and backgrounds.
Figure \ref{fig:gmsbcdftrilii} again shows the result, which indicate the
chargino mass reach is extended by 50~GeV for 10~fb$^{-1}$ and 70~GeV for
30~fb$^{-1}$.

\subsection{\rm {\D0} study of prompt slepton decay signals for the
Slepton co-NLSP Model Line}

\begin{figure}[tpb]
  \centerline{\epsfysize=3.0in\epsfbox{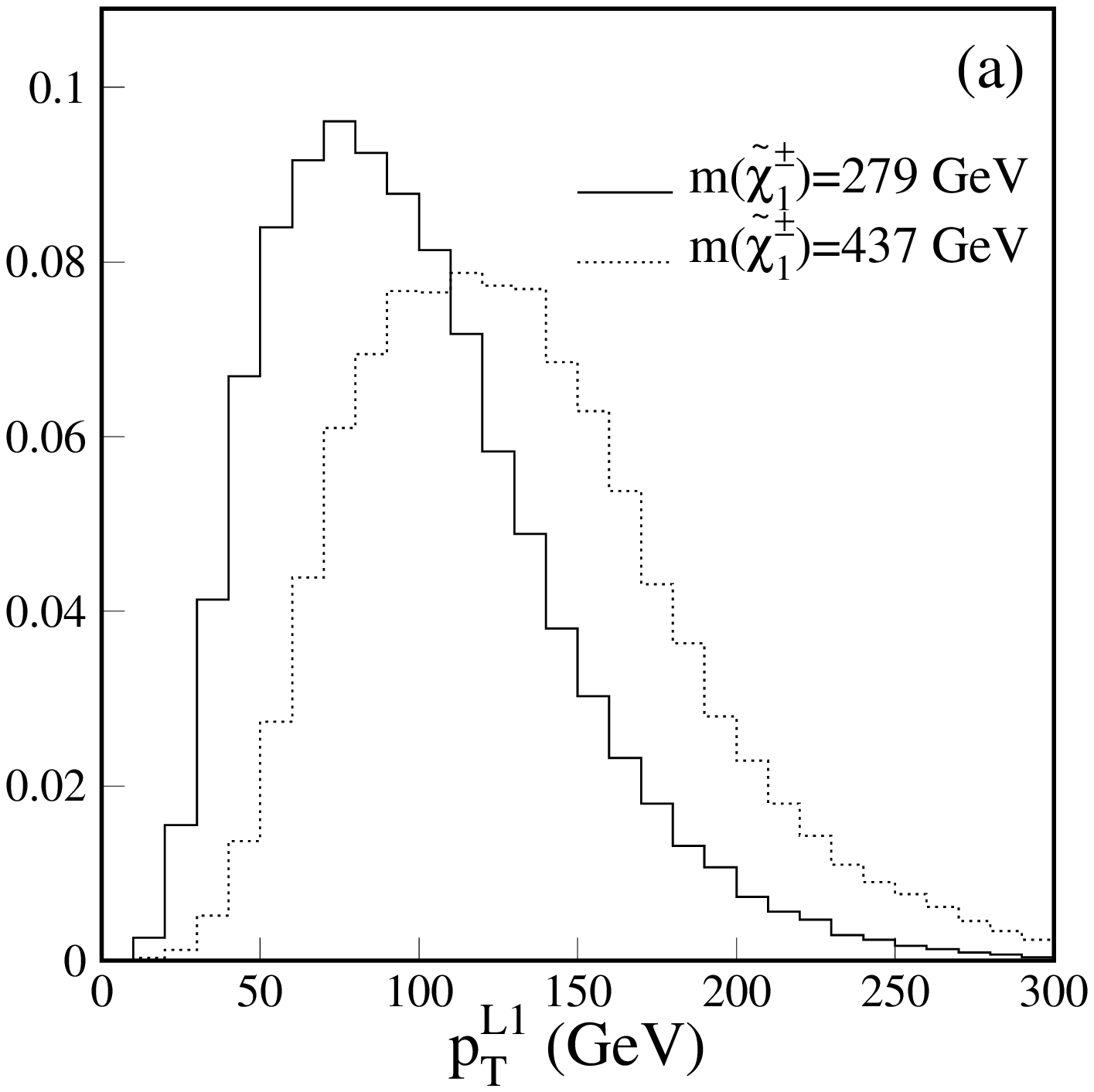}
              \epsfysize=3.0in\epsfbox{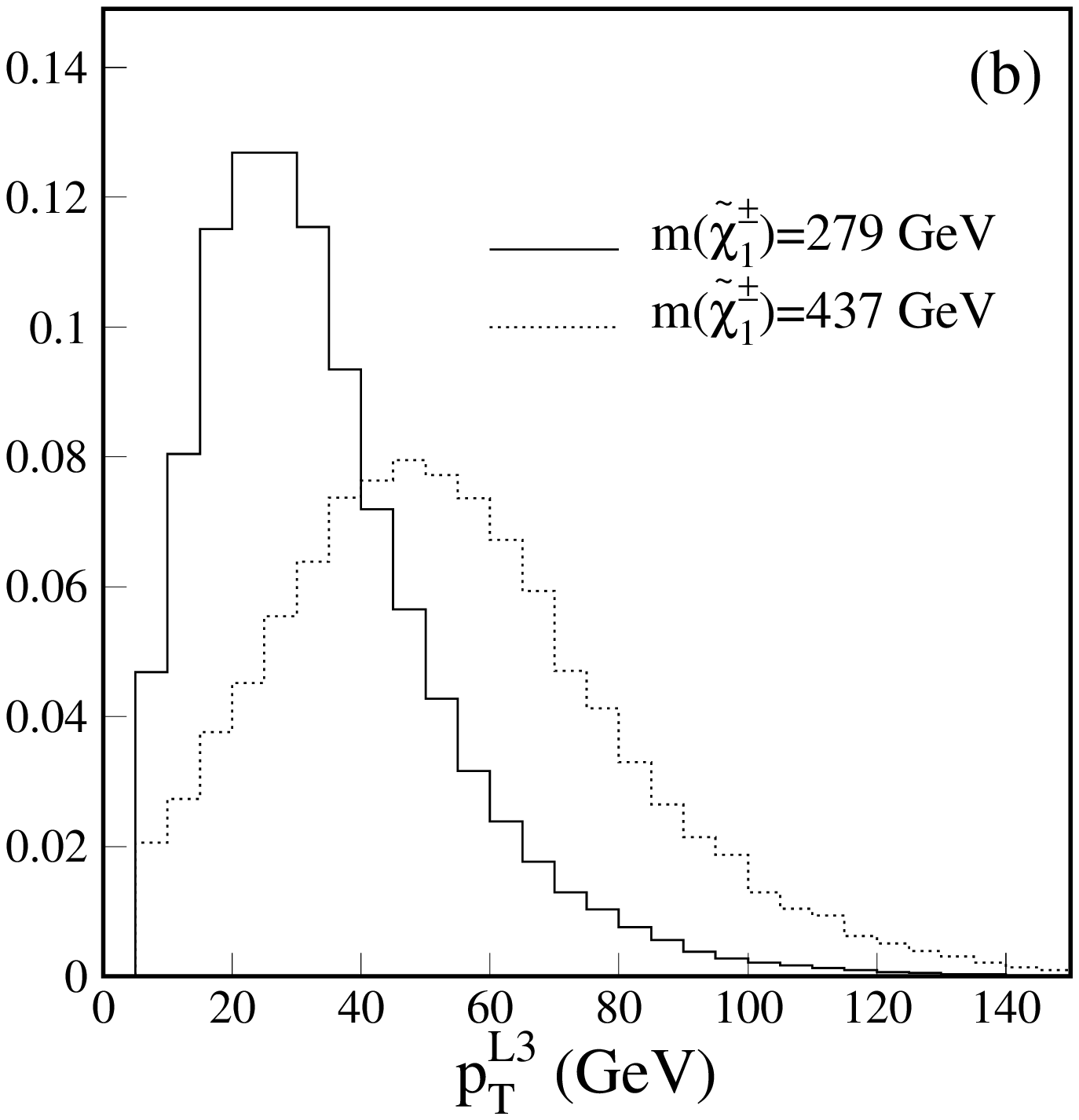}}
  \caption{The $p_T$ distributions of (a) the leading lepton and (b) the third
           lepton for the models with short-lived sleptons as
           co-NLSPs for $m_{\tilde\chi^\pm_1}=279,\ 437$~GeV
           ($\Lambda=40,\ 60$~TeV) in the \D0 study. Note that the $p_T$
requirement is
           15~GeV for the leading lepton and 5~GeV for the non-leading 
leptons.
           All distributions are normalized to unit area.}
  \label{fig:p3}
\end{figure}

\begin{figure}[tpb]
  \centerline{\epsfysize=3.5in\epsfbox{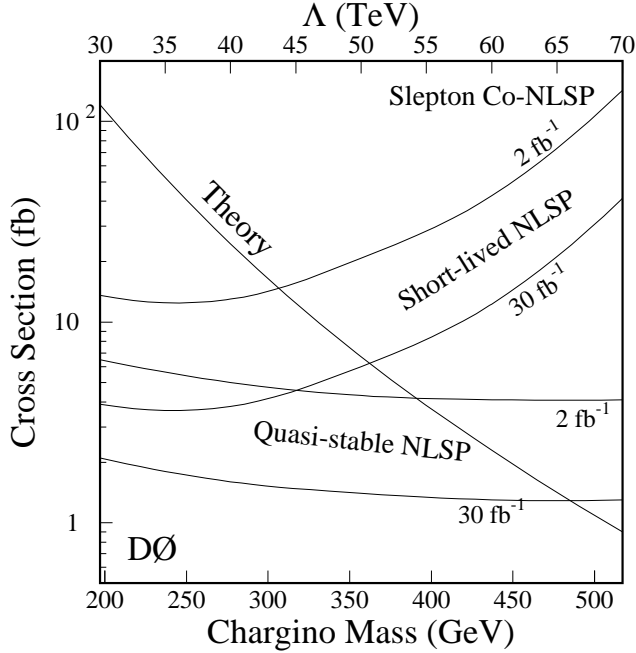}}
  \caption{The \D0 $5\sigma$ discovery cross section curves as functions
of
           mass of the lighter chargino and
$\Lambda$ for the Slepton co-NLSP Model Line, along with the
theoretical cross
           sections. The $5\sigma$ curves are shown for both short-lived
NLSPs (\protect\lllj selection criteria) and
           quasi-stable slepton co-NLSPs (\protect\lldedx selection) and
for integrated
           luminosities of 2, 30~fb$^{-1}$. }
  \label{fig:p3lim}
\end{figure}

If the decay $\tilde\ell\to\ell\tilde G$ is prompt ($\gamma c\tau\lsim 10$~cm),
\llmet\ events are expected from supersymmetry. Unfortunately, this final
state has large backgrounds from the Standard Model processes such as
$t\bar{t}$, $WW$, $WZ$ and $ZZ$ productions as well as from $W+{\rm jets}$
production with one of the jets misidentified as a lepton. However we note
that these events typically have multiple leptons in the final state and
most of them are in the central pseudorapidity region with good lepton
identification. Apart from those from $\tilde\ell$ decays, leptons are
also expected from $W^*$'s and $Z^*$'s produced in the cascade decays of
$\tilde\chi^\pm_1$ and $\tilde\chi^0_2$ of supersymmetry originated
events. Therefore, they can be selected using the \lllj\  criteria. The
$p_T$ distributions of the leading lepton and the third lepton of these
events are shown in Fig.~\ref{fig:p3}(a). 
Since most of the leading leptons are produced in the direct decays of
heavy $\tilde\ell$'s, its $p_T$ spectrum is relatively hard as shown in
the figure. The detection efficiencies and the expected significances are
summarized in Table~\ref{tab:p3}. The reduction in the relative cross
section of the tri-lepton producing $\widetilde\chi_1^+ \widetilde
\chi_1^-$  and \cn\ 
processes is responsible for the
decrease in efficiency as $\Lambda$ increases. For $\Lambda\gsim 50$~TeV
corresponding to $m_{\tilde \ell} \gsim 160$ GeV and $m_{\tilde
\chi_1^\pm} \gsim
360$ GeV, the $\tilde\ell\tilde\ell$ production cross section surpasses
that of the $\widetilde\chi_1^+ \widetilde \chi_1^-$
and \cn. With the $\tilde\ell\to\ell\tilde G$ decay,
$\tilde\ell\tilde\ell$ events will result in a high $p_T$ $\ell\ell \met$
final state. We note that the improvement by adding the \lljj\ 
selection is minimal in this case. The $5\sigma$ discovery curves are
compared with the theoretical cross sections in Fig.~\ref{fig:p3lim}. With
integrated luminosities of 2 and 30~fb$^{-1}$, the lighter chargino with
mass up to 310 and 360 GeV can be discovered, respectively.
\begin{table}[htbp]
\caption{The theoretical cross sections, $\tilde\chi^\pm_1$ and 
$\tilde\ell$ masses, detection efficiencies of the \protect\lllj\
selection criteria, and significances for different values of $\Lambda$
for points on the Slepton co-NLSP Model Line with prompt decays,
in the \D0 analysis. The efficiencies typically have a relative
statistical uncertainty of 4\%. The observable background cross
section is assumed to be 0.3~fb with a 20\% systematic uncertainty.}
\label{tab:p3}
\renewcommand{\arraystretch}{1.7}
\begin{tabular}{|c|ccccc|}\hline
   $\Lambda$ (TeV)                &  30  &   40 &  50  &  60  &  70  \\
   $m_{\tilde\chi^\pm_1}$ (GeV)   & 197  &  279 & 358  & 437  & 517  \\
   $m_{\tilde\ell}$ (GeV)         &  99  &  128 & 158  & 188  & 218  \\ \hline
   $\sigma_{th}$ (fb)             & 121  & 24.5 & 6.7  & 2.3  & 0.9  \\
\hline
                  $\epsilon$ (\%) & 14.7 & 15.4 &  9.6 & 4.7  &  1.4 \\
   \protect\rsb\ (2 fb$^{-1}$)    &  44  & 9.4  &  1.6 & 0.3 &  -- \\
   \protect\rsb\ (30 fb$^{-1}$)   & 152  & 32   &  5.5 & 0.9 &  0.1 \\
  \end{tabular}
\end{table}

\subsection{\rm {\D0} study of quasi-stable slepton signals for the
Slepton co-NLSP Model Line}

\begin{figure}[tpb]
  \centerline{\epsfysize=3.0in\epsfbox{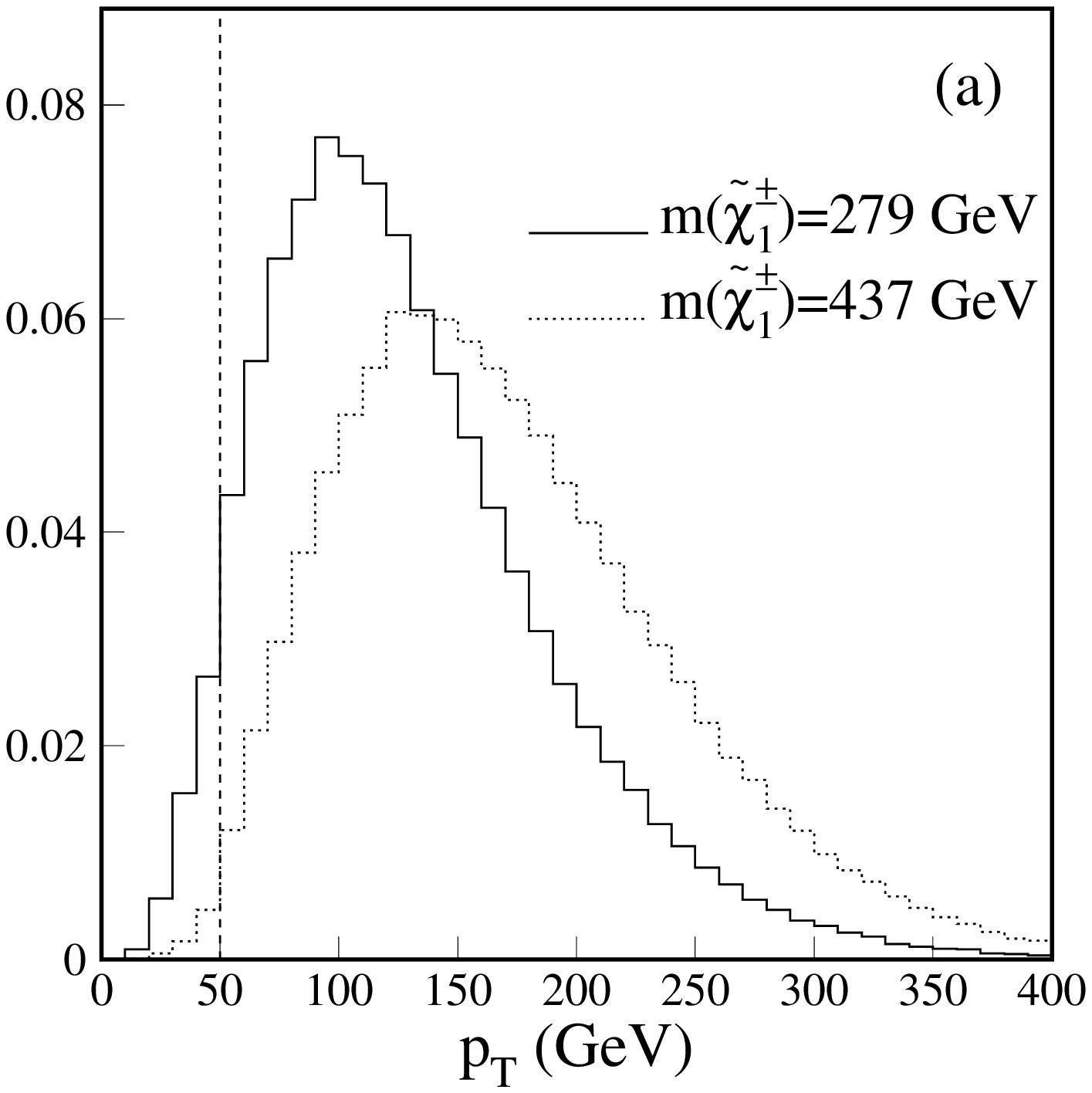}
              \epsfysize=3.0in\epsfbox{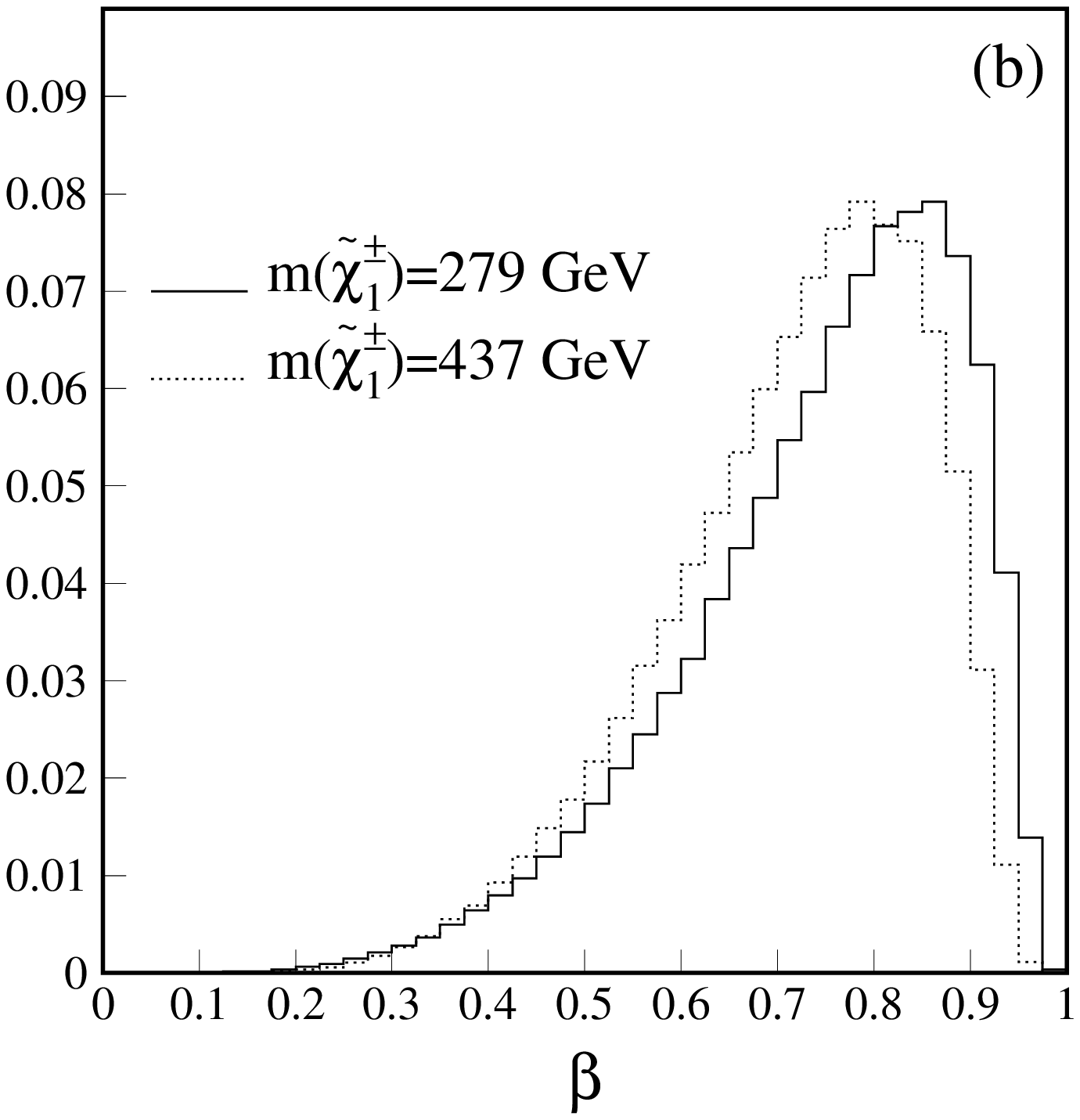}}
  \caption{The lepton $p_T$ (a) and the NLSP speed $\beta$ (b)
           distributions expected for models with quasi-stable
           sleptons as co-NLSPs, for
           $m_{\tilde\chi^\pm_1}=279,\ 437$~GeV ($\Lambda=40, 60$~TeV)
Here $\beta$ is measured in units of the speed of
           light $c$.
           All distributions are normalized to unit area.}
  \label{fig:p3h}
\end{figure}

If the slepton co-NLSPs have a long lifetime, it can decay outside the
detector ($\gamma c\tau\gsim 3$~m). In this case, the $\tilde\ell$ will
appear in the detector like a `muon' except that the ionization energy
loss will be large. This signature is identical to that of a quasi-stable
$\tilde\tau_1$ discussed above. Therefore, the signal events can be
identified using the same \lldedx\ selection. The expected $p_T$ and
$\beta$ distributions of the $\tilde\ell$ for $\Lambda=40,\ 60$~TeV are
shown in Fig.~\ref{fig:p3h}.
Note that the $\beta$ distribution is very similar to that shown
in Fig.~\ref{fig:p3h}(b) for the models with $\stauI$ as the sole NLSP.
The $\tilde\ell$s
typically have very large $p_T$ and are mostly central. For example, about
90\% of the $\tilde\ell$s are in central pseudorapidity region with the
tracking coverage for the case of $\Lambda=70$~TeV ($m_{\tilde \ell}
= 218$ GeV). Table~\ref{tab:p3h}
shows the detection efficiencies and the expected significances for different
$\Lambda$ values. The $5\sigma$ discovery curves are shown in
Fig.~\ref{fig:p3lim}. The lighter chargino mass discovery reach is
about 390 GeV for \ldt=2~fb$^{-1}$ and 480 GeV for \ldt=30~fb$^{-1}$.
\begin{table}[htbp]
\caption{The theoretical cross section, $\tilde\chi^\pm_1$ and $\tilde\ell$
masses, detection efficiency of the \protect\lldedx\ 
selection, and significances for different values of $\Lambda$ for the
models with quasi-stable $\tilde\ell$'s as co-NLSPs, in the
\D0 analysis. The relative statistical error on the efficiency is
typically 1\%. The background cross section is assumed to be 0.5~fb with
a systematic uncertainty of 20\%.}
\label{tab:p3h}
\renewcommand{\arraystretch}{1.7}
\begin{tabular}{|c|ccccc|}\hline
     $\Lambda$ (TeV)              &  30  &   40 &   50 &  60  &   70 \\
     $m_{\tilde\chi^\pm_1}$ (GeV) & 197  &  279 &  358 & 437  &  517 \\
     $m_{\tilde\ell}$ (GeV)       &  99  &  128 &  158 & 188  &  218 \\ \hline
     $\sigma_{th}$ (fb)           & 121  & 24.5 &  6.7 & 2.3  &  0.9 \\
\hline
     $\epsilon$ (\%)              & 34.8 & 45.2 & 52.6 & 54.9 & 55.1 \\
     \protect\rsb\ (2 fb$^{-1}$)  &  83  &  22  &  7.0 &  2.6 &  1.0 \\
     \protect\rsb\ (30 fb$^{-1}$) & 257  &  68  &  21  &  7.8 &  3.1 \\
  \end{tabular}
\end{table}


\section{Squark NLSP}

\label{sec:squarknlsp}
\setcounter{equation}{0}
\setcounter{footnote}{1}
\indent

A squark may arise as the NLSP in theories of gauge-mediated supersymmetry
breaking in which the strongly interacting messenger fields have
suppressed couplings to spontaneous supersymmetry breaking.
In addition,
any non-gauge couplings between the standard model superpartners
and the messenger or supersymmetry breaking sectors
have an {\it a priori} unknown effect on the superpartner spectrum.
If kinematically open, a squark NLSP can decay to the
Goldstino through emission of its partner quark,
\beq
\squark \to q \goldstino
\label{squarkdecay}
\eq
The lightest squark is likely to be mostly stop-like because
of left-right stop level repulsion, and
negative renormalization group evolution contribution to the
squared masses, both proportional to the square of the top Yukawa
coupling.
For a stop-like squark lighter than the top quark,
the only kinematically allowed
two-body decays are through flavor violating suppressed
modes to light quarks.
Since mixing between the second and third generations is
expected to be larger than between the first and third generations,
\beq
\tilde t \rightarrow c \widetilde G
\label{stoptwobody}
\eq
should be the dominant two-body decay mode.
Three body decay through a charged current interaction,
\beq
\tilde t \rightarrow bW \widetilde G
\label{stopthreebody}
\eq
is not suppressed by flavor violation, but
is phase-space suppressed.
Depending of the precise parameters, the three-body mode
$\tilde t \to bW \goldstino$ dominates over the two body mode
$\tilde t \to c \goldstino$ for values of the scharm-stop mixing
angle $\sin \theta_{\tilde c \tilde t} \lsim {\rm~few~}
\times 10^{-3}$ \cite{mesino}.
For a stop-like squark heavier than the top quark, the two-body
decay
$\tilde t \to t \goldstino$ generally dominates.
With the top decay $t \to bW$ this give the same
decay pattern as (\ref{stopthreebody}), but with different
kinematics.
Expressions for the two- and three-body squark decays to the
Goldstino are given in Appendix \ref{app:goldstino}.

For any reasonable supersymmetry breaking scale above the
electroweak scale, coupling to the Goldstino is weak enough
so that the squark decay length easily exceeds the hadronization
length scale.
A NLSP squark therefore always hadronizes before decaying.
Hadronization with a light antiquark leads to a
neutral or charged mesino
bound state,
$\mesino_{\widetilde Q \qbar} \equiv (\widetilde Q \qbar)$,
while hadronization with two light quarks leads to a
neutral or charged sbaryon
bound state
$\sbaryon_{\widetilde Q qq} \equiv (\widetilde Q qq)$, and likewise
for the antiparticle states.
The existence of these strongly interacting bound states can
lead to a number of novel signatures discussed below.


If the supersymmetry breaking scale is below a few 100 TeV, the decay
length 
for the decay (\ref{squarkdecay}) or
(\ref{stoptwobody}) and (\ref{stopthreebody}) is short enough
so that the decay products
appear to originate from the interaction region.
The experimental signatures for general prompt squark decay are
\beq
\begin{array}{llll}
\bullet~~{\rm Prompt~decays} & \squark \to g \goldstino~~:
               &  jj\met  & ~~~~~~~~~~~~~~   \\
\end{array}
\nonumber
\eeq
For a stop-like NLSP squark, the experimental signatures depend
on the relative importance of the two- or three-body
decay modes
\beq
\begin{array}{lll}
\bullet~~{\rm Prompt~decays} & \tilde t \to bW \goldstino~{\rm or}~
              \tilde t \to t \goldstino~~:
               &
		   	   bbWW \met ~{\rm or}~ tt \met
                \\
& &  \\
\bullet~~{\rm Prompt~decays} & \tilde t \to c \goldstino~~:  
              &
               cc \met   \\
\end{array}
\nonumber
\eeq
Observation of any of these signatures
interpreted as arising from decay to Goldstino pairs, would
of course imply a squark NLSP with a low supersymmetry breaking
scale.
Observation of either $bbWW \met$ or
$tt \met$ would imply the NLSP squark is stop-like.
In conjunction with
a bound on the $cc \met$ signature, this
would also provide a very stringent
upper limit on the scharm-stop mixing angle \cite{mesino}.

If the supersymmetry breaking scale is larger than a few 100 TeV
the squark decays (\ref{squarkdecay}) or
(\ref{stoptwobody}) and (\ref{stopthreebody}) within a hadronized
mesino or sbaryon can take place over a macroscopic distance. 
In this case displaced jets arise from the squark decay. 
If the displaced decays are short enough, Standard Model 
backgrounds from heavy quark production with displaced secondary
vertices are possible, but may be controlled by the use
of angular distributions.
Longer decay lengths can be distinguished by the magnitude of the 
jet impact parameter with the production vertex. 
Slowly moving charged mesinos and sbaryons which live long enough
to traverse the tracking region of a detector will yield
highly ionizing tracks (HITs),
much like a quasi-stable stau slepton discussed in sections 
\ref{subsec:CDFqsstau} and \ref{subsec:D0qsstau}.  
Hadronic interactions of an antimesino or sbaryon give rise to 
additional hadronic activity along the track and
also allow charge exchange with the detector material. 
Slowly moving antimesinos or sbaryons therefore give rise
to intermittent charge exchange associated with highly ionizing tracks
(CE-HITs). 
Mesinos and antisbaryons do not undergo as rapid charge exchange
as the antiparticle counterparts. 
But for a charged state there can still be additional hadronic activity 
along a highly ionizing track (H-HIT). This is due to inelastic hadronic
interactions of the bound state
in the calorimeter materials \cite{bjorken}.
Neutral mesino and sbaryon states, 
as well as the intermittent nature of antimesino or sbaryon 
CE-HITs, contribute to missing energy. 
The experimental signatures of a squark NLSP with macroscopic 
decay length to the Goldstino are therefore:
\beq
\begin{array}{lll}
\bullet~~{\rm Macroscopic~decays} & \squark \to g \goldstino~~:
               &  {\rm Displaced~jets}      \\
          &    &  {\rm CE-HITs}  \\
		  &    &  {\rm H-HITs}  \\
		  &    &  \met  \\
\end{array}
\nonumber
\eeq
A number of techniques to identify displaced jets, and the difficulty 
in identifying hadronic highly ionizing tracks 
are detailed below in sections \ref{subsec:dissquark} 
and \ref{subsec:qssquark}.
Observation of any signature associated with macroscopic
decay of an NLSP squark would imply a low SUSY breaking scale. 
Measurement of the decay length distribution would give a 
measure of the SUSY breaking scale. 
Flavor tagging of the final states would yield information 
about the NLSP squark flavor. 

An NLSP squark also presents the interesting possibility
of neutral mesino-antimesino oscillations analogous to 
meson-antimeson oscillations \cite{mesino}.
Rather than representing a very specialized process which 
could only be observed once the existence of an NLSP squark
is established, mesino oscillation can in fact provide 
a discovery channel, as detailed in section \ref{subsec:mesosc}.
Observation of mesino oscillation would provide a very sensitive
probe of sflavor violation in the squark sector. 

For the quantitative studies below, the signals are assumed to 
arise entirely from direct squark pair production mainly 
through $s$-channel gluon exchange. 
The squark NLSP Model Line is therefore simply defined to be 
a single squark with a mass which is allowed to vary. 
Squark pair production 
is very likely to be the largest SUSY production cross section 
in the squark NLSP scenario. 
It is, however, possible that gluino pair production could
also contribute to the signals if the gluino
is not too much heavier than the NLSP squark. 
In this case additional prompt partons would arise from gluino 
cascade decays, which are likely to be dominated by 
$\gluino \to q \squark$.

\subsection{Prompt Squark Decay}
\label{subsec:psquark}

The lightest squark is likely to be mostly stop-like because
of left-right stop level repulsion, and
negative renormalization group evolution contribution to the 
squared mass, both proportional to the square of the top Yukawa
coupling. 
Without any sflavor violation, the dominant decay mode for 
an NLSP stop squark heavier than about 90 GeV but lighter 
than the top quark is the three body decay  
$\widetilde t \to bW\GG$. 
Top squark pair production then yields the signature $bbWW \met$. 
Unfortunately this is the identical signature which arises from 
top quark pair production. 
One possibility would be to search for an excess in the top 
sample at large missing energy. 
This can be significantly improved by using additional 
kinematic information to distinguish top and stop decays.

\begin{figure}[tpb]
\centering
\epsfysize=2.8in
\hspace*{0in}
\epsfbox{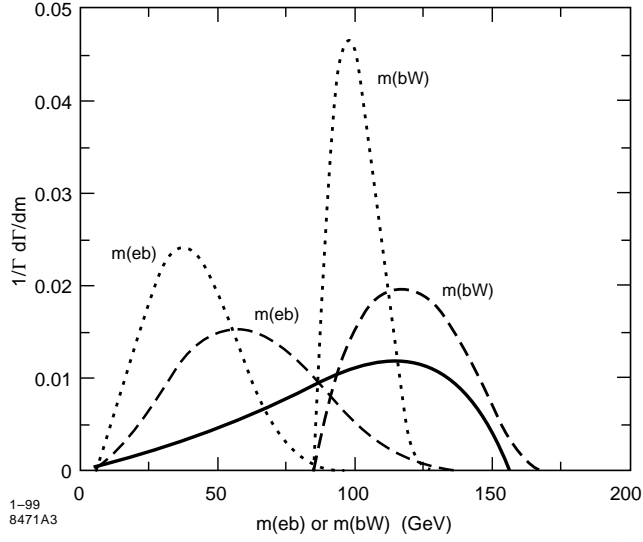}
\caption{$m_{eb}$ and $m_{bW}$
invariant mass spectra for stop squark decay $\stopq \to bW \goldstino$
for $m_{\stopq}=130$ GeV (dot) and
$m_{\stopq}=170$ GeV (dash) with $m_2=200$ GeV,
$\mu = 1000$ GeV, $\tan \beta = 1$, $m_{\tilde b_L}=300$ GeV,
and the left-right stop mixing angle $\sin \theta_t=-0.8$.
The $m_{eb}$ invariant mass spectrum for top quark decay
$t \to bW$ is shown for comparison (solid).}
\label{fig:peskinone}
\end{figure}

One observable which is available at the Tevatron which can 
distinguish top and stop decays is the invariant mass of the observed 
$b$-jet plus lepton system which results from a 
leptonic $W$ decay \cite{peskinstop}.
Because the Goldstino is derivatively coupled, the stop
decay amplitude is peaked at large values of the Goldstino
momentum, which implies small values of the invariant mass 
for the remaining (visible) decay products.  
In contrast, top quark decays favor 
large values of the invariant mass of the visible decay products. 
Fig. \ref{fig:peskinone} shows the lepton-$b$ invariant 
mass distribution for the top quark and two values
of the stop mass with typical parameters \cite{peskinstop}.
The $b$-$W$ invariant mass for stop decay is also shown. 
As is apparent, 
the stop excess in a top quark sample could be enhanced 
by a cut on the lepton-$b$-jet invariant mass.
In order to avoid combinatoric problems this would be applied
in the channel in which one
$W$ decays hadronically while the other decays leptonically.

It should be noted that at present, the available Pythia and ISAJET
Monte Carlo simulations 
of the three body decay $\tilde t \to bW\GG$ do not incorporate
a derivatively coupled Goldstino in the decay amplitude, and 
so would not reproduce the correct invariant mass distributions. 
This should be rectified in future studies of this process. 

\begin{figure}[tpb]
\centering
\epsfysize=2.8in
\hspace*{0in}
\epsfbox{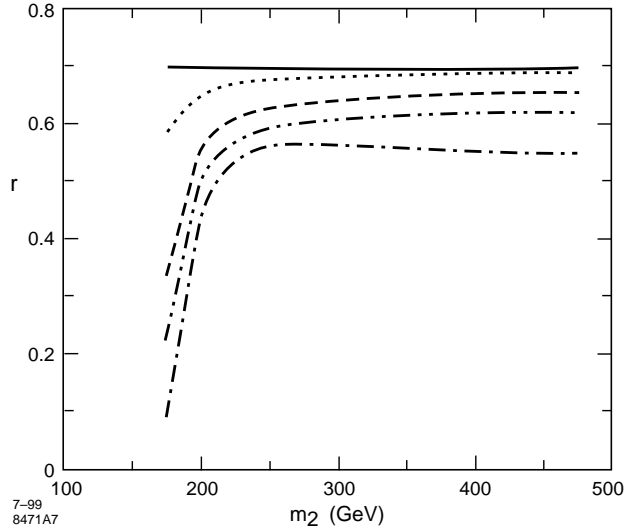}
\caption{Longitudinal $W$ boson polarization
$r\equiv \Gamma(W_0)/ \Gamma({\rm all})$ for stop squark
decay $\stopq \to bW \goldstino$ as a function of the $SU(2)_L$
Wino mass $m_2$ for
the left-right stop mixing angle
$\sin \theta_t = -0.98$ (dot), $-0.8$ (dash),
$-0.6$ (dot-dot-dash), and $0$ (dot-dash), with
$m_{\stopq}=170$ GeV, $\tan \beta =1$, $\mu=1000$ GeV, and
$m_{\tilde b_L}=300$ GeV.
The longitudinal polarization for top quark decay $t \to bW$
is shown for comparison (solid).}
\label{fig:peskintwo}
\end{figure}

Another observable which can distinguish 
top and stop decays is the $W$ longitudinal 
polarization \cite{peskinstop}. 
A $W$ boson arising from top decay 
is highly longitudinally polarized because of the large 
longitudinal Goldstone component coupling proportional to the 
top mass. 
The $W$ longitudinal polarization arising from stop decay depends
on the relative importance of various amplitudes. 
If the decay is dominated by diagrams in which the Goldstino
is emitted from the stop squark or intermediate off-shell 
bottom squark the polarization is similar to that for top quark 
decay. 
However, diagrams in which the Goldstino is radiated from an 
intermediate chargino which is mostly Wino contribute transverse
polarization \cite{peskinstop}. 
The latter diagrams are important if the chargino is not much
heavier than the stop squark or top quark. 
Fig. \ref{fig:peskintwo} shows 
the longitudinal $W$ polarization for top quark decays and stop
squark decays as a function of the Wino mass 
for various stop mixing angles and typical parameters \cite{peskinstop}.
%
This observable is model dependent and therefore sensitive
to the underlying superpartner spectrum. 
Because of the missing energy carried by the Goldstinos, 
a measurement of the $W$ longitudinal polarization in stop decay can not
make use of techniques which require the neutrino four-vector.
However, the polarization can be determined from the $W$ decay 
angle determined from four-vectors of the two jets 
assigned to the hadronic $W$ \cite{peskinstop}.

With non-trivial scharm-stop mixing the decay $\tilde t \to c \GG$
can dominate for stop squarks less massive than the top quark. 
Even without sflavor violation, standard quark mixing 
results in this decay mode dominating for a stop mass less
than about 90 GeV. 
Stop pair production $\tilde t \tilde t$ then leads to events with two
charm jets and $\missET$. 
A search for this mode can be handled as a special
case of a SUGRA search for $\widetilde t_1\rightarrow c \NI$
with the $\NI$ mass approaching zero.

\subsubsection{CDF study of Prompt Squark Decay}

CDF has performed a search for 
$\widetilde t_1 \widetilde t_1$ pair production with 
$\widetilde t_1\rightarrow c \NI$
in 88~pb$^{-1}$ of Run I data \cite{gmsbcdfccmet}.
The primary analysis cuts include
two jets with $E_T>15$~GeV, 40~GeV of $\missET$, and one of the jets
must have displaced tracks in the SVX.
The results include the region
where the $\NI$ mass approaches zero where the limit applies to
the low scale SUSY breaking decay $\widetilde t_1\rightarrow c \GG$.
The results rule out this scenario for a 
$\widetilde t_1$ mass between 40 and 85~GeV.

\begin{figure}[tpb]
\centering
\epsfysize=3.5in
\hspace*{0in}
\epsfbox{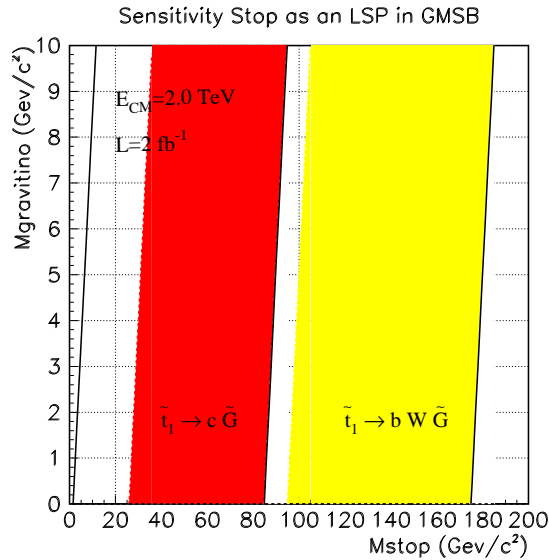}
\caption{Projected CDF Run II limits on a stop NLSP.}
\label{fig:gmsbcdfstop}
\end{figure}

To project to Run II, the cross sections are simply scaled to 2~fb$^{-1}$
and the efficiency for tagging displaced tracks is assumed to increase by
a factor of two.
The lowest $\widetilde t_1$ mass exclusion region boundary,
where the luminosity could
overcome the decreasing efficiency of the $\missET$ cut, is expected
to go down to approximately 25~GeV as shown in Fig. \ref{fig:gmsbcdfstop}.
The upper limit will remain at 85~GeV because that is where
the $\widetilde t_1 \to Wb\GG$ becomes kinematically available and dominates
in the absence some non-trivial stop-scharm mixing. 

CDF has also performed the search for $\widetilde t_1 \widetilde t_1$
pair production
in the lepton-plus-jets top data mode \cite{gmsbcdfstop}
which applies for heavier $\widetilde t_1$'s which decay by 
$\widetilde t_1 \to Wb\GG$.
The signal is discriminated from the
top quark background based on the shape of the transverse mass
distribution of the lepton and $\missET$.  This  analysis is
sensitive only for $\widetilde t_1$ masses greater than
90~GeV due to the lepton
$E_T$ and $\missET$ cuts.  This analysis is expected to be sensitive
up to the threshold where the decay $\widetilde t_1\rightarrow t \GG$
dominates.

\subsection{Mesino and Sbaryon Displaced Decay Signals}
\label{subsec:dissquark}

The NLSP squark decay length to the Goldstino depends very sensitively
on the supersymmetry breaking scale, and may take place over 
a macroscopic distance. 
For example, 
using the expressions for squark decay rate to the Goldstino given 
in Appendix \ref{app:goldstino}, for 
$m_{\tilde t} = 150$ GeV 
$\Gamma^{-1}(\tilde{t}_R \rightarrow bW \widetilde G) \simeq 
75~{\rm cm}~(\sqrt{F} / 100~{\rm TeV})^4$, while for 
$m_{\tilde t} = 190$ GeV
$\Gamma^{-1}(\tilde{t} \rightarrow t \widetilde G) \simeq 
0.75~{\rm cm}~(\sqrt{F} / 100~{\rm TeV})^4$.
The visible decay products of the squark decay can therefore
be displaced from the production vertex. 
Since the decay products always involve strongly interacting partons, 
an NLSP squark hadronized in a mesino or sbaryon bound state 
can give rise to displaced jets
with both large transverse
energy ($E_T$) and large missing transverse energy ($\met$).

Any search for squark decay involving displaced jets well within 
the tracking volume must contend
with backgrounds from analogous heavy ($c$- and $b$-) 
quark decays which can also give large $E_T$ displaced jets. 
This is particularly true for a metastable stop squark which 
decays to heavy quark flavors so that the background can not 
be reduced by simply anti-tagging on heavy flavor displaced jets. 
For sufficiently large squark decay length, the signal can 
be distinguished from the heavy quark background by 
the large beam axis impact parameter. 
An $\met$ cut can also significantly reduces 
background involving heavy quark hadronic decay modes. 
In addition, since the massive mesinos and sbaryons are non-relativistic,
the decay products are not significantly boosted in the lab frame, 
and are distributed roughly uniformly. 
This is in contrast to high $E_T$ displaced jets from heavy 
quark decay, which are highly boosted in the direction of the
relativistic heavy quark motion, namely away from the interaction vertex. 
So the angular distribution of displaced high momentum jets can 
greatly aid in the separation of a squark signal from heavy 
quark background. 
A useful observable in this regard is \cite{mesino}
\begin{equation}
\cos \varphi \equiv { \vec{p}_{\rm jet} \cdot \vec{n} \over 
                     | \vec{p}_{\rm jet} |  }
\end{equation}
where $\vec{p}_{\rm jet}$ is the three-momentum vector of the 
displaced jet 
and $\vec{n}$ is the unit normal from the beam axis to the 
origin of the displaced jet. 
The distribution of high $E_T$ displaced jets from direct heavy quark
production and  
decay is concentrated in $0 \lsim \cos \varphi \lsim 1$, with
$\cos\varphi \sim 0$ corresponding to high pseudorapidity. 
In contrast, the distribution from mesino or sbaryon decay 
is roughly uniformly distributed in $-1 \lsim \cos \varphi \lsim 1$. 
Mesino and sbaryon decay may therefore be distinguished by high $E_T$ 
jets with large $\met$ and with large negative impact parameters
(LNIPs), where the sign of the impact parameter is taken to be 
${\rm sgn}(\cos \varphi)$.
Negative impact parameters defined in this way result 
if the visible decay 
products recoil against the invisible Goldstino  
in a direction towards, rather than away from the 
production vertex. 
Equivalently, the jet appears to originate on the ``wrong side''
of the production vertex. 

Stop squarks which decay with a macroscopic decay 
length either by $\tilde t \to c \GG$
or $\tilde t \to bW \GG$ or $\tilde t \to t \GG$ could
be uncovered with an LNIP search. 
In the case of the charm final state an LNIP search could
be applied to the SUGRA search for stop pair production in 
events with two charm jets and $\met$. 
In the case of the top-like final states, the top quark background 
could be significantly reduced by requiring  
large impact parameters for the $W$ decay products
in a top quark sample. 
This could easily be implemented by requiring that none of the
leptons be associated with the interaction vertex in a di-lepton 
or lepton-hadron top quark sample. 

LNIPs 
provide an efficient means to search for any exotic massive 
metastable particles which decay to hadronic final 
states. 
In particular, a search for LNIPs would 
also be sensitive to a Higgsino-like neutralino NLSP, discussed 
in section \ref{sec:higgsino} and \ref{subsec:nphiggisino}, 
which decays with a macroscopic decay length 
by $\NI \to h\GG$ with $h \rightarrow bb$ or 
$\NI \to Z\GG$
with the $Z$ decaying hadronically. 

Displaced jets arising from neutral mesinos or sbaryons which 
decay within the tracking region would appear as incomplete tracks
leading to jets with very large impact parameter. 
A charged mesino or sbaryon which decays within the tracking 
region might be identified as a HIT stub intersecting 
a displaced jet.  
A search for decay lengths in this regime would require rather 
specialized analysis of events with large hadronic $E_T$
and $\met$.

\subsection{Quasi-stable Mesino and Sbaryon Signals}
\label{subsec:qssquark}

Quasi-stable NLSP squarks hadronized in mesino and sbaryons
with long enough life times to partially or completely 
traverse a detector lead to a number of interesting 
phenomena. 
It should first be noted that 
even though squarks carry significant momentum from the 
production vertex, there can only be very soft hadronic  
activity associated with squark hadronization. 
This is because massive squarks are non-relativistic, 
and in the heavy particle limit it is the relative velocity
which determines the magnitude of any jet activity 
associated with hadronization. 
Likewise, when a non-relativistic hadronized mesino
or sbaryon decays by decay of the constituent squark, 
the bound state light quark(s) do not lead to significant
jet activity since by decoupling the high momentum squark decay products
can not transfer significant 
energy to the low momentum light quark(s). 
The energy of the bound state quark(s) are only of order the 
QCD scale binding energy. 
These observations about 
jet activity are of course also true for any of 
the non-prompt squark 
decay signatures discussed in sections \ref{subsec:dissquark}
or \ref{subsec:mesosc}, and also for the 
prompt squark decay signatures discussed in section \ref{subsec:psquark}
since in that case the decay length still greatly exceeds the 
hadronization scale. 

Long-lived squarks may hadronize as either neutral or charged
mesinos or sbaryons. 
A neutral non-relativistic strongly interacting bound state
will experience only a few rather soft hadronic interactions
as it moves through a detector. 
The calorimeters will therefore only detect a small fraction 
of its energy compared with a usual relativistic jet of showering 
particles. 
This will result in an apparent $\met$ signature \cite{bjorken,hgluino}.
Since there are two NLSPs
produced in each event, and since there is minimal jet activity
associated with squark hadronization, 
the total missing transverse momentum vector can
point in a direction different from any jet in the event.

A charged non-relativistic quasi-stable mesino or sbaryon is very similar 
to a quasi-stable stau slepton. 
Because of the increased ionization of a slowly moving massive 
charged particle, highly ionizing tracks (HITs) result in the tracking 
region of a detector. 
However, a non-relativistic strongly interacting bound state which contains
light quarks can exchange isospin and charge 
with background material in a detector through hadronic 
interactions \cite{bjorken,hgluino}.
Such particles make transitions between neutral 
and charged states as they pass through matter. 
It is important to note that in the non-relativistic limit,
only bound states which contain 
light quarks (rather than anti-quarks), namely 
anti-mesinos and sbaryons, can significantly 
charge exchange with matter \cite{bjorken}.
Non-relativistic anti-mesinos and sbaryons moving through
a detector therefore yield the phenomenon
of intermittent charge exchange associated with highly ionizing tracks
(CE-HITs).
Depositions in the calorimeter alternate between highly ionizing charged
and neutral segments.

The oscillation length between charged and neutral anti-mesino
states
can be estimated in the heavy squark and chiral limits from 
the forward scattering isospin exchange amplitude in matter. 
In the chiral limit the neutral and charged states are degenerate
(ignoring electromagnetic mass splittings), and 
isospin exchange operators are of the form 
 $(4 \pi /\Lambda_{\chi})^2~\overline{\mesino} \vec{\tau} \Gamma 
\mesino\cdot\overline{N} \vec{\tau} \Gamma N$ 
where ${\cal M}$ and $N$ are the mesino and nucleon operators 
respectively, 
$\tau$ is the Pauli isospin matrix,
$\Gamma$ is a Dirac matrix representing the general 
Lorentz structure, 
$\Lambda_{\chi} \simeq 1.1$ GeV is the 
chiral symmetry breaking or cut off scale, and the estimate
of the coefficient follows from consistency of the cutoff chiral 
Lagrangian. 
This yields an oscillation length in matter of 
\begin{equation} 
 \lambda ( \overline{\cal M}^{\pm} \leftrightarrow 
           \overline{\cal M}^0 ) 
  \sim {1.2~{\rm m} \over \rho/({\rm gm}~{\rm cm}^{-3}) }
\end{equation} 
where ${\rho}$ is the material density. 
It should be noted that the actual oscillation length may 
differ from this estimate by up to a factor of a few. 

The exchange length per unit density
for a CE-HIT is small enough so that an antimesino or sbaryon
is unlikely to charge exchange in the inner tracking region of a 
detector. 
However charge exchange in the calorimeter region is possible. 
This can have the effect of converting a charged antimesino
or sbaryon which leaves a HIT in the inner tracking region
to a neutral state which does not register in the outer muon 
system. 
Conversely intermittent tracks in the muon system can 
arise from an initial neutral antimesino or sbaryon which 
does not register in the inner tracking region. 
The intermittent nature of CE-HITs can also reduce charge 
track trigger efficiency and contribute to $\met$.
Mis-identification of CE-HITs can also contribute to $\met$. 

A non-relativistic 
charged mesino or anti-sbaryon contain light antiquarks, 
and are therefore less likely to charge
exchange in matter than the anti-particle counterparts.
These slowly moving charged states experience a few rather soft
hadronic interactions as it moves through a detector, 
just as the neutral states. 
Because of the additional hadronic activity along the track, 
these particles yield hadronic highly ionizing tracks 
(H-HITs). 
In the inner tracking region H-HITs are likely to appear
simply as HITs. 
However, the additional hadronic activity could be
recorded by the calorimeter. 

Using CE-HITs as a discovery mode for a long lived NLSP squark
is problematic because of possibly significant backgrounds
from, for example, cosmics and because of the uncertainty 
in the charge exchange rate. 
More useful would be a HIT search based only on the inner tracking 
region. 
Both CE-HITs and H-HITs should appear as HITs in the inner tracker. 
Reconstruction of CE-HITs and H-HITs in the outer tracking 
region might be attempted in a sample of inner tracker HITs. 
TOF information utilizing the outer muon system might also be 
used, but with reduced efficiency compared with a stau HIT
due to charge exchange in the calorimeter. 
The $\met$ implied by the intermittency of CE-HITs, 
or the $\met$ from squarks hadronized in a neutral bound
state might also be useful in identifying events. 

It is interesting to note that a bound state NLSP squark could be 
identified as either an up- or down-type squark if the sign 
of the charged segments of CE-HITs could be determined. 
An up-type anti-squark bound in an antimesino with a 
quark is either negatively charged or neutral, 
$\overline{\cal M}^-$ or $\overline{\cal M}^0$.
Conversely, a down-type anti-squark bound in an antimesino
with a quark is either positively charged or negative, 
$\overline{\cal M}^+$ or $\overline{\cal M}^0$.
Since hadronization in (anti)sbaryon states is less likely
than in (anti)mesino states, and 
since CE-HITs arise mainly for antimesinos rather than mesinos, 
an excess of negative over positive 
CE-HITs would imply an up-type NLSP anti-squark, 
while an excess of positive over negative CE-HITs would imply a down-type
NLSP anti-squark. 

Finally, a bound state squark which decays within the detector
might be observed by a H-HIT which ends with a jet, 
an H-HIT to jet kink. 
Events of this type might be identified by 
HITs in the inner tracking region which end in a jet or 
deposited energy in the calorimeter. 

\subsection{Mesino Oscillations}
\label{subsec:mesosc}

A mesino bound state of a squark and antiquark, 
$\mesino_{\widetilde Q \qbar} \equiv (\widetilde Q \qbar)$, 
is a spin ${1 \over 2}$ Dirac fermion. 
A neutral mesino and its antiparticle differ by 
two units of (s)flavor, fermion number, $F$, and $R$-charge. 
All of these quantum numbers are, however, manifestly violated in 
any supersymmetric theory.
(S)quark flavor is violated by Yukawa couplings
and squark flavor may also be violated by scalar tri-linear
couplings and possibly the scalar mass-squared matrices. 
Fermion number and $R$-symmetry are violated by gaugino 
masses. 
Since no conserved quantum number distinguishes a mesino from antimesino,
these states can mix. 
So hadronization of squarks into neutral mesino bound states
allows for the interesting phenomenon of particle--antiparticle
oscillations which is impossible for an isolated charged particle. 

Mesino oscillations are analogous to meson oscillations. 
At the microscopic level the
$\Delta \tilde Q = \Delta q = \Delta F = \Delta R =2$
amplitudes which mix mesino and antimesino 
are typically dominated by tree level gluino exchange \cite{mesino}.
For a stop-like neutral mesino the gluino contribution to the
$\mesino_{\tilde{t} \overline u} \leftrightarrow
\overline{\mesino}_{\tilde{t}^* u}$ 
oscillation wavelength, $\beta \gamma \lambda$, 
is numerically \cite{mesino}
\begin{equation}
\lambda \simeq (\mbox{4 nm}) \,
\left( m_{\tilde g} \over 250 ~{\rm GeV} \right)^2 ~
{ f(m_{\tilde t}/m_{\tilde g}) \over 
  \sin^2 \theta_{\tilde{u} \tilde{t}}  }
\label{eq:osclength}
\end{equation}
where $\sin \theta_{\tilde{u} \tilde{t}}$
is the up squark-stop mixing angle$, f(y) = y (1-y^2)$, 
and by assumption $y < 1$ so 
$0 <f(y) < \sqrt[3]{2/3} \simeq 0.38$.
Oscillations on the scale of a detector could occur 
for $\sin \varphi_{\tilde{u} \tilde{t}}$ as small as $5 \times 10^{-5}$. 
Up-type sflavor violation involving the third generation is 
essentially 
unconstrained by present data, so extremely rapid oscillations
compared with the decay length and 
scale of a detector are conceivable.  

The time-integrated probability for an NLSP squark hadronized 
as a mesino to 
oscillate to an antimesino and decay as an antisquark 
depends on the oscillation frequency and decay rate
\beq
{\cal P}( \mesino \rightarrow \overline \mesino) = 
{x^2 \over 2(1 +  x^2) }
\label{PMM}
\eq
where $x = (2 \pi / \Gamma) / \lambda$
is the ratio of the decay length to oscillation length.
Rapid oscillations, 
$x \gg 1$, 
yield ${\cal P}( \mesino \rightarrow \overline \mesino) 
   = {1 \over 2}$,
while for slow oscillations, $x \ll 1$,  
${\cal P}( \mesino \rightarrow \overline \mesino) \to {1 \over 2} x^2$.

Neutral 
mesino--antimesino oscillations present the possibility of 
novel experimental signatures, even for decay lengths which are
too short to be resolved in real space. 
Oscillations may be revealed in any decay mode which tags
the sign of the (anti)squark in a neutral (anti)mesino. 
Squark--antisquark production 
events in combination with mesino--antimesino oscillation
can then lead to same-sign events. 
For example, the antisquark may hadronize as a neutral
antimesino which oscillates to a mesino before decaying, 
while the squark hadronizes as a charged mesino or sbaryon
which can not oscillate. 
Summing over all possibilities,
the time-integrated ratio of same- to
opposite-sign events is
\begin{equation}
R \equiv {N_{++} + N_{--} \over N_{+-} + N_{-+}}
= {2 {\cal P} f_0 (1 - {\cal P} f_0) 
   \over 1 - 2 {\cal P} f_0 + 2 {\cal P}^2 f_0^2}
\simeq 2 {\cal P} f_0 + 2 {\cal P}^2 f_0^2\,
\label{eq:Ratio}
\end{equation}
where ${\cal P} \equiv {\cal P}(\mesino \rightarrow \overline \mesino)$,
and $f_0$ is the neutral mesino hadronization fraction. 
Thus, for $x \gsim 1$ a significant fraction of squark--antisquark
events will yield same-sign events. 

The feasibility of determining the sign of an (anti)squark 
at decay depends on the decay products. 
For stop-like squark decays 
$\tilde t \rightarrow b W \widetilde G$ or 
$\tilde t \rightarrow t \widetilde G$ with 
$t \rightarrow bW$, the $W$-bosons
reliably tag the sign of the (anti)squark. 
The $W$-bosons signs may in turn 
be determined with the leptonic decay mode 
$W \rightarrow \ell \nu$ where $\ell=e,\mu$.
This requires isolating these primary 
leptons from any secondary leptons arising
from $b$-quark decay. 
Such distinctive, essentially background free
events have the topology of same-sign top-top
events in the di-lepton channel, 
and provide a possible discovery mode for SUSY. 
The largest background is probably from 
top-antitop production with the very small 
probability of misidentification of the primary 
leptons or mismeasurement of the charges. 
At the Fermilab Tevatron Run IIa with 2 fb$^{-1}$ of 
integrated luminosity,
a 175 GeV stop squark with 
dominant decay $\tilde t \rightarrow bW \widetilde G$ and 
oscillation parameter 
$x \sim 1$ would yield $\sim 10$ same-sign dilepton top-top
events, while $x \gg 1$ would yield $\sim 20$ events. 
A detection acceptance times efficiency $\gsim 30$\%
should give a detectable signal for these parameters. 

Observation of stop mesino oscillations requires, and 
very sensitively probes, up-type squark sflavor violation. 
For example, for a stop decay length $\Gamma^{-1} \sim 10$ cm, 
maximal  
$\mesino_{\tilde{t} \bar{u}} \leftrightarrow
\overline{\mesino}_{\tilde{t}^* {u}}$
mixing, $x \gsim 1$, occurs for all 
$\epsilon_{(N)13} \gtrsim
5 \times10^{-5}$.
Even for $\Gamma^{-1} \sim 2~\mu {\rm m}$
(which could not be resolved as a displaced vertex), 
maximal mixing occurs for any 
$\epsilon_{(N)13} \gtrsim 10^{-2}$.
The magnitude of squark sflavor violation depends on the 
scale at which (s)quark flavor is broken. 
If the flavor scale is not too much larger than the messenger
scale for transmitting supersymmetry breaking, interesting
levels of sflavor mixing are expected, and 
observable mesino oscillations can occur. 

The flavor violating two-body decay $\tilde t \rightarrow c 
\widetilde G$ can dominate if sflavor violation is large enough, 
as discussed above. 
This mode also dominates if the NLSP squark
is scharm-like, 
$\tilde c \rightarrow c \widetilde G$. 
Semi-leptonic decay of the $c$-quarks 
hadronized in $D^{0,\pm}$-mesons could
then be used to tag same-sign events 
in high $E_T$ 
charm-jets with large $\met$.
$D^0 \leftrightarrow \overline D^0$ oscillation is 
negligible and would not contaminate a mesino oscillation
signal at the discovery level in a relatively clean sample
of LNIPs.  
However, squark decays in this mode with a decay length that
is too short to resolve using LNIPs would be contaminated
by standard model production of 
$b$-jets which are not easily distinguished from charm-jets. 
This standard model background could be significant since 
$B^0 \leftrightarrow \overline B^0$ oscillations are non-negligible.
Self-tagging of the heavy flavor at production 
to determine its sign 
or measuring total jet charge after decay to isolate $D^{\pm}$
mesons which do not oscillate
could reduce this background, but requires large
statistics, and is probably 
not applicable at the discovery level. 

Observing oscillations for 
a squark NLSP which decays to other flavors is more problematic. 
For a sbottom-like squark which decays by 
$\tilde b \rightarrow b \widetilde G$, the $B^0$-meson backgrounds
discussed above are important. 
Decays to lighter quarks, 
$\widetilde Q \rightarrow (u,d,s)\widetilde G$, are difficult
to sign using self tagging. 
So a stop-like NLSP squark provides the best opportunity to observe
mesino oscillations, in particular in same-sign events
in the di-lepton top-top channel \cite{mesino}.

It is worth noting that if the mesino decay length
is macroscopic {\it and} the oscillation length
is fortuitously of the same order,  
$x \sim 1$, mesino oscillations could be observed in real space
in the signed decay length distributions. 
This would allow a direct measure of the oscillation length, 
and give an accurate determination of the up-squark-stop mixing 
angle. 

\subsubsection{CDF study of Mesino Oscillations}

The CDF top quark analysis results
may be used to investigate the like sign top-top mode
in the di-lepton channel.
In Run I with 109~pb$^{-1}$ of integrated luminosity,
9 events with opposite--sign leptons, $\missET$
and two jets were found, with an expected background of $2.4\pm 0.5$.
Two events with like--sign leptons were also found, with an expected
background of $0.61\pm 0.44$\cite{cdfgmsbtop}.  Using simple Poisson
statistics, it is possible to
exclude more than approximately 6 events of anomalous
like-sign top-top production.
To set the scale, note that 6 events is approximately the same as what
is expected from the top cross section of 5~pb.
So Run I would not be sensitive
to stop pair production with mesino oscillation for a stop
mass similar to the top quark mass since in this case the 
stop pair production 
cross section is an order of magnitude smaller than for 
top pair production. 


\section{Other NLSP Scenarios}\label{sec:waco}
\setcounter{equation}{0}
\setcounter{footnote}{1}
\indent

In principle any of the MSSM superpartners may be an NLSP.
It is therefore worth considering the signatures
associated with NLSP types not studied in detail
as part of the Run II workshop.
While perhaps unconventional and not as well studied,
because of some novel and unique signatures associated with
other NLSP types, these possibilities should not be overlooked in the
search for SUSY in Run II.

\subsection{Gluino NLSP}

The gluino is most likely to arise as the NLSP in theories of
GMSB with both suppressed
supersymmetry breaking for the strongly interacting messenger
fields and suppressed $U(1)_R$ breaking.
Since the gluino does not mix with any other states, it can
only decay to the Goldstino through emission of a gluon,
\beq
\gluino \to g \goldstino
\eq
Direct gluino pair production is the largest SUSY production cross 
in the gluino NLSP scenario. 

If the decay $\widetilde g \rightarrow g \widetilde G$ occurs promptly,
then the signature will be a pair of very hard gluon jets, accompanied
by large $\Eslash$.
A detailed study of this signal has not been performed. 
But there may be some 
reach, possibly significant, in this channel at Run II, provided
that the cuts on the $\Eslash$ and the $E_T$ of the two hardest jets
in an event are chosen suitably high to defeat backgrounds from
$Z\rightarrow \nu\overline \nu$ production, jet energy mismeasurements,
missed partons, etc.

A gluino NLSP with
macroscopic decay length can lead to other interesting signatures.
An NLSP gluino, which is an octet of $SU(3)_C$,
hadronizes with either a gluon or
light quark and antiquark to form what is
generally referred to as an $R$-hadron bound state,
$\Rhadron^0 \equiv ( \gluino g )$ or
$\Rhadron^{0,\pm} \equiv ( \gluino q \bar{q} )$.
If the decay length is macroscopic, but contained within the
tracking region, large $E_T$ displaced jets with large $\met$
result. 
These can be identified in an LNIP search described in 
section \ref{subsec:dissquark} in the context of a squark NLSP. 

For decay lengths comparable to or larger than a detector, 
the signatures of the charged or neutral $R$-hadron bound states
are similar to the quasi-stable mesino and sbaryon signatures
discussed in section \ref{subsec:qssquark}.
If the mass difference between $\Rhadron^{\pm}$ and $\Rhadron^0$
is less than the pion mass, then both are effectively stable
on the scale of the detector. 
A neutral  $\Rhadron^0$ contributes to $\met$. 
A charge $\Rhadron^{\pm}$ should appear 
as a HIT in the inner tracking region. 
In addition, an $R$-hadron can charge exchange with detector
material, yielding an intermittent charge exchange associated with a 
highly 
ionizing track (CE-HIT) as described in more detail for a squark NLSP in 
section \ref{subsec:qssquark}.
Since charge exchange is probably not significant in the 
inner tracking region of the detector, the best search 
strategy may be an attempt to identify CE-HIT signatures
in the outer tracking region in a sample of inner tracking region 
HITs. 
The $\met$ which arises neutral  neutral $\Rhadron^{0}$-hadrons
and the intermittent
nature of CE-HIT from charged $\Rhadron^{\pm}$-hadrons
may also be useful in identifying events. 
Unlike antimesino and sbaryon CE-HITs, the charged segments
of $R$-hadron CE-HITs are equally likely to have positive
or negative charge. 

\subsection{Singletino NLSP}

Some extensions of the MSSM contain
singlet superfields without Standard Model quantum numbers.
Gauge invariance implies that
such scalar singlet and fermionic singletino
fields can mix at the renormalizable
level only with the Higgs and Higgsino fields.
Singlet fields may in fact be motivated in GMSB
by the necessity of additional interactions in the
Higgs sector \cite{dnmodels,singletino}
discussed in Appendix A2i.
If the NLSP is a neutralino with a large fermionic singletino component,
it decays mainly through its mixing with the Higgsino to a Higgs
or $Z$ boson,
\beq
\NI \to (h,Z) \GG
\eq
The signatures for prompt decay are then very similar to the Higgsino-like
neutralino NLSP
discussed in section \ref{sec:higgsino}.
If the scalar singlet, $\phi$,
is lighter than $\widetilde\chi_1^0$, the NLSP decay
\beq
\NI \to \phi \GG
\eq
dominates.
However, this still generally gives rise to visible signatures since
$\phi$ is likely to decay mainly through its mixing with Higgs by
$\phi \to bb$ \cite{singletino}.

\subsection{Neutralino-Stau Co-NLSP}

The slepton co-NLSP scenario discussed in section \ref{sec:sleptonconlsp}
in which the splitting between the
sleptons is smaller than the lepton masses is natural if the
messenger interactions do not violate lepton flavor, as is the case
in the MGM.
It is also possible that some of the superpartners just happen
to be close enough in mass to give a co-NLSP scenario, even
though no symmetry enforces this near degeneracy.
This in fact occurs in the MGM at low to moderate $\tan \beta$
with $N=2$ generations of messengers and the messenger scale not too
far above the supersymmetry breaking scale.
In this case the lightest neutralino, $\NI$, and the right handed
sleptons, $\tilde{l}_R$, just happen to be nearly degenerate \cite{mgm}.
A near degeneracy between $\NI$ and $\stau_1$ can also occur
in a somewhat smaller region of parameter space for
$N=3$ with large $\tan \beta$.
For  $|m_{\tilde \chi_1^0} - m_{\tilde\tau_1}| < m_{\tau}$,
either the decays $\NI \rightarrow \tau \stau_1$ or
$\stau \rightarrow \tau \NI$
are blocked kinematically, depending on the mass ordering.
In this case $\NI$ and $\widetilde \tau_1$
share the role of NLSP, since they each have no kinematically allowed
decays except into the Goldstino.
In this scenario, all SUSY decay
chains terminate either in
\beq
\NI \rightarrow \gamma \GG~~~~{\rm or}~~~~\widetilde \tau_1
\rightarrow \tau \GG
\eq
Cascades passing through $\NI \NI$ or $\stau_1 \stau_1$ should
give similar signatures to those of the
Bino-like Neutralino NLSP discussed in section \ref{sec:bino} or the
stau NLSP discussed in section \ref{sec:stauNLSP} respectively.
Cascades passing through $\NI \stau_1$ would give a mixed signature,
and could be searched for in a $\gamma X  \met$ event sample.
The sensitivity in each these modes of course
depends on the precise branching ratios of the cascade decays.

\subsection{Sneutrino NLSP}

Finally, for completeness the possibility of a sneutrino NLSP
may also be considered.
Both $\widetilde \nu$ and $\nu$ interact only weakly, and so escape the detector
without depositing energy.
An NLSP decay $\widetilde \nu \to \nu \GG$ is therefore unobservable.
In this case the only signatures available would be from cascade decays,
$X \met$,
to the NLSP $\tilde{\nu}$.


\section{Direct Goldstino Pair Production}\label{sec:directGG}
\setcounter{equation}{0}
\setcounter{footnote}{1}
\indent

The overall mass scale for Standard Model superpartners could be
related to the electroweak scale, but may still be too heavy to allow
direct superparticle production in Run II.
However, with low scale supersymmetry breaking the gravitino
is very light compared with collider energies, and is
always kinematically accessible.
In principle then, final states could arise with the Goldstino
components of the gravitino as the only supersymmetric particles.
With $R$-parity conservation supersymmetric particles are produced
in pairs, so the simplest example of this type is Goldstino pair production.
At a hadron collider the dominant contribution is likely
to come from $t$- and $u$-channel virtual squark exchange \cite{gmsbcdfGG}.
The weakly interacting Goldstinos escape the detector, yielding
an invisible event.
However, initial--state radiation may be used to tag such events,
giving a signature of one jet and
$\missET$ or one photon and $\missET$.
The supersymmetry breaking scale must be extremely low (right at the
electroweak scale) in order for this process to yield an
observable signal.
If this interesting scenario were realized in nature, essentially
all the superpartners would probably be very strongly coupled
to the supersymmetry breaking sector.

\subsubsection{{\rm CDF studies of direct Goldstino production}}

CDF has searched for the one jet and $\missET$ signature in the Run I data
\cite{cdfprldirectgoldstino}.
The trigger is based on $\missET$ which has a nominal 35~GeV threshold.
A large set of clean--up cuts are necessary to remove the copious backgrounds
from cosmic rays and from QCD events with severely mismeasured $\missET$.
In addition, the leading jet must have $E_T>80$~GeV and there must be no
leptons with $E_T>10$~GeV, which could indicate a $W\rightarrow \ell\nu$ decay.
The final $\missET$ requirement is optimized to be 175~GeV.

With this set of cuts, 19 events remain while $22\pm 7$
are expected from Standard Model sources, mostly from $Z\rightarrow \nu\nu$
plus an initial--state radiation jet.
Comparing to cross section estimates
for Goldstino pair production \cite{gmsbcdfGG},
the result
excludes approximately $\sqrt{F}>217$~GeV which corresponds to
$m_{\tilde G}\ge 1.1\times 10^{-5}$~eV.
For Run II, the important backgrounds have been extrapolated
using the ratio of the cross sections at 1.8~TeV and 2.0~TeV.
The result is an expected 20\% increase in the backgrounds.
Assuming that the systematic uncertainty in the background
remains at 30\%, the expected limit with 2~fb$^{-1}$
of integrated luminosity is $\sqrt{F}>260$~GeV or
$m_{\tilde G}\ge 1.6\times 10^{-5}$~eV.
The rather modest increase in the expected bound
on the supersymmetry breaking
scale (or equivalently the Goldstino decay constant) over
the Run I limit follows from the derivative coupling of the Goldstino
which implies the cross section is a rapidly falling function
of the supersymmetry breaking scale,
proportional to  $(1/\sqrt{F})^8$
or $(1/m_{\tilde G})^4$.


\section{Summary}\label{sec:summary}
\setcounter{equation}{0}
\setcounter{footnote}{1}
\indent

In this report the potential of the Fermilab Tevatron Run II for
discovery and study of low-scale supersymmetry breaking 
has been assessed. 
Low scale gauge mediated supersymmetry breaking
provides a particularly attractive theoretical solution to the problem of
flavor-violation in the supersymmetric Standard Model. 
The existence of a nearly massless Goldstino to which the 
NLSP can decay, provides an 
attractive and rich set of experimental possibilities for Run II,
with a variety of potentially spectacular signals.
The various classes of signals depend on the identity of the NLSP 
and its decay length. 

The experimental signatures which have been identified 
in this report as being useful for
the Tevatron Run II and future upgrades are summarized in Table 
\ref{tab:summary} and below:
\begin{table}[hbtp]
\caption{Experimental signatures for different NLSP scenarios.
LNIP $\equiv$ Large Negative Impact Parameter.
MIT $\equiv$ Minimum Ionizing Track (muon candidate).
HIT $\equiv$ Highly Ionizing Track (anomalously large $dE/dx$).
CC-HIT $\equiv$ Charge Changing Highly Ionizing Track.
CE-HIT $\equiv$ Charge Exchange Highly Ionizing Track.
H-HIT $\equiv$ Hadronic Highly Ionizing Track.
TOF $\equiv$ large Time of Flight measurement.
$X$ $\equiv$ Additional partons in the final state. If the
decay length is comparable to the size of the detector, then
signatures from two or three columns can appear simultaneously.
}
\label{tab:summary}
\begin{tabular}{llll}
\hline \hline \\
   \multicolumn{1}{c}{NLSP}
  & \multicolumn{1}{c}{Prompt Decay}
  & \multicolumn{1}{c}{Macroscopic}
  & \multicolumn{1}{c}{Long-lived}  \\
 & & \multicolumn{1}{c}{Decay Length} & \\
 & & & \\
\hline  \\  & \\
Bino-$\NI$         & $\gamma \gamma ~X ~\Eslash$
                   & (Displaced $\gamma$) $~X~ \Eslash ~~~$
                                   & $X ~\Eslash$    \\
                                   & & TOF & \\
                                   & & & \\
Higgsino-$\NI$     & $(\gamma,h,Z)(\gamma,h,Z) ~X~ \Eslash ~~~$
                   & (Displaced $\gamma$~,& $ X~ \Eslash$ \\
                                   & [$ \gamma b ~X~ \Eslash$,
~~$\gamma b j X \Eslash$,
                                   & ~Displaced $Z$~,
                                   &           \\
                                   & $\gamma j j ~X~ \Eslash$,~~
$\gamma  X \Eslash$,
                                   & ~LNIP $b$-jets ) $~X~ \Eslash$
                                   &    \\
                                   & $b \overline b ~X~ \Eslash$,
~~                   $ bbb ~X~ \Eslash$,
               & TOF  &  \\
                                   & $\gamma\ell\ell ~X~ \Eslash$,
~~ $\ell\ell\ell\ell ~X~ \Eslash$] ~~~
&
&
\\ 
                                   & & & \\
$\stau_1$ & $\tau^{\pm} \tau^{\pm} ~X~ \Eslash$ & HIT $\rightarrow \tau$~~
                                                          kinks & HITs \\
        & $\ell^{\pm} \ell^{\pm} ~X~ \Eslash$
                   & HIT $\rightarrow e,\mu$~~kinks & Same-Charge HITs \\
& $\tau\tau\tau ~X~ \Eslash $ & & Same-Charge MITs \\ 
		   &  $\tau\tau\ell ~X~\Eslash $ 
						   & & $\ell\ell\ell X
\Eslash$ \\
                   &  $\tau\ell\ell ~X~\Eslash$ & & $\ell\ell\ell\ell X
\Eslash$ \\
                   &  $\ell\ell\ell ~X~\Eslash$ & & CC-HITs \\
                   &  $\tau\tau\ell\ell ~X~\Eslash$ & & TOF \\
                   &  $\tau\ell\ell\ell ~X~\Eslash$ & &  \\
                                   & & & \\
$\slepton$ ~co-NLSP ~~~               & (as for Stau NLSP, but
                                   & HIT $\rightarrow e,\mu,\tau$~~kinks
                                   & HITs \\
                                   & with different profiles, &
& $\ell\ell\ell X \Eslash$            \\
                       &
lepton democracy)
& &
$\ell\ell\ell\ell X \Eslash$ \\
                                   & 
$\ell\ell\ell\ell ~X~\Eslash$ &
& TOF \\
                                   & & & \\
$\squark$          & $ jj~X$
                   & Displaced jets  & CE-HITs \\
                                   & $cc~X~\Eslash$
                                   & H-HIT $\to$ jet kinks
                                   & H-HITs \\
                                   & $bb~X~\Eslash$
                                   & LNIPs & $\met$ \\
                                   & $tt~X~\Eslash$ 
								   & Mesino Oscillations &  TOF \\
                                    & Same-Charge $tt~X~\Eslash$ \\
                                   & & & \\
$\gluino$          & $jj~X~\Eslash$
                   & Displaced jets
                                   & CE-HITs \\
                                   &  & LNIPs & H-HITs \\
                                   & & &$\Eslash$ \\
                                   & & &TOF \\
                                   & & & \\
\hline \hline
\end{tabular}
\end{table}

\begin{enumerate}
\item Bino-Like NLSP 
\begin{itemize}
\item For prompt decays of a Bino-like $\widetilde \chi_1^0$ NLSP, there
will be a very substantial reach in the $\gamma\gamma ~X~ \met$ channel,
where $X$ can be anything, but likely includes jets.
\item Cascade decays to the Bino-like $\widetilde \chi_1^0$ NLSP 
can include a neutral Higgs boson, $h$, with fairly high probability. 
The SUSY signature of two hard photons and $\met$ could then be used 
as a unique method of obtaining a sample of Higgs bosons. 
\item For macroscopic decay lengths of a Bino-like $\widetilde \chi_1^0$
NLSP, the resulting displaced photons
can be resolved and provide a useful signal. 
The properties of the \D0\ 
preradiator allow for a particularly sensitive probe of decay lengths down
to the
few centimeter level.
\end{itemize} 
\item Higgsino-Like NLSP
\begin{itemize}
\item Prompt decays of a Higgsino-like $\NI$ NLSP can yield $\gamma$, $h$, 
and $Z$ bosons, giving rise to signatures with photons, $b$-jets, jets, 
and reconstructed leptonic $Z$ bosons in 
combinations that depend strongly on the underlying SUSY parameters.
This allows for a particularly rich set of possibilities for 
event selections. 
Many of the signatures are interesting on general
grounds, since they can arise in other, unrelated, new physics scenarios.
\item The presence of Higgs bosons from $\NI$ decays can lead to
a interesting source of tagged Higgs events. 
\item For decays of a Higgsino-like $\NI$ NLSP with macroscopic
decay length, but contained within the 
detector, displaced photons, displaced $Z$ bosons, or displaced Higgs bosons
arise.
The displaced hadronic final states, including $b$-jets from displaced 
Higgs decay, 
yield tracks with large negative impact parameters (LNIPs)
with reconstructed displaced jets pointing towards, rather than away from, the 
beam axis. 
\end{itemize}
\item Stau NLSP
\begin{itemize}
\item Prompt decays of a $\widetilde \tau_1$ NLSP give rise to
events with same-charge taus (either manifested as hadronic one-prong or
three-prong decays, or as leptonic decays). 
Depending on the underlying SUSY parameters
a variety of different multi-lepton and multi-tau event selections
are possible. 
In some cases, it is best to
require an additional one or two hard jets,
since these occur in SUSY cascade decays but not in relevant backgrounds.
All these signatures
depend crucially on tau identification efficiency, which will 
need to be evaluated once the detectors are operating. 
\item Stau decays that take place within the instrumented region 
of a detector yield events with a decay kink. 
Such a non-relativistic stau leaves a highly ionizing track (HIT)
with a kink to a hadronic tau jet or an $e$ or $\mu$ from a 
leptonic tau decay.
\item Quasi-stable staus which traverse the entire detector before
decaying will appear either as HITs or an excess of fake
``muons", i.e. minimum ionizing tracks (MITs). Both CDF and \D0 have
found a significant reach in this search. In addition, the CDF
time-of-flight (TOF) detector has been found to be quite useful in this
search since the staus are non-relativistic. 
Stau pairs 
will often have the same charge, allowing another useful handle on the events.
\item The charge changing three-body cascade decays of selectrons and
smuons, for example, $\widetilde e_R^+ \rightarrow
e^+ \tau^+ \widetilde \tau_1^-$, can also have a macroscopic decay length
if the selectron (or smuon) is nearly degenerate with the stau.
The electron and tau released in the decay
are typically very soft (and could easily be missed)
with the final state stau traveling in the forward direction. 
This gives rise to charge-changing HITs (CC-HITs), since the
non-relativistic selectron (or smuon) can convert 
to a stau of the opposite charge within the tracking region. 
\end{itemize}
\item Slepton Co-NLSP
\begin{itemize}
\item If the three lightest sleptons $\widetilde \tau_1$, $\widetilde e_R$,
and $\widetilde \mu_R$ are degenerate in mass to within 1.8 GeV, then
all three play the role of the NLSP. 
If the decays of the slepton co-NLSPs
are prompt, a variety of signatures involving taus and leptons result. 
The event topologies are very similar to those in
the Stau NLSP case, but the flavor and multiplicity profiles can be 
quite different.
In particular, there is a greater tendency for lepton democracy in the
events.
\item As in the case of a Stau NLSP, macroscopic or long decay lengths
for slepton co-NLSPs can give rise to HIT$\rightarrow $ lepton kinks,
HITs through the detector, an excess of fake ``muons", and an anomalous
TOF.
\end{itemize}
\item Squark NLSP
\begin{itemize}
\item Prompt decay of a stop-like squark NLSP to a top-like
final state gives a signature with a top quark event topology.
Large $\met$, lepton-$b$-jet invariant mass, and $W$
boson polarization can be used to partially separate these
from top quark backgrounds. 
Prompt decay to a charm final state
can be searched for in a standard SUGRA analysis for stop pair production. 
\item Decay of a squark NLSP over a macroscopic distance, but 
contained within the tracking region, gives rise to displaced jets
with large $E_T$ and $\met$.  
The angular distribution of the displaced jets is roughly uniform, 
and may be searched for in events with large negative impact parameters
(LNIPs) with the reconstructed displaced jets pointing towards,
rather than away from, the beams axis. 
\item 
Quasi-stable squarks (anti-squarks) will hadronize to form mesinos and
sbaryons (anti-mesinos and anti-sbaryons). The slowly-moving anti-mesino 
and sbaryon bound states can exchange isospin and charge with background
material in the course of traversing a detector.  Non-relativistic bound
states of these types therefore yield intermittent charge-exchange
associated with highly ionizing tracks (CE-HITs) which alternate between  
highly ionizing charged and neutral segments in the calorimeter.
Mis-identification of CE-HITs can contribute to $\met$. Quasi-stable     
mesinos or anti-sbaryons do not as readily charge exchange. Squark and
gluino bound states can also yield fairly soft hadronic activity along a
highly ionizing track (H-HIT). This is due to inelastic hadronic
interactions of the bound state with the calorimeter materials. 
Even though non-relativistic hadronized NLSP squarks can
carry significant momentum, they deposit little energy in the
calorimeters.  Both CE-HITs and H-HITs are likely to appear as HITs in the
inner tracking region.
\item NLSP squarks hadronized in mesino bound states can undergo 
mesino-antimesino oscillations.
For stop-like squarks this yields events with a same sign top-top topology.
Mesino oscillations might also be observed directly as oscillations
in the signed decay length distributions. 
\end{itemize}
\item Gluino NLSP
\begin{itemize}
\item Prompt decay of a gluino NLSP will lead to events
with two very hard gluon jets and very large $\Eslash$.
\item For decays of a gluino NLSP with macroscopic
decay length, but contained within the 
detector, large $E_T$ displaced gluon jets with large $\met$
result. 
These can be searched for in a sample of LNIPs. 
\item Quasi-stable gluino NLSPs hadronize as $R$-hadrons and
can charge exchange with matter, resulting in CE-HITs. 
Charged non-relativistic $R$-hadrons should appear as HITs in the 
inner tracking region and have anomalous TOF. 
\end{itemize}
\end{enumerate}
We look forward to the implementation of these signatures in searches
for low-scale supersymmetry using real Run II data.


\addcontentsline{toc}{section}{Appendix A: Minimal gauge
mediation and variations}
\section*{Appendix A: Minimal gauge
mediation and variations}\label{sec:minimalgmsb}
\setcounter{equation}{0}
\setcounter{footnote}{1}
\setcounter{subsubsection}{0}
\renewcommand{\theequation}{A.\arabic{equation}}

The precise definition of the minimal model of
gauge-mediated supersymmetry breaking (MGM)
used in many of the studies presented in this report
is outlined below.
This model, however, can not represent a full theory of gauge-mediated
supersymmetry breaking.
Some of the important variations and extensions
of the minimal model which affect the
phenomenology are therefore also described below.
For reviews of the MGM see Ref. \cite{mgm,gmsb}.


\subsubsection{Minimal model of gauge mediation}\label{app:mgm}
\indent

Gauge-mediated supersymmetry breaking results if
some of the fields in the messenger sector which feel
supersymmetry breaking also transform under the
Standard Model gauge group.
Soft masses for visible sector superpartners then arise
radiatively.
The successful supersymmetric prediction of gauge coupling
unification is not affected if the messenger fields
form unifiable representations.
In the MGM the messenger fields are taken to be $N$ generations of
chiral supermultiplets,
$\Phi_i$ and $\overline{\Phi}_i$, each transforming as
${\bf 5} \oplus {\bf \overline 5} \in SU(5) \supset
SU(3)_C \times SU(2)_L \times U(1)_Y$.
The messenger fields couple to a Standard Model
singlet chiral superfield, $S$, through the superpotential coupling
\beq
W = \lambda S \Phi_i \overline \Phi_i,
\label{SPP}
\eeq
where $\lambda$ is a dimensionless coupling.
Both the scalar component of the singlet superfield,
$S$, and auxiliary component, $F_S$, are assumed to acquire
expectation values.
The origin of the these expectation values are
not specified in the MGM, and may simply be taken as
background spurions.
In a full theory the $F_S$ auxiliary expectation value
felt by the messengers may not coincide with the intrinsic
supersymmetry breaking order parameter $F$ which determines
the Goldstino decay constant.
For this reason it is useful to define a factor $C_G \geq 1$
which relates the Goldstino decay constant with $F_S$:
\beq
F = C_G F_S .
\eeq
For a fixed value of $F_S$, the decay rate of the NLSP to
its Standard Model partner and Goldstino is therefore proportional
to $1/C_G^2$.
In a full theory the auxiliary expectation values
would presumably arise ultimately from
non-perturbative gauge dynamics in the supersymmetry breaking
sector.

The fermionic components of
$\Phi_i$ and $\overline\Phi_i$ obtain Dirac masses from the $S$ scalar
expectation value
\beq
m_{\psi_i} = \lambda S \equiv M_m .
\label{messengerfermionmasses}
\eeq
The auxiliary expectation value gives rise to a holomorphic
soft supersymmetry breaking mass for the scalar messengers.
The resulting scalar  mass squared matrix
\beq
\pmatrix{ m^2_{\psi_i} & \lambda F_S \cr
          \lambda F_S & m^2_{\psi_i} }
        \label{scalarmatrix}
\eeq
has  mass eigenvalues
\beq
m_{\pm i} = m_{\psi_i} \sqrt{1 \pm \Lambda/m_{\psi_i}} ,
\label{messengerscalarmasses}
\eeq
where
\beq
\Lambda \equiv {F_S \over S}
\label{defineLambda}
\eeq
The scalar expectation value $S$ sets the overall mass scale
for the messengers, while the auxiliary expectation value,
$F_S$, sets the supersymmetry breaking scale through the
mass splitting $m_{\psi_i} - m_{\pm i}$
between the messenger fermions and scalars.

Supersymmetry breaking in the messenger fermion and
scalar spectrum is transmitted radiatively to the visible sector
superpartners by gauge interactions.
The gauginos couple directly to the messengers and
acquire a mass radiatively from a single
messenger
loop,\footnote{Note that here and in the following
the Standard Model normalization $\alpha_1 = g^{\prime 2}/4\pi$
is employed.
In the literature, the  GUT normalization, which absorbs
the factor of $k_1 = {5 \over 3}$ into $\alpha_1$, is often
used.  Due care should be exercised
when comparing formulas from different sources.}
\beq
M_a =  k_a N \Lambda ~{\alpha_a \over  4 \pi}
\label{gauginomass}
\eq
where $a=1,2,3$ for the Bino, Wino, and gluino respectively,
and $k_1={5 \over 3}$, $k_2=k_3=1$.
The scalars gain mass at two loops from a gauge loop coupling
the scalars to a messenger loop
\beq
m_{\phi}^2 = 2 N \Lambda^2 \left[
  { 5 \over 3} \left( {Y \over 2} \right)^2
       \left( {\alpha_1 \over  4 \pi} \right)^2
 + C_2 \left( {\alpha_2 \over  4 \pi} \right)^2
 + C_3 \left( {\alpha_3 \over  4 \pi} \right)^2
  \right] ,
  \label{ascalarmass}
\eq
where $Y$ is the ordinary weak hypercharge normalized as
$Q=T_3 + {1 \over 2} Y$,
$C_2={3\over 4}$ for weak isodoublet scalars and zero
for weak isosinglets, and
$C_3={4 \over 3}$ for squarks and zero for other scalars.
Since gaugino masses arise at one loop while
scalar masses squared at two loops, the gaugino and scalar
masses are the same order.
In addition, since the supersymmetry breaking is transmitted by the
Standard Model gauge interactions, the superpartner
masses are roughly in proportion to their gauge couplings
squared.
The factors of $N$ in (\ref{gauginomass}) and (\ref{ascalarmass})
count the multiplicity of messenger generations in the messenger
loops.
Gaugino masses scale like $N$ while scalar masses scale
like $\sqrt{N}$.
Finally, it is worth noting that the scalar masses squared
(\ref{ascalarmass}) are fortuitously positive.
A negative result would have rendered this class of models
untenable.

The visible sector gaugino and scalar radiative masses
(\ref{gauginomass}) and (\ref{ascalarmass})
are generated at the messenger scale $M_m$
with the appropriate gauge couplings evaluated at that scale.
These masses must therefore be evolved to the electroweak
scale by renormalization group running.

The dimensionful Higgs sector parameters
which determine the electroweak scale
must be linked in some way to supersymmetry breaking.
These parameters violate $U(1)_{PQ}$ in the superpotential
mass term which determines the Higgsino Dirac mass:
\beq
W = \mu H_u H_d ,
\eq
and $U(1)_{R-PQ}$ in the Higgs soft mass parameter
\beq
V = -m_{ud}^2 H_u H_d ~+~h.c.
\eq
Unlike the gaugino and scalar masses, these terms
are not generated simply by gauge couplings to the messenger sector.
These mass parameters must arise in a viable model, implying there
must be additional interactions between the visible and
messenger sectors which violate $U(1)_{PQ}$ and
$U(1)_{R-PQ}$ symmetries.
In the MGM the detailed form of these interactions is not
specified and $\mu$ and $m_{ud}^2$
are taken to be free parameters.
For phenomenological studies it is most useful to eliminate
these parameters in favor of $\tan \beta$ and
$m_Z$ by imposing the constraints of electroweak symmetry
breaking.
Radiative electroweak symmetry breaking, in which
the up-type Higgs mass squared is driven negative by
the stop squark soft mass through
the large top quark Yukawa coupling under renormalization
group evolution, naturally occurs in the MGM.
Most numerical codes which evolve the soft masses from the
messenger scale using the renormalization
group equations and solve the constraints of electroweak symmetry
breaking, such as ISAJET in versions 7.34 and later,
employ the full one-loop corrected Higgs potential.
In the MGM the boundary condition for the soft Higgs masses
at the messenger scale are identical to the left handed sleptons
which have the same gauge quantum numbers.

Soft tri-linear $A$-terms are not generated directly at one-loop
by gauge interactions with the messenger sector fields.
These terms therefore vanish at the messenger scale, but
are generated under renormalization
group evolution below the messenger scale.
For this reason $A$-terms are
typically smaller in magnitude than the soft masses
in the MGM unless the messenger scale is quite large.

An appealing feature of gauge-mediated supersymmetry breaking
is the natural lack of supersymmetric flavor violation.
Since the Standard Model gauge interactions are flavor independent,
the radiatively induced superpartner soft masses
do not violate flavor.
In a full theory this would be the case if the
scale for determining the flavor structure of the Standard Model
Yukawa couplings is well above the messenger scale.
If this scale is at or below the messenger scale,
however, interesting levels of supersymmetric flavor
mixing could occur.


\subsubsection{Beyond minimal gauge mediation}
\label{subsec:variations}
\indent

The minimal model of gauge mediation described above can not
represent the full theory of supersymmetry breaking.
Even so it does capture some of the important features expected
in many classes of theories of gauge-mediated supersymmetry breaking.
Important features of more complete theories can, however, differ
significantly from the minimal model.
Some of the more important variations and extensions
of the minimal model and the effects
on accelerator phenomenology are described below.

\vspace{.15cm}
\noindent i) {\it{Non-gauge corrections to Higgs masses}}
\vspace{.15cm}

The most serious deficiency of the MGM is the lack of specific
interactions which
break the $U(1)_{PQ}$ and $U(1)_{R-PQ}$ symmetries and
give rise to the Higgs sector
mass parameters $\mu$ and $m_{ud}^2$ described above.
Any realistic theory must contain additional interactions
between the Higgs superfields and messenger sector which give rise
to these terms.
In the MGM the Higgs soft masses at the messenger scale
are assumed to be equal to the left handed slepton soft masses
since these fields have the same gauge quantum numbers
$
m_{H_u}^2(M_m) = m_{H_d}^2(M_m) = m_{\slepton_L}^2(M_m)
$.
However, it is possible that the additional interactions
required to break $U(1)_{PQ}$ and $U(1)_{R-PQ}$ symmetries
also contribute, either constructively or destructively,
to the gauge mediated soft Higgs masses.
If this is the case, the MGM relations between Higgs sector
parameters $\mu$ and $m_{ud}^2$
and the electroweak parameters $\tan \beta$ and $m_Z$,
implied by the constraints of electroweak symmetry breaking,
are modified by the additional contributions to
$m_{H_u}^2$ and $m_{H_d}^2$. From the phenomenological perspective
this has the practical effect that the $\mu$ parameter, which determines
the Higgsino Dirac mass, may be regarded as
a free parameter.

The most important phenomenological effect of the Higgsino
Dirac mass $\mu$ is on the
Higgsino content of the lightest neutralino,
$\widetilde{\chi}_1^0$.
In the MGM, with only gauge-mediated contributions to the
Higgs soft masses, the constraints of electroweak symmetry
breaking typically imply $\mu \sim {\rm few} \times M_1$.
In this case the lightest neutralino is Bino-like.
However, taking $\mu$ as a free parameter leads to the possibility
of a Higgsino like $\widetilde{\chi}_1^0$.
Some of the signatures associated with a Higgsino-like
neutralino NLSP
are discussed in section \ref{sec:higgsino}.

\vspace{.15cm}
\noindent ii) {\it{$U(1)_R$-symmetry suppression of
gaugino masses}}
\vspace{.15cm}

Gaugino masses require the breaking of both $U(1)_R$ symmetry
and supersymmetry, while scalar masses require only
supersymmetry breaking.
In the MGM $U(1)_R$ and supersymmetry in the messenger sector
are broken at the same scale by
the auxiliary expectation value $F_S$.
It is possible however with multiple spurions that
$U(1)_R$ is an approximate symmetry at the supersymmetry
breaking scale, and is only broken at a slightly lower
scale \cite{mgm}.
In this case the gaugino masses are suppressed with respect
to the scalar masses.
The suppression can be parameterized with a parameter
$\Rslash \leq 1$, which relates the gaugino masses $M_a$
to the MGM predictions:
\beq
\left.
M_a = \Rslash ~ M_a \right|_{\rm MGM} .
\eq
With suppressed $U(1)_R$ breaking in the messenger sector the
gauginos are somewhat lighter (in comparison with the MGM) than the
scalars with similar quantum numbers.
It should be noted that while gaugino masses require continuous
$U(1)_R$ breaking they are invariant under discrete $Z_2$ $R$-parity.
The breaking of continuous $U(1)_R$ symmetry therefore
does not necessarily have anything
at all to do with the conservation or possible violation of
$R$-parity.

\vspace{.15cm}
\noindent iii) {\it{Messenger threshold corrections}}
\vspace{.15cm}

In the MGM all the messenger sector superfields are
assumed to be approximately degenerate.
The off-diagonal supersymmetry breaking contributions
to the scalar mass squared matrix (\ref{scalarmatrix})
are assumed to be small compared with the diagonal supersymmetric
masses, amounting to $\Lambda \ll M_m$.
This could be modified in several ways.
First, it may be that there are significant hierarchies
between the masses of different messenger fields.
If there are two distinct messenger
scales $M$ and $M^{'}$, then there will be
additional contributions
to scalar squared masses proportional to
$ (\alpha_a M_a^2  / 4 \pi) \ln (M/M^{'}) $ for each
gauge group $(a=1,2,3)$.
Second, it could be that the off-diagonal supersymmetry
breaking contributions to scalar messenger masses are comparable to the
diagonal supersymmetric masses.
This effect can be computed in terms of
a parameter $x = \Lambda/M_m$.
The multiplicative threshold correction for the contributions to
Standard Model gaugino masses is then
$g(x) =1 + {1 \over 6} x^2 + \cdots$ while the multiplicative
correction for scalar masses squared is
$f(x) = 1 + {1 \over 36} x^2 + \cdots$.
Full expressions for these threshold corrections can be found in
Refs. \cite{Dimopoulos:1996gy}
and \cite{Martin:1997zb}.
It should be noted that these corrections tend to be quite small
in cases with moderate hierarchies, and
do not violate the degeneracy of scalar partner masses with the
same Standard Model gauge quantum numbers.

\vspace{.15cm}
\noindent iv) {\it{Non-$SU(5)$ multiplet messengers}}
\vspace{.15cm}

In the MGM the messenger fields are assumed to form complete
multiplets of $SU(5) \supset
SU(3)_C \times SU(2)_L \times U(1)_Y$.
This has the advantage of naturally
maintaining the successful supersymmetric
prediction of gauge-coupling unification.
However, it is also possible to view this apparent
unification of gauge couplings as accidental, in whole or in part.
The possibility of messenger fields which do not
transform as complete $SU(5)$ multiplets may therefore be considered.
In this case
the expressions for the Standard Model
gaugino and scalar partner masses at the messenger
mass scale can be generalized to
\beq
M_a =  k_a N_a \Lambda {\alpha_a \over 4 \pi} \qquad\qquad
\eeq
where $a=1,2,3$, and
\beq
m_{\phi}^2 = 2  \Lambda^2 \left[
  N_1 { 5 \over 3} \left( {Y \over 2} \right)^2
       \left( {\alpha_1 \over  4 \pi} \right)^2
 + N_2 C_2 \left( {\alpha_2 \over  4 \pi} \right)^2
 + N_3 C_3 \left( {\alpha_3 \over  4 \pi} \right)^2
  \right]
  \label{scalarmass} .
\eeq Here $N_a$ is the Dynkin index
for the
messenger fields for the appropriate
gauge group, in a normalization
in which $N_a=1$ for $a=1,2,3$ for a complete
${\bf 5 \oplus \overline 5} \in SU(5)$.
In general, for vector messenger representations
which can form Dirac states, $N_2$ and $N_3$ are required to be integers by
non-Abelian gauge invariance.
However, for the Abelian case of weak hypercharge,
\beq
N_1 = {6 \over 5} ~\sum_i \left( { Y_i \over 2} \right)^2
\eq
where the sum $\sum_i$ is over all messenger pairs.
In general, $N_1$ must be an integer multiple of $1/5$. To see this,
note that $(Y_i/2)^2 = (Q_i- T_{3i})^2$ where $Q_i$ is the electric
charge and $T_{3i}$ is the weak isospin.
Now, if fractional electric
charges are confined by QCD (which is always the case
for unifiable representations), then $3 Q_i$ must be equal to the
$SU(3)_C$ triality of the representation mod 3.
This follows because it is always possible to
combine the messenger field with some combination of ordinary
Standard Model quark and anti-quark states to get a confined color singlet state
with vanishing triality mod 3.
Now, $SU(3)_C$ representations with non-zero triality have
dimensions that are integer multiples of 3. Furthermore, $T_{3i}$ is an
integer(half-integer) for odd(even) dimensional representations of $SU(2)_L$.
So if fractional charges are confined by QCD, then
for any representation the quantity $(Y_i/2)^2=(Q_i-T_{3i})^2$
multiplied by the dimension of the representation must be an integer
multiple of $1/6$.
This implies that $N_1$ is always an integer multiple of $1/5$
as long as QCD confines all fractional charges.

Clearly, choosing $N_1$, $N_2$ and $N_3$ independently
(instead of $N=N_1 = N_2 = N_3$ as in the MGM) can give rise to
a much more general set of possibilities for the
gauge-mediated superpartner spectrum.
Some of the possibilities for non-$SU(5)$ multiplet messengers have been
studied in Ref. \cite{Martin:1997zb}.
Another possibility is
that the messenger fields do form
complete $SU(5)$ multiplets,
but that the supersymmetry breaking auxiliary expectation
value(s) does not couple in an $SU(5)$-invariant manner \cite{mgm}.

\vspace{.15cm}
\noindent v) {\it{Gauge-multiplet messengers}}
\vspace{.15cm}

The massive messenger sector
fields of the MGM are scalars and fermions that comprise
chiral supermultiplets.
However, it is also possible that some of the
messengers are gauge bosons and gaugino fields
which gain mass at the messenger scale.
These fields comprise gauge multiplets corresponding to
spontaneously broken gauge
groups in the messenger or supersymmetry breaking sectors
beyond the Standard Model gauge group.
Such gauge-multiplet messengers always arise
by virtue of the super-Higgs mechanism if chiral multiplets
in the messenger sector which gain an expectation value
transform under both the Standard Model
gauge group and these additional gauge group(s).
In this case $SU(3)_C \times SU(2)_L \times U(1)_Y \subset G$,
where $G$ also contains (some of) the
messenger or supersymmetry breaking sector gauge groups
which are spontaneously broken at the messenger scale.

The spectrum of the heavy gauge supermultiplets may break supersymmetry.
Radiative masses for the
visible sector gauginos and scalars are then generated
analogously to the case of chiral messenger multiplets
\cite{Giudice:1997ni}.
The most important feature of gauge multiplet messengers is that
{\it negative} contributions to the visible sector
scalar masses squared are induced.
This effect poses a serious problem for
such models since squark and/or slepton expectation values would
result leading to color and/or charge breaking.
Models which contain gauge-multiplet messengers
must also in general contain chiral multiplet messengers
to offset the negative contributions to scalar masses squared.
In this case the scalars are likely to be somewhat
lighter (in comparison with the MGM)
than the gauginos with similar quantum numbers.

\vspace{.15cm}
\noindent vi) {\it{Non-holomorphic messenger masses}}
\vspace{.15cm}

In the MGM, supersymmetry breaking in the messenger spectrum
appears in the superpotential coupling (\ref{SPP}),
resulting in the off-diagonal holomorphic messenger soft masses
in the scalar mass squared matrix (\ref{scalarmatrix}).
The diagonal scalar messenger masses in (\ref{scalarmatrix}) are not
affected by the auxiliary expectation value and are identical
to the messenger fermion masses.
This holomorphic form of supersymmetry breaking in the
messenger sector has the special property ${\cal S}{\rm Tr}~m^2=0$
where the supertrace is over all messenger particles,
${\cal S}{\rm Tr} \equiv \sum (-1)^f$ and $f=0(1)$ for bosons(fermions),
and each spin degree of freedom is counted separately.
This has the important effect that the messenger soft masses
to not mix with (or induce) visible sector scalar superpartner
soft masses under renormalization group evolution.
Only the finite, radiatively induced, two-loop soft scalar masses
(\ref{ascalarmass}) result from messenger sector holomorphic
supersymmetry breaking.
The lack of renormalization group mixing can also be understood
in terms of the messenger sector global symmetries carried by the
holomorphic soft terms.
The visible sector scalar superpartner soft masses do not transform
under these symmetries.
Since operators with different global quantum numbers can not mix,
visible sector soft masses are therefore not induced under renormalization
group evolution.

With a general messenger sector, supersymmetry breaking may appear
in the messenger spectrum in a more general way than through
the holomorphic superpotential coupling (\ref{SPP}).
The general form of the messenger scalar mass squared matrix
may then be taken to be
\beq
\pmatrix{
m_{\psi_i}^2 + \delta m_+^2 & \lambda F_S + \delta m^2 \cr
\lambda F_S + \delta m^2 & m_{\psi_i}^2 + \delta m_-^2 }
\label{nonhmatrix}
\eeq
where $\delta m^2$ is a holomorphic soft mass squared parameter
while $\delta m^2_+$ and $\delta m^2_-$
are non-holomorphic soft mass squared parameters.
The inclusion of these diagonal non-holomorphic soft masses
leads to ${\cal S}{\rm Tr} ~m^2= 2(\delta m_+^2 + \delta m_-^2)  \neq 0$.
Because of this,
non-holomorphic messenger soft masses can induce \cite{Poppitz:1997xw}
visible sector
soft masses squared at two loops under renormalization group
evolution \cite{Jack:1994rk}.
This occurs because no symmetry forbids non-holomorphic soft masses
in the messenger sector from mixing with visible sector soft masses.
Because of the renormalization group contribution, two-loop soft
squared masses
induced by non-holomorphic terms in the messenger sector
are larger than the finite
two-loop soft squared masses induced by holomorphic terms
by a factor $\ln(M_m / m_m)$,
where $M_m$ is the messenger scale at which the non-holomorphic
messenger masses are generated and $m_m$ is the mass of the messengers.
Likewise, since gaugino masses require $U(1)_R$ breaking which
only appears in (\ref{nonhmatrix}) from the off-diagonal holomorphic
soft masses,
the non-holomorphic renormalization group contribution to the visible
sector scalar squared masses are also larger than the gaugino masses by
the
same $\ln(M_m / m_m)$ factor.
So in theories with non-holomorphic messenger masses and $M_m \gg m_m$
the magnitude of the visible sector scalar squared masses are larger than
the
associated
gaugino masses with similar gauge quantum numbers.
It is also worth noting that for ${\cal S}{\rm Tr} ~m^2= 2(\delta m_+^2 +
\delta m_-^2) > 0$
in (\ref{nonhmatrix}) the induced
visible sector soft masses squared are negative, which is
phenomenologically unacceptable.
Non-holomorphic messenger masses are therefore
only viable for $\delta m_+^2 + \delta m_-^2 < 0$.

Non-holomorphic messenger soft masses can arise in theories
in which the Standard Model gauge group is embedded directly
in the supersymmetry breaking sector.
In such theories it is possible that some of the moduli
(or sigma model degrees of freedom)
in the supersymmetry breaking sector transform under the Standard
Model gauge group and act as messengers.
Such moduli naturally receive non-holomorphic soft masses from
Kahler potential couplings due to non-trivial curvature
of the Kahler manifold.
The messenger scale, $M_m$, in such theories is set by the mass of other
chiral supermultiplet and gauge supermultiplet
messenger fields which gain a mass via the super-Higgs
mechanism due to gauge symmetry breaking expectation values.
If the supersymmetry sector is non-renormalizable the moduli
with non-holomorphic soft masses
gain a mass $m_m$, at a hierarchically smaller scale than
the messenger scale, $m_m \ll M_m$.
The logarithmic enhancement of the visible sector soft squared masses,
$\ln(M_m / m_m)$, can then be sizeable.
In most non-renormalizable gauge-mediated supersymmetry breaking
theories of this type the magnitude of the logarithm is large
enough to require some degree of fine tuning to obtain
electroweak symmetry breaking consistent with current bounds
on gaugino masses.
Ignoring this problem of fine tuning, and assuming
${\cal S}{\rm Tr} ~m^2 <0$ so that visible sector soft masses squared
are positive,
the gauginos are generally lighter than the associated scalars with
similar quantum numbers by the logarithmic factor
$\ln(M_m / m_m)$.

\vspace{.15cm}
\noindent vii) {\it{$D$-term contributions to scalar
masses}}
\vspace{.15cm}

In addition to the radiatively generated scalar soft  masses
arising from $F$-term auxiliary expectation values in the
messenger sector, it is possible for scalar soft masses
to receive contributions directly from a
non-zero $D$-term expectation value for $U(1)_Y$
hypercharge.
This leads to a shift in the soft scalar masses proportional to the weak
hypercharge according to
\beq
\Delta m^2_\phi = -Y_\phi g' \langle D_Y \rangle.
\eeq
where $Y_{\phi}$ is the weak $U(1)_Y$ hypercharge of the scalar
$\phi$ and  $D_Y$ is the $U(1)_Y$ auxiliary field.
More generally, if the unbroken gauge symmetry at the
supersymmetry-breaking scale contains an additional Abelian factor(s)
$U(1)_X$, then similar contributions can arise
\beq
\Delta m^2_\phi = -X_\phi g_X \langle D_X \rangle,
\eeq
where $X_\phi$ is the $U(1)_X$ charge of $\phi$ and $D_X$ is the $U(1)_X$
auxiliary field.
If present, these $D$-term contributions will not affect
the MSSM gaugino mass parameters, but can leave a ``fingerprint" on the
MSSM scalar mass spectrum which can be quite distinct from that of the MGM.
Note that since ${\rm Tr}~Y=0$ and ${\rm Tr}~X=0$ in order to ensure
vanishing of gravitational anomalies and quadratic
divergences, some scalar field(s) necessarily
receive a negative mass squared contribution from $D$-term
expectation values.

Non-vanishing $D$-terms may arise either at
tree-level or radiatively.
Tree-level $D$-term expectation values for unbroken
$U(1)$ gauge symmetries generally arise
if there is chiral matter in the supersymmetry breaking sector
which gains a scalar expectation value and transforms under the $U(1)$.
In non-renormalizable models with large scalar moduli
expectation values these $D$-terms are generally hierarchically
smaller than the supersymmetry breaking scale and unimportant.
In renormalizable models however the non-vanishing $D$-terms
are generally only suppressed compared with the supersymmetry
breaking scale by some power of a Yukawa coupling in this sector.
$D$-term expectation values can also be generated radiatively
at one loop from messenger fields which transform
under the $U(1)$.
This occurs if
$\delta m_+ \neq \delta m_-$ in the scalar mass squared matrix
(\ref{nonhmatrix}).

Both tree-level and radiatively generated $D$-term expectation
values automatically vanish if there is an
unbroken discrete
symmetry under which $D$ transforms.
In the MGM messenger parity provides such a discrete symmetry,
and $D_Y=0$ at leading order.

\vspace{.15cm}
\noindent viii) {\it{Strongly coupled messengers}}
\vspace{.15cm}

The calculation of gaugino and scalar superpartner masses
in the MGM assumes that the messenger
dynamics can be treated perturbatively.
However, it is likely that the
ultimate source of supersymmetry breaking involves non-perturbative
dynamics which may be strongly coupled.
It is then natural in certain classes
of renormalizable theories in which the messenger
and supersymmetry breaking scales coincide
to consider the possibility that the messengers are strongly
coupled \cite{Dimopoulos:1996vz}.
Assuming the Standard Model gauginos are elementary at
the messenger scale, the induced gaugino masses may be
estimated using the standard rules of naive dimensional
analysis for strongly coupled theories \cite{Cohen:1997rt}
\beq
M_a \sim N ~{\alpha_a \over 4 \pi} ~ {4 \pi F \over M}
\eq
where the supersymmetry breaking scale $\sqrt{F}$ and cutoff or
messenger scale $M$ are related by $F \sim 4 \pi M^2$.
For elementary Standard Model scalar superpartners, the
induced scalar masses are of the order \cite{Cohen:1997rt}
\beq
m_{\phi}^2 \sim N ~
  \left({ \alpha_a \over 4 \pi}\right)^2
  \left( {4 \pi F \over M} \right)^2
\eq
As in the perturbative case, the scalar and gaugino masses
are the same order.
However, relative to these masses, the Goldstino decay constant $F$ is
roughly a factor
of $4 \pi$ {\it smaller} than for the perturbative case.

Another interesting possibility is that some of the Standard Model
matter supermultiplets are composite and/or directly coupled to the
messenger and supersymmetry breaking sectors.
In this case the scalar superpartners of these supermultiplets
could naturally gain a mass at the supersymmetry breaking scale.
For example if the first two generations are composite,
with supersymmetry breaking and compositeness scales of the same
order, it is likely that only the gauginos and
third generation scalars would be accessible
at accelerator energies \cite{CKNheavy}.

\addcontentsline{toc}{section}{Appendix B: Decays to the Goldstino}
\section*{Appendix B: Decays to the Goldstino}\label{app:goldstino}
\setcounter{equation}{0}
\setcounter{footnote}{1}
\setcounter{subsubsection}{0}
\renewcommand{\theequation}{B.\arabic{equation}}
\indent

The spontaneous breaking of global supersymmetry leads to the existence of
a massless Goldstone fermion, the Goldstino.
The lowest order derivative coupling for
emission or absorption of a single on-shell Goldstino is
fixed by the supersymmetric
Goldberger-Treiman low energy theorem to be
proportional to
\beq
{1 \over F} \partial_{\mu} G^{\alpha}  j^{\mu}_{~\alpha}
~+~h.c.
\eq
where $\sqrt{F}$ is the supersymmetry breaking scale, and
$j^{\mu}_{~\alpha}$ the supercurrent.
This allows the model independent
decay rate of a sparticle to its partner plus the
Goldstino to be calculated in terms of the supersymmetry
breaking scale.
Since cascade decays of heavy superpartners through tree-level
interactions are generally very rapid,
decay by Goldstino emission is generally only relevant for
the lightest Standard Model superpartner(s).

For a slepton or squark the decay rate
to its partner plus the Goldstino in the absence of any
supersymmetric flavor mixing is
\cite{gwidth,mgm}
\beq
\Gamma( \tilde{f} \to f \goldstino) = { m_{\tilde{f}}^5 \over 16 \pi F^2}
\left( 1 -{ m_f^2 \over m_{\tilde{f}}^2 } \right)^4 .
\label{sleptonwidth}
\eeq
Because the Goldstino is derivatively coupled, the two body decay rate to
any massive final state suffers a $\beta^4$ suppression near
threshold, where $\beta$ is the massive final state velocity
in the decay rest frame.

Non-vanishing supersymmetric flavor mixing
in general introduces flavor violation in decay to the Goldstino.
For example, the general decay rate of an up-type squark to
a quark and the Goldstino is
\beq
\Gamma( \widetilde{Q}_a \rightarrow q_i \widetilde{G} ) =
{ m_{\tilde{Q}}^5 \over 16 \pi F^2}
   \left( |U_{Lai}|^2 + |U_{Rai}|^2 \right)
  \left( 1 - { m_{q_i}^2 \over  m_{\tilde{Q}}^2 } \right)^4
  \label{sqflavor}
\eq
where here $U_{Lai}$ and $U_{Rai}$ are the
up-squark mixing matrices between the mass eigenstates
$a=1, \dots,6$ and left and right flavor eigenstates $i=1,2,3$.
These are defined with respect to the up-quark mass eigenstates,
$\widetilde{Q}_a = U_{Lai} \widetilde{Q}_{Li} +
     U_{Rai} \widetilde{Q}_{Ri}$, with
         $U_{Lia}^{\dagger} U_{Laj} = \delta_{ij}$,
         $U_{Ria}^{\dagger} U_{Raj} = \delta_{ij}$, and
         $U_{Lia}^{\dagger} U_{Raj} = 0$.
There are no $L-R$ mixing contributions to (\ref{sqflavor})
since the Goldstino is massless.
Similar expressions hold for the down-type squarks and sleptons.
For squarks lighter than the top quark, two-body decays to
the Goldstino and top quark are kinematically forbidden.
In this case three-body decays to the $W$-boson, quark, and Goldstino
can become relevant depending on the magnitude of
supersymmetric flavor mixing \cite{mesino}.
In general these decays proceed through an off-shell squark,
quark, or chargino.
In the limit $m_{\cci} \gg m_{\tilde{Q}_a}$ and
$m_{\tilde{Q}_b} \gg m_{\tilde{Q}_a}$,
the three-body decay of the $a$-th up-type squark
takes place predominantly through an off-shell quark plus
diagrams related by gauge invariance.
The general expression for the three-body decay simplifies
in this limit to \cite{mesino}
$$
\Gamma(  \widetilde{Q}_a \rightarrow q_j W\widetilde{G} ) = 
{ \alpha_2 m_{\tilde{Q}}^5 \over 128 \pi^2 F^2}
  \Biggl [  \left|U_{Lai}
                 V^{\dagger}_{ij}  \right|^2
          ~I\left( {m_W^2 / m_{\tilde{Q}}^2} ,
                      {m_{q_i}^2 / m_{\tilde{Q}}^2} \right)
$$
\beq ~~~~~~~~~~~~~~~~~~~~~~~~+ \left | U_{Rai} V^{\dagger}_{ij} \right |^2
          ~J\left( {m_W^2 / m_{\tilde{Q}}^2} ,
                      {m_{q_i}^2 / m_{\tilde{Q}}^2} \right)
  \Biggr ],
\eeq
where here $V_{ij}$ is the CKM quark mixing matrix, and
the phase space integrals are
\beq
I(a,b) = \int_a^1 dx
   { (1-x)^4 (x-a)^2 \over 12 x^3 a}
     \left(
           {    6x^3(3a+x) \over (x-b)^2}
         + { 4x^2 (4a-x) \over (x-b) }
     + x^2 + 2xa + 3a^2
           \right) 
\eeq
\beq
J(a,b) = \int_a^1 dx
    { b (1-x)^4(x-a)^2(2a+x) \over 2 x^2 a (x-b)^2} .
\eeq
These three-body decays are particularly relevant for a stop-like
squark lighter than the top quark, which can decay by
$\tilde t \to b W \GG$.

Neutralinos are in general mixtures of gaugino and Higgsino eigenstates,
and so can decay to neutral gauge bosons and Higgs bosons if kinematically
accessible through Goldstino emission.
The decay rates to the photon final state through the gaugino components,
and $Z$-boson final state through the gaugino and Higgsino components
are \cite{gwidth,mgm}
\beq
\Gamma(\nnone \to \gamma \goldstino) = | \cos \theta_W N_{1 \tilde{B}} +
                                \sin \theta_W N_{1 \tilde{W}} |^2
    ~{ m_{\nnone}^5 \over 16 \pi F^2}
\eq
\beq
\Gamma(\nnone \to Z \goldstino) =   \left( |
                                \sin \theta_W N_{1 \tilde{B}} -
                                \cos \theta_W N_{1 \tilde{W}} |^2
              + {1 \over 2}  |  \cos \beta N_{1 {d}} -
                                \sin \beta N_{1 {u}}  |^2  \right)
    { m_{\nnone}^5 \over 16 \pi F^2}
    \left( 1 - { m_{Z}^2 \over m_{\nnone}^2} \right)^4
\eq
where $N_{ij}$ are the neutralino eigenvectors which
diagonalize the neutralino mass matrix $M_D = N^* M N^{-1}$
\cite{haberkane}.
The decay rates to Higgs boson final states through the Higgsino
components are \cite{gwidth,mgm}
\beq
\Gamma( \nnone \to h^0  \goldstino) = {1 \over 2}   |
                                \sin \alpha N_{1 {d}} -
                                \cos \alpha N_{1 {u}} |^2
    ~{ m_{\nnone}^5 \over 16 \pi F^2}
    \left( 1 - { m_{h^0}^2 \over m_{\nnone}^2} \right)^4
\eq
\beq
\Gamma( \nnone \to H^0 \goldstino) =  {1 \over 2}  |
                                \cos \alpha N_{1 {d}} +
                                \sin \alpha N_{1 {u}} |^2
    ~{ m_{\nnone}^5 \over 16 \pi F^2}
    \left( 1 - { m_{H^0}^2 \over m_{\nnone}^2 } \right)^4
\eq
\beq
\Gamma( \nnone \to A^0  \goldstino) =  {1 \over 2}  |
                                \sin \beta N_{1 {d}} +
                                \cos \beta N_{1 {u}} |^2
    ~{ m_{\nnone}^5 \over 16 \pi F^2}
    \left( 1 - { m_{A^0}^2 \over m_{\nnone}^2 } \right)^4
\eq
where $\alpha$ is the $h^0-H^0$ Higgs mixing angle.
In the Higgs decoupling limit in which only $h^0$ remains
light, this angle is related to $\tan \beta$ by
$\sin \alpha \simeq - \cos \beta$ and
$\cos \alpha \simeq \sin \beta$ \cite{hhguide}.

Decay of neutralinos through Goldstino emission to three-body final states
can take place through an off-shell intermediate
gauge boson \cite{gwidth}.
The most important of these are Dalitz decays of neutralinos
through an off-shell photon
to fermion final states,
$\nnone \to \gamma^* \goldstino$ with $\gamma^* \to f \bar{f}$.
The neutralino Dalitz decay rate to a Goldstino is
\beq
{ \Gamma( \nnone \rightarrow f \bar{f} \GG) \over
\Gamma(\nnone \rightarrow \gamma\GG ) } =
{\alpha Q_f^2 N_c^f\over 3\pi}
\left ({\rm ln}(m^2_{\nnone}/m^2_f) -{15 \over 4} \right )
\eeq
where $Q_f$ and $m_f$ are the fermion
electric charge and mass respectively, and $N_c^f = 1$ for leptons
and 3 for quarks.
For light quarks $m_f$ should be replaced with an infrared cutoff
of order $\Lambda_{QCD}$.

Charginos are in general mixtures of Wino and charged Higgsino
eigenstates, and so can decay to $W^{\pm}$-boson and $H^{\pm}$
charged Higgs final states through Goldstino emission.
The chargino decay rates are \cite{gwidth}
\beq
\Gamma(\cci \to W^{\pm} \goldstino) =  {1 \over 2} \left(
          |V_{i \tilde{W}}|^2 + |U_{i \tilde{W}}|^2
             + \sin^2 \beta  |V_{i \tilde{H}}|^2
         + \cos^2 \beta  |U_{i \tilde{H}}|^2 \right)
{ m_{\cci}^5 \over 16 \pi F^2}
    \left( 1 - { m_{W^{\pm}}^2 \over m_{\cci}^2} \right)^4
\eq
\beq
\Gamma(\cci \to H^{\pm} \goldstino) =  {1 \over 2} \left(
              \cos^2 \beta  |V_{i \tilde{H}}|^2
         + \sin^2 \beta  |U_{i \tilde{H}}|^2 \right)
{ m_{\cci}^5 \over 16 \pi F^2}
    \left( 1 - { m_{H^{\pm}}^2 \over m_{\cci}^2} \right)^4 ,
\eq
where here $U_{ij}$ and $V_{ij}$ are the chargino eigenvectors which
diagonalize the chargino mass matrix by the bi-unitary
transformation $M_D = U^* M V^{-1}$ \cite{haberkane}.

Gluinos do not mix with other states, and so decay through Goldstino
emission only to gluon final states, with decay rate \cite{gwidth}
\beq
\Gamma(\gluino \to g \goldstino) =
    { m_{\gluino}^5 \over 16 \pi F^2}.
\label{gluinodecaywidth}
\eq

\addcontentsline{toc}{section}{Appendix C: List of Model Lines Studied}
\section*{Appendix C: List of Model Lines 
Studied}\label{app:modellinesummary}
\setcounter{equation}{0}
\setcounter{footnote}{1}
\setcounter{subsubsection}{0}
\renewcommand{\theequation}{C.\arabic{equation}}
\indent

The Model Lines selected for this study are listed below. In each case,
$\Lambda$ is varied with the other parameters or ratios held fixed as
shown. Different workers may obtain slightly different masses and
branching fractions due to using different Standard Model gauge
couplings, top mass, etc.
\begin{itemize} 
\item Bino-like Neutralino NLSP Model Line: 
$N=1$; $M_m/\Lambda = 2$; $\tan\beta = 2.5$; $\mu > 0$ fixed by EWSB.
\item Higgsino-like Neutralino NLSP Model Line I: 
$N=2$; $M_m/\Lambda = 3$; $\tan\beta = 3$; $\mu = -{3\over 4} M_1$.
\item Higgsino-like Neutralino NLSP Model Line II: 
$N=2$; $M_m/\Lambda = 3$; $\tan\beta = 3$; $\mu = {1\over 3} M_1$.
\item Stau NLSP Model Line: 
$N=2$; $M_m/\Lambda = 3$; $\tan\beta = 15$; $\mu > 0$ fixed by EWSB.
\item Slepton co-NLSP Model Line: 
$N=3$; $M_m/\Lambda = 3$; $\tan\beta = 3$; $\mu > 0$ fixed by EWSB.
\item Squark NLSP Model Line: Single squark with varying mass.
\end{itemize}


\addcontentsline{toc}{section}{Appendix D: Glossary of Acronyms}
\section*{Appendix D: Glossary of Acronyms}\label{app:acronyms}
\setcounter{equation}{0}
\setcounter{footnote}{1}
\setcounter{subsubsection}{0}
\renewcommand{\theequation}{D.\arabic{equation}}
\indent

Acronyms used in this report which are
peculiar to signals of low-scale supersymmetry breaking:
\begin{itemize}

\item {\bf CC-HIT} Charge-changing highly ionizing track. These can arise
when a charged slepton decays to a nearly degenerate 
stau of the opposite charge, by
emitting a very soft lepton and tau, with the stau traveling in the
forward direction. This decay can have a 
macroscopic length for non-relativistic sleptons, 
and so leads to a highly ionizing track which appears to change sign. 
\item {\bf CE-HIT} Charge-exchange highly ionizing track. 
Strongly interacting squark or gluino bound states 
can exchange 
isospin and charge with matter in the course
of traversing a detector. 
Non-relativistic bound states of this type therefore give rise to 
calorimeter deposits which alternate between 
highly ionizing charged and neutral segments associated with
highly ionizing tracks (HITs) in the tracking and vertex detectors. 
\item {\bf DCA} The distance of closest approach to the primary vertex for
a displaced photon
or charged track.
\item {\bf GMSB} Gauge-mediated supersymmetry breaking, the most concrete
and perhaps the best-motivated example of low-scale supersymmetry
breaking.
\item {\bf HIT} Highly ionizing track left by a long-lived
charged particle such as a stau, selectron, smuon, 
squark in a mesino or sbaryon hadronic bound state, 
or gluino in a $R$-hadron bound state.
These massive particles can travel slowly enough to
deposit energy at a rate $dE/dx$ which is much larger than for a MIP.
\item {\bf H-HIT} A highly ionizing track associated with additional
soft hadronic
activity. These can result from inelastic hadronic interactions of
slowly-moving squark or gluino bound states in the calorimeters.
\item {\bf LNIP} Large negative impact parameter.  
The sign of the impact parameter is defined to be ${\rm sgn}(\cos \varphi)$,
where $\varphi$ is the angle between the reconstructed momentum vector
of a displaced parton and the unit normal from the beam axis to
the origin of the displaced jet. (See section
\ref{subsec:dissquark}.)
Negative values arise from massive slowly moving long-lived unstable
particles
which decay to visible partons in a direction towards, rather than away from
(as for a light relativistic unstable particle),
the production vertex.
\item {\bf MGM} Minimal gauge mediation. GMSB with $N$ generations
of messengers and the $\mu$-parameter determined (up to a sign) by EWSB. 
\item {\bf MIP} A minimum ionizing particle, such as a muon or a
long-lived stau with $\beta\gamma \gsim 0.85$.
\item {\bf NLSP} The next-to-lightest supersymmetric particle
(or particles), which decay to the nearly massless Goldstino.
\end{itemize}


\end{document}